\documentclass[12pt]{reportj}
\usepackage{deluxetablej}
\usepackage{hyperref} 
\hypersetup{colorlinks=true, linkcolor=blue, filecolor=magenta,urlcolor=blue}
\usepackage{cleveref} 
\makeatletter
\def\@to{to}
\makeatother
\usepackage{times}
\usepackage{graphicx}
\usepackage{xcolor}
\usepackage{tocloft}
\usepackage{rotating}
\usepackage{lscape}
\usepackage{fancyheadings}
\newcommand{\isrnum}{2025-03} 
\newcommand{\bd}{BD~$+$75$^{\circ}$325}

\emergencystretch=\maxdimen
\usepackage{datetime}
\newdateformat{ddmonthyyyy}{\THEDAY\ \monthname[\THEMONTH]\ \THEYEAR}

\renewcommand\thesection{\arabic{section}}
\renewcommand\thesubsection{\thesection.\arabic{subsection}}

\def\ssection#1{\setcounter{subsection}{0} \refstepcounter{section} \section*{\hbox to \hsize{\large\bf \arabic{section}. #1\hfill }}\label{sec} \addcontentsline{toc}{section}{\arabic{section}. #1}}
\def\ssubsection#1{\setcounter{subsubsection}{0} \refstepcounter{subsection}\subsection*{\hbox to \hsize{\normalsize\bfseries\itshape \arabic{section}.\arabic{subsection} #1\hfill}}\label{subsec} \addcontentsline{toc}{subsection}{\arabic{section}.\arabic{subsection} #1}}
\def\ssubsubsection#1{\refstepcounter{subsubsection}\subsection*{\hbox to \hsize{\normalsize\it \arabic{section}.\arabic{subsection}.\arabic{subsubsection} #1\hfill}}\label{subsubsec} \addcontentsline{toc}{subsubsection}{\arabic{section}.\arabic{subsection}.\arabic{subsubsection} #1}}

\def\ssectionstar#1{\section*{\hbox to \hsize{\large\bf #1\hfill}} \addcontentsline{toc}{section}{#1}}
\def\ssubsectionstar#1{\subsection*{\hbox to \hsize{\normalsize\bfseries\itshape #1\hfill}} \addcontentsline{toc}{subsection}{#1}}
\def\ssubsubsectionstar#1{\subsection*{\hbox to \hsize{\normalsize\it  #1\hfill}} \addcontentsline{toc}{subsection}{#1}}

\renewcommand{\cftaftertoctitle}{%
\mbox{}\hfill{\normalfont Page}}
\begin{document}

~\\

\vspace{-2.4cm}
\noindent\includegraphics*[width=0.295\linewidth]{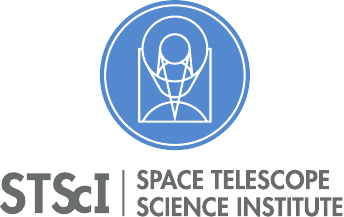}

\vspace{-0.4cm}

\begin{flushright}
    {\bf Instrument Science Report STIS {\isrnum}(v1)}
    
    \vspace{1.1cm}
    
    {\bf\Huge Recalibrating the Sensitivities of the STIS First-Order, Medium-Resolution Modes}
    
    \rule{0.25\linewidth}{0.5pt}
    \vspace{0.5cm}
    
    Alex Fullerton\\

    \footnotesize{Space Telescope Science Institute, Baltimore, MD\\}
    \vspace{0.5cm}
     \ddmonthyyyy{13 August 2025} 
\end{flushright}

\vspace{0.1cm}

\noindent\rule{\linewidth}{1.0pt}
\noindent{\bf A{\footnotesize BSTRACT}}

{\it \noindent 
The sensitivities of STIS first-order, medium resolution modes were 
redetermined from on-orbit observations and CALSPEC models 
(version 11) of the primary white-dwarf spectrophotometric standard 
stars G191-B2B, GD~71, and GD~153.
The sensitivity of an additional configuration was updated 
by comparing observations of the secondary standard {\bd}
with the STIS low-resolution spectrum that has been calibrated
consistently with the version 11 models.
The procedures used to derive the sensitivities and verify the 
PHOTTAB reference files prior to their activation in CRDS (on May 1, 2025)
are described.
Results are presented in graphical form in an extensive appendix. 
Issues and uncertainties are discussed briefly, along with 
recommendations for future work. \\
}
\vspace{-1.2cm}
\noindent\rule{\linewidth}{1.0pt}

\renewcommand{\cftaftertoctitle}{\thispagestyle{fancy}}
\tableofcontents

\begin{figure}[t]
  \centering
  \includegraphics[width=\textwidth]{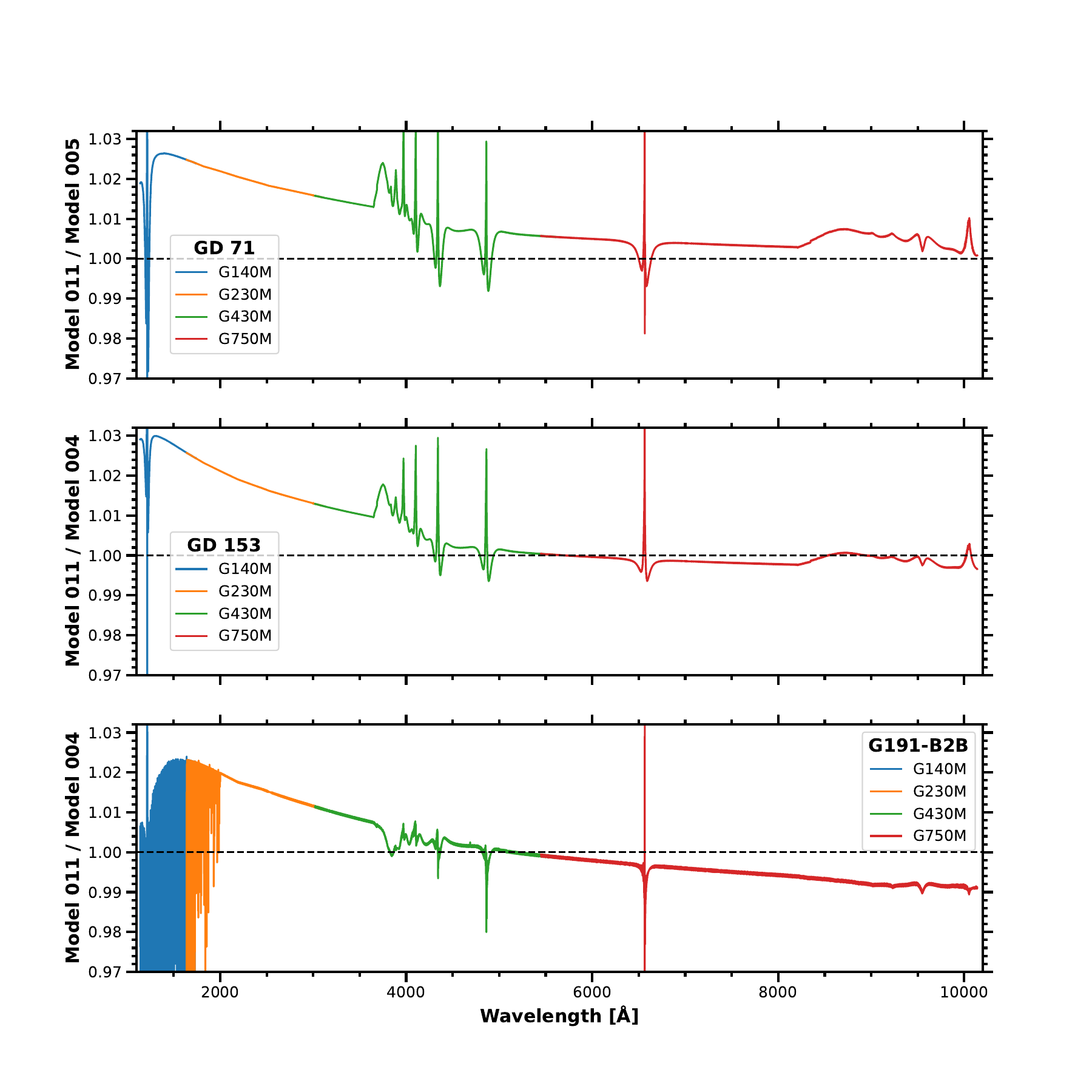}
  \caption{Comparison of the CALSPEC models used to determine sensitivities.
            The wavelength coverage provided by the G230MB grating is not 
            indicated because it is identical to that of G230M.
            In contrast to previous models, the CALSPEC version 11 model for 
            G191-B2B includes line blanketing in the ultraviolet region.
            }
    \label{fig:models}
\end{figure}

\vspace{-0.3cm}
\ssection{Introduction}\label{sec:intro}
The suite of STIS first-order, medium-resolution gratings provide spectroscopic coverage 
between 1150 and 10,000~{\AA} with resolving powers that range from $\sim$10,000 near 
Lyman~$\alpha$ to $\sim$5,000 in the near-infrared.
With a few exceptions, the spectrophotometric calibration of these first-order M-modes rely 
on sensitivities determined by comparison with CALSPEC models of the primary white-dwarf 
standards G191-B2B, GD~71, and GD~153.
These sensitivities were last updated in 2005 -- 2006 by referencing spectra of the standards 
obtained in 2000 -- 2002 to version 4 (for G191-B2B and GD 153) or 5 (for GD 71) of their 
respective CALSPEC models (Proffitt 2006).
In the interim, further improvements have been made to the models of the standards: see 
Bohlin, Hubeny, \& Rauch (2020) for details.
Figure~\ref{fig:models} shows that version 11 of the CALSPEC models differ from their predecessors 
by as much as 3\% in some wavelength regions.
The magnitude of these differences motivated recalibration of all STIS spectroscopic and imaging modes:
see, e.g, Carlberg et al. (2022; E140M grating), Siebert et al. (2024; echelle modes pre-SM4), 
Hernandez et al. (2024; echelle modes post-SM4), Jones et al. (2025; first-order L modes),
and Dos Santos et al. (in prep; near-UV imaging modes).

This report describes the  recalibration of the sensitivity of the first-order M-modes. 
Although straightforward in principle, this work is complicated by the comparatively narrow 
wavelength interval covered by a particular setting of an M-grating (which is labelled by 
the central wavelength of the interval: the CENWAVE), especially in the vicinity of broad 
photospheric features like the Lyman $\alpha$ line of atomic hydrogen.   
At the longest wavelengths, the pattern of interference fringes caused by coatings of the 
CCD detector also complicates the reliable derivation of sensitivities. 
Table~\ref{tab:mmodes} identifies the 61 primary, 19 secondary, and 1 
``available but unsupported'' M-mode configurations of STIS.
\clearpage
\begin{deluxetable}{llll}
\tablewidth{0pt}
\tablecaption{CENWAVES available with the STIS 1st-order M-gratings\label{tab:mmodes}}
\tabletypesize{\footnotesize}
\tablehead{\colhead{Grating} & \colhead{Detector} & \colhead{Type} & \colhead{CENWAVE [\AA]}}
\startdata
G140M  & FUV-MAMA & Primary     & 1173, 1222, 1272, 1321, 1371, 1420, 1470, 1518, 1567        \\
       &          &             & 1616, 1665, 1714                                            \\
       &          & Secondary   & {\bf 1218, 1387, 1400, 1540, 1550, 1640}                    \\
       &          &                                                                           \\
G230M  & NUV-MAMA & Primary     & 1687, 1769, 1851, 1884, 1933, 2014, 2095, 2176, 2257        \\
       &          &             & 2338, 2419, 2499, 2579, 2659, 2739, 2818, 2898, 2977, 3055  \\
       &          & Secondary   & {\bf 2600, 2800, 2828}                                      \\
       &          &                                                                           \\
G230MB & CCD      & Primary     & 1713, 1854, 1995, 2135, 2276, 2416, 2557, 2697, 2836        \\
       &          &             & 2967, 3115                                                  \\
       &          & Secondary   & 2794                                                        \\
       &          &                                                                           \\                                                                                
G430M  & CCD      & Primary     & 3165, 3423, 3680, 3936, 4194, 4451, 4706, 4961, 5216, 5471  \\
       &          & Secondary   & 3305, 3843, 4781, 5093                                      \\
       &          &                                                                           \\
G750M  & CCD      & Primary     & 5734, 6252, 6768, 7283, 7795, 8311, 8825, 9336, 9851        \\
       &          & Secondary   & 6094, 6581, 8561, {\bf 9286}, {\bf 9806}                    \\
       &          & Unsupported & {\bf 10363}                                                 \\
\enddata
\vspace{-0.35in}
\tablecomments{CENWAVES that were not updated are indicated in bold face. 
               See Section~{\ref{sec:data}}.
              }
\end{deluxetable}

\newpage

\lhead{}
\rhead{}
\cfoot{\rm {\hspace{-1.9cm} Instrument Science Report STIS {\isrnum}(v1) Page \thepage}}

\vspace{-0.3cm}
\ssection{Observations}\label{sec:data}
Table~\ref{tab:calprograms} summarizes the calibration programs that form the basis of the spectrophotometric 
calibration of the STIS M-modes.
Following the installation of STIS in 1997 February, observations of the primary white dwarf standards 
GD~71 and GD~153 during Cycle 7 were used to calibrate spectra obtained with the G140M and G230M gratings, 
while observations of the secondary standard {\bd} were used for the remaining gratings.
As described by Proffitt (2006), there was a concerted effort in 2000 -- 2001 to ensure internal consistency 
by relying only on primary standards, in particular G191-B2B.
The bulk of the recalibration effort in 2005 -- 2006 relied on observations from programs 8421 and 8916 obtained in
Cycles 8 and 10, respectively, with significant additions from Cycle 7 programs 7096, 7097, 7657 and a single 
spectrum from program 7937.  

Details of the spectra that were used to produce the current calibration are provided in 
{\hyperref[sec:AppA]{Appendix A}}.
This suite of spectra is nearly identical to the data used to produce the previous calibration,  
though some spectra with low signal-to-noise ratios (S/N) were dropped from consideration.
All observations were made at the central position of the $52\times2$ aperture following a standard 
target acquisition; and only observations with CCDGAIN = 1.0 were included for configurations that 
rely on the CCD detector.
\begin{deluxetable}{llll}
  \tablewidth{0pt}
  \tablecaption{M-Mode Calibration Programs\label{tab:calprograms}}
  \tabletypesize{\footnotesize}
  \tablehead{\colhead{Program~/~PI} & 
             \colhead{Targets}  & 
             \colhead{Gratings} &
             \colhead{CENWAVEs} 
            }
  \startdata
  7094~/~Bohlin    & \bd           & G230MB, G430M, G750M & Selected Primary             \\
  7096~/~Bohlin    & GD 71, GD 153 & G140M, G230M         & Selected Primary, Secondary  \\
  7097~/~Bohlin    & GD 71, GD 153 & G140M, G230M         & 1272, 1518, 1851, 2659       \\
  7656~/~Leitherer & \bd           & G230MB, G430M, G750M & All primary                  \\
  7657~/~Leitherer & GD 71         & G140M, G230M         & Selected Primary, Secondary  \\
  7810~/~Leitherer & \bd           & G430M, G750M         & 4781, 6581, 8561             \\
  7937~/~Keyes     & G191-B2B      & G140M                & 1272                         \\       
  8421~/~Leitherer & G191-B2B      & All                  & Selected Primary             \\
  8916~/~Brown     & G191-B2B      & All                  & Primary, Selected Secondary  \\
  \enddata
\end{deluxetable}

There are several limitations associated with the data sets that are available.
\begin{enumerate}
  \item 
  With few exceptions, the calibration of a specific grating/CENWAVE configuration relies 
  on a single target.   
  Consequently, it is not generally possible to explore systematic differences 
  due, e.g., to subtle mismatches with or errors in the underlying stellar models.  
  Jones et al. (2025) show how the sensitivity curves derived from individual standard
  stars can deviate from the ensemble average obtained from the more extensive 
  and diversified collection of observations obtained with the STIS low-resolution gratings.

  \item 
  The tilt of the G140M grating was changed in 1999 March to position spectra below the 
  repeller wire of the FUV-MAMA.  
  Consequently, standard star spectra obtained before 
  (programs 7096, 7097, 7657, 7937) and after this transition (programs 8421, 8916) produce 
  separate calibrations. 
  During the earlier era, the mode-select mechanism (MSM) was set at $+3''$, which is referred to 
  throughout this document as the ``up'' configuration (e.g., G140M\_up).
  After 1999 March, the MSM was set to $-3''$, which is termed the ``dn'' (down; G140M\_dn) configuration. 
  Proffitt (2006) describes the rationale for this change, and also notes that as a result, 
  observations of only a single target are generally available for G140M modes: GD~71 during 
  the ``up'' era, and G191-B2B during the ``dn'' era.

  \item
  Programs 8421 and 8916 did not observe G191-B2B with all grating/CENWAVE configurations,
  which leads to several additional complications.
  \begin{itemize}
    \item Primary standards were not observed with G140M/1616 at the ``up'' location. 
          Following the precedent set by Proffitt (2006), spectra of GD~71 and G191-B2B obtained 
          at the ``dn'' location have instead been used to determine the sensitivity for this 
          primary CENWAVE at the ``up'' location.
          While this approach is expedient, there is no particular justification for it.
    \item Similarly, there are no observations of a primary standard for G140M/1550 at the ``dn'' 
          location, so a spectrum of GD~71 at the ``up'' location has been used to determine 
          the sensitivity for this secondary CENWAVE.
   \item The secondary standard {\bd} was observed for all primary and a few secondary 
          modes of the gratings that feed the CCD detector (Table~\ref{table:bd}).
          Although these spectra have good S/N, they were not incorporated in this recalibration
          because the currently available CALSPEC model for {\bd} is not on the same flux scale
          as the version 11 models of the primary standards. 
          With one exception, the observations of {\bd} were instead used to check for systematic 
          differences in the sensitivity curves computed from the primary standards; see 
          Section~\ref{sec:discuss}.
          For this purpose, sensitivity curves were produced from the {\bd} spectra by using the 
          CALSPEC data product {\tt bd\_75d325\_stis\_006.fits}, which is derived from L-mode STIS 
          spectra that been placed on the version 11 flux system defined by Bohlin, Hubeny, \& Rauch (2020).
          The sole exception is for the secondary mode G430M/4781, which adopted the sensitivity curve based
          on {\bd} because observations of a primary standard have never been obtained.
    \item Since observations of primary or secondary standard stars could not be identified for 
          the secondary CENWAVES\footnote{The observation of {\bd} with G750M/10363 was excluded 
          since this mode in not supported.} indicated in bold face in Table~\ref{tab:mmodes}, 
          these configurations were not updated.
          Proffitt (2006) followed the same approach in dealing with these modes.
  \end{itemize}

\end{enumerate}

\vspace{-0.3cm}
\ssection{Procedure}\label{sec:procedure}

\ssubsection{Approach}\label{ssec:background}
The sensitivity $S$ of a given grating/CENWAVE configuration of STIS at time $t$ 
is given by:
\begin{displaymath}
S(t)= C_{std}(t) / \mathcal{F}_{std}(t)~~~{\rm cm}^2\,{\rm \AA}\,{\rm erg}^{-1},
\end{displaymath}
where $C_{std}$ is the observed count-rate spectrum of a spectrophotometric standard star (with units of
counts/s for a specified extraction aperture) and $\mathcal{F}_{std}$ is either a
calibrated spectrum of the same standard extracted with the same aperture expressed in physical units 
(which are {erg/s/cm$^2$/\AA} for HST spectrographs) {\bf or} a synthetic model spectrum.
Both these quantities are available in the standard data products produced by {\tt calstis}, which 
have file names ending in either {\tt x1d} or {\tt sx1} for spectra recorded by a MAMA or CCD detector, 
respectively.

In general the instrumental sensitivity decreases with time.  
To compensate, standard {\tt calstis} processing applies several time-dependent corrections to the
extracted flux $\mathcal{F}_{std}(t)$ to return it to the value it would have had at ``beginning of 
life'' (BoL; time $t_0$).  
Since these corrections are not applied to the count-rate spectrum $C_{std}(t)$, it
must be multiplied by a correction factor $f_{corr} \ge 1$ to recover approximately the value it 
would have had at $t_0$.
This correction factor is determined by re-processing the standard star data without (``sans'') 
the time-dependent sensitivity (TDS) correction and -- for spectroscopic modes that use the CCD 
detector -- without the charge-transfer efficiency (CTE) and computing the ratio
\begin{displaymath} 
f_{corr} = \mathcal{F}_{std}^{\,{with}~{TDS}}  / \mathcal{F}_{std}^{\,{sans}~{TDS}}. 
\end{displaymath}
Since all the spectra used to recompute the sensitivity were obtained comparatively early in the life 
of STIS, these correction factors are small.

With this correction, sensitivity information is determined by dividing the count-rate of a standard 
star at BoL by its updated CALSPEC model spectrum, which in this case corresponds to ``version 11."  
Multiple sensitivity curves for the same grating/CENWAVE configuration obtained from either the same 
standard or different standards are merged, and a weighted fit of a low-order spline is fit to the 
ensemble to obtain a smoothly varying representation of the sensitivity.

\ssubsection{Description of Steps}\label{ssec:steps}
The current approach follows closely the procedure described by Proffitt (2006), though
the individual steps were re-implemented in a series of Python scripts.

\begin{figure}[ht]
  \centering
  \includegraphics[width=\textwidth]{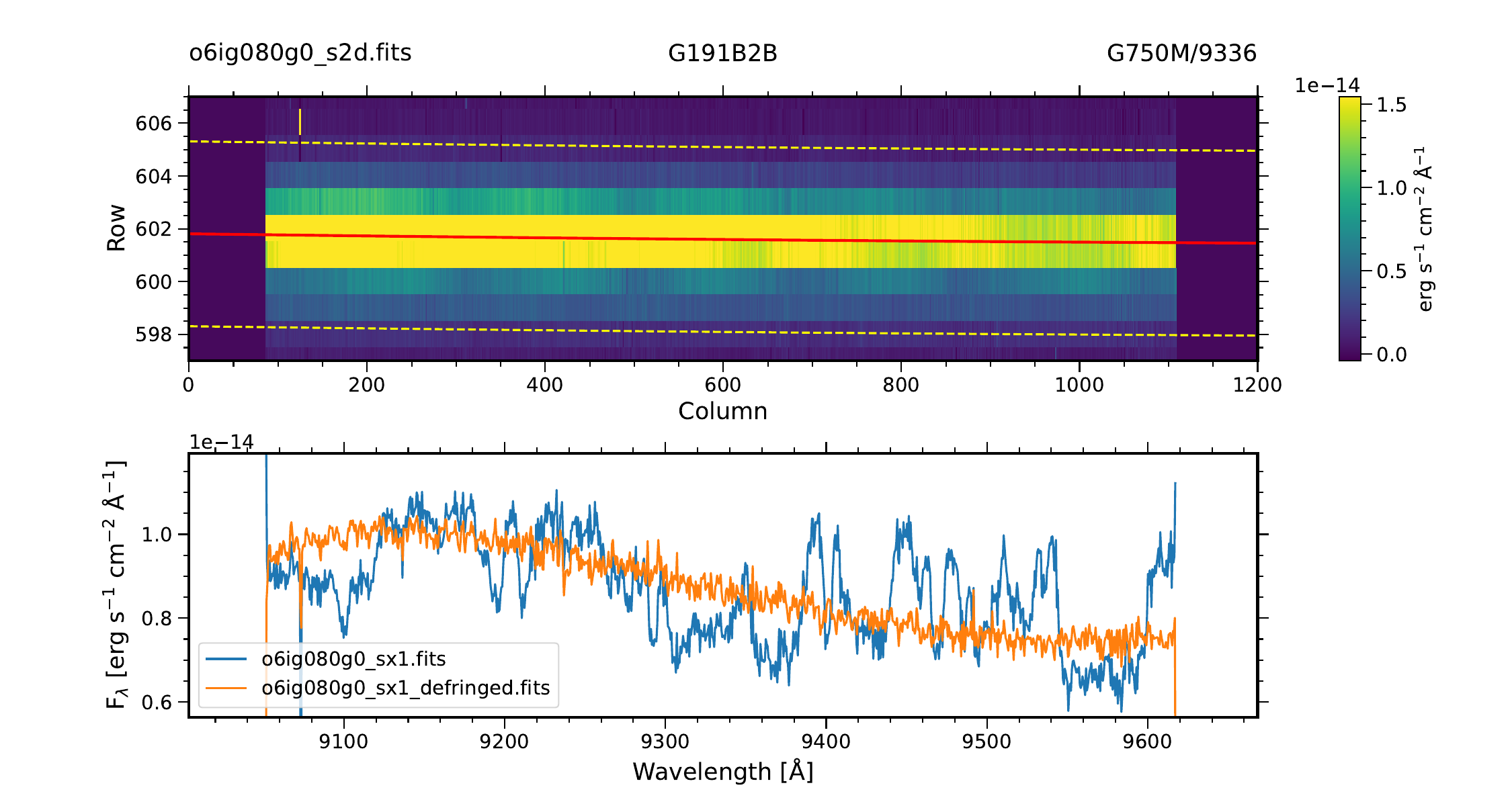}
  \caption{Illustration of the fringe-corrected spectrum for G750M/9336.
           Top: Rectified, fringe-corrected s2d image.  
                The center of the spectral trace is indicated by a solid red line.  
                The y-pixel limits of the 7-pixel extraction box are indicated 
                by dashed yellow lines.
           Bottom: Comparison of the extracted spectra before and after defringing.
           }
  \label{fig:defringe}
\end{figure}

\ssubsubsection{Assemble and Prepare the Data}
\begin{enumerate}
  \item 
  The relevant data sets were identified by querying the Mikulski Archive for Space Telescopes (MAST)
  and downloaded to a local directory.  
  A simple database containing relevant information from the FITS headers of the files was constructed to 
  streamline the selection of data sets for different grating/CENWAVE configurations during 
  subsequent processing.
  
  \item
  The data sets were reprocessed through {\tt calstis} with the time-dependent sensitivity file
  set to {\tt TDSTAB = 'N/A'} and {\tt CTECORR = 'OMIT'} for CCD data to create the 
  $\mathcal{F}_{std}^{\,{sans}~{TDS}}$ versions of the extracted spectra required to 
  compute $f_{corr}$.

  \item 
  Spectra obtained with the G750M grating for CENWAVEs $>$ 7000~{\AA} required additional processing 
  to mitigate the effects of fringing.  
  \begin{itemize}
    \item The appropriate fringe-flat data were downloaded from MAST.
    \item A normalized ``fringe flat'' image was produced by (a) running 
          {\tt stistools.defringe.normspflat} on the {\tt raw} fringe-flat data; and 
          (b) using {\tt stistools.defringe.mkfringeflat} to shift and scale it 
          to best match the fringes in the {\tt sx2} image of the standard star.
          Shift parameters were varied from $-$1.0 to 3.0 in steps of 0.10, and scale factors 
          were varied from 0.8 to 1.5 in steps of 0.04.  
    \item A fringe-corrected, rectified {\tt s2d} image file was created by using
          {\tt stistools.defringe.defringe } to apply the optimal, normalized ``fringe flat'' 
          to the input {\tt sx2} image of the standard star created in Step 2.
          The resultant {\tt s2d} image was calibrated in flux units by using the current
          operational {\tt PHOTTAB} calibration reference file (which is based on the 
          sensitivity curves derived in 2006), but did not include time-dependent 
          sensitivity or charge-transfer efficiency corrections.  
    \item A fringe-corrected, fluxed spectrum was extracted from the rectified {\tt s2d} image 
          by summing over the 7-pixel extraction aperture specified by the {\tt XTRACTAB}
          reference file {\tt n7p1031qo\_1dx.fits}.
          The location of the spectrum was determined by fitting a Gaussian function 
          to selected columns of the image to identify the central row in the spatial direction,
          then tracing the spectrum in the dispersion direction by fitting a linear function 
          to the positions defined by the peaks of the Gaussians. 
          Since the image had already been rectified, the trace only varied by a small 
          fraction of a pixel across the detector.
          Spectra were extracted by centering the extraction aperture on the position of the
          trace at each column of the image, and summing over rows that fell within the 
          aperture.
          The summation included the contributions from fractional pixels at both extremes
          of the aperture. 
          Errors and data-quality flags were extracted in a similar manner from their 
          respective arrays in the {\tt s2d} file.
          Figure~\ref{fig:defringe} provides an example of a fringe-corrected data set. 
    \item A fringe-corrected, {\it count-rate spectrum} was {\it derived} from the extracted, 
          fluxed spectrum and the sensitivity curve used to calibrate them (i.e., from 2006):  
          $C_{std}  = S \times \mathcal{F}_{std}$.    
  \end{itemize}          
\end{enumerate}  

\ssubsubsection{Compute, Merge, and Fit Sensitivity Curves}
\begin{enumerate}
  \item Sensitivity curves were computed for each standard star data set in several steps.
        First, the correction factor $f_{corr}$ was determined from the extracted spectra 
        processed with and without time-dependent corrections, as described in 
        Section~\ref{ssec:background}.  
        In the case of G750M spectra with CENWAVE $>$ 7000~{\AA}, $f_{corr}$ was estimated 
        from the ``fringed'' {\tt sx1} files computed with and without the time-dependent 
        corrections. 
        The correction factor was applied to the count-rate spectrum ``sans TDS'' and 
        pixels flagged with serious data quality issues were eliminated.
        Second, the appropriate CALSPEC version 11 model was linearly interpolated to the 
        same wavelength grid. 
        In principle, the model spectrum should also have been convolved
        with the  line-spread function for the instrumental configuration, though in practice the 
        mismatch in resolution did not play a significant role in subsequent fitting. 
        Finally, the sensitivity curve was computed by dividing the corrected count-rate spectrum
        by the interpolated CALSPEC model and stored in a FITS file, along with the 
        specified error vectors.

  \item Merged sensitivity curves were created for all grating/CENWAVE combinations that had multiple 
        observations of one or more standard star.  
        For these cases, the individual sensitivity curves and their errors were retrieved from their 
        respective FITS files, concatenated, and sorted into order of ascending wavelength. 
        Each data point was assigned a relative weight computed from the inverse square of its error and
        normalized so that the weights of the ensemble summed to 1. 
        The merged sensitivity curve, its error and weight vectors, and an indication of the data 
        set from which each point originated were written to a FITS binary table. 

  \item A cubic spline was fit to the merged sensitivity data by weighted least squares to derive a 
        smooth sensitivity curve for each grating/CENWAVE configuration. 
        Before beginning the procedure, wavelength intervals that coincided with strong 
        spectral features (e.g., Lyman $\alpha$) were masked to avoid biasing the fits. 

        Splines were computed by using the {\tt scipy.interpolate.CubicSpline} function with 
        ``natural'' boundary conditions to ensure the second derivative of the sensitivity curve is 
        0 at the end points.
        Although splines were generally calculated for a grid of 6 equally-spaced knots, 
        the number and location of the knots were occasionally refined after inspection of the 
        results of a given iteration of fitting to follow large-scale undulations in the merged 
        sensitivity data more reliably; see, e.g., the location of the knots in 
        Figure~\ref{fig:G230MBC2557a}.
        
        Least-squares fitting of the sensitivity data at the spline knots was accomplished by using 
        the program {\tt scipy.optimize.least\_squares}.  
        Primarily to remove outliers caused by mismatches between the spectral resolution of the 
        data and the CALSPEC models, data points located more than 3$\sigma$ from the fit were 
        removed by 3 iterations of sigma clipping.

        To ensure reproducibility, all the information pertinent to this fitting procedure 
        was documented in a single multi-extension FITS file that contained: 
        \begin{itemize}
          \item the fitted sensitivity curve computed for the wavelength grid used previously; 
          \item the location and value of the spline knots; 
          \item the merged sensitivity data; 
          \item the merged sensitivity data after sigma-clipping (i.e., the data used to produce the final fit); and 
          \item a provenance table listing salient details of the individual data sets used in the fit.
        \end{itemize}
\end{enumerate}

\ssubsubsection{Prepare New Reference Files}

Consolidated sensitivity files {\tt \{grating\}\_sens.fits} were regenerated with the same 
format as those produced in 2005 -- 2006 but with the revised sensitivity curves for the 
instrumental configurations indicated in Table~\ref{tab:mmodes}.
This replacement was entirely straightforward because the new curves had already been computed 
on the same wavelength grid as used previously.
Header keywords were updated appropriately.

However, the sensitivity files are not used directly by {\tt calstis}, but instead 
are constituents of the {\tt PHOTTAB} reference files that reside in the Calibration
Reference Data System (CRDS).
Revised ``release candidate'' {\tt PHOTTAB} files were constructed by retrieving the 
current suite of files from CRDS (Table~3) and executing the 
heritage IDL procedure {\tt update\_pht\_with\_sens.pro} to convert the new sensitivities 
to throughputs.
The new throughputs were inserted into the appropriate locations of the existing 
{\tt PHOTTAB} files for each grating/CENWAVE combination.
The {\tt DATE} and {\tt DESCRIP} keywords and the {\tt HISTORY} text block were 
updated in the primary header, and the new file renamed to indicate its status
as a release candidate before being copied to the local reference file directory 
for subsequent use in the verification process.
\begin{deluxetable}{lccl}
  \tablewidth{0pt}
  \tablecaption{Revised {\tt PHOTTAB} Reference Files\label{tab:phtfiles}}
  \tabletypesize{\scriptsize}
  \tablehead{\colhead{Detector}               & 
             \colhead{Release-Candidate File} & 
             \colhead{USEAFTER Date}          &
             \colhead{Gratings} 
            }
  \startdata
  FUV-MAMA  & 77o1827ho\_rc5\_pht.fits & OCT 01 1996 & E140H  E140M  G140L {\bf G140M\_up}                     \\
  NUV-MAMA  & 77o18276o\_rc5\_pht.fits & OCT 01 1996 & E230H  E230M  G230L {\bf G230M} PRISM                   \\
  CCD       & 74e15479o\_rc5\_pht.fits & OCT 01 1996 & G230LB {\bf G230MB} G430L {\bf G430M} G750L {\bf G750M} \\
            &                          &             &                                                         \\
  FUV-MAMA  & 77o1827ko\_rc5\_pht.fits & MAR 15 1999 & E140H  E140M  G140L {\bf G140M\_dn}                     \\
            &                          &             &                                                         \\ 
  FUV-MAMA  & 77o18270o\_rc5\_pht.fits & JUL 10 2001 & E140H  E140M  G140L {\bf G140M\_dn}                     \\
  NUV-MAMA  & 77o1826to\_rc5\_pht.fits & JUL 10 2001 & E230H  E230M  G230L {\bf G230M} PRISM                   \\
            &                          &             &                                                         \\ 
  FUV-MAMA  & 6471930qo\_rc5\_pht.fits & MAY 11 2009 & E140H  E140M  G140L {\bf G140M\_dn}                     \\
  NUV-MAMA  & 77o1827fo\_rc5\_pht.fits & MAY 11 2009 & E230H  E230M  G230L {\bf G230M} PRISM                   \\
            &                          &             &                                                         \\
  FUV-MAMA  & 64719317o\_rc5\_pht.fits & JUL 01 2012 & E140H  E140M  G140L {\bf G140M\_dn}                     \\
            &                          &             &                                                         \\
  FUV-MAMA  & 6471930po\_rc5\_pht.fits & JUL 01 2016 & E140H  E140M  G140L {\bf G140M\_dn}                     \\
            &                          &             &                                                         \\ 
  NUV-MAMA  & 77o18278o\_rc5\_pht.fits & MAR 31 2018 & E230H  E230M  G230L {\bf G230M} PRISM                   \\
  \enddata
  \vspace{-0.2in}
  \tablecomments{The release-candidate files were renamed after delivery to CRDS as indicated in 
  the \href{https://www.stsci.edu/contents/news/stis-stans/june-2025-stan.html\#article1}{June 2025 STAN}.  
  They were activated on May 1, 2025.}
\end{deluxetable}

\vspace{-0.3cm}
\ssection{Results and Verification}\label{sec:results}
The figures in {\hyperref[sec:AppB]{Appendix B}} illustrate the merged sensitivity data 
and the final spline fit that defines the revised sensitivity curve for the 
grating/CENWAVE combinations indicated in Table~\ref{tab:mmodes}.

The revised sensitivity curves were verified by using {\tt calstis} to 
recalibrate the input data for each spectrophotometric standard with the release
candidate {\tt PHOTTAB} files and visually inspecting a comparison with version 11 
of its CALSPEC model.
This comparison is shown in the top panel of the figures, and confirms that 
the new {\tt PHOTTAB} file correctly calibrates the standard-star spectra in all cases.
This approach provides sufficient verification of the quality of the 
recalibration, even though the same data were used to derive and verify the 
sensitivity.

Additional panels illustrate the differences between the previous and current 
sensitivity curves directly (third panel from the top) or fractionally (bottom panel).
The bottom panel also shows the ratio of the CALSPEC models for the relevant standard 
star used in the previous and present calibration efforts. 
All other things being equal, the ratios of the sensitivity curves and the model 
spectra should be mirror reflections about unity, since the same data were used in 
both efforts.
Cases where this expectation is not met indicate differences in details of 
the spline fitting or some other aspect of the data handling (e.g., defringing).
This comparison provided a useful diagnostic of processing issues during the 
iterative fitting procedure.

\begin{figure}[ht]
  \centering
  \includegraphics[width=\textwidth]{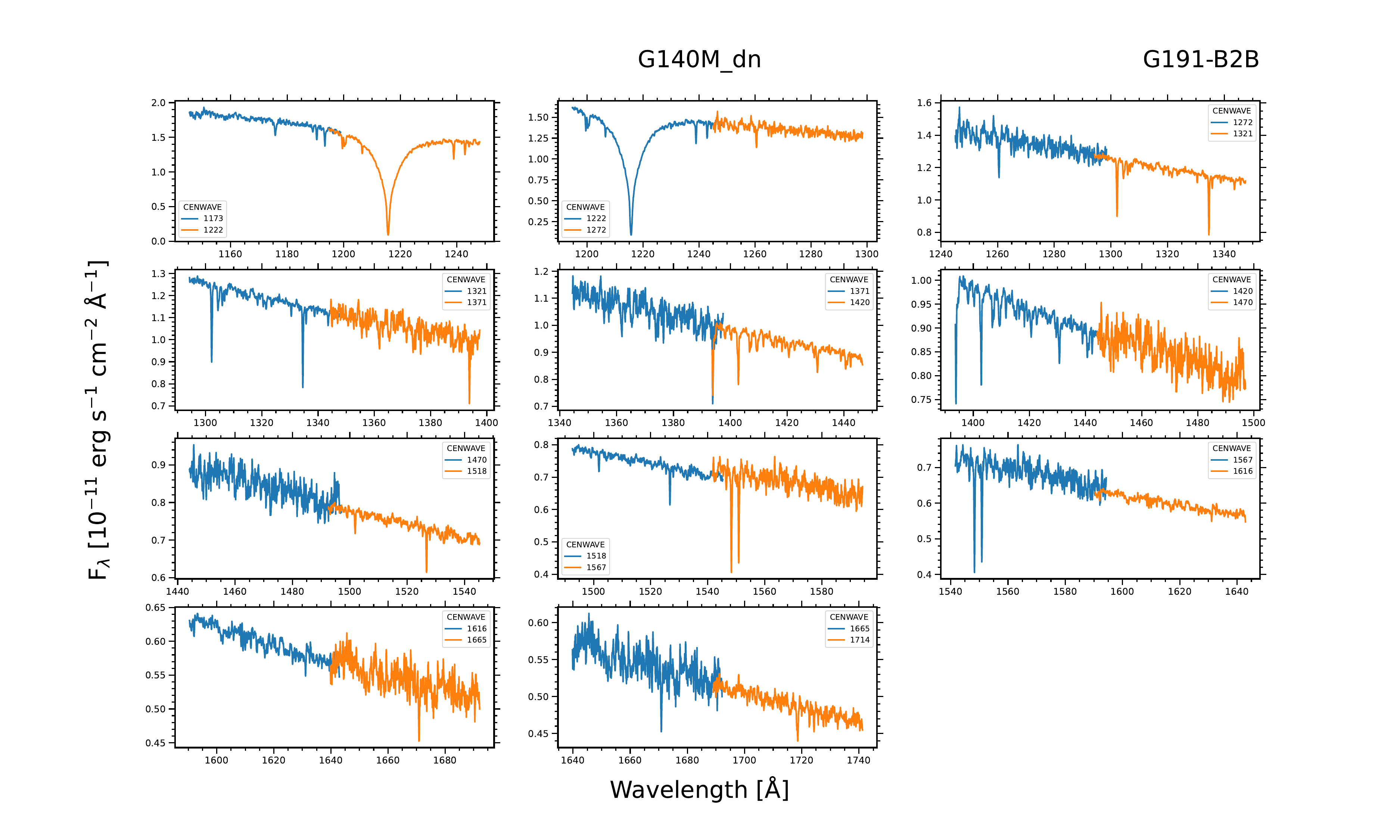}
  \caption{G140M\_dn spectra of G191-B2B recalibrated with the newly derived 
           sensitivity curves. 
           Although individual CENWAVE settings were calibrated independently,
           there is good internal consistency in the region where consecutive  
           settings overlap.
           Significant differences in the S/N of the spectra obtained at adjacent 
           CENWAVE settings are apparent. 
           }
  \label{fig:overlapG140M}
\end{figure}

\begin{figure}[ht]
  \centering
  \includegraphics[width=\textwidth]{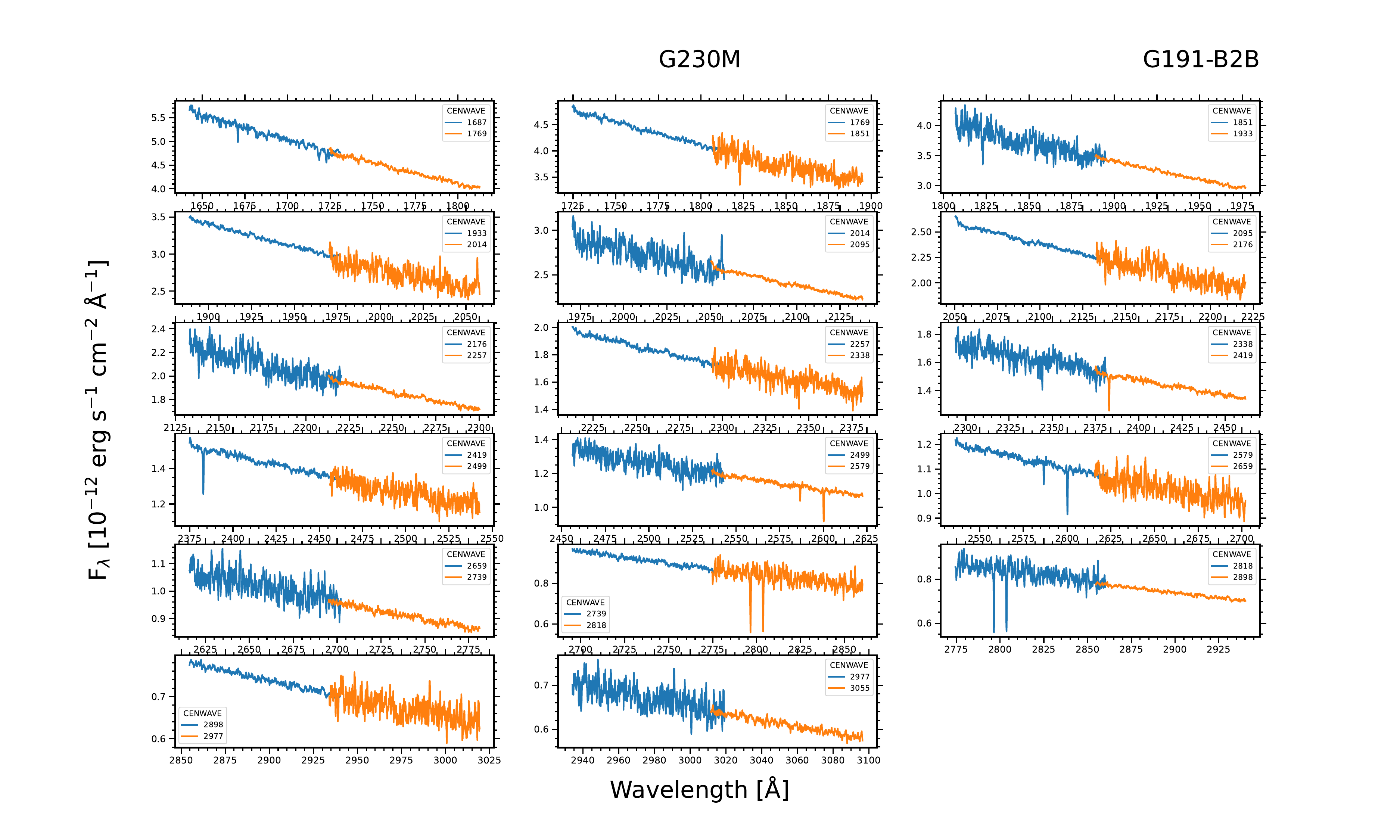}
  \caption{Same as Figure~\ref{fig:overlapG140M} for G230M.}
  \label{fig:overlapG230M}
\end{figure}

\begin{figure}[ht]
  \centering
  \includegraphics[width=\textwidth]{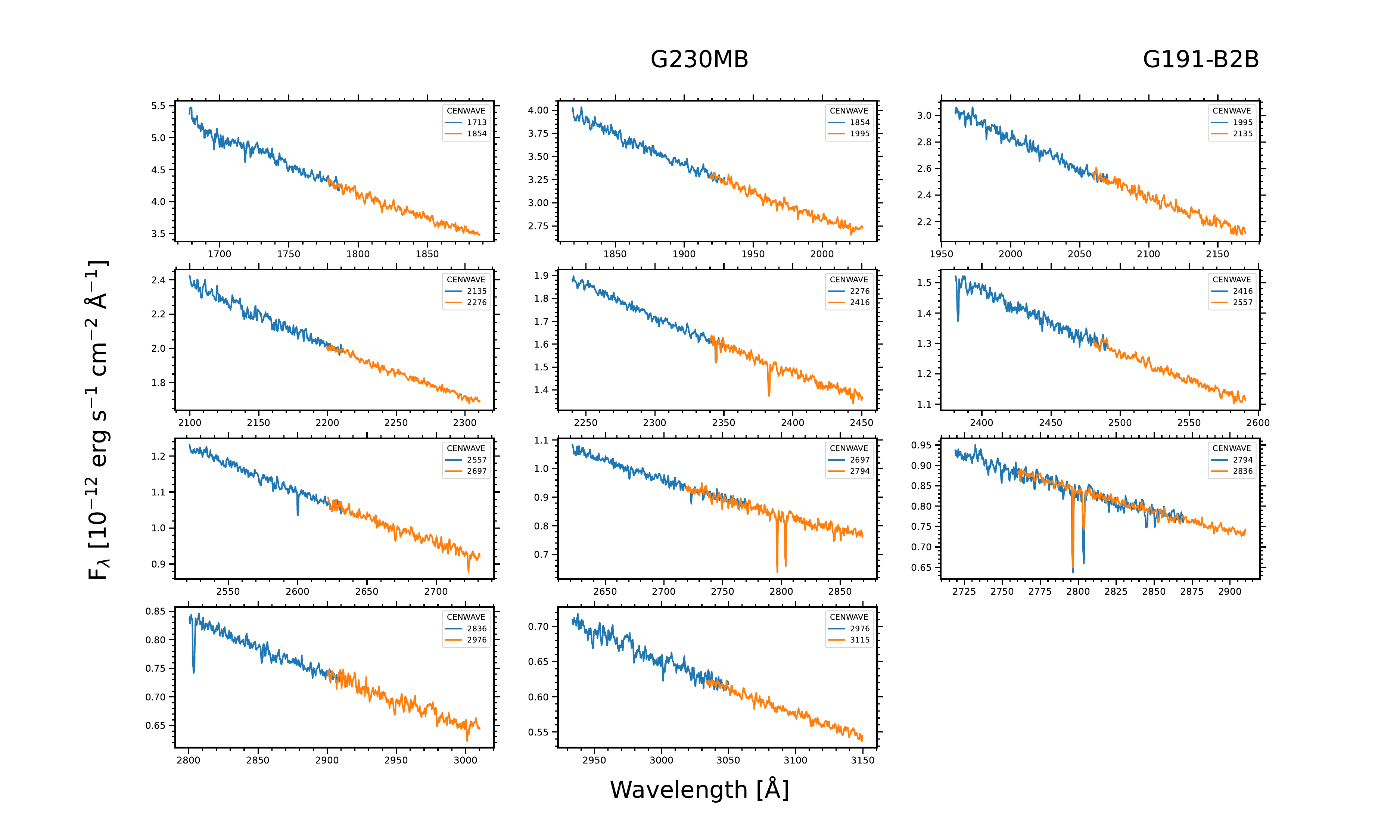}
  \caption{Same as Figure~\ref{fig:overlapG140M} for G230MB.
           Differences in the appearance of narrow spectral features 
           in consecutive orders (e.g., Mg~{\sc ii} $\lambda\lambda$ 2796, 2803)
           are artifacts of data screening and smoothing.
          }
  \label{fig:overlapG230MB}
\end{figure}

\begin{figure}[hb]
  \centering
  \includegraphics[width=\textwidth]{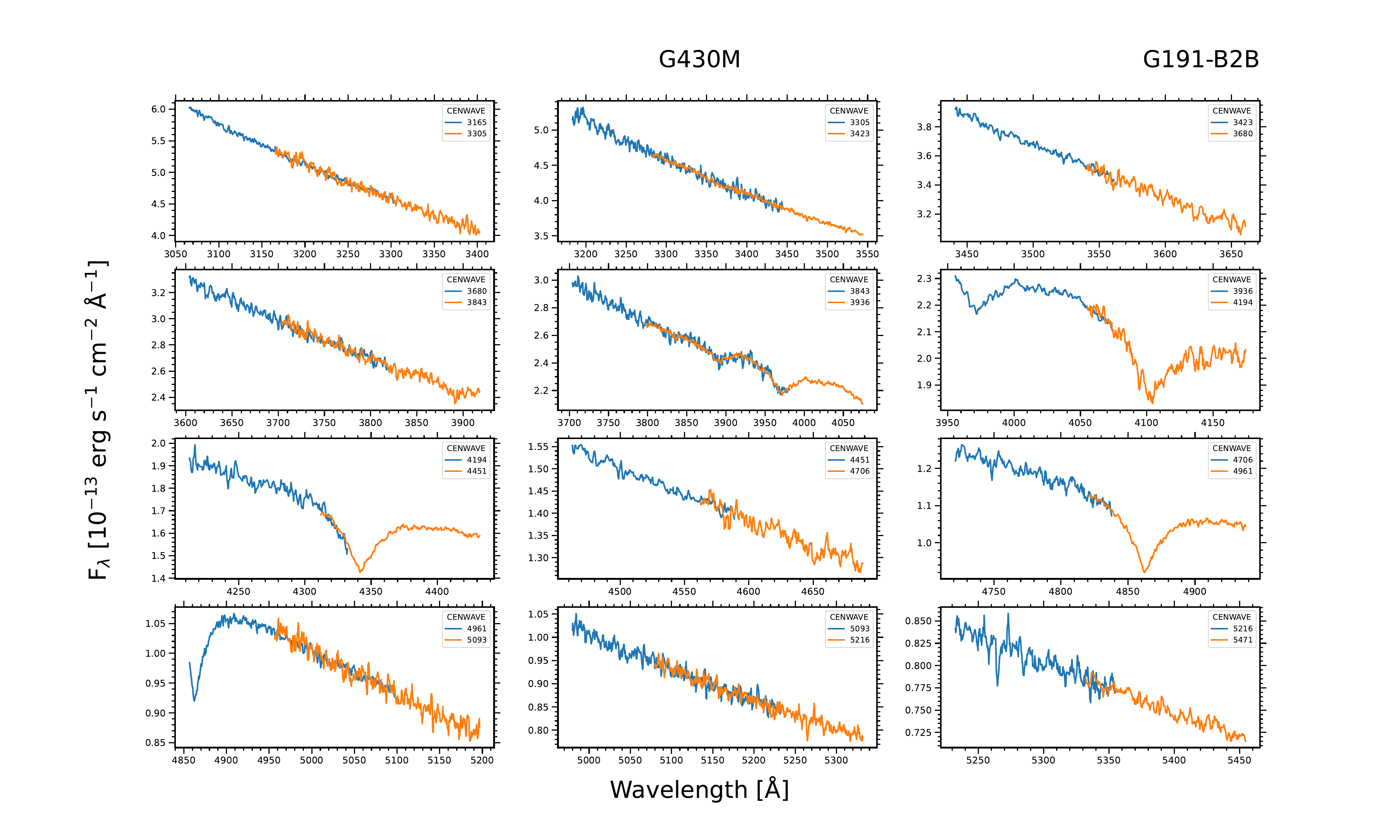}
  \caption{Same as Figure~\ref{fig:overlapG140M} for G430M.}
  \label{fig:overlapG430M}
\end{figure}

\begin{figure}[ht]
  \centering
  \includegraphics[width=\textwidth]{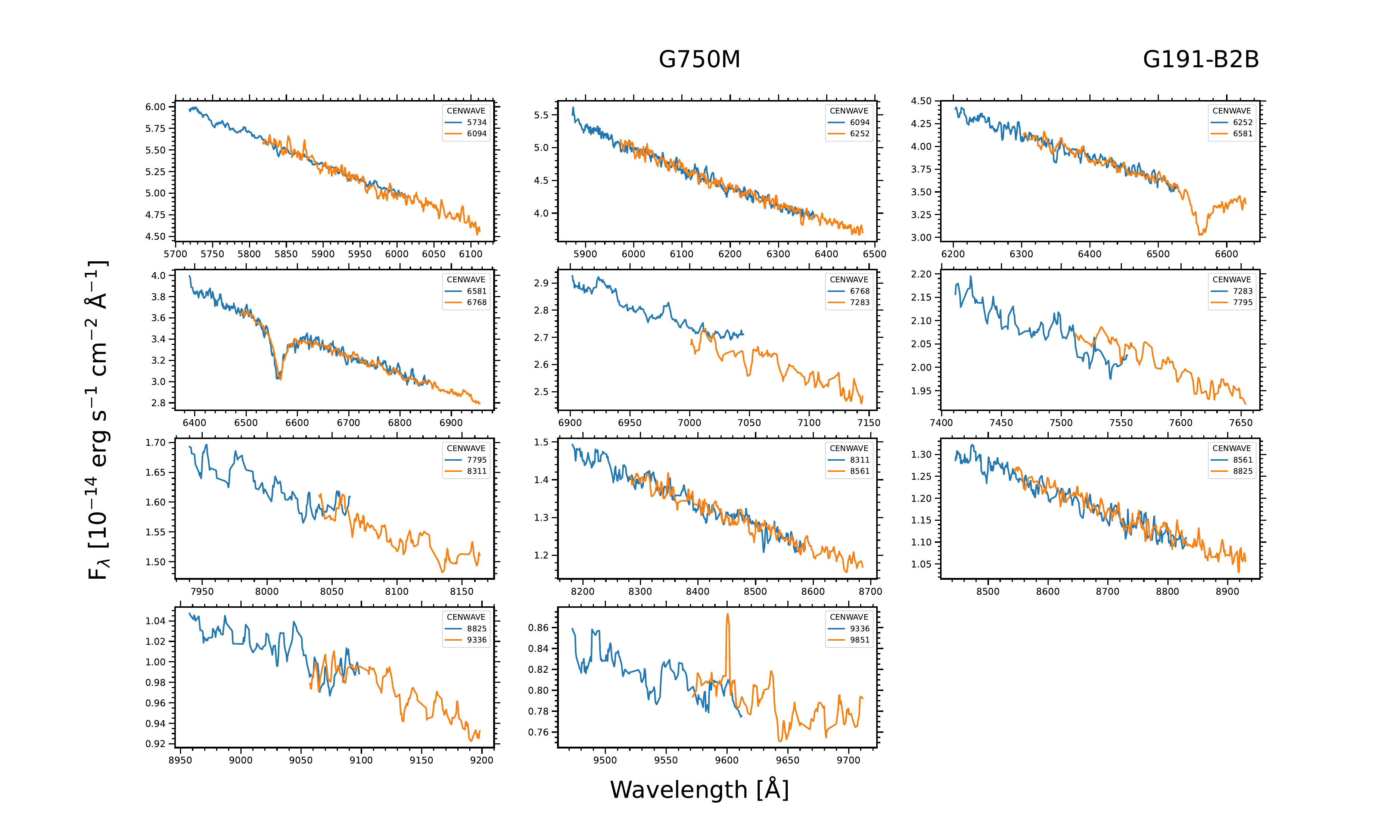}
  \caption{Same as Figure~\ref{fig:overlapG140M} for G750M.
           The slight mismatch in the overlap region between CENWAVES
           6768~{\AA} and 7283~{\AA} results from the use of the
           defringing procedure for CENWAVES greater than 7200~{\AA}.
          }
  \label{fig:overlapG750M}
\end{figure}

Figures~\ref{fig:overlapG140M} -- \ref{fig:overlapG750M} demonstrate the internal 
consistency of the recalibration of G191-B2B by showing that regions of adjacent 
CENWAVEs that overlap in wavelength are well matched in flux.

The final versions of the new PHOTTAB reference files were activated in the HST 
Calibration Reference Data System (CRDS) on May 1, 2025.  
Their names are documented in the 
{\href{https://www.stsci.edu/contents/news/stis-stans/june-2025-stan.html#article1}
      {June 2025 issue of the STScI Analysis Newsletter}
}.

\clearpage
\begin{figure}[ht]
  \centering
  \includegraphics[width=\textwidth]{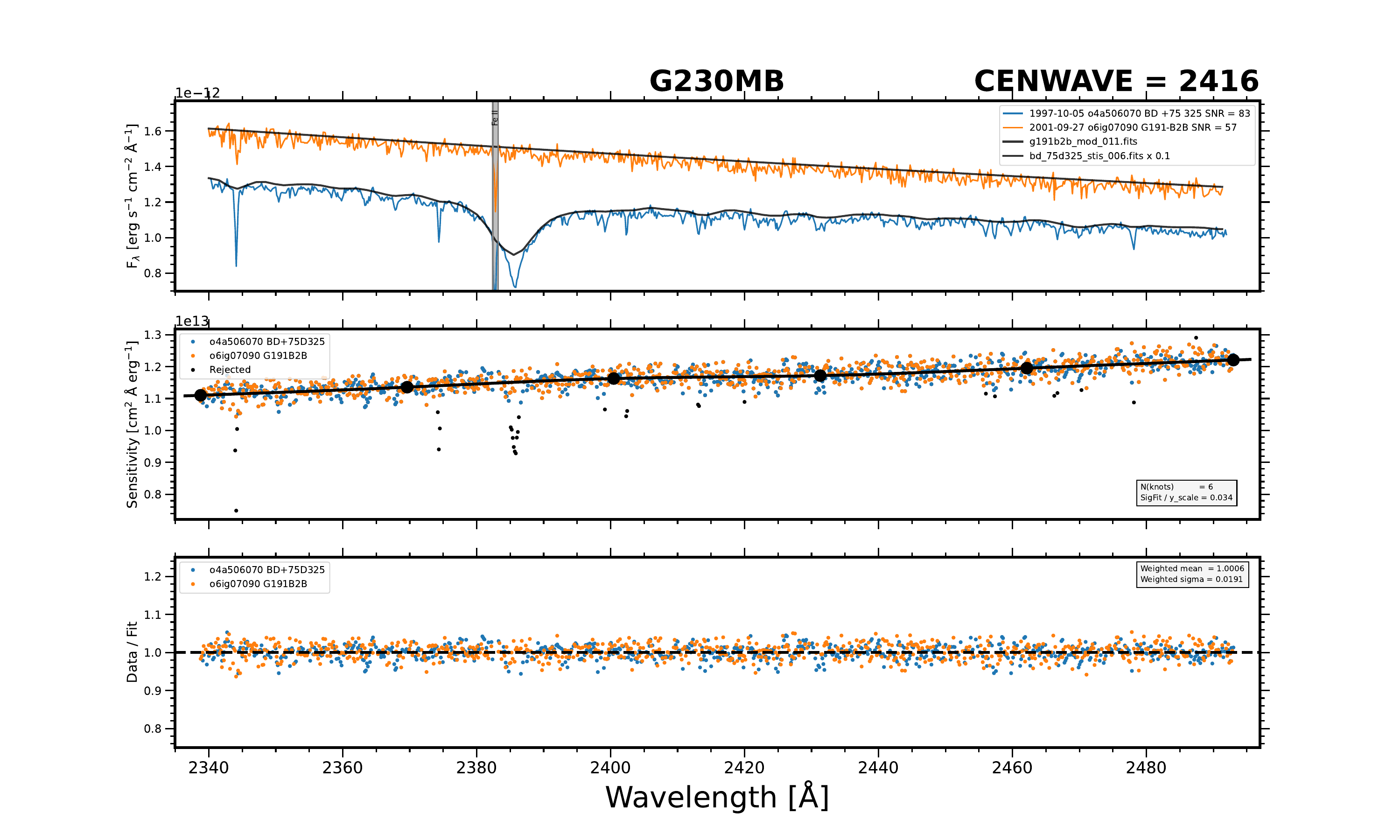}
  \caption{Comparison of the sensitivity curves for G230MB/2416 derived from data 
           of similar quality for G191-B2B (orange) and {\bd} (blue). 
           Top: Observed and standard (model or observed) flux distributions 
                used to derive the new sensitivity.  
                The flux of {\bd} has been reduced by a factor
                of 10 for display purposes.
           Middle: The sensitivity curves used in the fitting procedure.
           Bottom: The sensitivity data divided by the fitted curve.  
          }
  \label{fig:cfprimsecA}
\end{figure}

\vspace{-0.3cm}
\ssection{Discussion and Recommendations}\label{sec:discuss}

Figures \ref{fig:G140MupC1173b} -- \ref{fig:G750MC9851a} confirm that the revised sensitivities 
derived for the STIS M-modes reproduce the input CALSPEC model spectra accurately.
The S/N of the spectra of the primary white dwarf standards used as input is the limiting factor.
Although the S/N varies between 9 and 142 for the data sets used in the recalibration,
the median value is 47.7.
Consequently, fluxes derived from the current re-calibration should be accurate to $\sim$2\%,
with the possible exception of localized excursions in the vicinity of Lyman~$\alpha$.

Since the same data sets used to produce the previous calibration were used in the present work,
differences in the two sets of sensitivity curves can be attributed to details of the spline fitting
(e.g., number and placement of nodes) or data handling (e.g., use of {\tt stistools.defringe}).
However, the more fundamental limitations and issues with the data noted by Proffitt (2006) remain.
\begin{enumerate}
  \item 
  The calibrations rely heavily on a single standard -- G191-B2B -- so there is limited scope to 
  explore issues that might be associated with the CALSPEC model of its spectrum.
  \begin{itemize}
    \item
          Figure~\ref{fig:cfprimsecA} compares the sensitivity data derived from observations of 
          {\bd} for G230MB/2416.  
          The agreement is quite good, despite the significant difference in brightness of the two 
          sources, so there is no particular concern about the fidelity of either model at these
          wavelengths.  
    \item However, the small but systematic offsets between observations of G191-B2B and GD~71 with the 
          G140M$\_$up/1272 configuration indicated by the separation of colors in the sensitivity data
          panels (second panel from the top) in Figs~\ref{fig:G140MupC1272a} and \ref{fig:G140MupC1272b}
          suggest that minor issues with the models might persist.
          These discrepancies may cause subtle differences in the calibrations 
          for observations made at the G140M\_up and G140M\_dn locations on the detector, which rely 
          on GD~71 and G191-B2B, respectively.
          Fortunately, these differences are only an issue for the small fraction of M-mode observations 
          obtained during the ``up'' era.
          Potential differences with the popular G140M/1222 configuration due to mismatches with the 
          photospheric Lyman~$\alpha$ line also require further investigation; compare, e.g., 
          Figs~\ref{fig:G140MupC1222b} and \ref{fig:G140MdnC1222a} for GD~71 and G191-B2B, respectively.
  \end{itemize}

  \item
  As indicated in Table~\ref{tab:mmodes}, observations of the primary or secondary spectrophotometric 
  standard stars do not appear to exist for some M-modes. 
  Since it is not possible to recalibrate these modes on the same photometric system as version 
  11 CALSPEC models of the primary standards, they have been left at their previous values, some 
  of which have uncertain provenance.
\end{enumerate}

Table~\ref{tab:demand} summarizes the historical use of the modes that have not been recalibrated 
by indicating the number of calibration and General Observer spectra that have been obtained over 
the lifetime of STIS.
With two notable exceptions, these modes are not in high demand, and their use has generally been
confined to GO cycles that are more than a decade old.
This combination of low demand and stale calibration suggests that it might be time to remove them 
from the list of available configurations. 
Alternately, if a GO selects one of these modes, it might be beneficial  
to alert them to the stale calibration (e.g., of G140M/1218 and G140M/1540) and 
perhaps recommend the selection of an adjacent, more recently calibrated configuration.
\begin{deluxetable}{cccccl}
  \tablewidth{0pt}
  \tablecaption{Historical Use of M-Modes That Were Not Recalibrated\label{tab:demand}}
  \tabletypesize{\footnotesize}
  \tablehead{\colhead{Grating}     & 
             \colhead{CENWAVE}     & 
             \colhead{Cal Spectra} &
             \colhead{GO Spectra}  &
             \colhead{Cycles}      &
             \colhead{Comment}
            }
  \startdata
  G140M &  1218 &  0 &  40 &  8 -- 27 & Secondary Mode   \\
  G140M &  1387 &  0 &   0 &          & Secondary Mode   \\
  G140M &  1400 &  0 &   7 &  9 -- 12 & Secondary Mode   \\
  G140M &  1540 &  0 &  84 & 17 -- 30 & Secondary Mode   \\
  G140M &  1640 &  0 &   0 &          & Secondary Mode   \\
        &       &    &     &          &                  \\
  G230M &  2600 &  0 &   0 &          & Secondary Mode   \\
  G230M &  2800 &  0 &   5 & 19 -- 25 & Secondary Mode   \\
  G230M &  2828 &  0 &   0 &          & Secondary Mode   \\
        &       &    &     &          &                  \\
  G750M &  9286 &  0 &   0 &          & Secondary Mode   \\
  G750M &  9806 &  3 &   0 &          & Secondary Mode   \\
  G750M & 10363 &  7 &   3 &  7 -- 8  & Unsupported Mode \\
\enddata
\end{deluxetable}

Acquiring the data required to enable recalibration of the complete suite of primary M-Modes 
should be considered a priority as a close-out activity. 
Achieving uniform S/N near 100 would be beneficial in reducing uncertainties in the calibration,
as would observing multiple standards for modes that include Lyman~$\alpha$. 
A direct comparison of the sensitivity derived near ``beginning of life" (e.g., with the data
sets currently available) and ``end of life'' would improve the integrity of the calibration over
the lifetime of STIS, and would enhance the legacy value of the final archive.
\newpage 

\vspace{-0.3cm}
\ssectionstar{Acknowledgements}
\vspace{-0.3cm}
This work benefitted substantially from the advice, encouragement, and patience 
of the entire STIS Branch, most particularly Svea Hernandez, Tala Monroe, 
Joleen Carlberg, and Matt Siebert.

\vspace{-0.3cm}
\ssectionstar{Change History for STIS ISR {\isrnum} }\label{sec:History}
\vspace{-0.3cm}
Version 1: {\ddmonthyyyy{13 August 2025}} -- Original Document 

\vspace{-0.3cm}
\ssectionstar{References}\label{sec:References}
\vspace{-0.3cm}

\noindent
Bohlin, R. C., Hubeny, I., \& Rauch, T. 
\href{https://iopscience.iop.org/article/10.3847/1538-3881/ab94b4/pdf}
{2020, AJ, 160, 21}          \\
\noindent
Carlberg, J. K. et al. 2022 
\href{https://www.stsci.edu/files/live/sites/www/files/home/hst/instrumentation/stis/documentation/instrument-science-reports/_documents/2022-04.pdf}
{STIS Instrument Science Report 2022-04} \\
\noindent
Dos Santos, L. A. et al. STIS Instrument Science Report, in prep   \\
\noindent
Hernandez, S. et al. 2024    
\href{https://www.stsci.edu/files/live/sites/www/files/home/hst/instrumentation/stis/documentation/instrument-science-reports/_documents/STIS_ISR_2024-04.pdf}
{STIS Instrument Science Report 2024-04}  \\
\noindent
Jones, A. M. et al., 2025
\href{https://www.stsci.edu/files/live/sites/www/files/home/hst/instrumentation/stis/documentation/instrument-science-reports/_documents/STIS_ISR_2025-02.pdf}
{STIS Instrument Science Report 2025-02}   \\
\noindent
Siebert, M. R. et al. 2024 
\href{https://www.stsci.edu/files/live/sites/www/files/home/hst/instrumentation/stis/documentation/instrument-science-reports/_documents/STIS_ISR_2024-02.pdf}
{STIS Instrument Science Report 2024-02}   \\
\noindent
Proffitt, C. R. 2006 
\href{https://www.stsci.edu/files/live/sites/www/files/home/hst/instrumentation/stis/documentation/instrument-science-reports/_documents/200604.pdf}
{STIS Instrument Science Report 2006-04}    \\

\clearpage
\newpage
\appendix
\vspace{-0.3cm}
\ssectionstar{Appendix A: Observational Material}\label{sec:AppA}
\vspace{-0.3cm}
Tables \ref{table:g140m_up} -- \ref{table:g750m} provide details of the specific 
data sets of the primary white-dwarf standard stars used to update the 
spectrophotometric calibration of the STIS M-Modes.  
Successive columns in these tables record:
\begin{itemize}
   \item the CENWAVE setting;
   \item the name of the target;
   \item the calibration program identification;
   \item the rootname of the data set;
   \item the observation date;
   \item the exposure time in seconds; and 
   \item an estimate of the signal-to-noise ratio, which was produced by masking the
         FLUX vector of the extracted spectrum to remove pixels with serious data quality 
         flags; dividing by the similarly masked ERROR vector; and computing the median
         value of the ratio.
\end{itemize}
Table~\ref{table:bd} lists the data sets available for the secondary standard {\bd}.
It includes the grating in addition to the information provided in the other tables.
\begin{deluxetable}{clccccr}
\tablewidth{0pt}
\tablecaption{Observations of Primary Standards with the G140M grating 
              (``up" location)\label{table:g140m_up}
              }
\tabletypesize{\footnotesize}
\tablehead{ \colhead{CENWAVE}     & 
            \colhead{TARGNAME}    & 
            \colhead{PID}         &
            \colhead{DATA SET}    & 
            \colhead{DATEOBS}     &
            \colhead{EXPTIME [s]} &
            \colhead{S/N}
            }
\startdata 
1173 & GD 71    & 7657 & o4dd13020 & 1999-03-04 &  971 & 48 \\
     &          &      &           &            &      &    \\
1222 & GD 71    & 7657 & o4dd04020 & 1997-11-22 &  720 & 48 \\
     & GD 71    & 7657 & o4dd13030 & 1999-03-04 &  720 & 48 \\
     &          &      &           &            &      &    \\
1272 & GD 71    & 7657 & o4dd04030 & 1997-11-22 &  720 & 69 \\
     & G191-B2B & 7937 & o530020f0 & 1999-02-23 &  120 & 61 \\
     & GD 71    & 7657 & o4dd13040 & 1999-03-04 &  600 & 63 \\
     &          &      &           &            &      &    \\
1321 & GD 71    & 7657 & o4dd04040 & 1997-11-22 &  600 & 59 \\
     & GD 71    & 7657 & o4dd13050 & 1999-03-04 &  600 & 59 \\
     &          &      &           &            &      &    \\
1371 & GD 71    & 7657 & o4dd04050 & 1997-11-22 &  600 & 52 \\
     & GD 71    & 7657 & o4dd13060 & 1999-03-04 &  652 & 53 \\
     &          &      &           &            &      &    \\
1420 & GD 71    & 7657 & o4dd13070 & 1999-03-04 & 1080 & 60 \\
     &          &      &           &            &      &    \\
1470 & GD 71    & 7657 & o4dd04060 & 1997-11-22 &  612 & 38 \\
     &          &      &           &            &      &    \\
1518 & GD 153   & 7097 & o43j030i0 & 1997-07-25 &  500 & 26 \\
     & GD 71    & 7657 & o4dd04070 & 1997-11-22 &  720 & 34 \\
     &          &      &           &            &      &    \\
1550 & GD 71    & 7657 & o4dd13080 & 1999-03-04 & 1080 & 35 \\
     &          &      &           &            &      &    \\
1567 & GD 71    & 7657 & o4dd04080 & 1997-11-22 &  720 & 27 \\
     &          &      &           &            &      &    \\
1665 & GD 71    & 7657 & o4dd04090 & 1997-11-22 & 2348 & 31 \\
     &          &      &           &            &      &    \\
1714 & GD 71    & 7657 & o4dd040a0 & 1997-11-22 & 2332 & 27 \\
\enddata
\end{deluxetable}

\begin{deluxetable}{clccccr}
\tablewidth{0pt}
\tablecaption{Observations of Primary Standards with the G140M grating 
              (``dn" location)\label{table:g140m_dn}
              }
\tabletypesize{\footnotesize}
\tablehead{ \colhead{CENWAVE}     & 
            \colhead{TARGNAME}    & 
            \colhead{PID}         &
            \colhead{DATA SET}    & 
            \colhead{DATEOBS}     &
            \colhead{EXPTIME [s]} &
            \colhead{S/N}
            }
\startdata 
1173 & G191-B2B & 8421 & o5i003010 & 2000-02-28 &  600 &  81 \\
     & G191-B2B & 8916 & o6ig06010 & 2001-11-21 &   22 &  15 \\
     &          &      &           &            &      &     \\
1222 & G191-B2B & 8421 & o5i003020 & 2000-02-28 &  600 & 108 \\
     & G191-B2B & 8916 & o6ig06020 & 2001-11-21 &   15 &  17 \\
     &          &      &           &            &      &     \\
1272 & G191-B2B & 8916 & o6ig06030 & 2001-11-21 &   12 &  19 \\
     &          &      &           &            &      &     \\
1321 & G191-B2B & 8421 & o5i003030 & 2000-02-28 &  395 & 100 \\
     & G191-B2B & 8421 & o5i003040 & 2000-02-28 &  205 &  72 \\
     & G191-B2B & 8916 & o6ig06040 & 2001-11-21 &   12 &  18 \\
     &          &      &           &            &      &     \\
1371 & G191-B2B & 8916 & o6ig06060 & 2001-11-21 &    4 &   9 \\
     & G191-B2B & 8916 & o6ig06070 & 2001-11-21 &   16 &  18 \\
     &          &      &           &            &      &     \\
1420 & G191-B2B & 8421 & o5i003050 & 2000-02-28 &  600 &  92 \\
     & G191-B2B & 8916 & o6ig06080 & 2001-11-21 &   20 &  17 \\
     &          &      &           &            &      &     \\
1470 & GD 71    & 7657 & o4dd14030 & 1999-03-26 &  600 &  37 \\
     & G191-B2B & 8916 & o6ig06050 & 2001-11-21 &   30 &  17 \\
     &          &      &           &            &      &     \\
1518 & GD 71    & 7657 & o4dd14040 & 1999-03-26 &  643 &  32 \\
     & G191-B2B & 8421 & o5i003060 & 2000-02-28 &  600 &  64 \\
     & G191-B2B & 8916 & o6ig06090 & 2001-11-21 &   45 &  17 \\
     &          &      &           &            &      &     \\
1567 & GD 71    & 7657 & o4dd14050 & 1999-03-26 &  720 &  26 \\
     & G191-B2B & 8916 & o6ig060a0 & 2001-11-21 &   70 &  17 \\
     &          &      &           &            &      &     \\
1616 & GD 71    & 7657 & o4dd14060 & 1999-03-26 & 2327 &  36 \\
     & G191-B2B & 8421 & o5i003070 & 2000-02-28 &  304 &  27 \\
     & G191-B2B & 8421 & o5i003080 & 2000-02-28 & 1496 &  60 \\
     & G191-B2B & 8916 & o6ig060b0 & 2001-11-21 &  120 &  17 \\
     &          &      &           &            &      &     \\
1665 & G191-B2B & 8916 & o6ig060c0 & 2001-11-21 &  165 &  16 \\
     &          &      &           &            &      &     \\
1714 & GD 71    & 7657 & o4dd14020 & 1999-03-26 & 2086 &  25 \\
     & G191-B2B & 8421 & o5i003090 & 2000-02-28 & 1111 &  37 \\
     & G191-B2B & 8421 & o5i0030a0 & 2000-02-28 &  827 &  32 \\
     & G191-B2B & 8916 & o6ig060d0 & 2001-11-21 &  240 &  17 \\
\enddata
\end{deluxetable}

\begin{deluxetable}{clccccr}
\tablewidth{0pt}
\tablecaption{Observations of Primary Standards 
              with the G230M grating\label{table:g230m}}
\tabletypesize{\footnotesize}
\tablehead{ \colhead{CENWAVE}     & 
            \colhead{TARGNAME}    & 
            \colhead{PID}         &
            \colhead{DATA SET}    & 
            \colhead{DATEOBS}     &
            \colhead{EXPTIME [s]} &
            \colhead{S/N}
            }
\startdata 
1687 & GD 153   & 7096 & o3zx090a0 & 1997-05-27 & 2500 &  27 \\
     & GD 71    & 7657 & o4dd10020 & 1999-01-08 & 2121 &  28 \\
     & G191-B2B & 8421 & o5i004010 & 2000-03-04 & 2520 &  64 \\
     & G191-B2B & 8916 & o6ig060e0 & 2001-11-21 &  180 &  17 \\
     &          &      &           &            &      &     \\
1769 & GD 71    & 7657 & o4dd01020 & 1997-11-10 & 2096 &  35 \\
     & G191-B2B & 8421 & o5i004020 & 2000-03-04 & 2880 &  83 \\
     & G191-B2B & 8916 & o6ig060f0 & 2001-11-21 &   75 &  13 \\
     &          &      &           &            &      &     \\
1851 & GD 153   & 7097 & o43j040h0 & 1997-07-21 &  550 &  19 \\
     & GD 71    & 7657 & o4dd01030 & 1997-11-10 & 2338 &  43 \\
     & GD 71    & 7657 & o4dd10040 & 1999-01-08 & 2536 &  45 \\
     & G191-B2B & 8916 & o6ig060g0 & 2001-11-21 &   60 &  14 \\
     &          &      &           &            &      &     \\
1884 & GD 71    & 7657 & o4dd11090 & 1999-03-02 & 2357 &  46 \\
     &          &      &           &            &      &     \\
1933 & GD 71    & 7657 & o4dd01040 & 1997-11-10 & 2382 &  48 \\
     & GD 71    & 7657 & o4dd10050 & 1999-01-09 & 2536 &  50 \\
     & G191-B2B & 8421 & o5i004030 & 2000-03-04 & 2880 & 106 \\
     & G191-B2B & 8916 & o6ig060h0 & 2001-11-21 &   50 &  14 \\
     &          &      &           &            &      &     \\
2014 & GD 71    & 7657 & o4dd01050 & 1997-11-10 & 2382 &  50 \\
     & GD 71    & 7657 & o4dd10060 & 1999-01-09 & 2401 &  51 \\
     & G191-B2B & 8916 & o6ig060i0 & 2001-11-21 &   45 &  14 \\
     &          &      &           &            &      &     \\
2095 & GD 71    & 7657 & o4dd01060 & 1997-11-10 & 2382 &  55 \\
     & GD 71    & 7657 & o4dd11040 & 1999-03-01 & 2492 &  57 \\
     & G191-B2B & 8421 & o5i004040 & 2000-03-04 & 2880 & 120 \\
     & G191-B2B & 8916 & o6ig060j0 & 2001-11-21 &   60 &  17 \\
     &          &      &           &            &      &     \\
2176 & GD 71    & 7657 & o4dd02030 & 1997-11-23 & 1000 &  39 \\
     & GD 71    & 7657 & o4dd11030 & 1999-03-01 & 1526 &  48 \\
     & G191-B2B & 8916 & o6ig060k0 & 2001-11-21 &   32 &  14 \\
     &          &      &           &            &      &     \\
2257 & GD 71    & 7657 & o4dd02040 & 1997-11-23 & 1136 &  41 \\
     & G191-B2B & 8421 & o5i005010 & 2000-02-28 & 1000 &  77 \\
     & G191-B2B & 8916 & o6ig060l0 & 2001-11-21 &   35 &  14 \\
     &          &      &           &            &      &     \\
2338 & GD153    & 7096 & o3zx09080 & 1997-05-27 &  975 &  33 \\
     & GD 71    & 7657 & o4dd11050 & 1999-03-01 & 1200 &  41 \\
     & G191-B2B & 8916 & o6ig060m0 & 2001-11-21 &   40 &  15 \\
     &          &      &           &            &      &     \\
2419 & GD 71    & 7657 & o4dd02050 & 1997-11-23 & 1000 &  36 \\
     & GD 71    & 7657 & o4dd11060 & 1999-03-01 & 1016 &  36 \\
     & G191-B2B & 8421 & o5i005020 & 2000-02-28 & 1090 &  74 \\
     & G191-B2B & 8916 & o6ig060n0 & 2001-11-21 &   40 &  14 \\
     &          &      &           &            &      &     \\
2499 & GD 71    & 7657 & o4dd02060 & 1997-11-23 & 1136 &  37 \\
     & GD 71    & 7657 & o4dd11070 & 1999-03-01 & 1200 &  38 \\
     & G191-B2B & 8916 & o6ig060o0 & 2001-11-21 &   45 &  14 \\
     &          &      &           &            &      &     \\
2579 & GD 71    & 7657 & o4dd02070 & 1997-11-23 & 1000 &  32 \\
     & GD 71    & 7657 & o4dd11080 & 1999-03-01 &  972 &  32 \\
     & G191-B2B & 8421 & o5i005030 & 2000-02-28 & 1200 &  70 \\
     & G191-B2B & 8916 & o6ig060p0 & 2001-11-21 &   50 &  14 \\
     &          &      &           &            &      &     \\
2659 & GD153    & 7097 & o43j050k0 & 1997-07-19 &  586 &  21 \\
     & GD 71    & 7657 & o4dd02080 & 1997-11-23 & 1139 &  32 \\
     & G191-B2B & 8916 & o6ig060q0 & 2001-11-21 &   55 &  14 \\
     &          &      &           &            &      &     \\
2739 & GD 71    & 7657 & o4dd02020 & 1997-11-23 & 1200 &  32 \\
     & G191-B2B & 8421 & o5i005040 & 2000-02-28 & 1200 &  64 \\
     & G191-B2B & 8916 & o6ig060r0 & 2001-11-21 &   60 &  14 \\
     &          &      &           &            &      &     \\
2818 & GD 71    & 7657 & o4dd03030 & 1998-01-04 & 2328 &  43 \\
     & G191-B2B & 8916 & o6ig060s0 & 2001-11-21 &   66 &  14 \\
     &          &      &           &            &      &     \\
2898 & GD 71    & 7657 & o4dd03040 & 1998-01-04 & 2382 &  39 \\
     & G191-B2B & 8421 & o5i005050 & 2000-02-28 & 2880 &  84 \\
     & G191-B2B & 8916 & o6ig060t0 & 2001-11-21 &   80 &  14 \\
     &          &      &           &            &      &     \\
2977 & GD153    & 7096 & o3zx09070 & 1997-05-27 & 2705 &  31 \\
     & GD 71    & 7657 & o4dd10030 & 1999-01-08 & 2492 &  35 \\
     & G191-B2B & 8916 & o6ig060u0 & 2001-11-21 &  110 &  14 \\
     &          &      &           &            &      &     \\
3055 & GD 71    & 7657 & o4dd03050 & 1998-01-04 & 2382 &  27 \\
     & G191-B2B & 8421 & o5i005060 & 2000-02-28 & 2880 &  60 \\
     & G191-B2B & 8916 & o6ig060v0 & 2001-11-21 &  155 &  14 \\
\enddata
\end{deluxetable}

\begin{deluxetable}{clccccr}
\tablewidth{0pt}
\tablecaption{Observations of Primary Standards 
              with the G230MB grating\label{table:g230mb}}
\tabletypesize{\footnotesize}
\tablehead{ \colhead{CENWAVE}     & 
            \colhead{TARGNAME}    & 
            \colhead{PID}         &
            \colhead{DATA SET}    & 
            \colhead{DATEOBS}     &
            \colhead{EXPTIME [s]} &
            \colhead{S/N}
            }
\startdata 

1713 & G191-B2B & 8421 & o5i006010 & 2000-02-27 &  800 & 33 \\
     & G191-B2B & 8916 & o6ig07030 & 2001-09-27 & 1200 & 40 \\
     &          &      &           &            &      &    \\
1854 & G191-B2B & 8916 & o6ig07040 & 2001-09-27 &  840 & 55 \\
     &          &      &           &            &      &    \\
1995 & G191-B2B & 8421 & o5i006020 & 2000-02-27 &  402 & 50 \\
     & G191-B2B & 8916 & o6ig070b0 & 2001-09-27 &  380 & 47 \\
     &          &      &           &            &      &    \\
2135 & G191-B2B & 8916 & o6ig070a0 & 2001-09-27 &  360 & 56 \\
     &          &      &           &            &      &    \\
2276 & G191-B2B & 8421 & o5i006030 & 2000-02-27 &  480 & 77 \\
     & G191-B2B & 8916 & o6ig07080 & 2001-09-27 &  280 & 57 \\
     &          &      &           &            &      &    \\
2416 & G191-B2B & 8916 & o6ig07090 & 2001-09-27 &  280 & 56 \\
     &          &      &           &            &      &    \\
2557 & G191-B2B & 8421 & o5i006040 & 2000-02-27 &  480 & 73 \\
     & G191-B2B & 8916 & o6ig07060 & 2001-09-27 &  300 & 56 \\
     &          &      &           &            &      &    \\
2697 & G191-B2B & 8916 & o6ig07050 & 2001-09-27 &  300 & 54 \\
     &          &      &           &            &      &    \\
2794 & G191-B2B & 8916 & o6ig070e0 & 2001-09-27 &  300 & 51 \\
     &          &      &           &            &      &    \\
2836 & G191-B2B & 8421 & o5i006060 & 2000-02-27 &  480 & 71 \\
     & G191-B2B & 8421 & o5i006050 & 2000-02-27 &  480 & 71 \\
     & G191-B2B & 8916 & o6ig070f0 & 2001-09-27 &  300 & 55 \\
     & G191-B2B & 8916 & o6ig100c0 & 2002-04-22 &  150 & 38 \\
     &          &      &           &            &      &    \\
2976 & G191-B2B & 8916 & o6ig070g0 & 2001-09-27 &  220 & 50 \\
     &          &      &           &            &      &    \\
3115 & G191-B2B & 8421 & o5i006070 & 2000-02-27 &  480 & 80 \\
     & G191-B2B & 8916 & o6ig070h0 & 2001-09-27 &  200 & 50 \\
\enddata
\end{deluxetable}

\begin{deluxetable}{clccccr}
\tablewidth{0pt}
\tablecaption{Observations of Primary Standards 
              with the G430M grating\label{table:g430m}}
\tabletypesize{\footnotesize}
\tablehead{ \colhead{CENWAVE}     & 
            \colhead{TARGNAME}    & 
            \colhead{PID}         &
            \colhead{DATA SET}    & 
            \colhead{DATEOBS}     &
            \colhead{EXPTIME [s]} &
            \colhead{S/N}
            }
\startdata 
3165 & G191-B2B & 8421 & o5i007010 & 2000-02-28 & 400 & 123 \\
     & G191-B2B & 8916 & o6ig070i0 & 2001-09-28 & 120 &  65 \\
     &          &      &           &            &     &     \\
3305 & G191-B2B & 8916 & o6ig070c0 & 2001-09-27 &  80 &  52 \\
     &          &      &           &            &     &     \\
3423 & G191-B2B & 8421 & o5i007020 & 2000-02-28 & 400 & 116 \\
     & G191-B2B & 8916 & o6ig070j0 & 2001-09-28 & 120 &  62 \\
     &          &      &           &            &     &     \\
3680 & G191-B2B & 8916 & o6ig070k0 & 2001-09-28 & 120 &  64 \\
     &          &      &           &            &     &     \\
3843 & G191-B2B & 8916 & o6ig07070 & 2001-09-27 &  80 &  57 \\
     &          &      &           &            &     &     \\
3936 & G191-B2B & 8421 & o5i007030 & 2000-02-28 & 438 & 142 \\
     & G191-B2B & 8916 & o6ig08020 & 2001-09-24 &  60 &  50 \\
     &          &      &           &            &     &     \\
4194 & G191-B2B & 8916 & o6ig08030 & 2001-09-24 &  70 &  53 \\
     &          &      &           &            &     &     \\
4451 & G191-B2B & 8421 & o5i007040 & 2000-02-28 & 480 & 137 \\
     & G191-B2B & 8916 & o6ig070m0 & 2001-09-28 &  80 &  54 \\
     &          &      &           &            &     &     \\
4706 & G191-B2B & 8916 & o6ig070d0 & 2001-09-27 &  90 &  52 \\
     &          &      &           &            &     &     \\
4961 & G191-B2B & 8421 & o5i007050 & 2000-02-28 & 480 & 121 \\
     & G191-B2B & 8916 & o6ig07020 & 2001-09-27 & 100 &  53 \\
     &          &      &           &            &     &     \\
5093 & G191-B2B & 8916 & o6ig08010 & 2001-09-24 & 100 &  53 \\
     &          &      &           &            &     &     \\
5216 & G191-B2B & 8916 & o6ig070l0 & 2001-09-28 & 120 &  54 \\
     &          &      &           &            &     &     \\
5471 & G191-B2B & 8421 & o5i007060 & 2000-02-28 & 480 &  97 \\
     & G191-B2B & 8916 & o6ig070n0 & 2001-09-28 & 150 &  52 \\
\enddata
\end{deluxetable}
\begin{deluxetable}{clccccr}
\tablewidth{0pt}
\tablecaption{Observations of Primary Standards 
              with the G750M grating\label{table:g750m}}
\tabletypesize{\footnotesize}
\tablehead{ \colhead{CENWAVE}     & 
            \colhead{TARGNAME}    & 
            \colhead{PID}         &
            \colhead{DATA SET}    & 
            \colhead{DATEOBS}     &
            \colhead{EXPTIME [s]} &
            \colhead{S/N}
            }
\startdata 
5734 & G191-B2B & 8421 & o5i008010 & 2000-01-30 & 480 & 142 \\
     & G191-B2B & 8916 & o6ig08040 & 2001-09-24 &  70 &  52 \\
     &          &      &           &            &     &     \\
6094 & G191-B2B & 8916 & o6ig08050 & 2001-09-24 &  80 &  52 \\
     &          &      &           &            &     &     \\
6252 & G191-B2B & 8916 & o6ig08060 & 2001-09-24 &  90 &  54 \\
     &          &      &           &            &     &     \\
6581 & G191-B2B & 8916 & o6ig07010 & 2001-09-27 & 120 &  59 \\
     &          &      &           &            &     &     \\
6768 & G191-B2B & 8421 & o5i008020 & 2000-01-30 & 480 & 120 \\
     & G191-B2B & 8916 & o6ig08070 & 2001-09-24 & 100 &  52 \\
     &          &      &           &            &     &     \\
7283 & G191-B2B & 8916 & o6ig08080 & 2001-09-24 & 140 &  52 \\
     &          &      &           &            &     &     \\
7795 & G191-B2B & 8421 & o5i008030 & 2000-01-30 & 480 &  80 \\
     & G191-B2B & 8916 & o6ig080a0 & 2001-09-24 & 210 &  51 \\
     &          &      &           &            &     &     \\
8311 & G191-B2B & 8916 & o6ig080c0 & 2001-09-24 & 360 &  48 \\
     &          &      &           &            &     &     \\
8561 & G191-B2B & 8916 & o6ig080e0 & 2001-09-24 & 400 &  46 \\
     &          &      &           &            &     &     \\
8825 & G191-B2B & 8421 & o5i008050 & 2000-01-30 & 480 &  49 \\
     &          &      &           &            &     &     \\
9336 & G191-B2B & 8916 & o6ig080g0 & 2001-09-25 & 540 &  41 \\
     &          &      &           &            &     &     \\
9851 & G191-B2B & 8421 & o5i008070 & 2000-01-30 & 480 &  24 \\
     & G191-B2B & 8916 & o6ig080i0 & 2001-09-25 & 690 &  28 \\
\enddata
\end{deluxetable}

\begin{deluxetable}{lcccccr}
\tablewidth{0pt}
\tablecaption{Observations of the Secondary Standard BD~$+75^{\circ}325$\label{table:bd}}
\tabletypesize{\footnotesize}
\tablehead{\colhead{GRATING}     & 
           \colhead{CENWAVE}     & 
           \colhead{PID}         & 
           \colhead{DATA SET}    & 
           \colhead{DATEOBS}     & 
           \colhead{EXPTIME [s]} & 
           \colhead{SNR}}
\startdata
G230MB & 1713 & 7656 & o4a506020 & 1997-10-05 & 120 &  34 \\
G230MB & 1854 & 7656 & o4a506030 & 1997-10-05 & 138 &  61 \\
G230MB & 1995 & 7656 & o4a506040 & 1997-10-05 &  72 &  58 \\
G230MB & 2135 & 7656 & o4a506050 & 1997-10-05 &  72 &  56 \\
G230MB & 2276 & 7656 & o4a506060 & 1997-10-05 &  72 &  84 \\
G230MB & 2416 & 7656 & o4a506070 & 1997-10-05 &  72 &  83 \\
G230MB & 2557 & 7656 & o4a506080 & 1997-10-05 &  72 &  80 \\
G230MB & 2697 & 7656 & o4a506090 & 1997-10-05 &  72 &  78 \\
G230MB & 2836 & 7656 & o4a5060a0 & 1997-10-05 &  72 &  79 \\
G230MB & 2976 & 7656 & o4a5060b0 & 1997-10-05 &  72 &  84 \\
G230MB & 3115 & 7656 & o4a5060c0 & 1997-10-05 &  72 &  89 \\
       &      &      &           &            &           \\
G430M  & 3165 & 7656 & o4a5050l0 & 1997-10-13 &  60 & 136 \\
G430M  & 3423 & 7656 & o4a5050m0 & 1997-10-13 &  60 & 129 \\
G430M  & 3680 & 7656 & o4a5050n0 & 1997-10-13 &  60 & 133 \\
G430M  & 3936 & 7656 & o4a5050o0 & 1997-10-14 &  60 & 150 \\
G430M  & 4194 & 7656 & o4a5050p0 & 1997-10-14 &  60 & 145 \\
G430M  & 4451 & 7094 & o3wy020b0 & 1997-05-18 & 120 & 195 \\
G430M  & 4451 & 7656 & o4a5050q0 & 1997-10-14 &  60 & 137 \\
G430M  & 4706 & 7656 & o4a5050r0 & 1997-10-14 &  60 & 125 \\
G430M  & 4781 & 7810 & o4lu01020 & 1998-03-27 & 144 & 189 \\
G430M  & 4961 & 7656 & o4a5060d0 & 1997-10-05 &  60 & 121 \\
G430M  & 5216 & 7656 & o4a5060e0 & 1997-10-05 &  60 & 113 \\
G430M  & 5471 & 7656 & o4a5060f0 & 1997-10-05 &  60 &  96 \\
       &      &      &           &            &           \\  
G750M  & 5734 & 7656 & o4a505030 & 1997-10-13 &  72 & 154 \\
G750M  & 6252 & 7656 & o4a505040 & 1997-10-13 &  72 & 144 \\
G750M  & 6581 & 7810 & o4lu01030 & 1998-03-27 & 144 & 188 \\
G750M  & 6768 & 7656 & o4a505050 & 1997-10-13 &  72 & 112 \\
G750M  & 7283 & 7656 & o4a505060 & 1997-10-13 &  72 & 112 \\
G750M  & 7795 & 7656 & o4a505080 & 1997-10-13 &  72 &  89 \\
G750M  & 8311 & 7656 & o4a5050a0 & 1997-10-13 &  72 &  63 \\
G750M  & 8561 & 7810 & o4lu01040 & 1998-03-27 & 144 &  85 \\
G750M  & 8825 & 7656 & o4a5050c0 & 1997-10-13 &  72 &  55 \\
G750M  & 9336 & 7656 & o4a5050e0 & 1997-10-13 &  72 &  45 \\
G750M  & 9851 & 7656 & o4a5050g0 & 1997-10-13 &  72 &  28 \\
\enddata
\end{deluxetable}

\begin{figure}[t]
  \hspace{-0.5in}
  \includegraphics[width=1.1\textwidth]{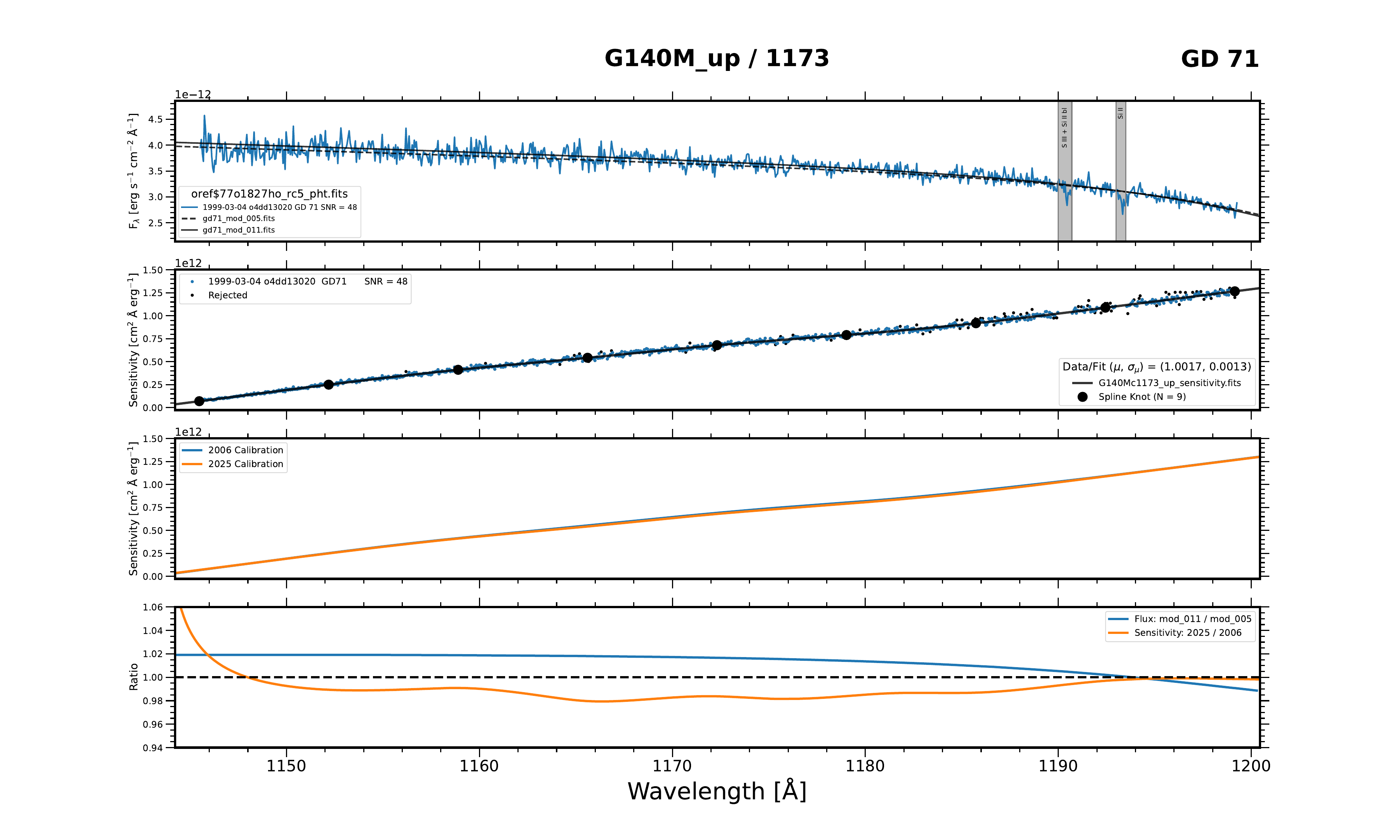}
  \footnotesize
  \caption{Calibration of GD 71 for G140M\_up/1173.}
  \label{fig:G140MupC1173b}
\end{figure}

\begin{figure}[b]
  \hspace{-0.5in}
  \includegraphics[width=1.1\textwidth]{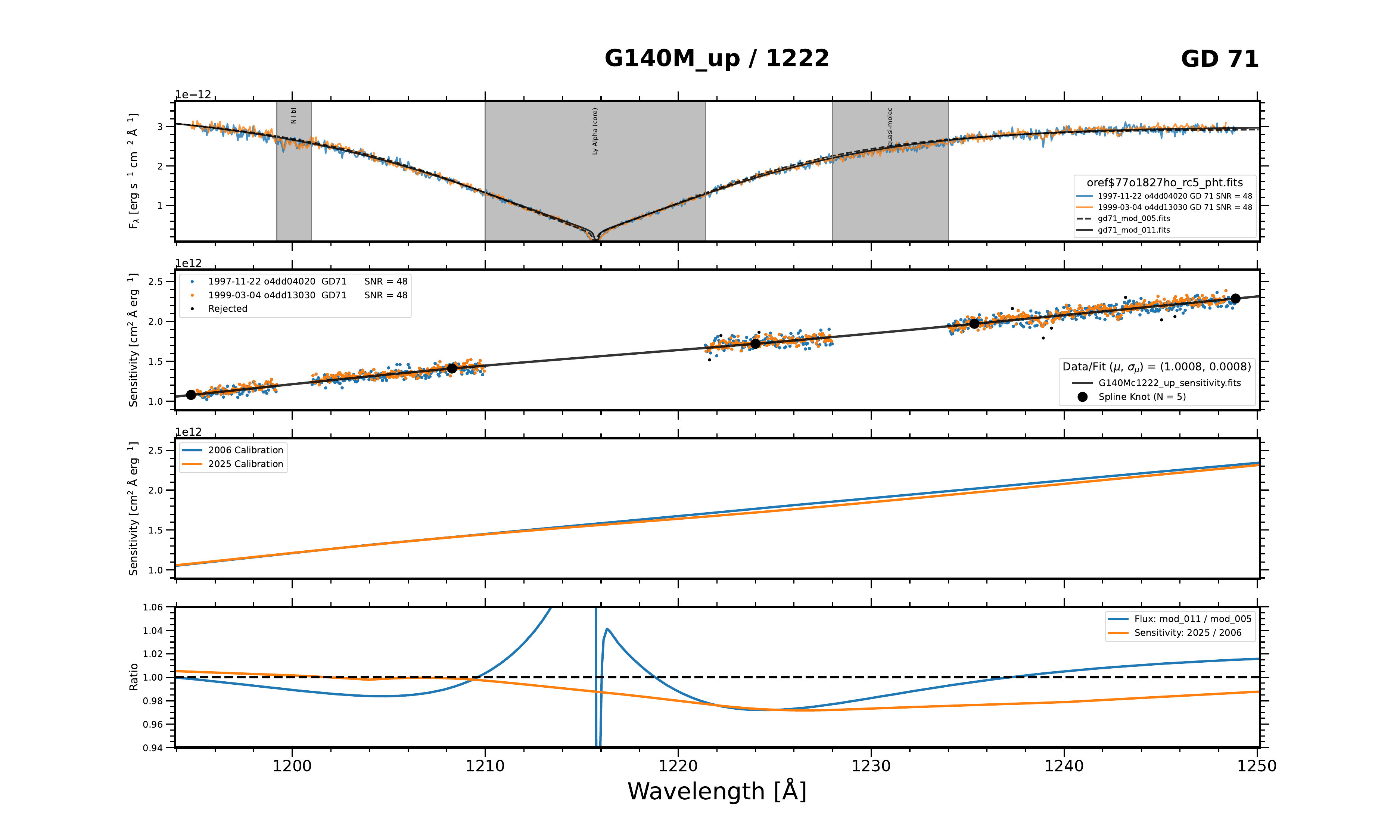}
  \footnotesize
  \caption{Calibration of GD 71 for G140M\_up/1222.}
  \label{fig:G140MupC1222b}
\end{figure}

\begin{figure}[t]
  \hspace{-0.5in}
  \includegraphics[width=1.1\textwidth]{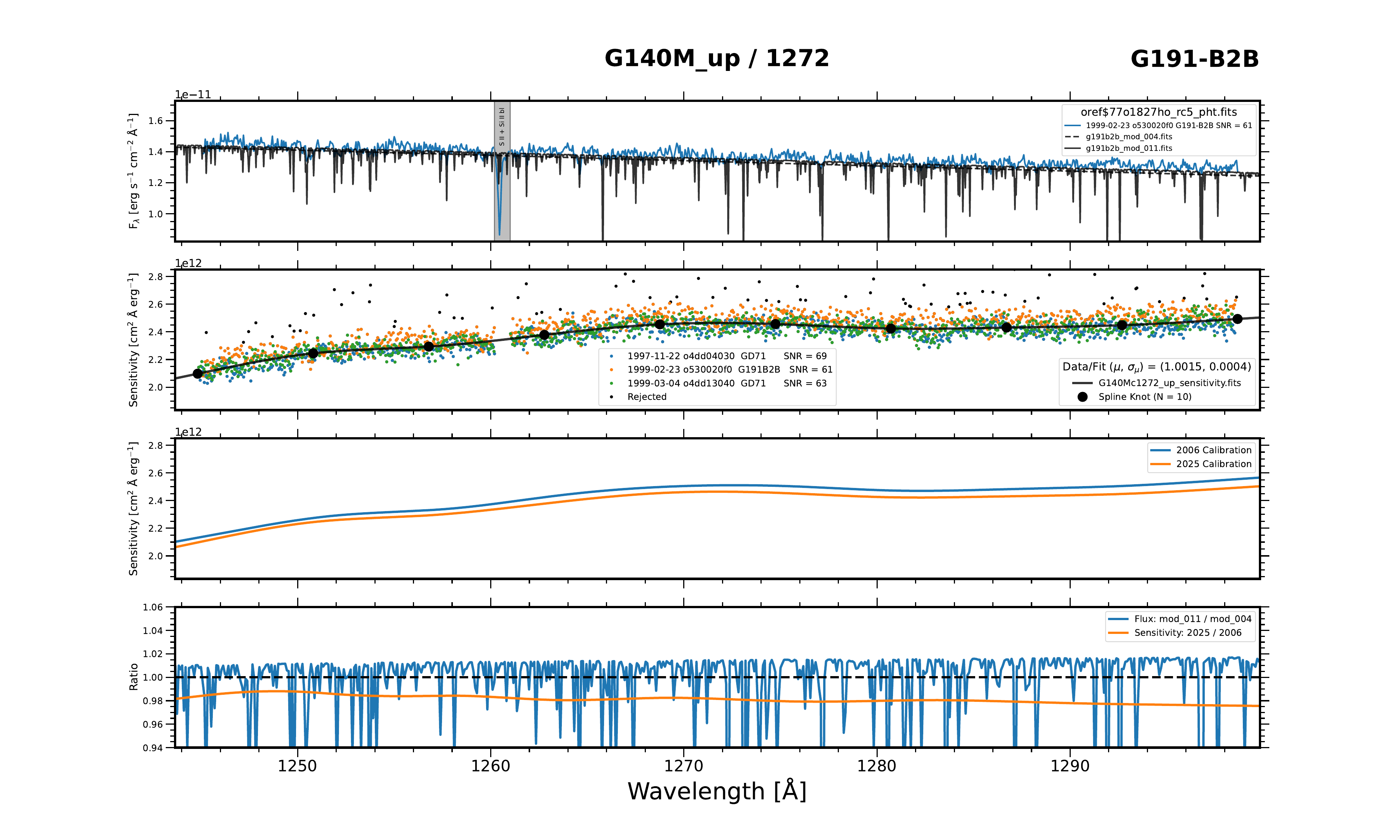}
  \footnotesize
  \caption{Calibration of G191-B2B for G140M\_up/1272.}
  \label{fig:G140MupC1272a}
\end{figure}

\begin{figure}[b]
  \hspace{-0.5in}
  \includegraphics[width=1.1\textwidth]{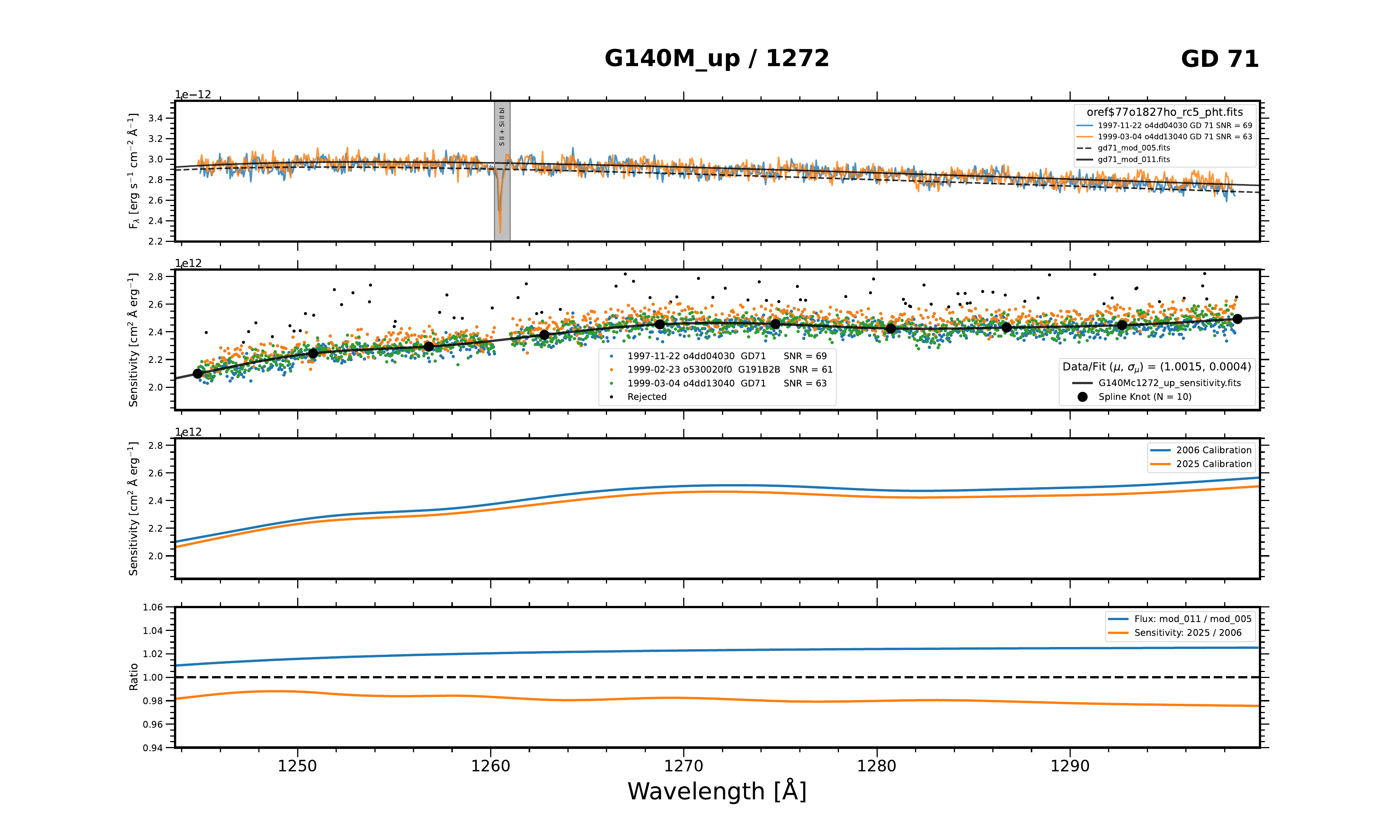}
  \footnotesize
  \caption{Calibration of GD 71 for G140M\_up/1272.}
  \label{fig:G140MupC1272b}
\end{figure}

\begin{figure}[t]
  \hspace{-0.5in}
  \includegraphics[width=1.1\textwidth]{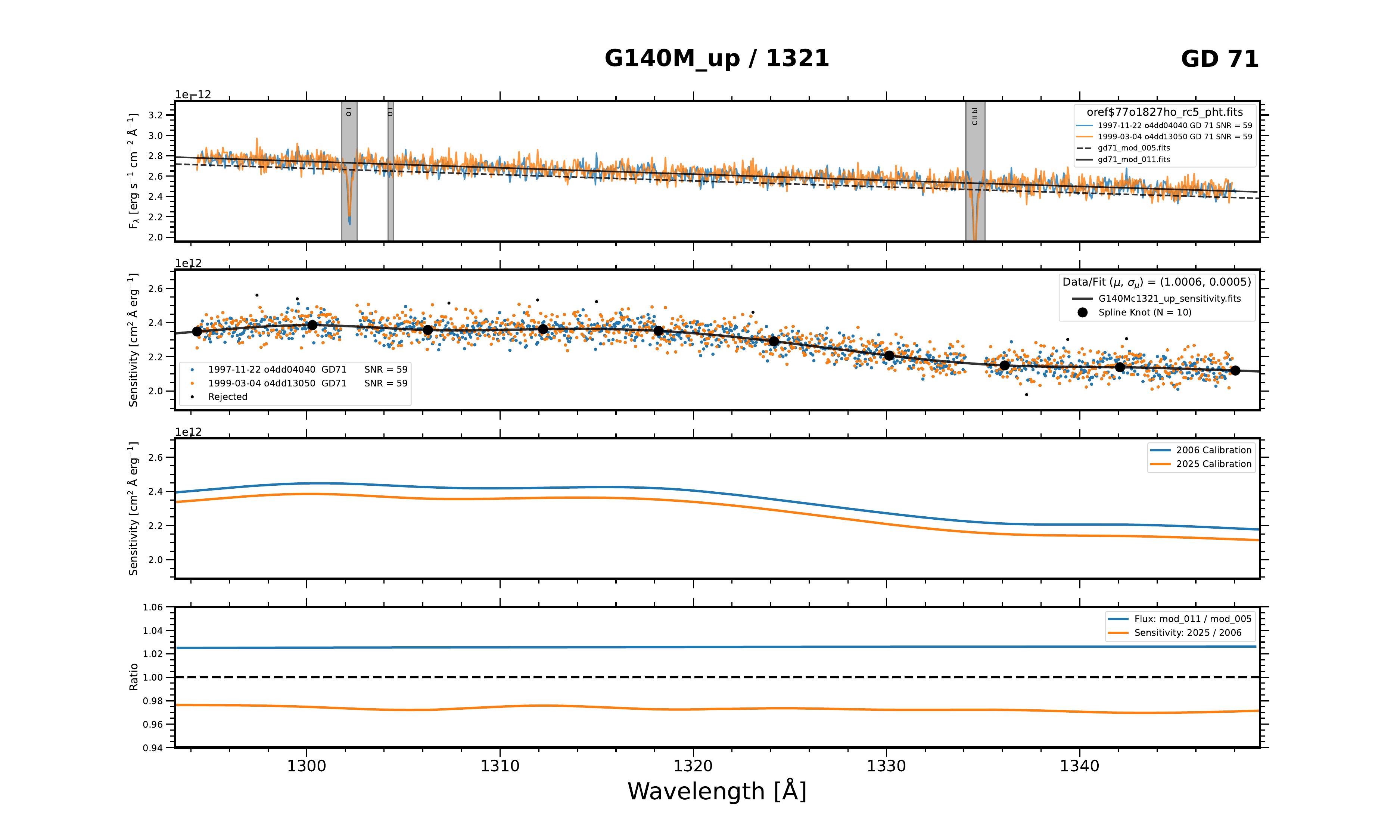}
  \footnotesize
  \caption{Calibration of GD 71 for G140M\_up/1321.}
  \label{fig:G140MupC1321b}
\end{figure}

\begin{figure}[b]
  \hspace{-0.5in}
  \includegraphics[width=1.1\textwidth]{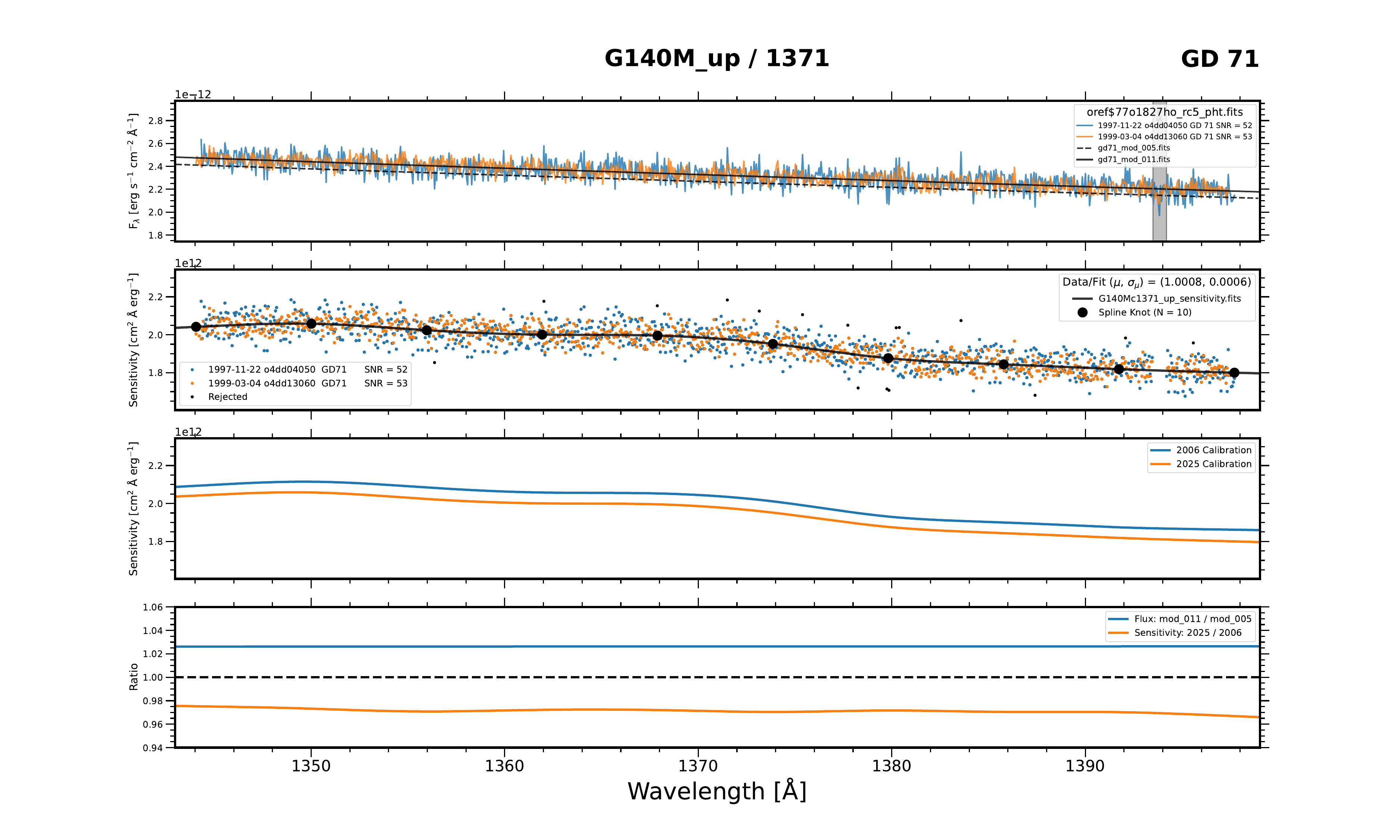}
  \footnotesize
  \caption{Calibration of GD 71 for G140M\_up/1371.}
  \label{fig:G140MupC1371b}
\end{figure}

\begin{figure}[t]
  \hspace{-0.5in}
  \includegraphics[width=1.1\textwidth]{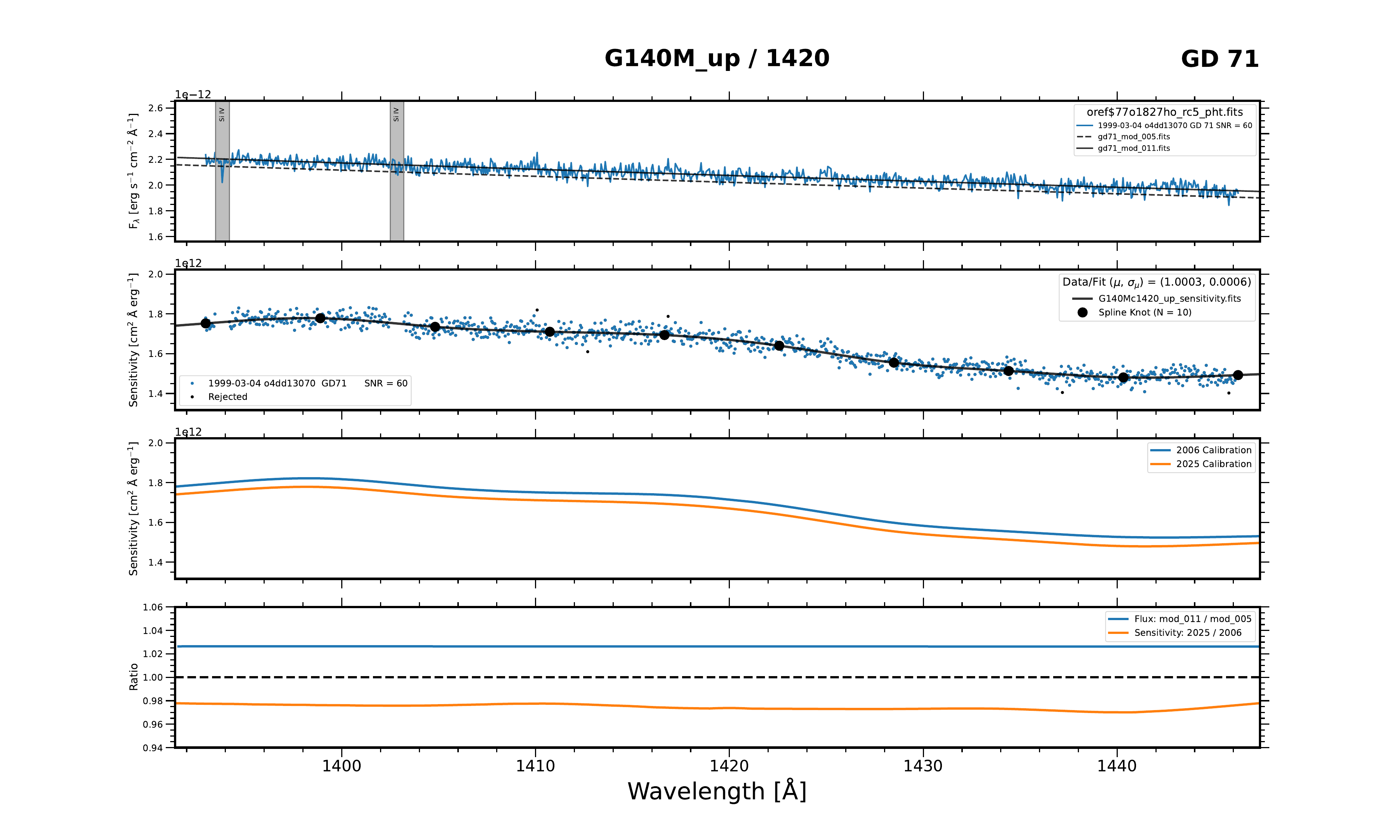}
  \footnotesize
  \caption{Calibration of GD 71 for G140M\_up/1420.}
  \label{fig:G140MupC1420b}
\end{figure}

\begin{figure}[b]
  \hspace{-0.5in}
  \includegraphics[width=1.1\textwidth]{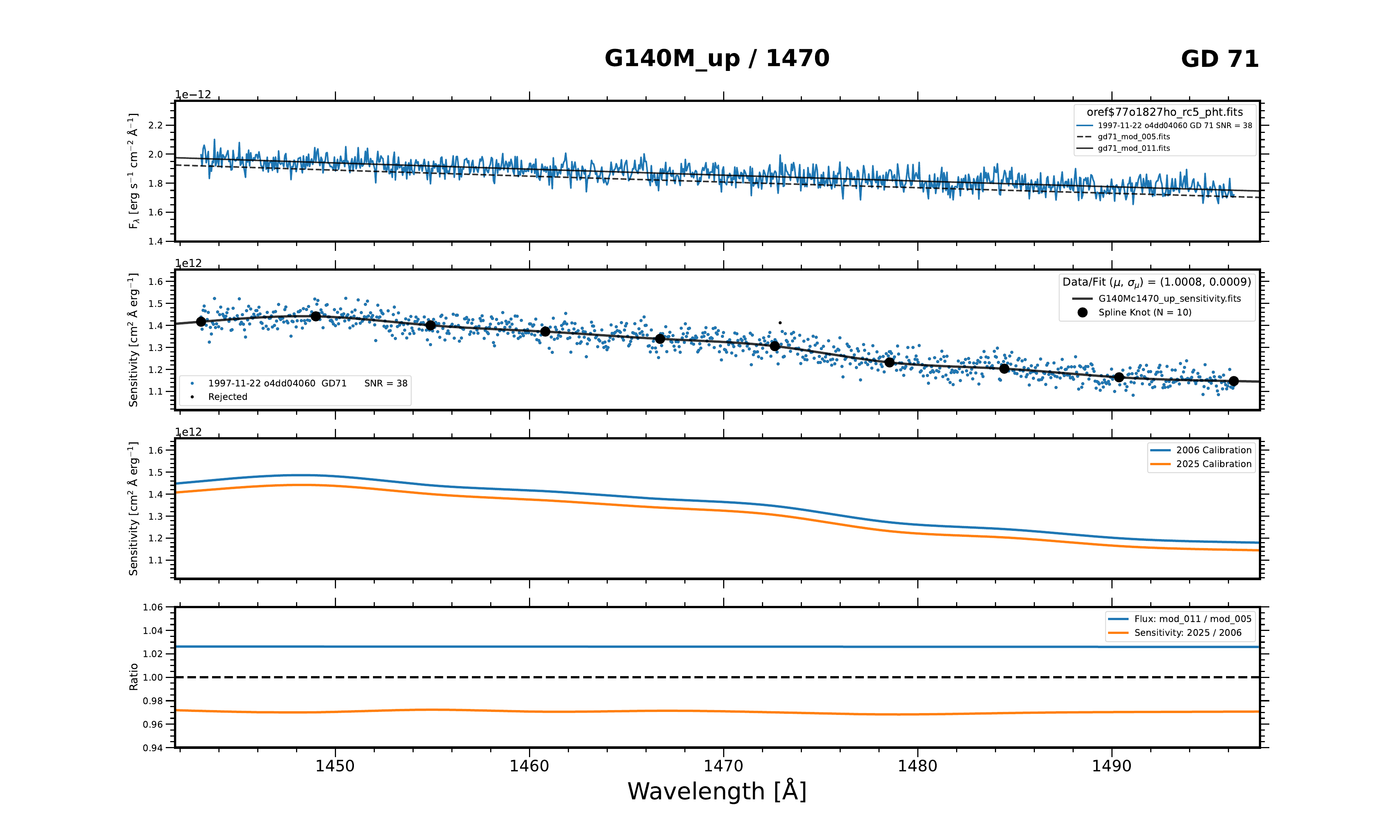}
  \footnotesize
  \caption{Calibration of GD 71 for G140M\_up/1470.}
  \label{fig:G140MupC1470b}
\end{figure}

\begin{figure}[t]
  \hspace{-0.5in}
  \includegraphics[width=1.1\textwidth]{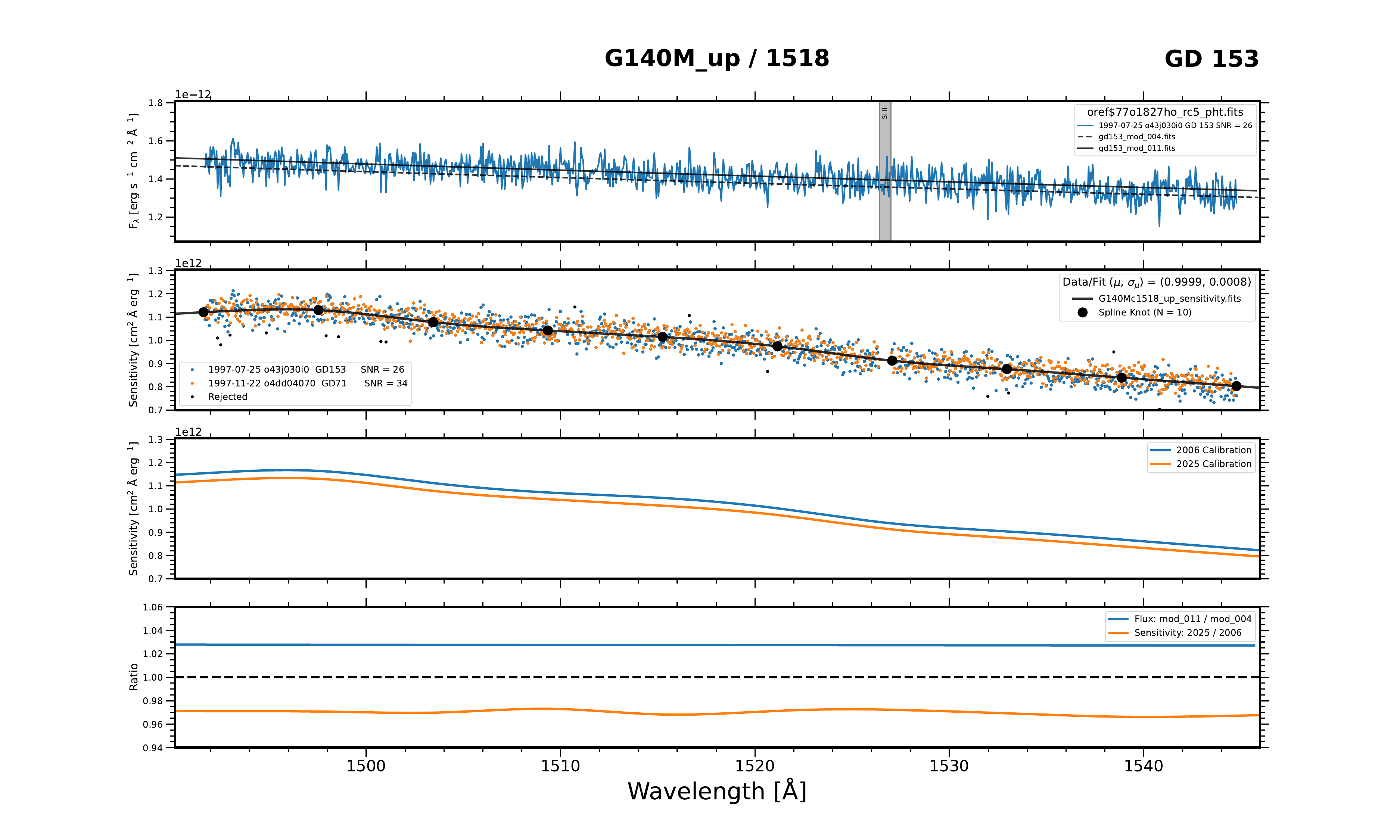}
  \footnotesize
  \caption{Calibration of GD 153 for G140M\_up/1518.}
  \label{fig:G140MupC1518c}
\end{figure}

\begin{figure}[b]
  \hspace{-0.5in}
  \includegraphics[width=1.1\textwidth]{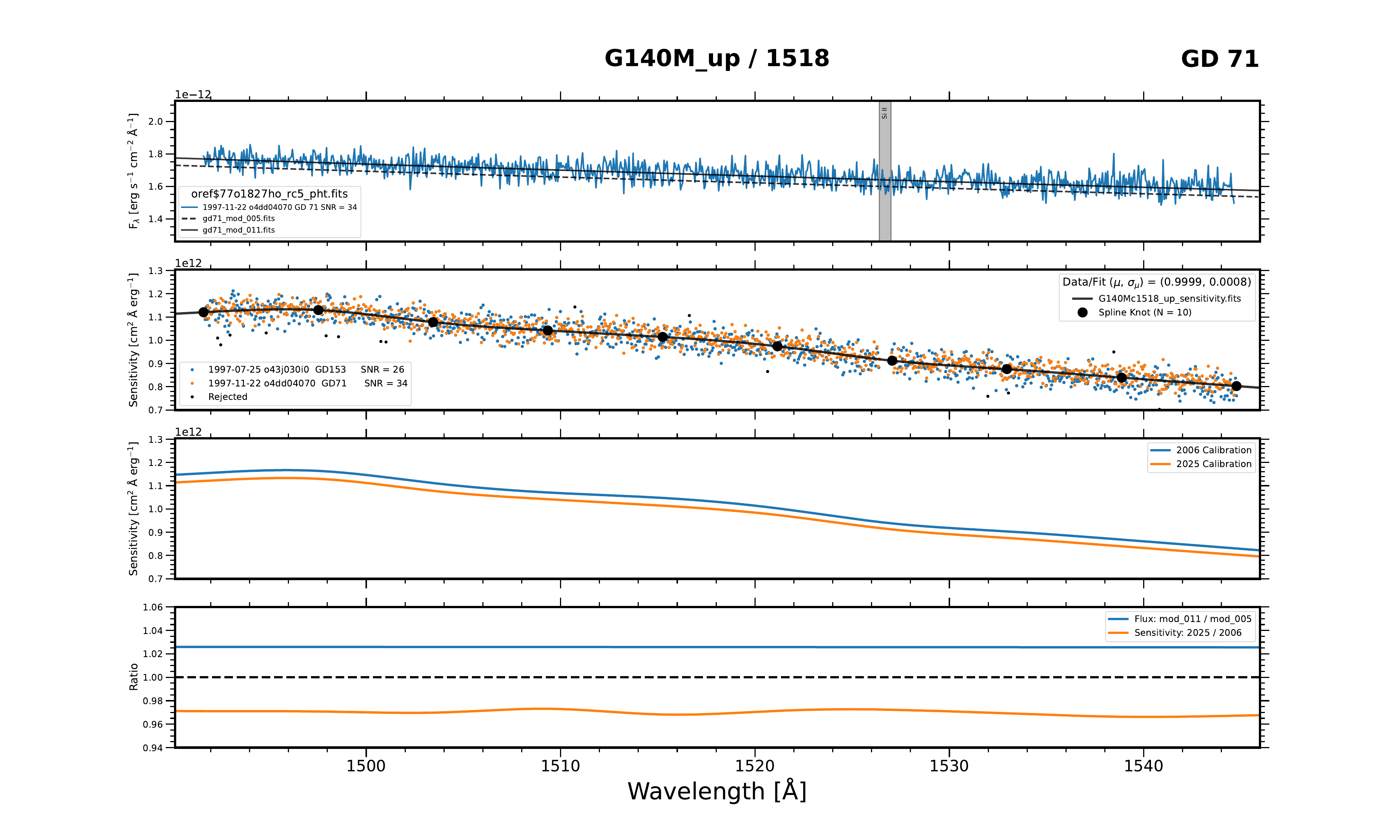}
  \footnotesize
  \caption{Calibration of GD 71 for G140M\_up/1518.}
  \label{fig:G140MupC1518b}
\end{figure}

\begin{figure}[t]
  \hspace{-0.5in}
  \includegraphics[width=1.1\textwidth]{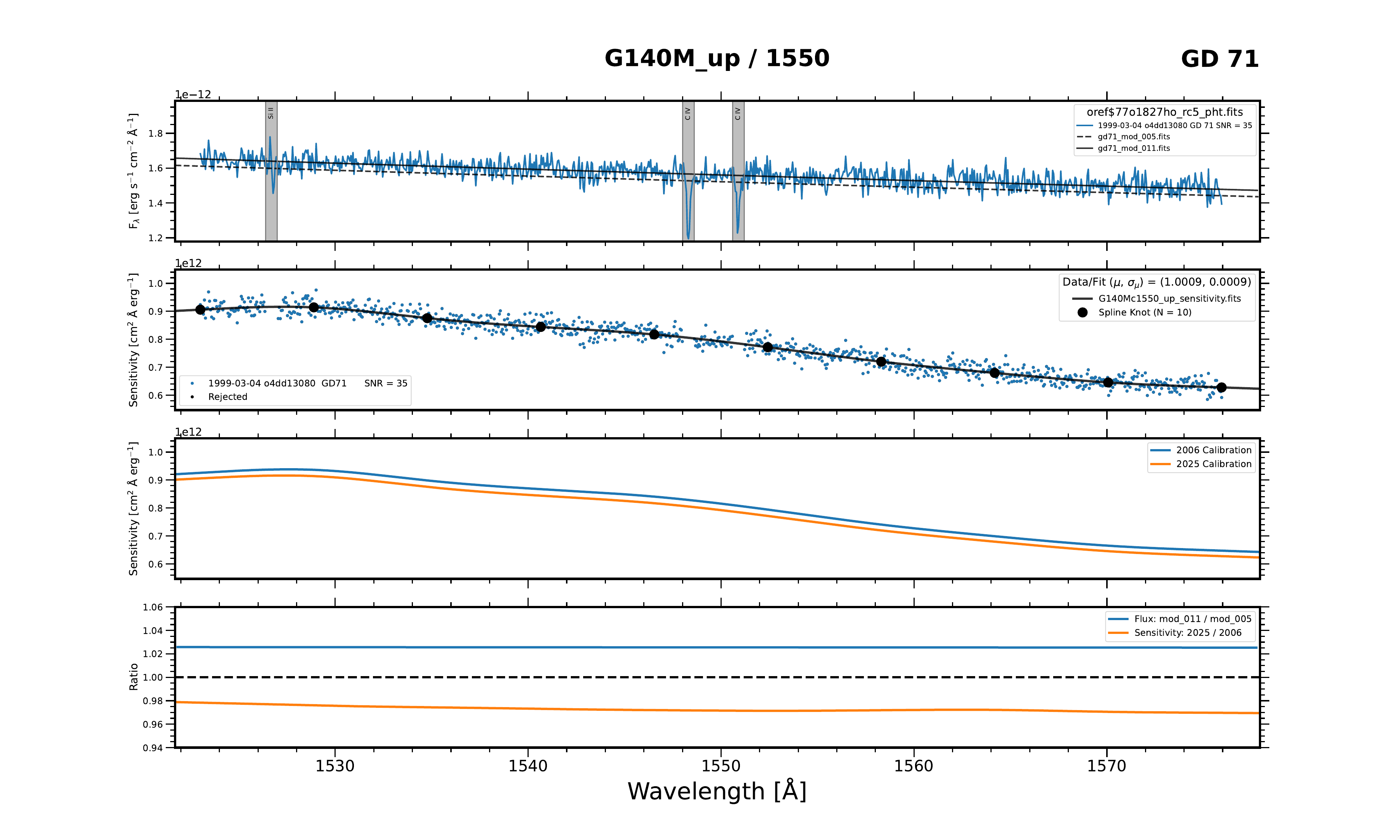}
  \footnotesize
  \caption{Calibration of GD 71 for G140M\_up/1550.}
  \label{fig:G140MupC1550b}
\end{figure}

\begin{figure}[b]
  \hspace{-0.5in}
  \includegraphics[width=1.1\textwidth]{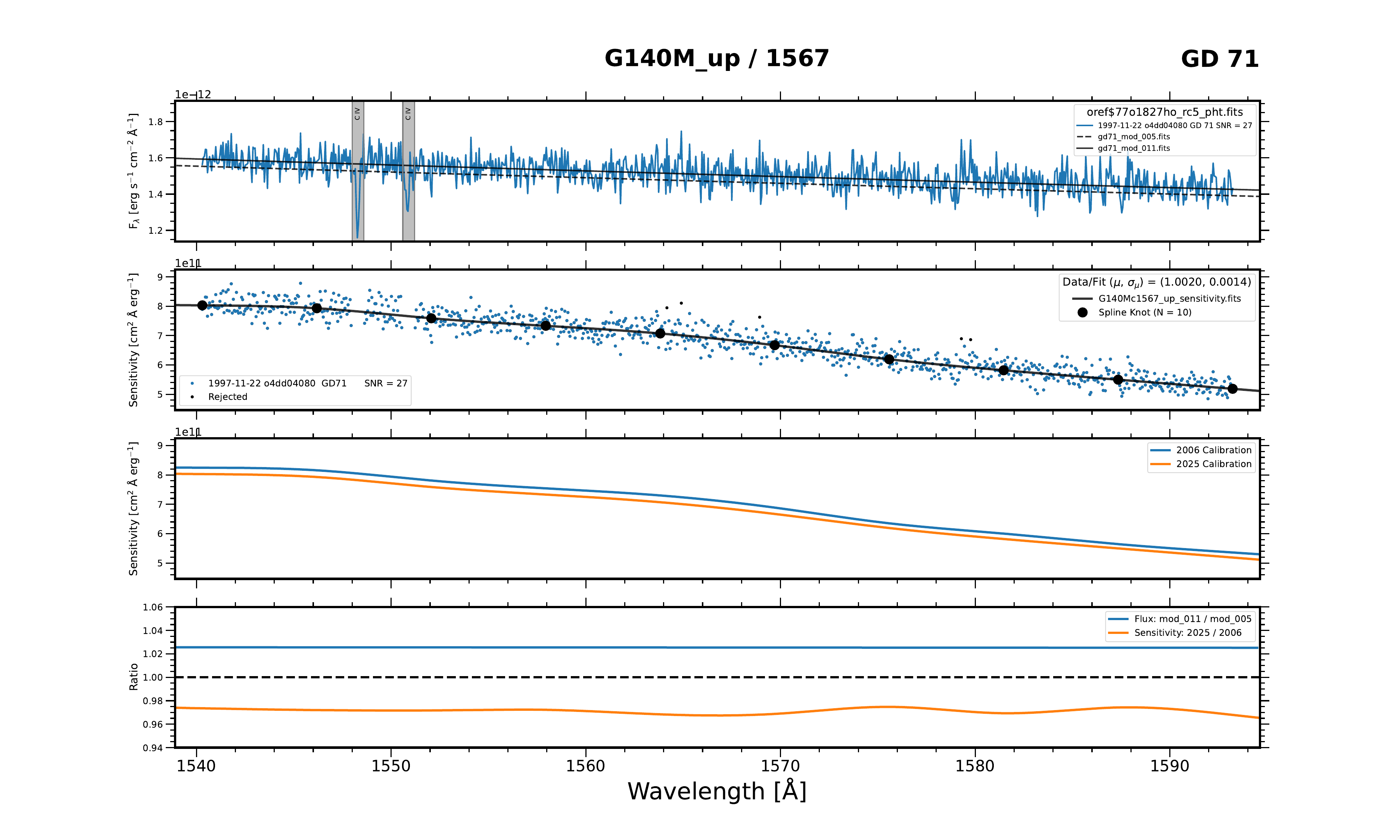}
  \footnotesize
  \caption{Calibration of GD 71 for G140M\_up/1567.}
  \label{fig:G140MupC1567b}
\end{figure}

\begin{figure}[t]
  \hspace{-0.5in}
  \includegraphics[width=1.1\textwidth]{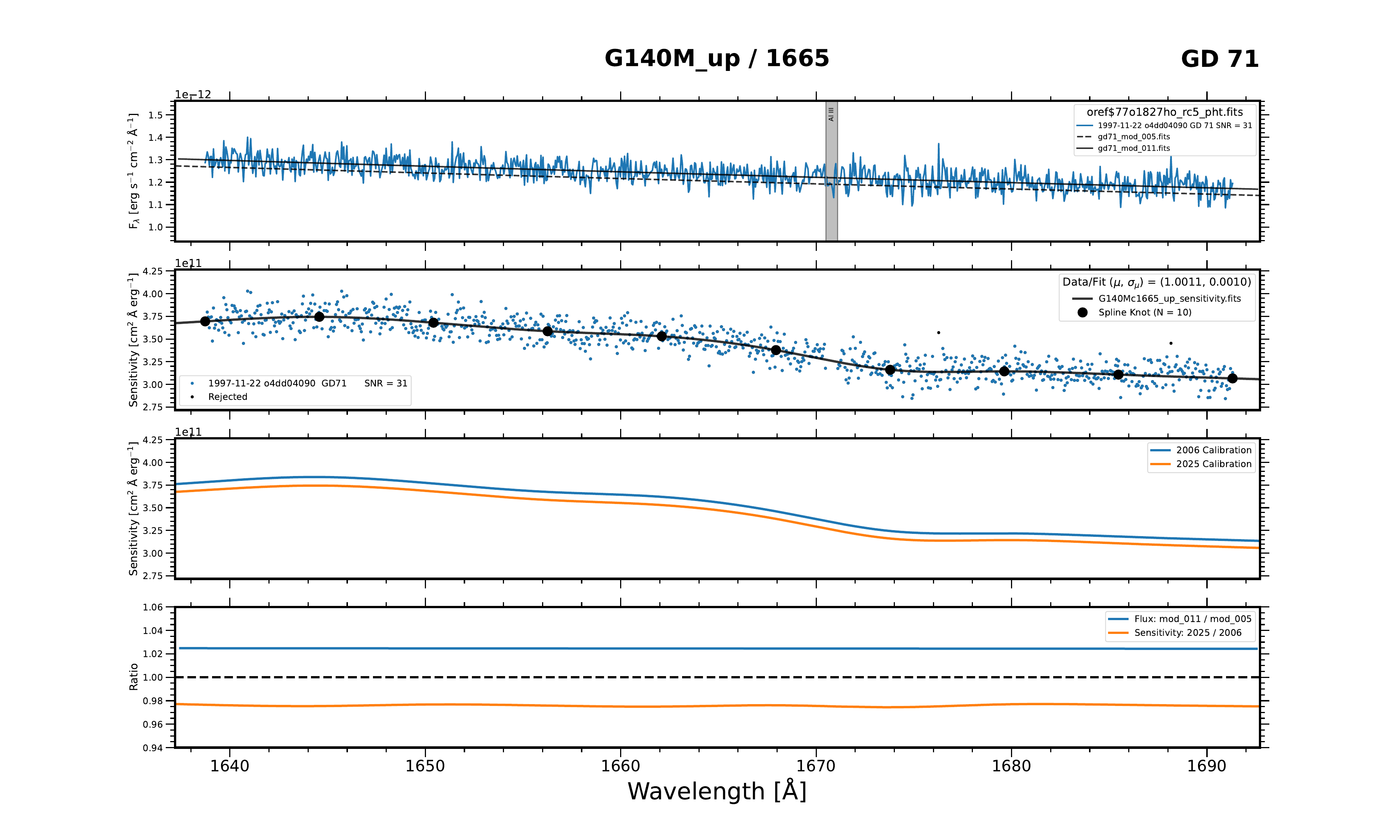}
  \footnotesize
  \caption{Calibration of GD 71 for G140M\_up/1665.}
  \label{fig:G140MupC1665b}
\end{figure}

\begin{figure}[b]
  \hspace{-0.5in}
  \includegraphics[width=1.1\textwidth]{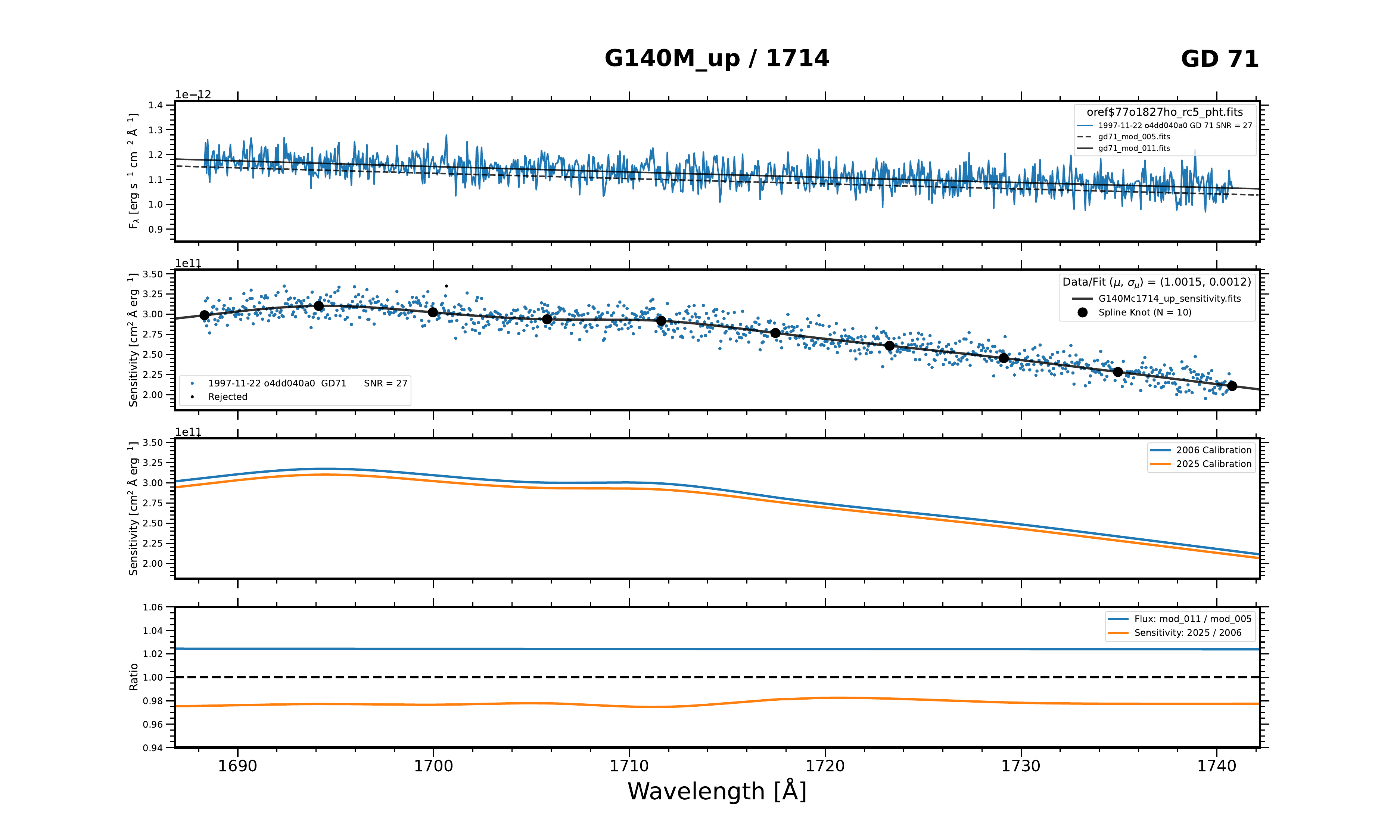}
  \footnotesize
  \caption{Calibration of GD 71 for G140M\_up/1714.}
  \label{fig:G140MupC1714b}
\end{figure}

\begin{figure}[t]
  \hspace{-0.5in}
  \includegraphics[width=1.1\textwidth]{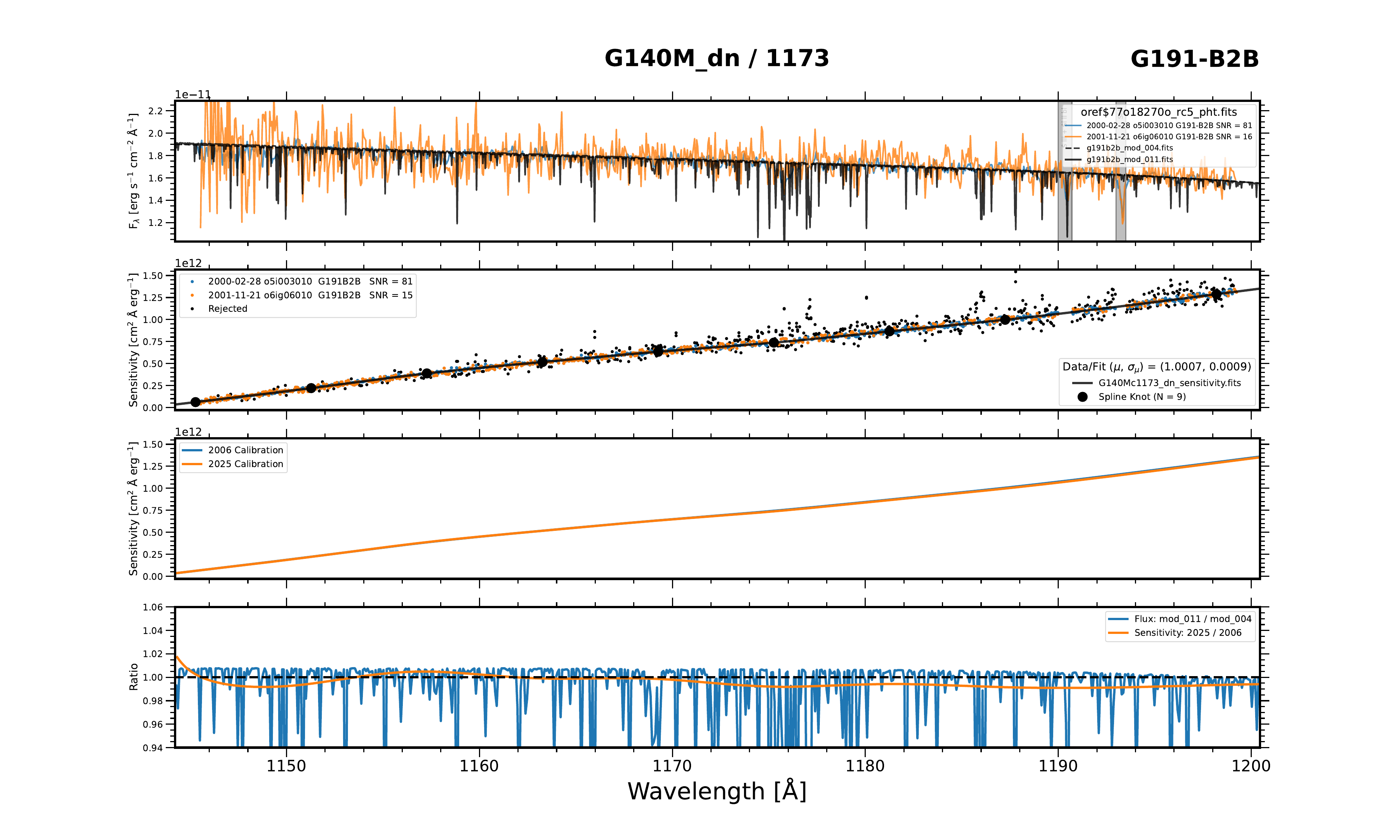}
  \footnotesize
  \caption{Calibration of G191-B2B for G140M\_dn/1173.}
  \label{fig:G140MdnC1173a}
\end{figure}

\begin{figure}[b]
  \hspace{-0.5in}
  \includegraphics[width=1.1\textwidth]{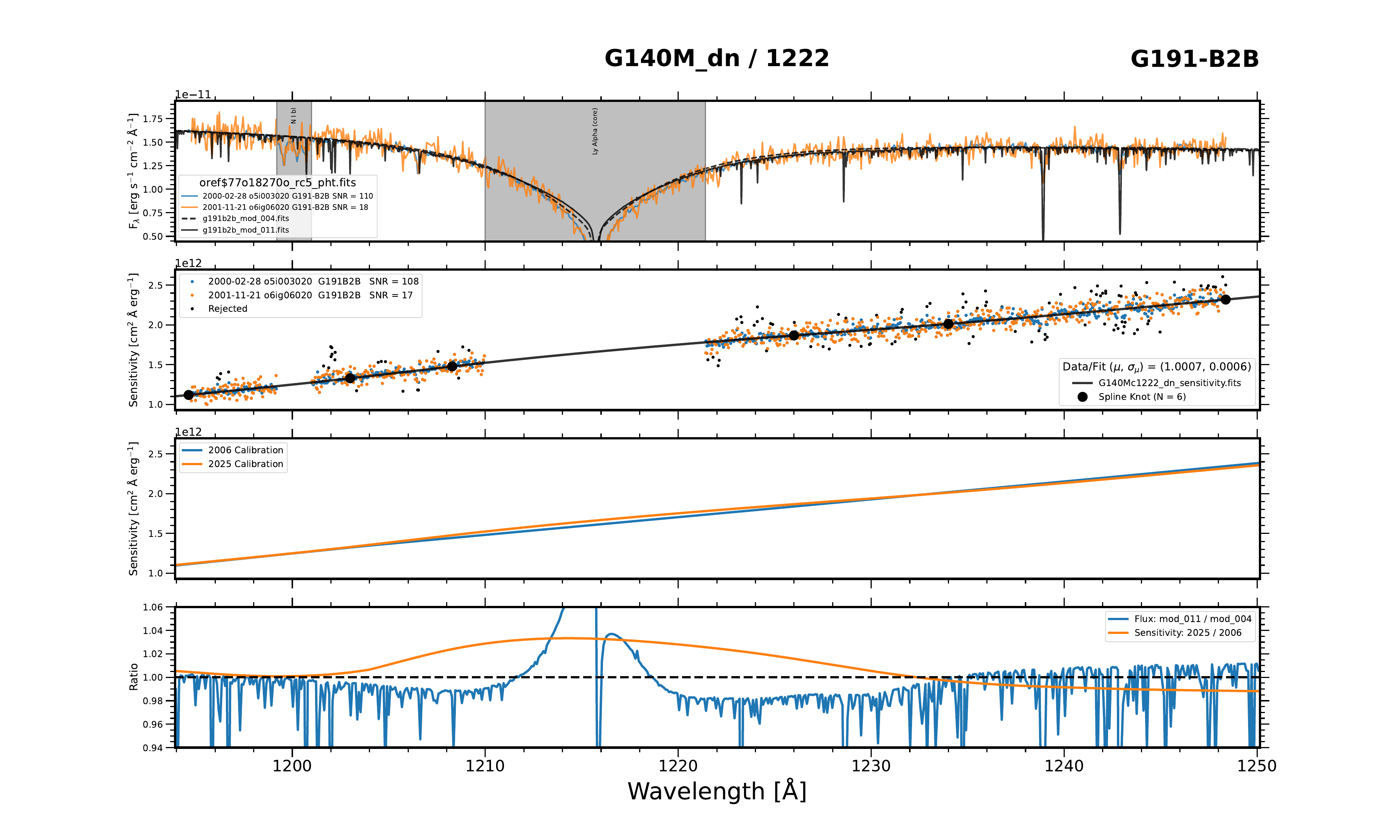}
  \footnotesize
  \caption{Calibration of G191-B2B for G140M\_dn/1222.}
  \label{fig:G140MdnC1222a}
\end{figure}

\begin{figure}[t]
  \hspace{-0.5in}
  \includegraphics[width=1.1\textwidth]{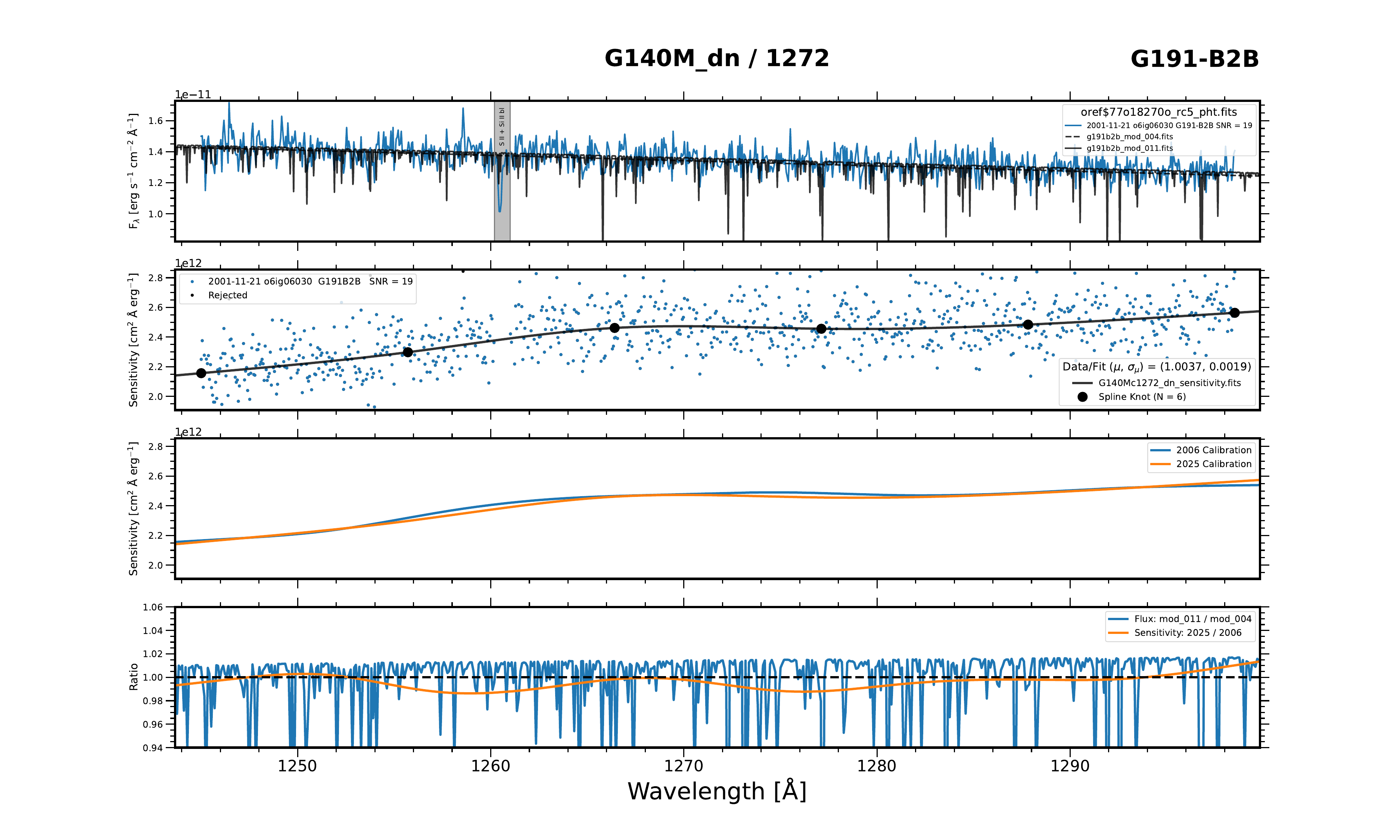}
  \footnotesize
  \caption{Calibration of G191-B2B for G140M\_dn/1272.}
  \label{fig:G140MdnC1272a}
\end{figure}

\begin{figure}[b]
  \hspace{-0.5in}
  \includegraphics[width=1.1\textwidth]{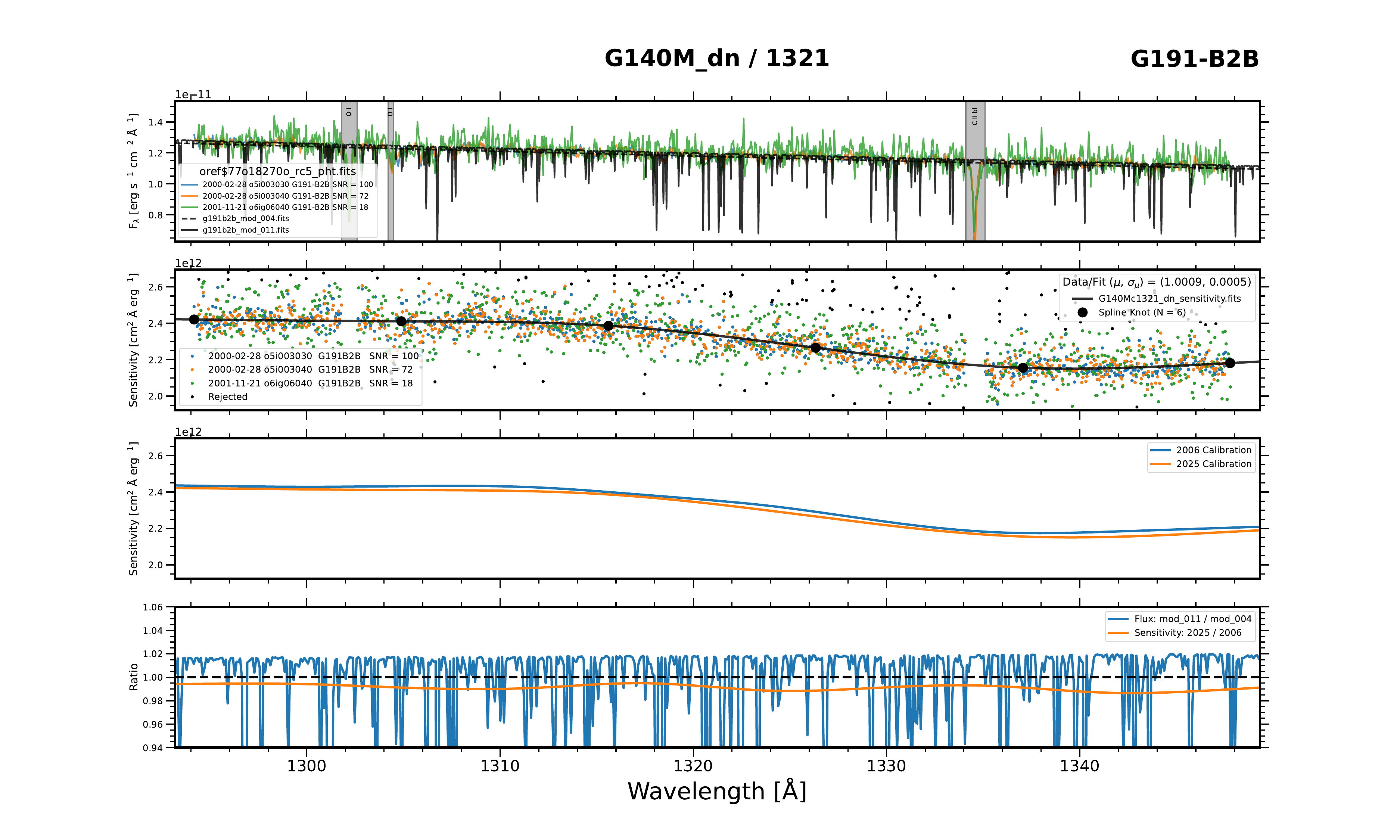}
  \footnotesize
  \caption{Calibration of G191-B2B for G140M\_dn/1321.}
  \label{fig:G140MdnC1321a}
\end{figure}

\begin{figure}[t]
  \hspace{-0.5in}
  \includegraphics[width=1.1\textwidth]{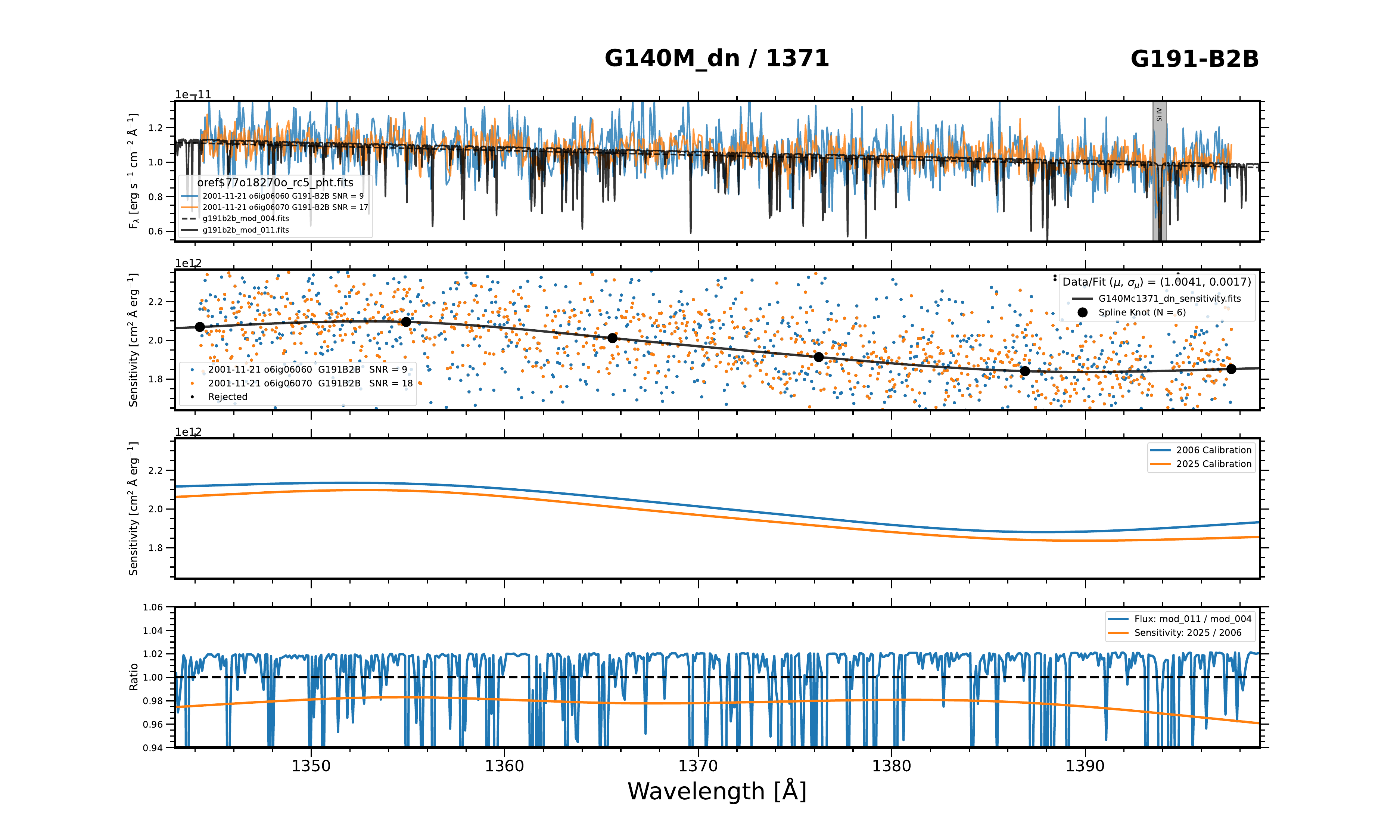}
  \footnotesize
  \caption{Calibration of G191-B2B for G140M\_dn/1371.}
  \label{fig:G140MdnC1371a}
\end{figure}

\begin{figure}[b]
  \hspace{-0.5in}
  \includegraphics[width=1.1\textwidth]{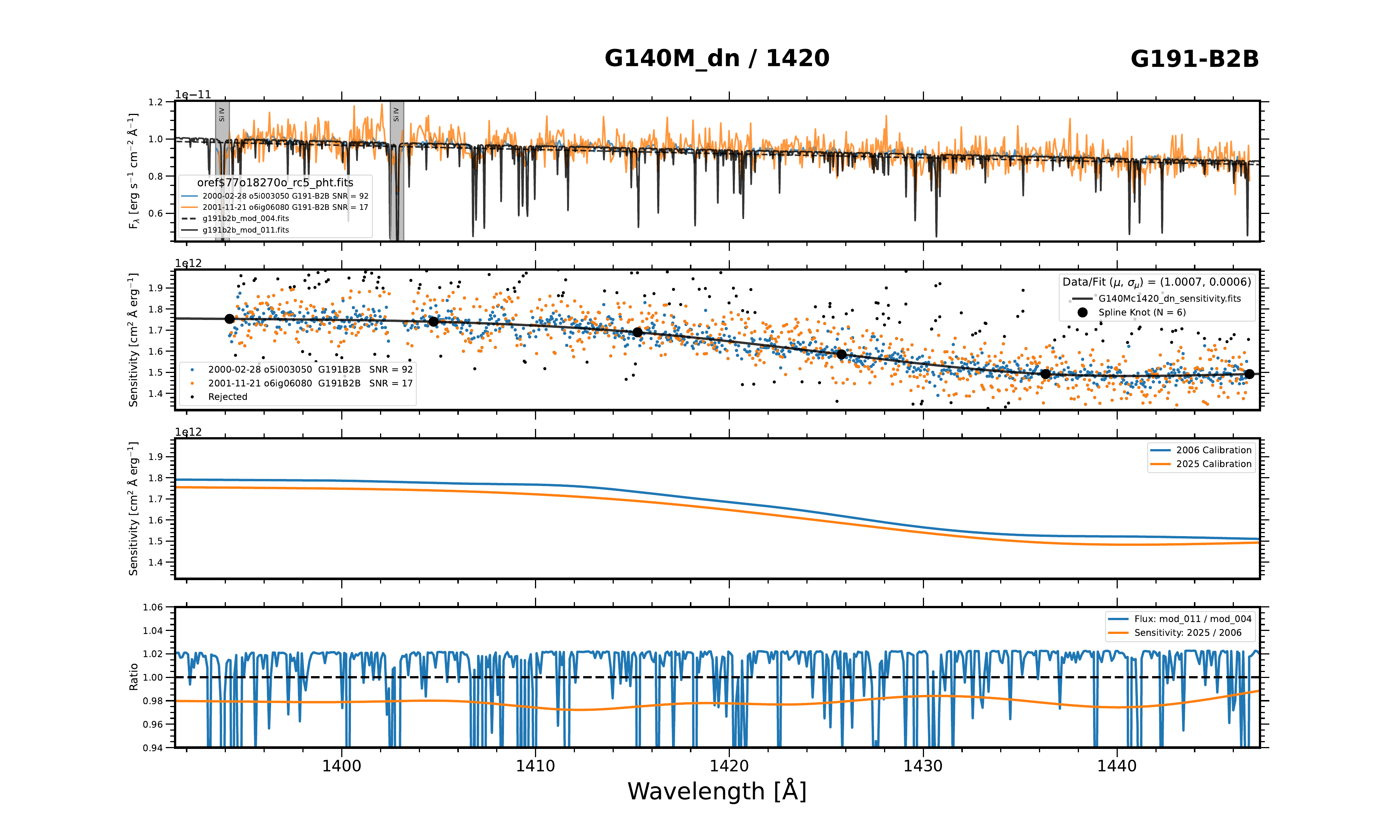}
  \footnotesize
  \caption{Calibration of G191-B2B for G140M\_dn/1420.}
  \label{fig:G140MdnC1420a}
\end{figure}

\begin{figure}[t]
  \hspace{-0.5in}
  \includegraphics[width=1.1\textwidth]{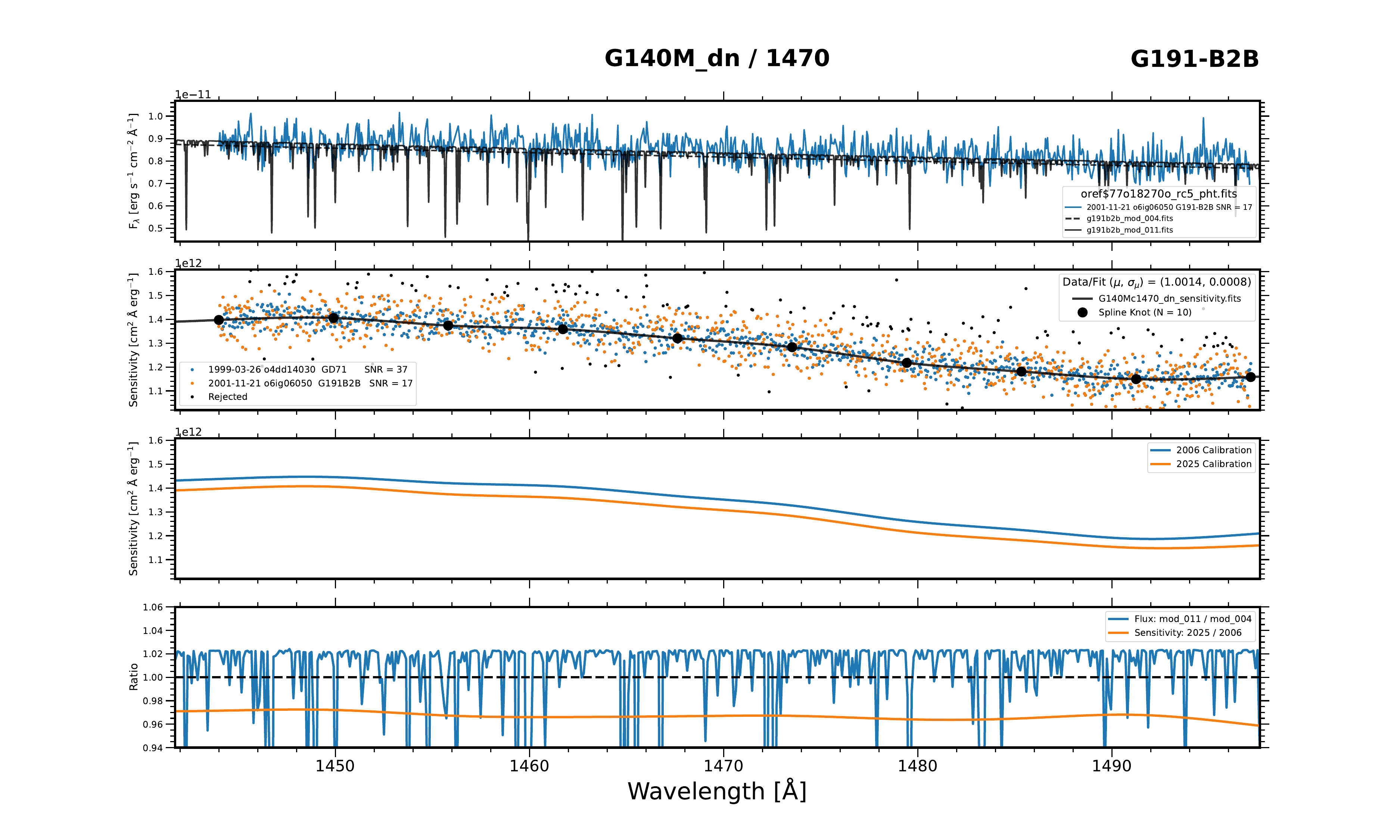}
  \footnotesize
  \caption{Calibration of G191-B2B for G140M\_dn/1470.}
  \label{fig:G140MdnC1470a}
\end{figure}

\begin{figure}[b]
  \hspace{-0.5in}
  \includegraphics[width=1.1\textwidth]{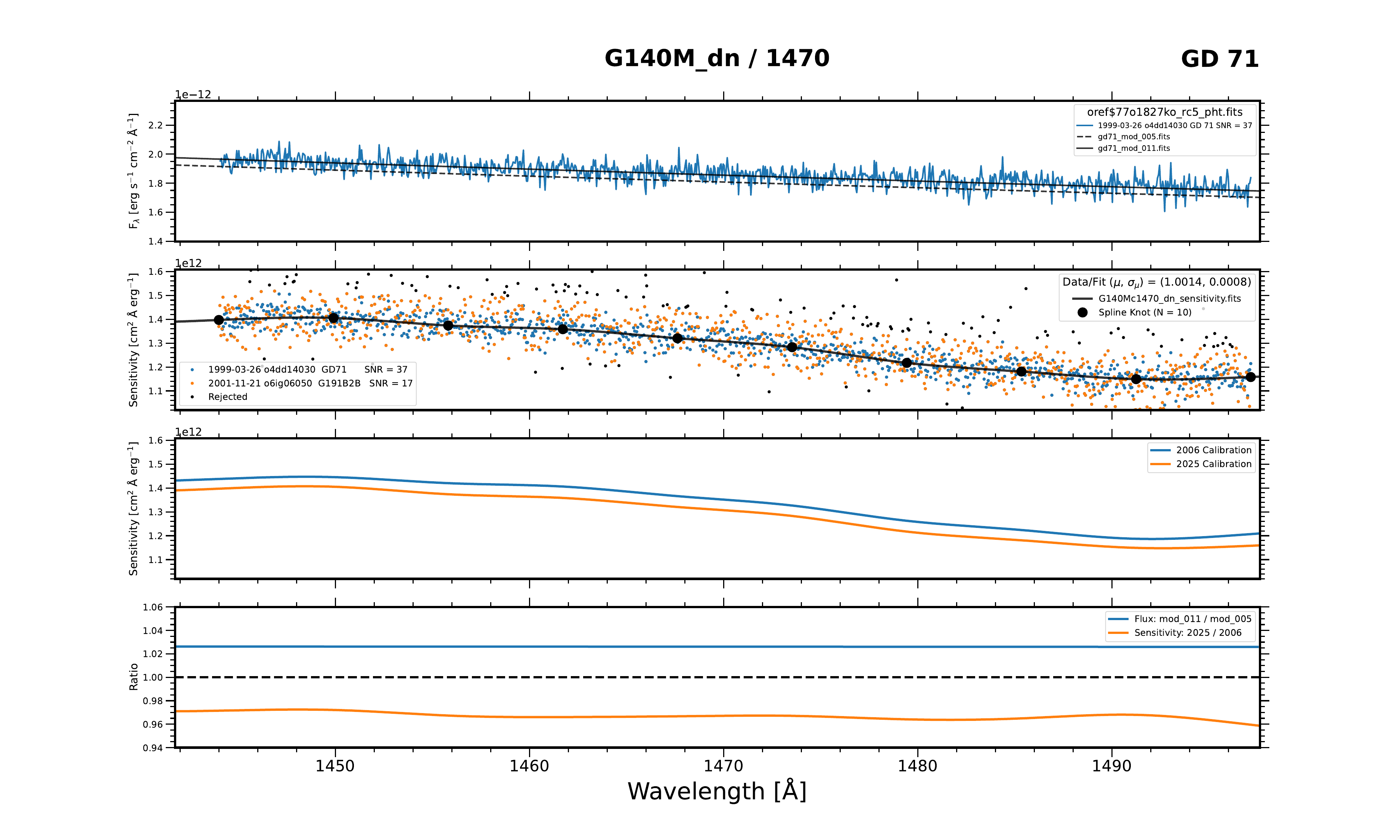}
  \footnotesize
  \caption{Calibration of GD 71 for G140M\_dn/1470.}
  \label{fig:G140MdnC1470b}
\end{figure}

\begin{figure}[t]
  \hspace{-0.5in}
  \includegraphics[width=1.1\textwidth]{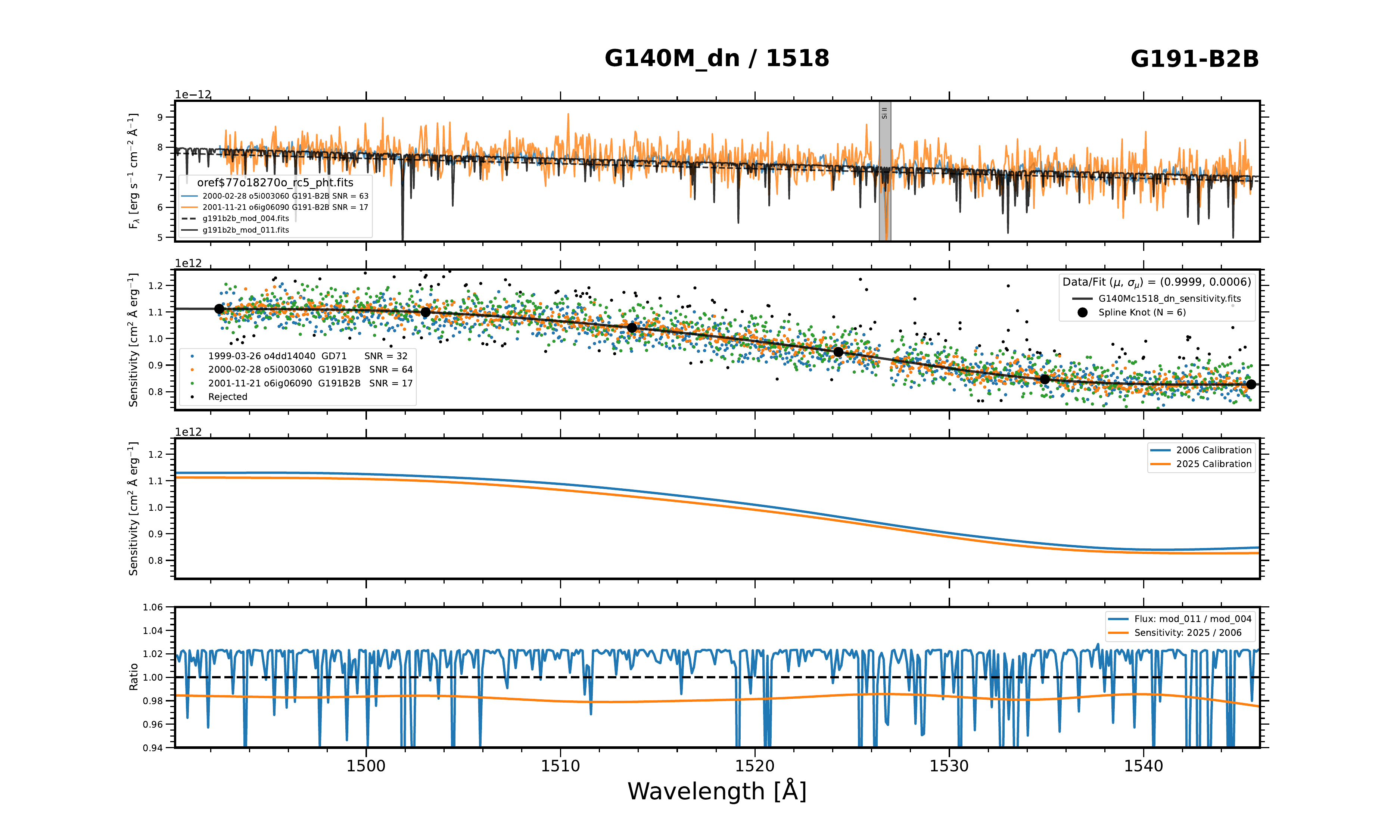}
  \footnotesize
  \caption{Calibration of G191-B2B for G140M\_dn/1518.}
  \label{fig:G140MdnC1518a}
\end{figure}

\begin{figure}[b]
  \hspace{-0.5in}
  \includegraphics[width=1.1\textwidth]{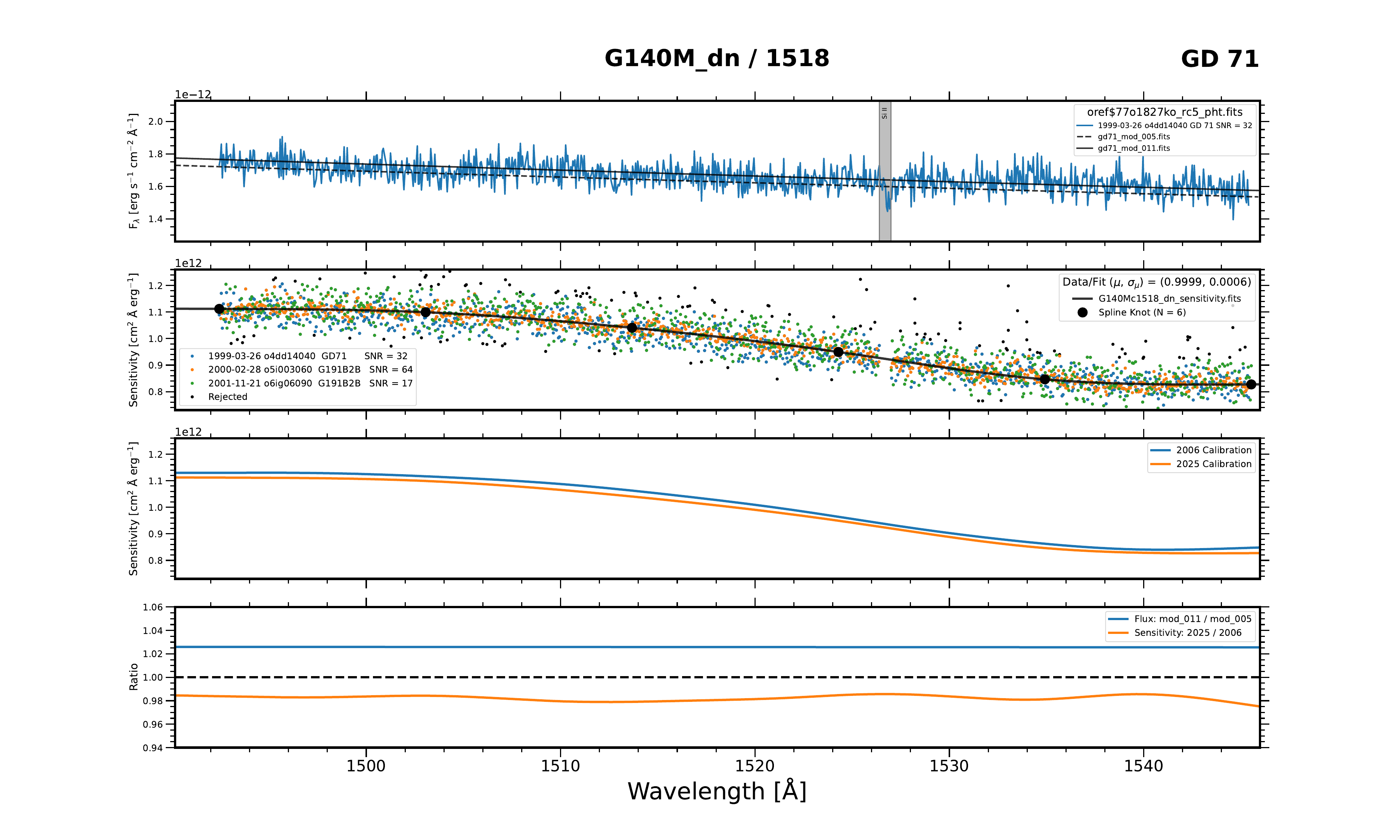}
  \footnotesize
  \caption{Calibration of GD 71 for G140M\_dn/1518.}
  \label{fig:G140MdnC1518b}
\end{figure}

\begin{figure}[t]
  \hspace{-0.5in}
  \includegraphics[width=1.1\textwidth]{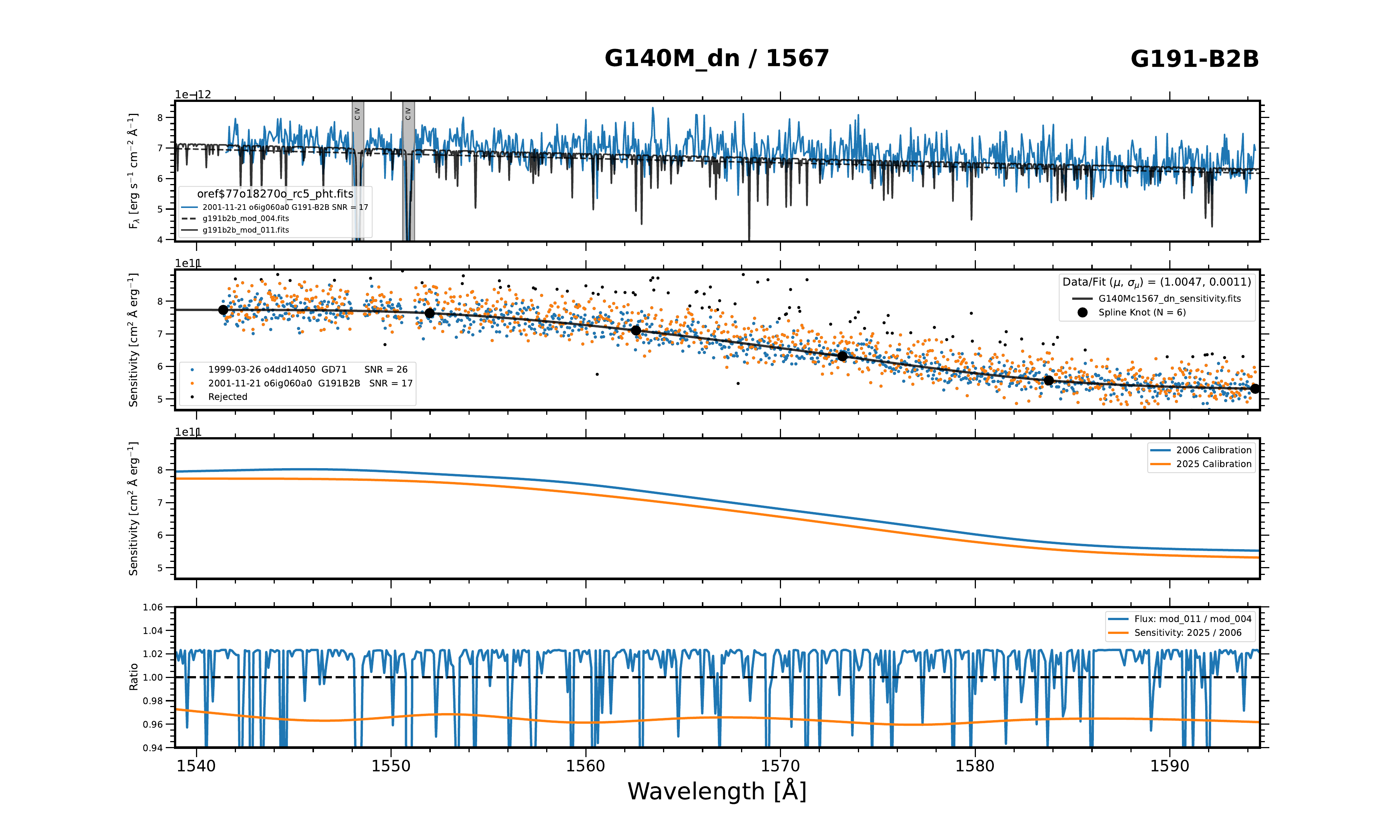}
  \footnotesize
  \caption{Calibration of G191-B2B for G140M\_dn/1567.}
  \label{fig:G140MdnC1567a}
\end{figure}

\begin{figure}[b]
  \hspace{-0.5in}
  \includegraphics[width=1.1\textwidth]{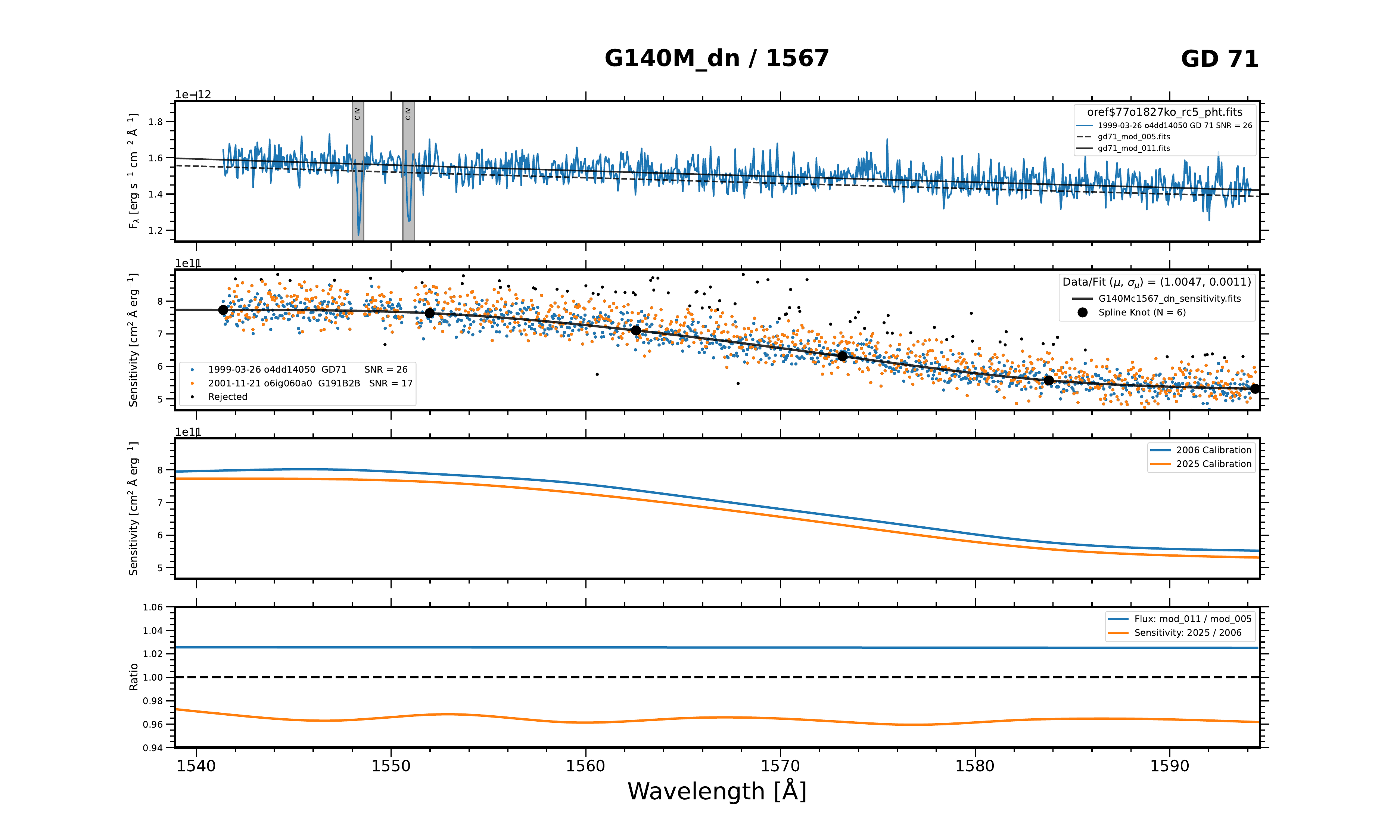}
  \footnotesize
  \caption{Calibration of GD 71 for G140M\_dn/1567.}
  \label{fig:G140MdnC1567b}
\end{figure}

\begin{figure}[t]
  \hspace{-0.5in}
  \includegraphics[width=1.1\textwidth]{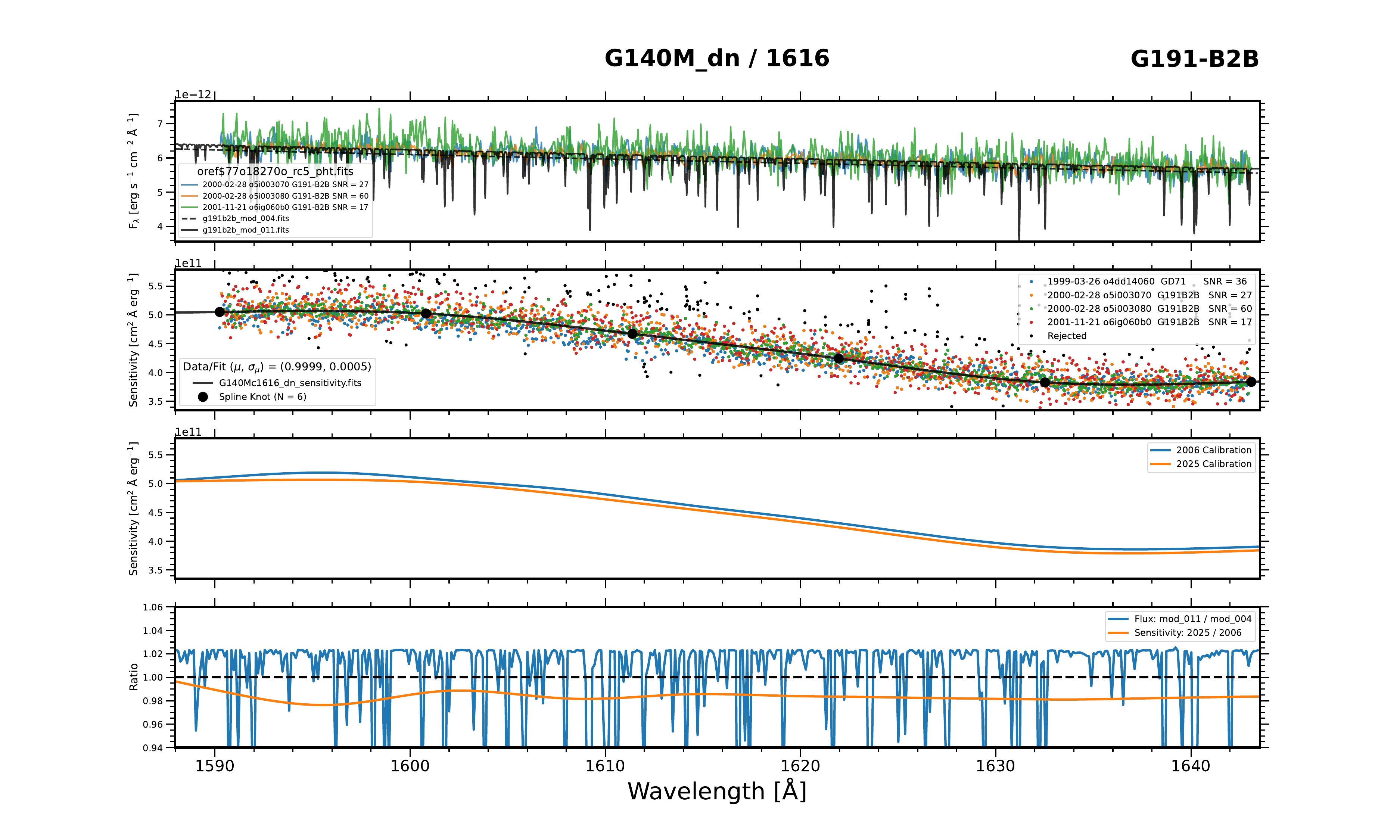}
  \footnotesize
  \caption{Calibration of G191-B2B for G140M\_dn/1616.}
  \label{fig:G140MdnC1616a}
\end{figure}

\begin{figure}[b]
  \hspace{-0.5in}
  \includegraphics[width=1.1\textwidth]{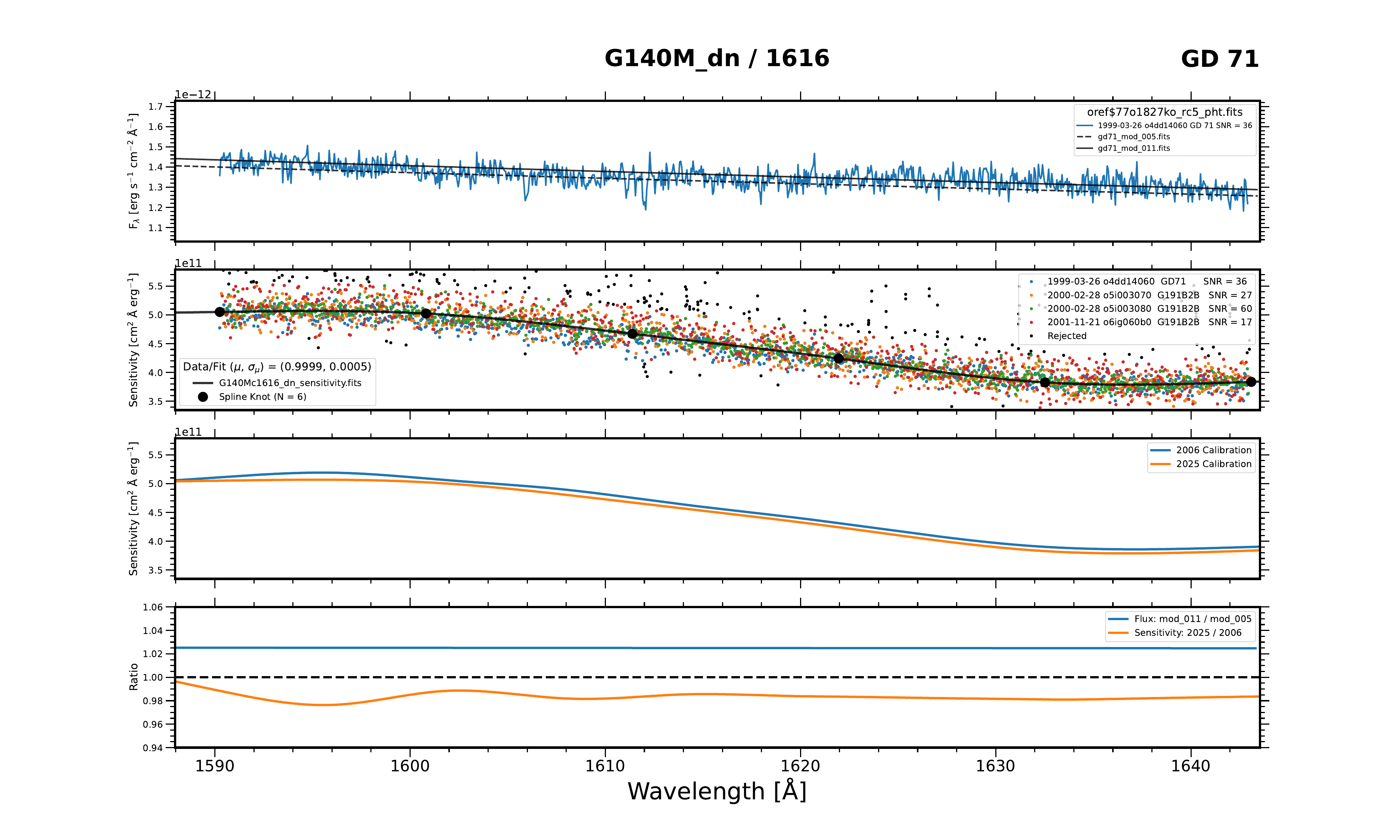}
  \footnotesize
  \caption{Calibration of GD 71 for G140M\_dn/1616.}
  \label{fig:G140MdnC1616b}
\end{figure}

\begin{figure}[t]
  \hspace{-0.5in}
  \includegraphics[width=1.1\textwidth]{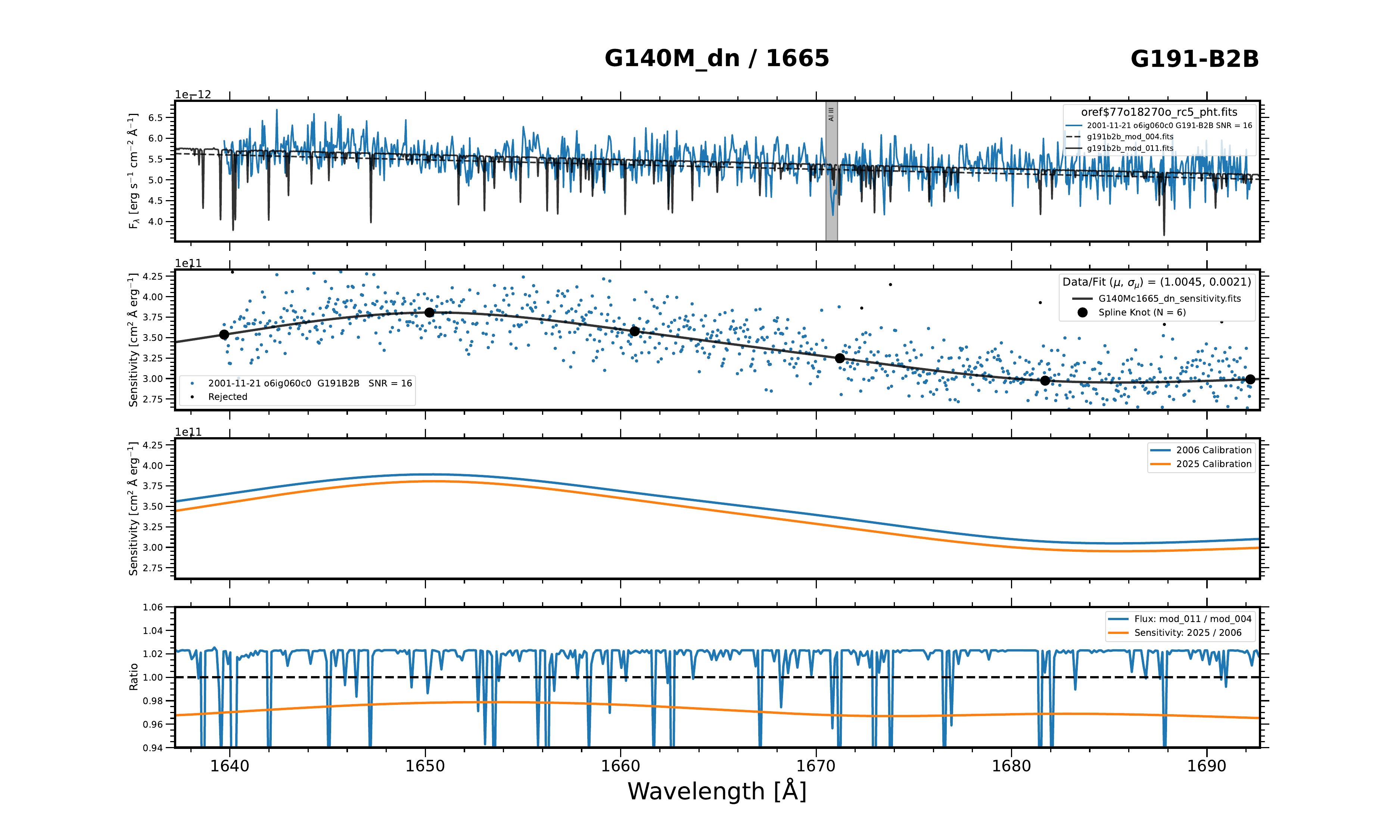}
  \footnotesize
  \caption{Calibration of G191-B2B for G140M\_dn/1665.}
  \label{fig:G140MdnC1665a}
\end{figure}

\begin{figure}[b]
  \hspace{-0.5in}
  \includegraphics[width=1.1\textwidth]{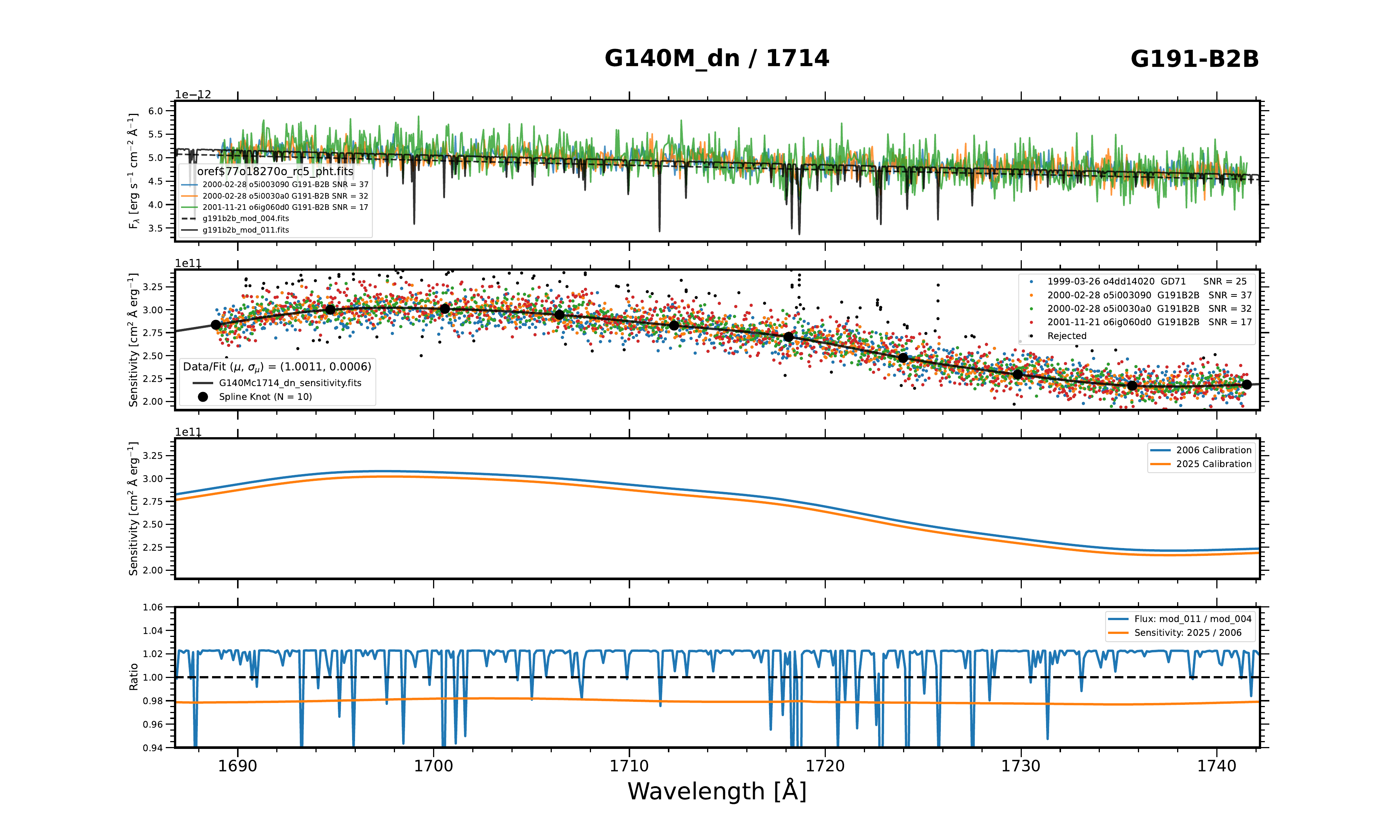}
  \footnotesize
  \caption{Calibration of G191-B2B for G140M\_dn/1714.}
  \label{fig:G140MdnC1714a}
\end{figure}

\begin{figure}[t]
  \hspace{-0.5in}
  \includegraphics[width=1.1\textwidth]{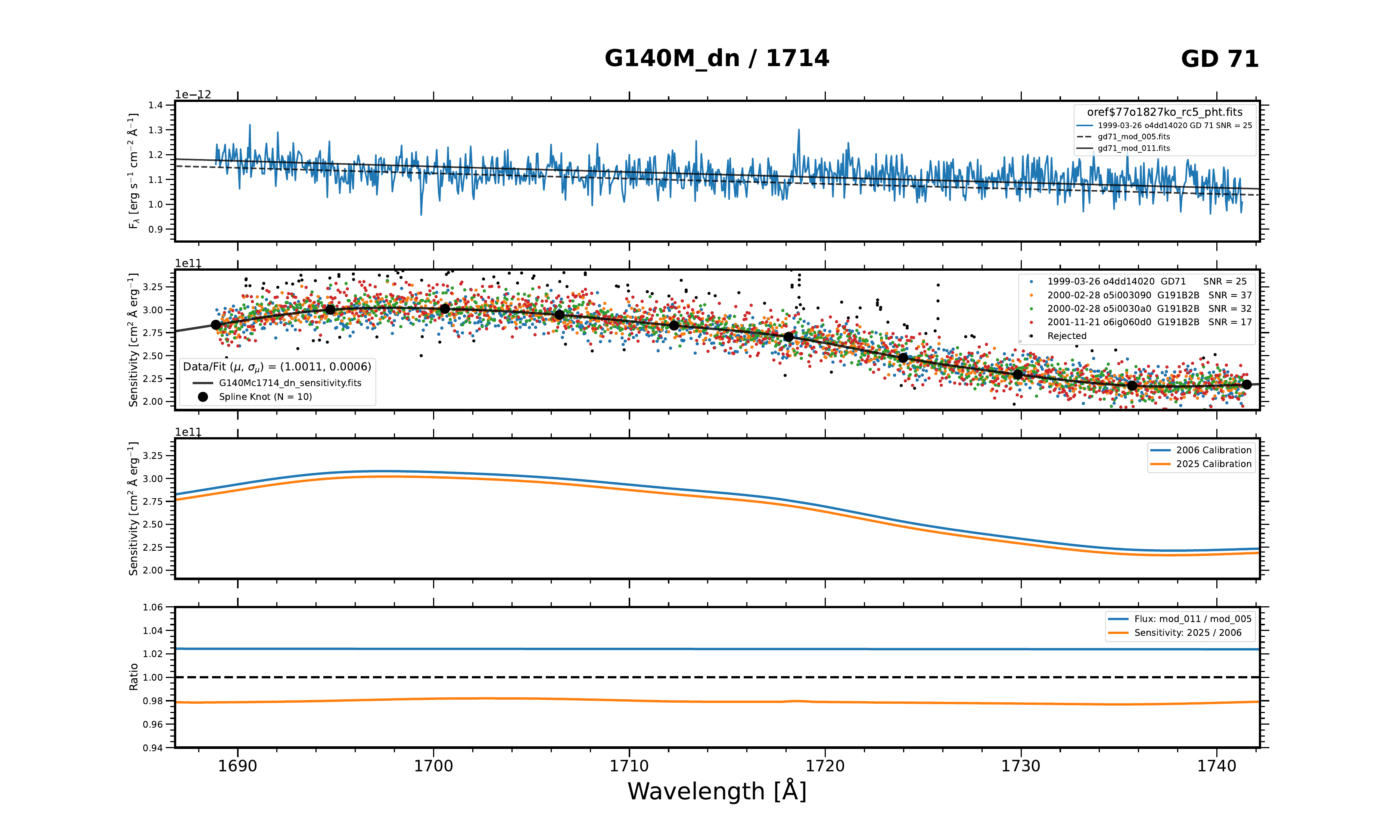}
  \footnotesize
  \caption{Calibration of GD 71 for G140M\_dn/1714.}
  \label{fig:G140MdnC1714b}
\end{figure}

\begin{figure}[t]
  \hspace{-0.5in}
  \includegraphics[width=1.1\textwidth]{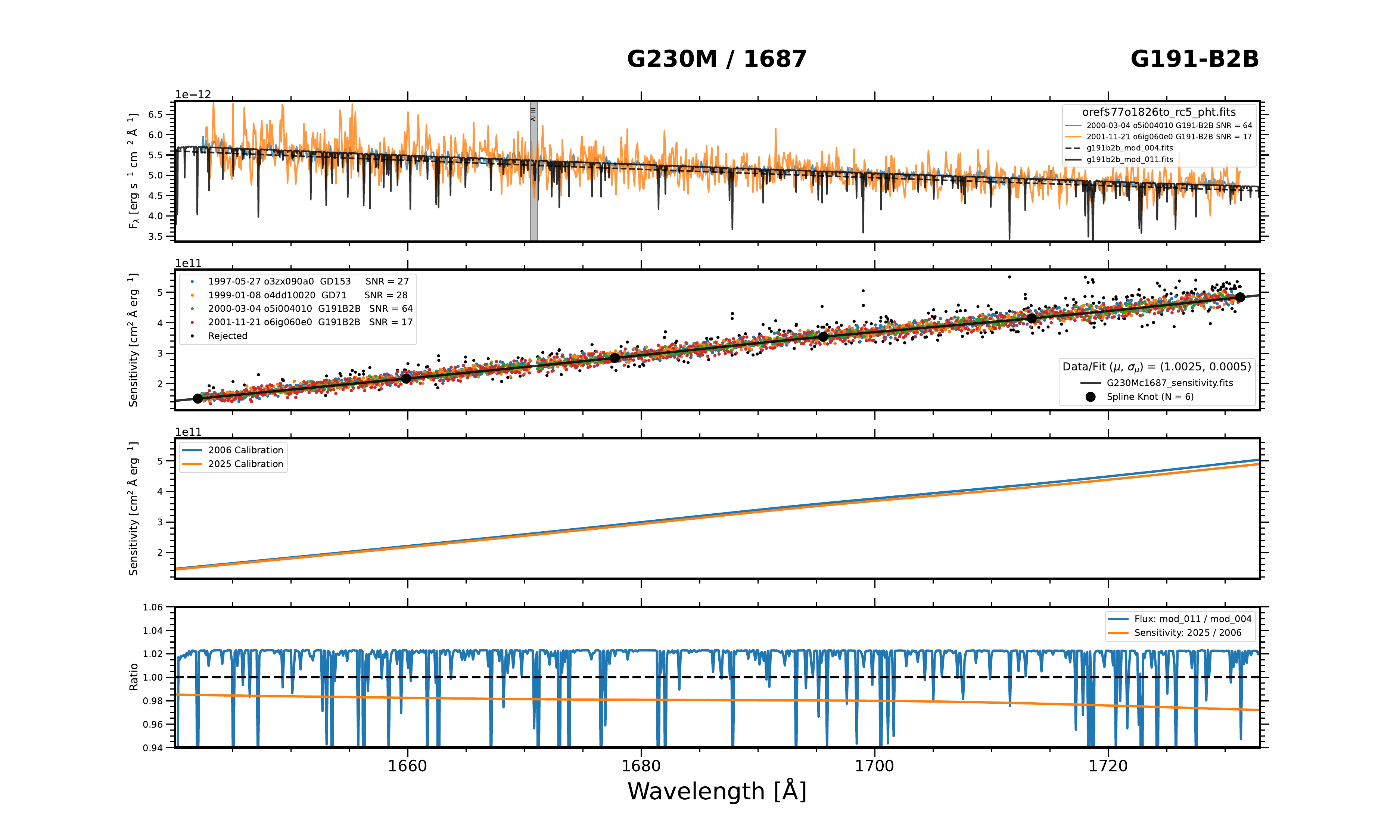}
  \footnotesize
  \caption{Calibration of G191-B2B for G230M/1687.}
  \label{fig:G230MC1687a}
\end{figure}

\begin{figure}[b]
  \hspace{-0.5in}
  \includegraphics[width=1.1\textwidth]{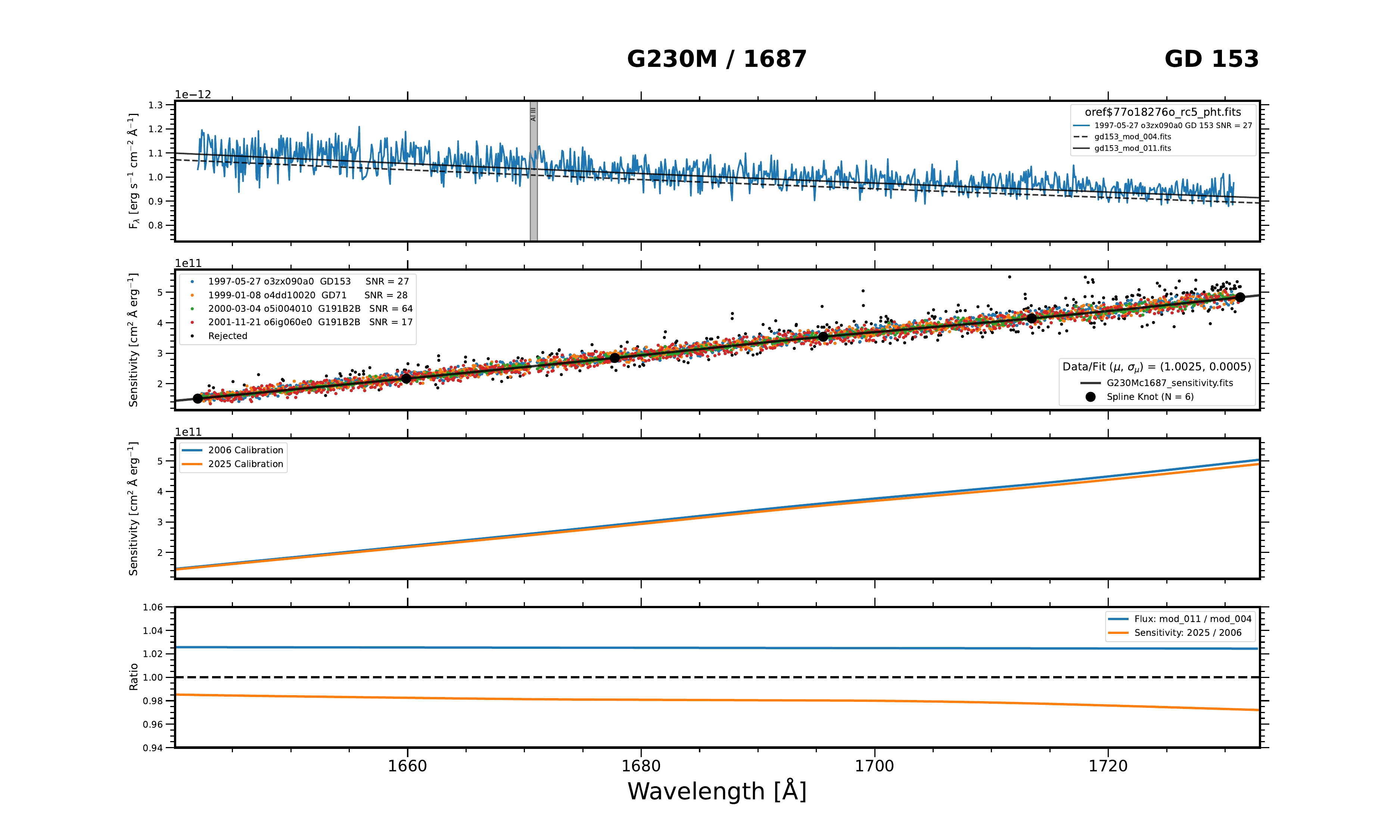}
  \footnotesize
  \caption{Calibration of GD 153 for G230M/1687.}
  \label{fig:G230MC1687c}
\end{figure}

\begin{figure}[t]
  \hspace{-0.5in}
  \includegraphics[width=1.1\textwidth]{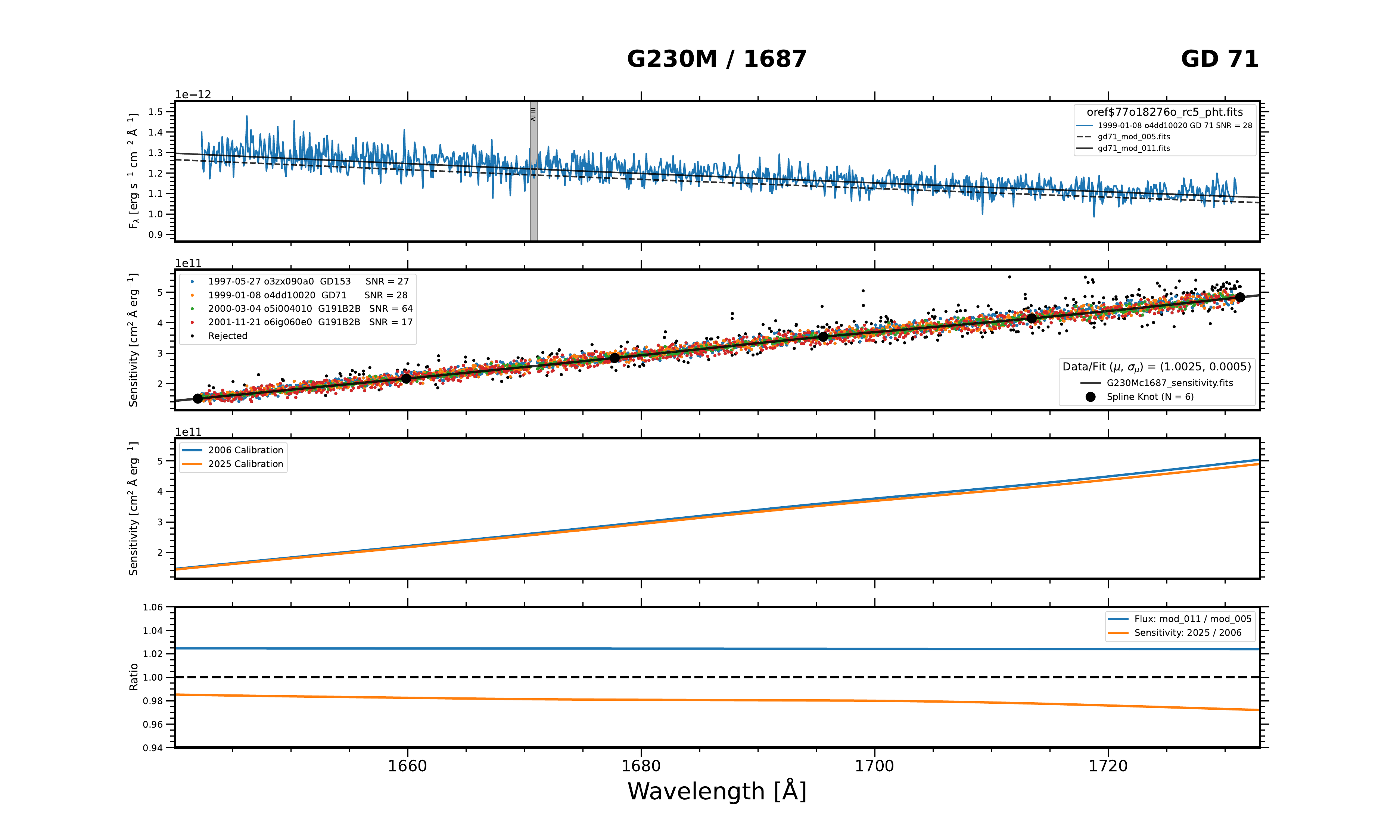}
  \footnotesize
  \caption{Calibration of GD 71 for G230M/1687.}
  \label{fig:G230MC1687b}
\end{figure}

\begin{figure}[b]
  \hspace{-0.5in}
  \includegraphics[width=1.1\textwidth]{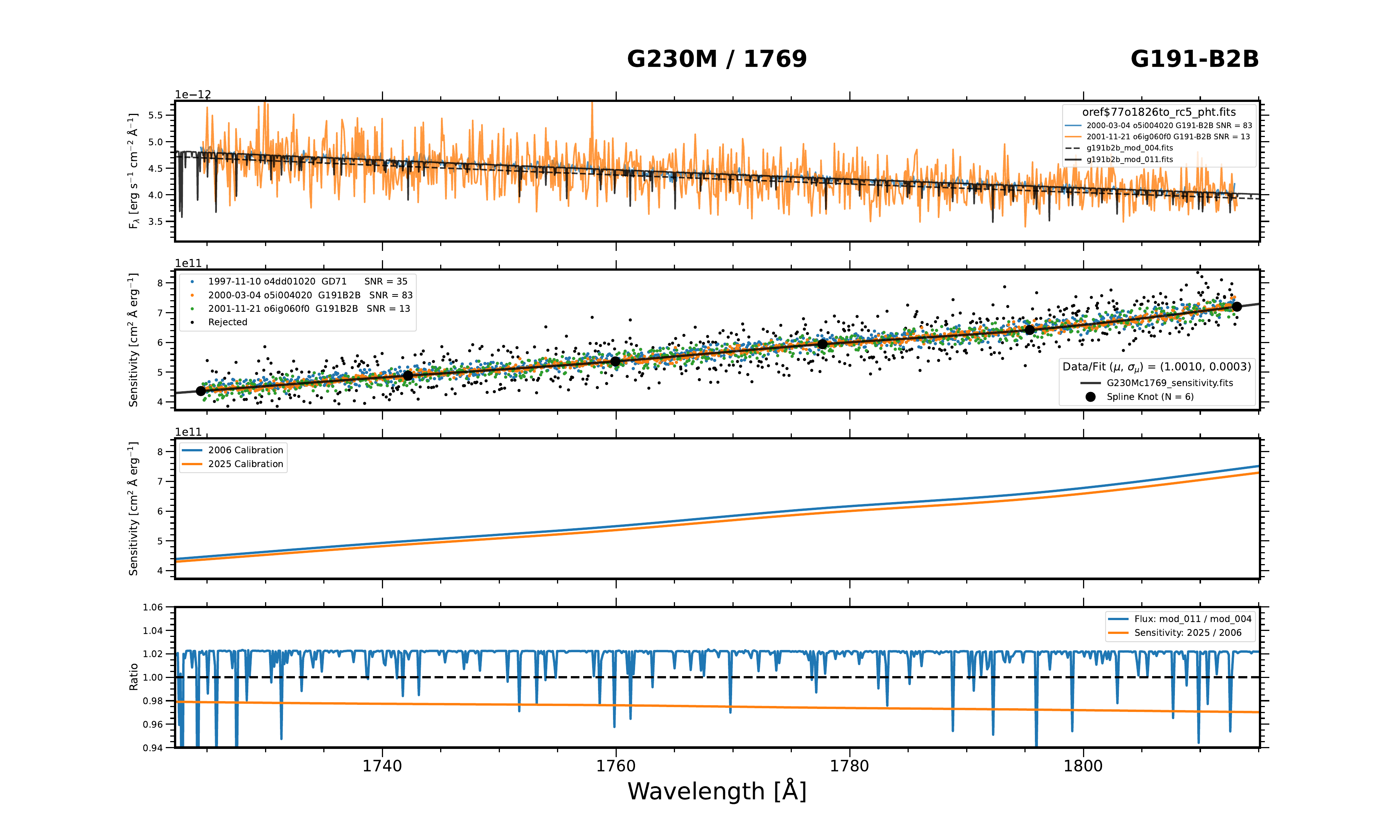}
  \footnotesize
  \caption{Calibration of G191-B2B for G230M/1769.}
  \label{fig:G230MC1769a}
\end{figure}

\begin{figure}[t]
  \hspace{-0.5in}
  \includegraphics[width=1.1\textwidth]{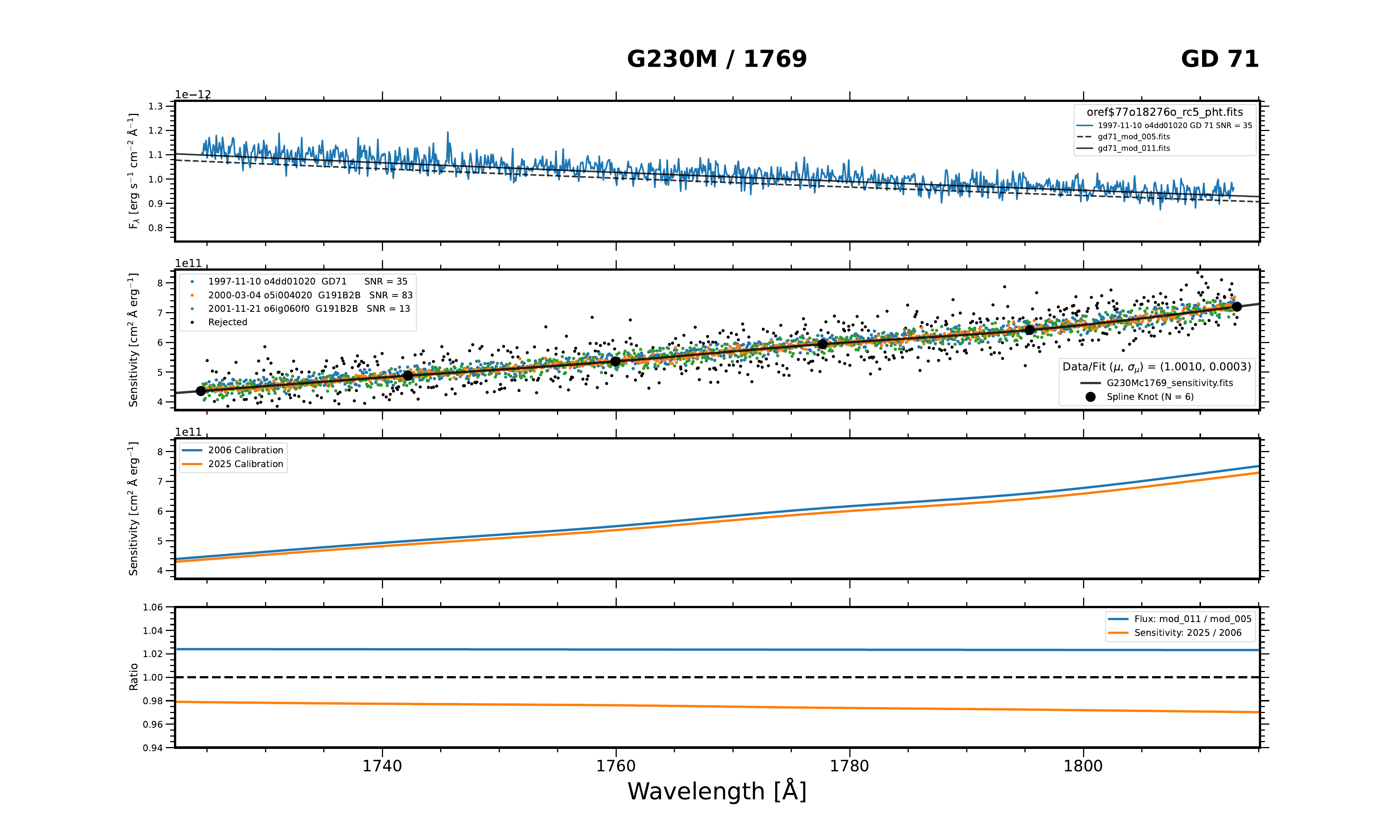}
  \footnotesize
  \caption{Calibration of GD 71 for G230M/1769.}
  \label{fig:G230MC1769b}
\end{figure}

\begin{figure}[b]
  \hspace{-0.5in}
  \includegraphics[width=1.1\textwidth]{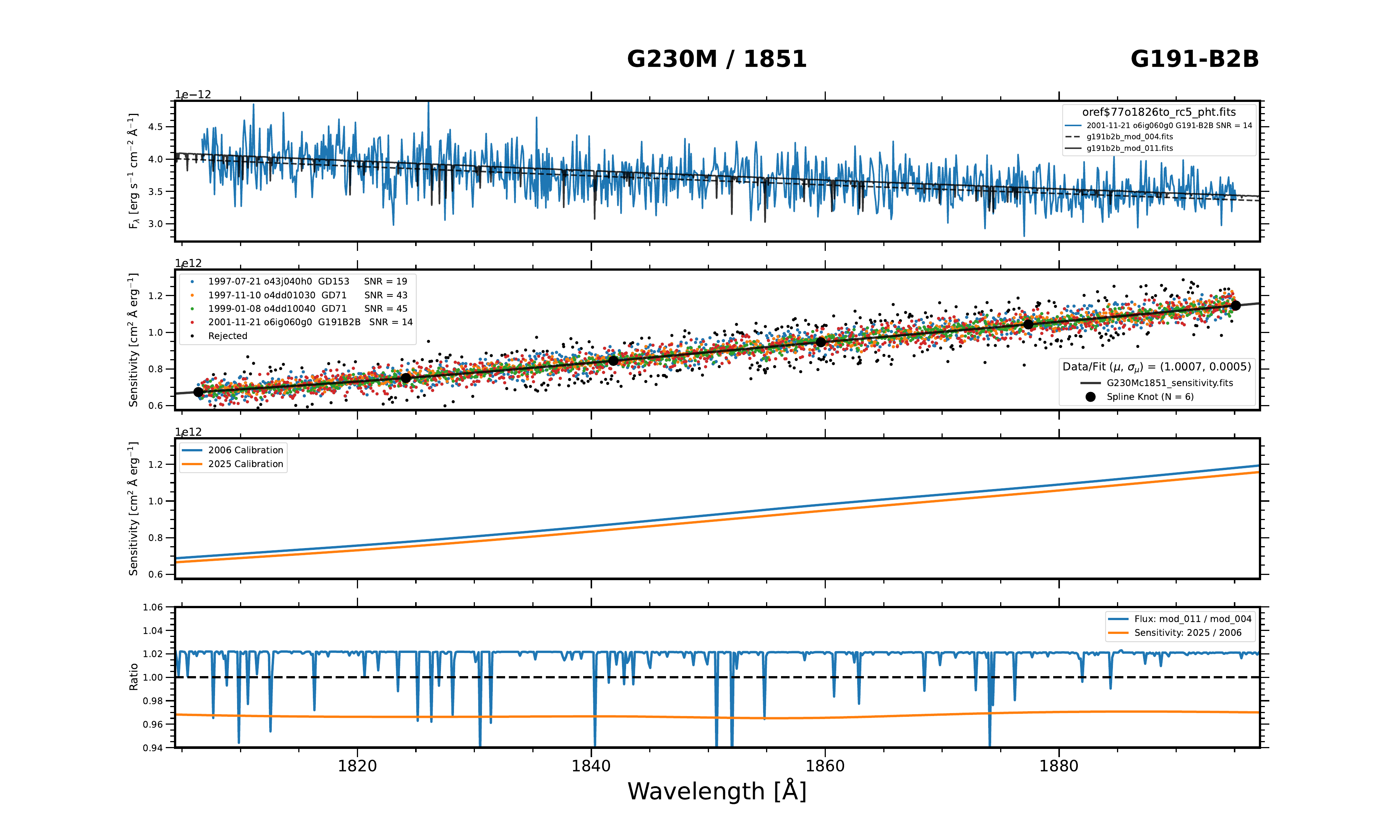}
  \footnotesize
  \caption{Calibration of G191-B2B for G230M/1851.}
  \label{fig:G230MC1851a}
\end{figure}

\begin{figure}[t]
  \hspace{-0.5in}
  \includegraphics[width=1.1\textwidth]{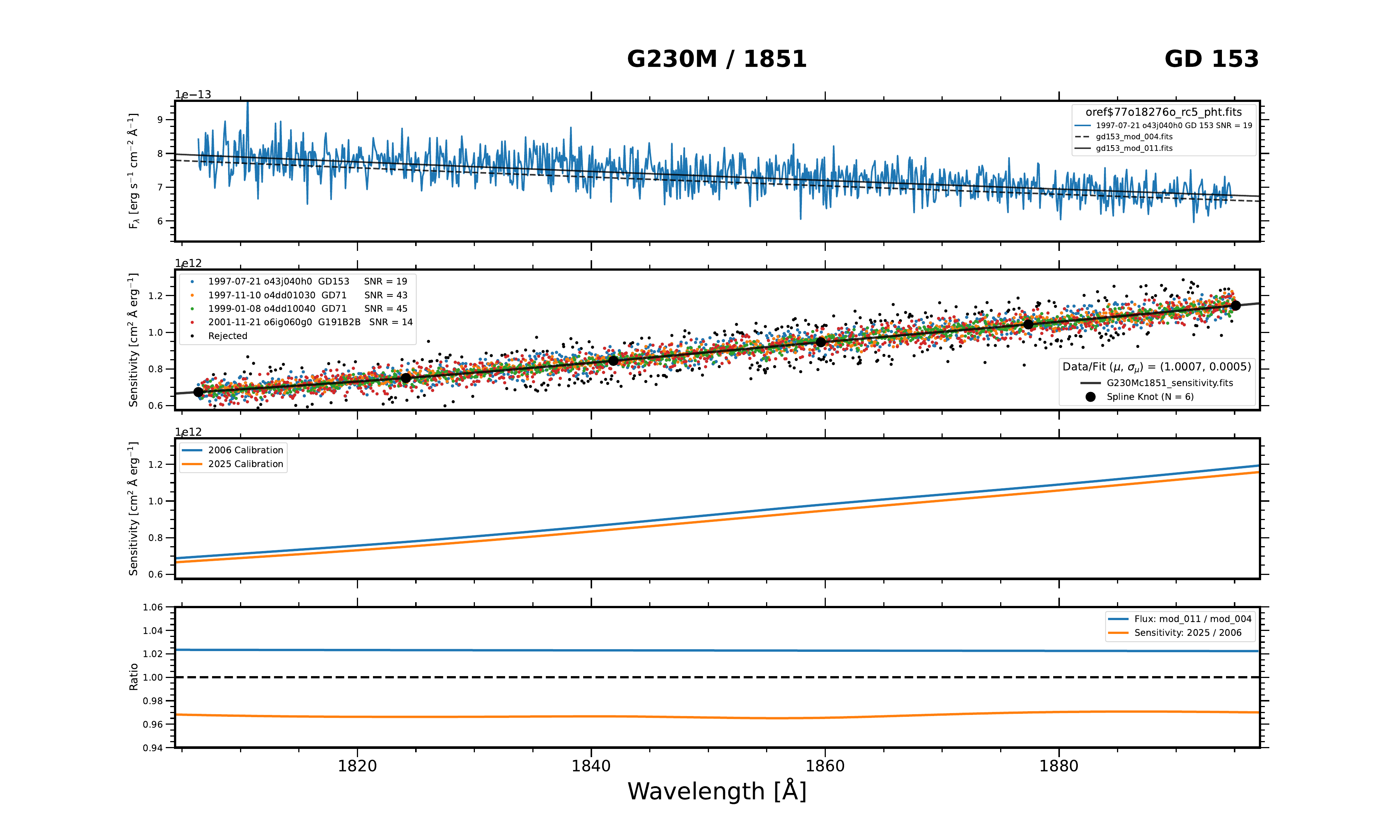}
  \footnotesize
  \caption{Calibration of GD 153 for G230M/1851.}
  \label{fig:G230MC1851c}
\end{figure}

\begin{figure}[b]
  \hspace{-0.5in}
  \includegraphics[width=1.1\textwidth]{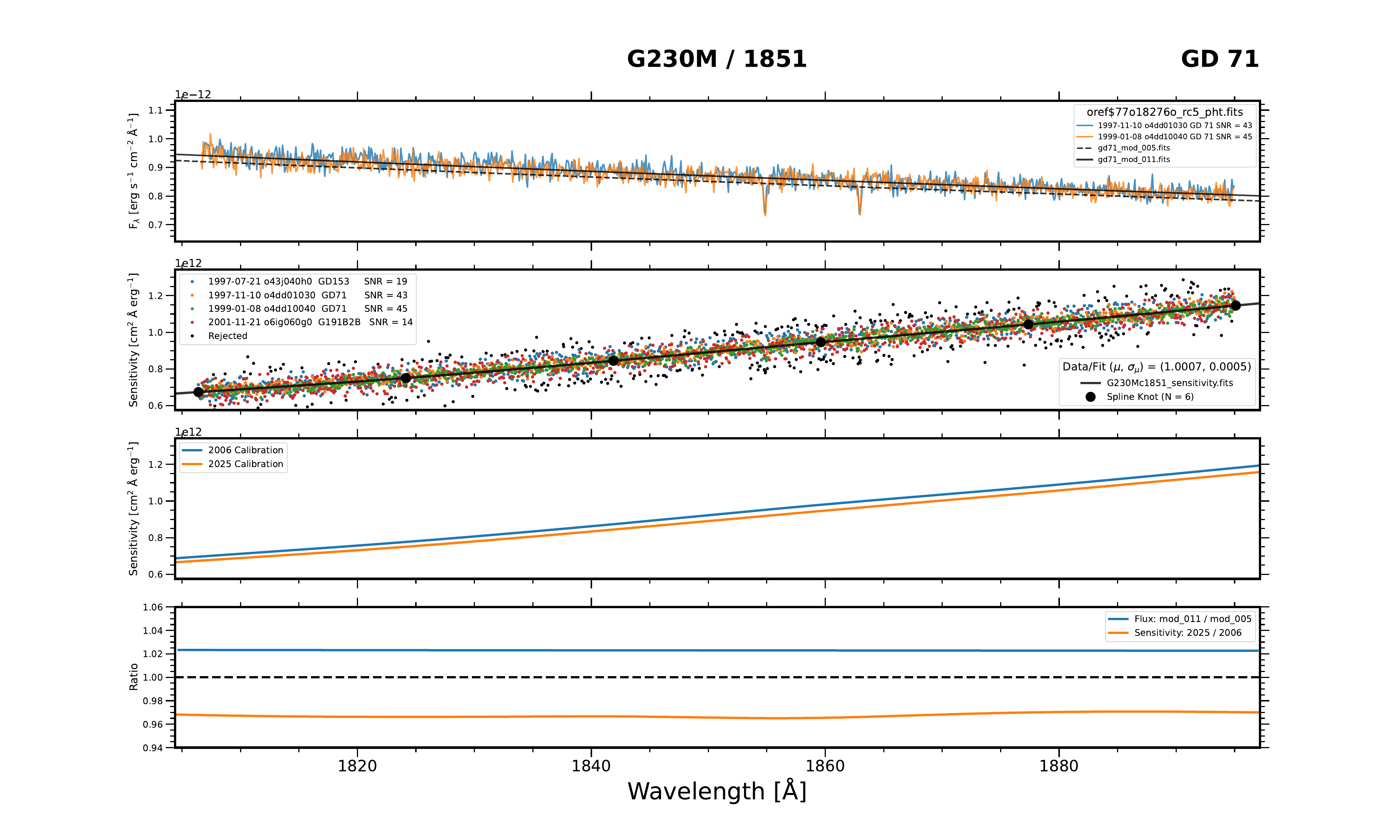}
  \footnotesize
  \caption{Calibration of GD 71 for G230M/1851.}
  \label{fig:G230MC1851b}
\end{figure}

\begin{figure}[b]
  \hspace{-0.5in}
  \includegraphics[width=1.1\textwidth]{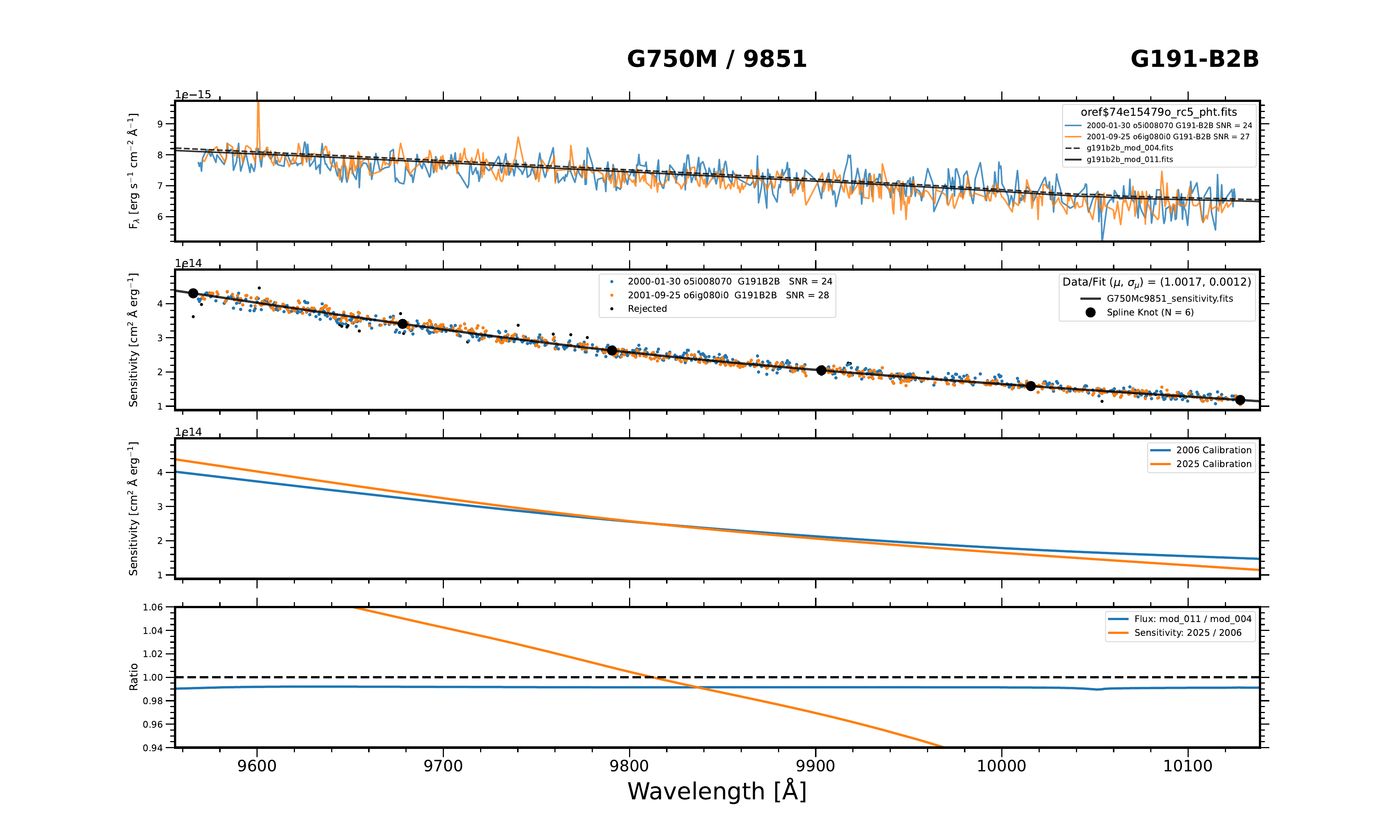}
  \footnotesize
  \caption{Calibration of G191-B2B for G750M/9851.}
  \label{fig:G750MC9851a}
\end{figure}

\ssectionstar{Appendix B: Summary Plots for Sensitivity Fits}\label{sec:AppB}
\vspace{-0.3cm}
Figures \ref{fig:G140MupC1173b} -- \ref{fig:G750MC9851a} illustrate the fits to 
the sensitivity curves derived from CALSPEC version 11 model spectra\footnote{For G430M/4781, 
the consolidated STIS L-mode spectrum of {\bd} was used.}
and provide various comparisons with the previous calibration for each grating, 
CENWAVE, and spectrophotometric standard star. 
The plots are arranged by grating in the following order: G140M\_up, G140M\_dn, 
G230M, G230MB, G430M, and G750M.  
For each grating, the figures are ordered by CENWAVE (smallest to largest) and  
standard star, whenever multiple standards contributed to the fits.\par

\noindent
Each figure consists of 4 panels.
\begin{enumerate}
   \item
   The top panel compares the data reprocessed with the new PHOTTAB files to 
   both the newly adopted (version 11) and previously adopted (version 4 or 5) CALSPEC models. 
   Regions that were excluded from the fits due to the presence of stellar or interstellar features 
   are shaded in gray.
   The reprocessed data better match the v11 model spectra, as expected.

   \item 
   The second panel from the top shows the cubic spline fit to the merged sensitivity data derived 
   from all available standard star data.  
   Individual points are color coded to indicate their origin, and data points that were removed from 
   the fit by sigma clipping (3 iterations, clipped at 3$\sigma$; see Section~\ref{sec:procedure}) are 
   shown in black.  
   The spline knots used to define the fit are represented by large black circles. 
   The weighted mean ($\mu$) and standard error of the mean ($\sigma_\mu$) of the {\it ratio} of the data 
   to the fit provide a coarse indication of the quality of the fit. 

  \item
  The third panel compares the sensitivity curve derived in 2006 with the newly revised version.

  \item
  The bottom panel compares the ratio of the (new/old) flux distributions and sensitivity curves.   
  Since both calibrations used essentially the same standard star data, the flux and sensitivity
  ratios should generally be offset from unity by equal but opposite amounts.  
  The G750M CENWAVEs that required defringing constitute the most significant exception owing to the 
  different defringing procedure adopted in the present work.
\end{enumerate}
 
\clearpage
 
\clearpage
 
\clearpage
 
\clearpage
 
\clearpage
 
\clearpage
 
\clearpage
 
\clearpage
 
\clearpage
 
\clearpage
 
\clearpage
 
\clearpage
 
\clearpage
 
\clearpage
 
\clearpage
 
\newpage

\begin{figure}[t]
  \hspace{-0.5in}
  \includegraphics[width=1.1\textwidth]{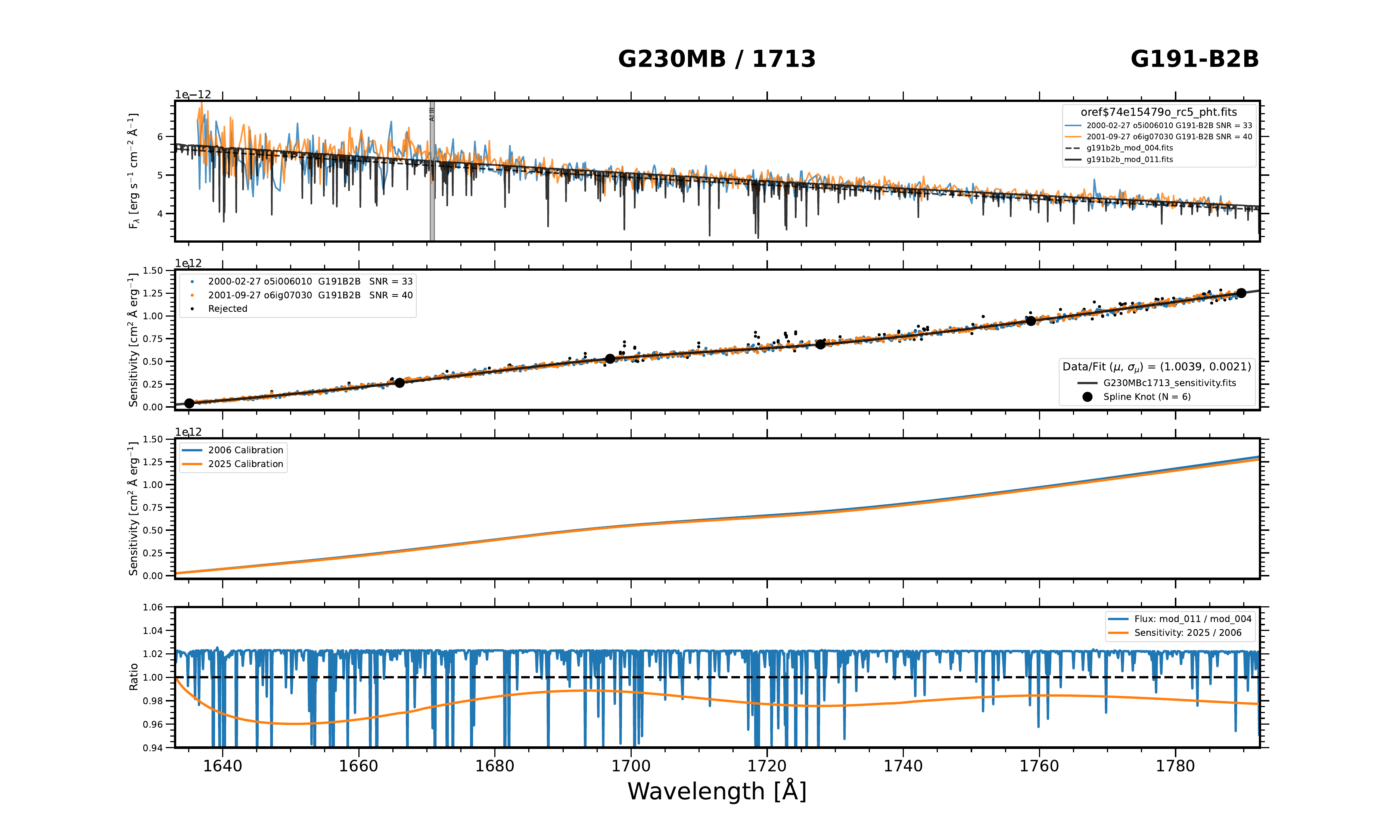}
  \footnotesize
  \caption{Calibration of G191-B2B for G230MB/1713.}
  \label{fig:G230MBC1713a}
\end{figure}
 
\begin{figure}[b]
  \hspace{-0.5in}
  \includegraphics[width=1.1\textwidth]{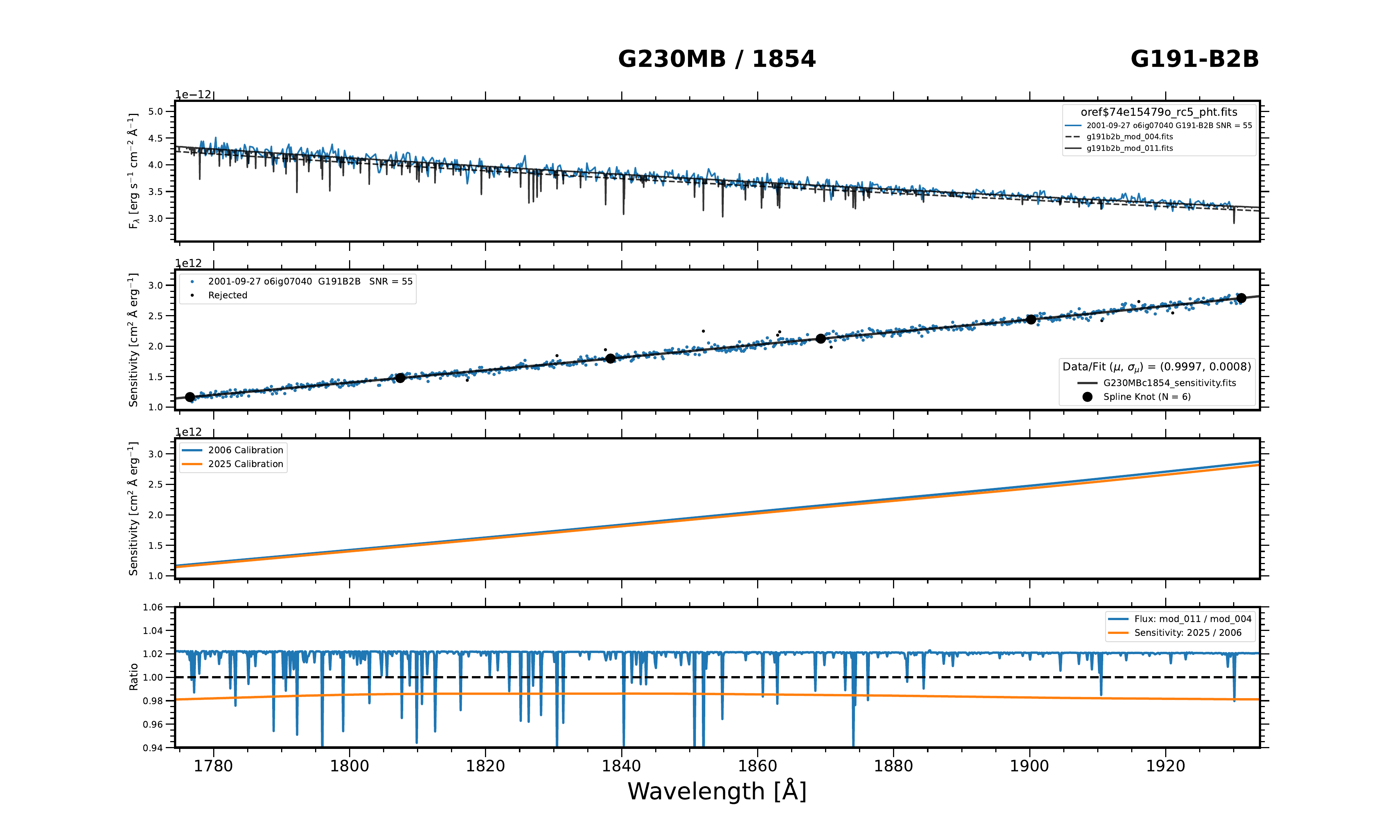}
  \footnotesize
  \caption{Calibration of G191-B2B for G230MB/1854.}
  \label{fig:G230MBC1854a}
\end{figure}
 
\clearpage
\begin{figure}[t]
  \hspace{-0.5in}
  \includegraphics[width=1.1\textwidth]{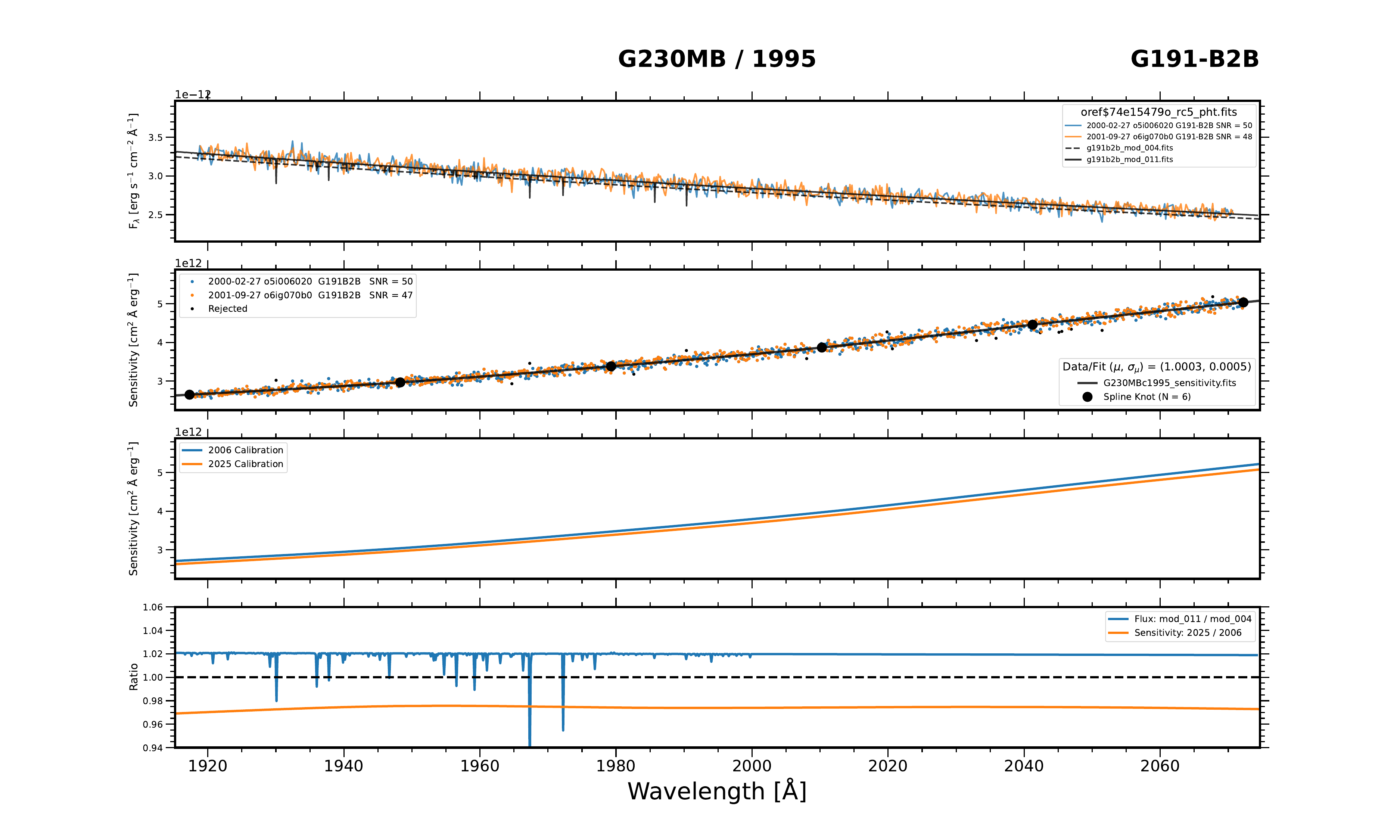}
  \footnotesize
  \caption{Calibration of G191-B2B for G230MB/1995.}
  \label{fig:G230MBC1995a}
\end{figure}
 
\begin{figure}[b]
  \hspace{-0.5in}
  \includegraphics[width=1.1\textwidth]{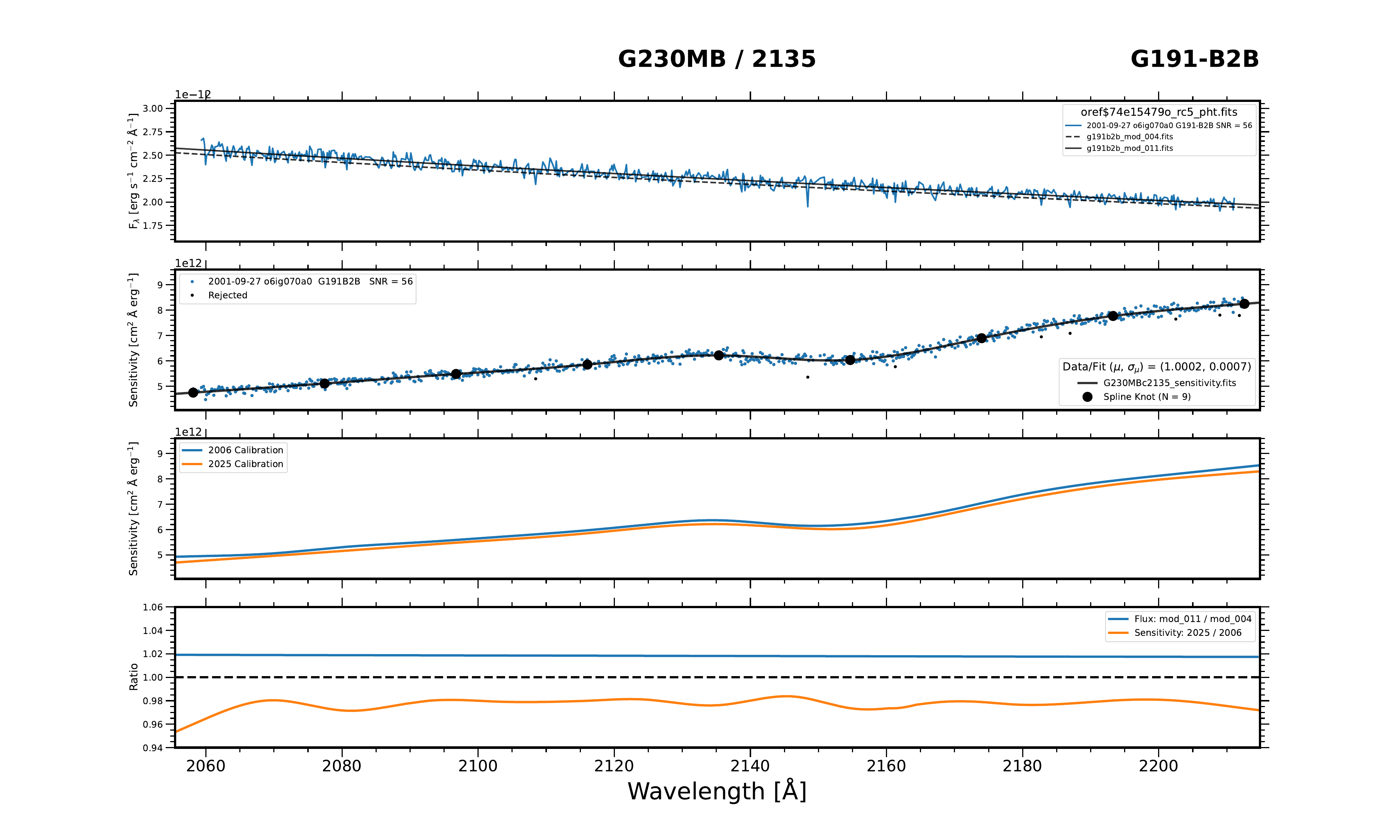}
  \footnotesize
  \caption{Calibration of G191-B2B for G230MB/2135.}
  \label{fig:G230MBC2135a}
\end{figure}
 
\clearpage
\begin{figure}[t]
  \hspace{-0.5in}
  \includegraphics[width=1.1\textwidth]{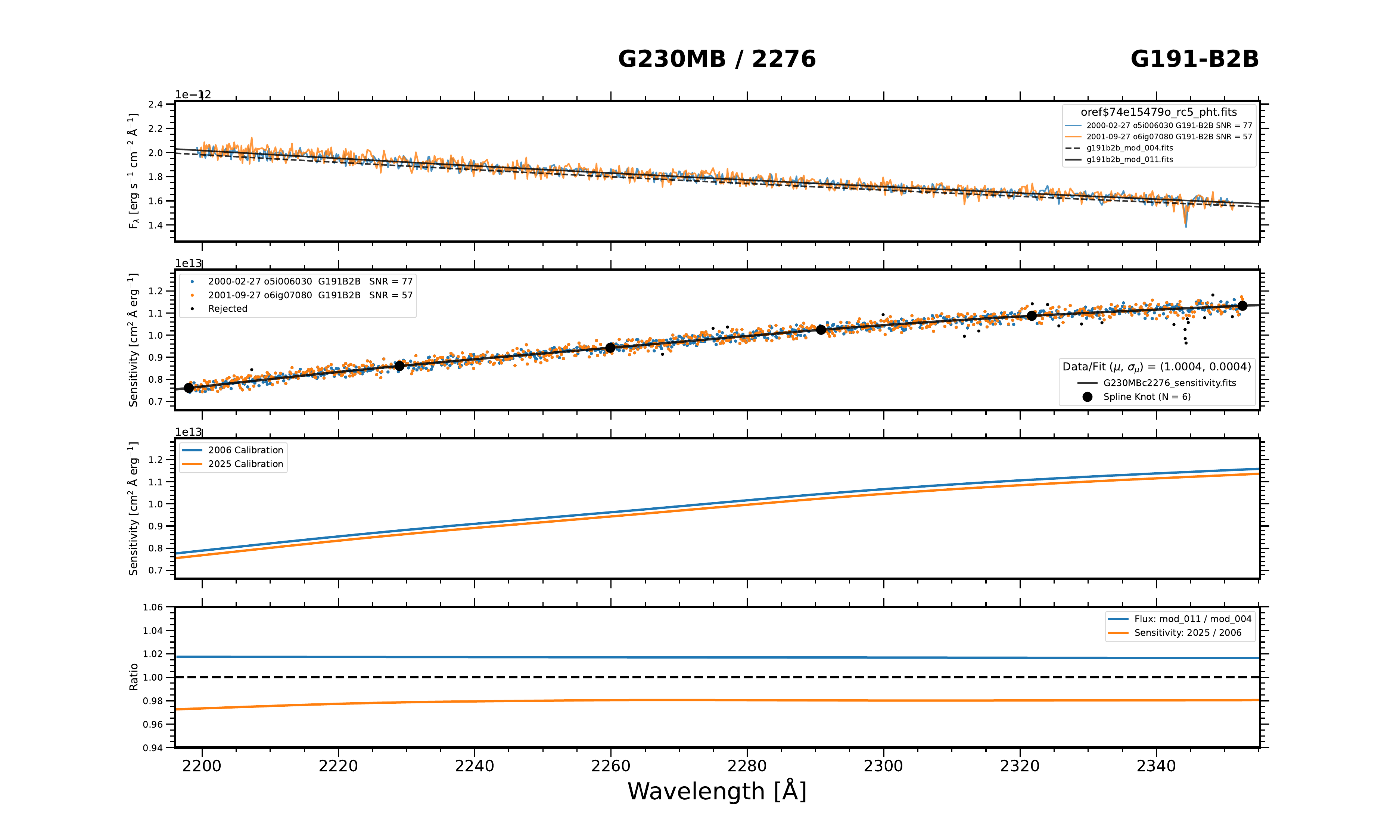}
  \footnotesize
  \caption{Calibration of G191-B2B for G230MB/2276.}
  \label{fig:G230MBC2276a}
\end{figure}
 
\begin{figure}[b]
  \hspace{-0.5in}
  \includegraphics[width=1.1\textwidth]{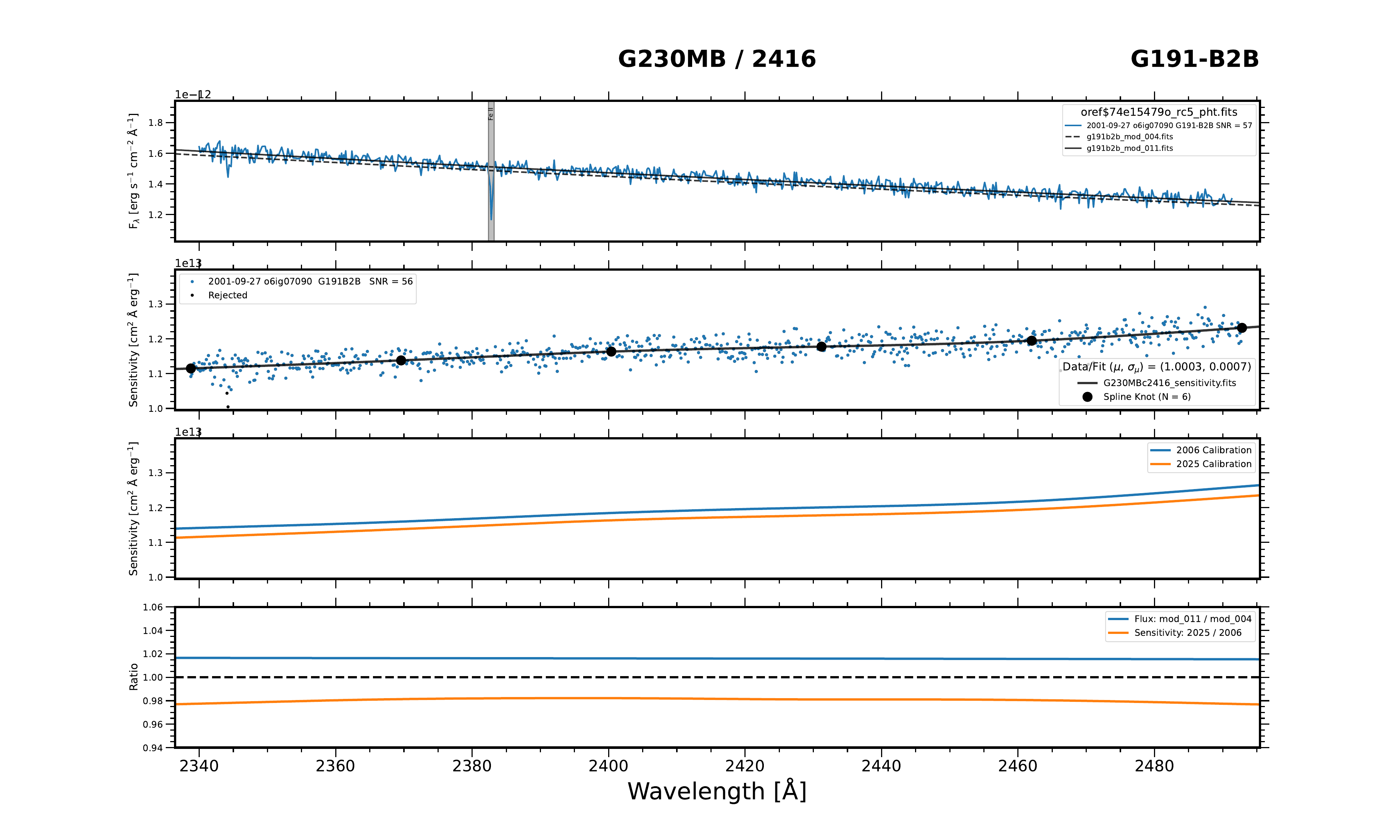}
  \footnotesize
  \caption{Calibration of G191-B2B for G230MB/2416.}
  \label{fig:G230MBC2416a}
\end{figure}
 
\clearpage
\begin{figure}[t]
  \hspace{-0.5in}
  \includegraphics[width=1.1\textwidth]{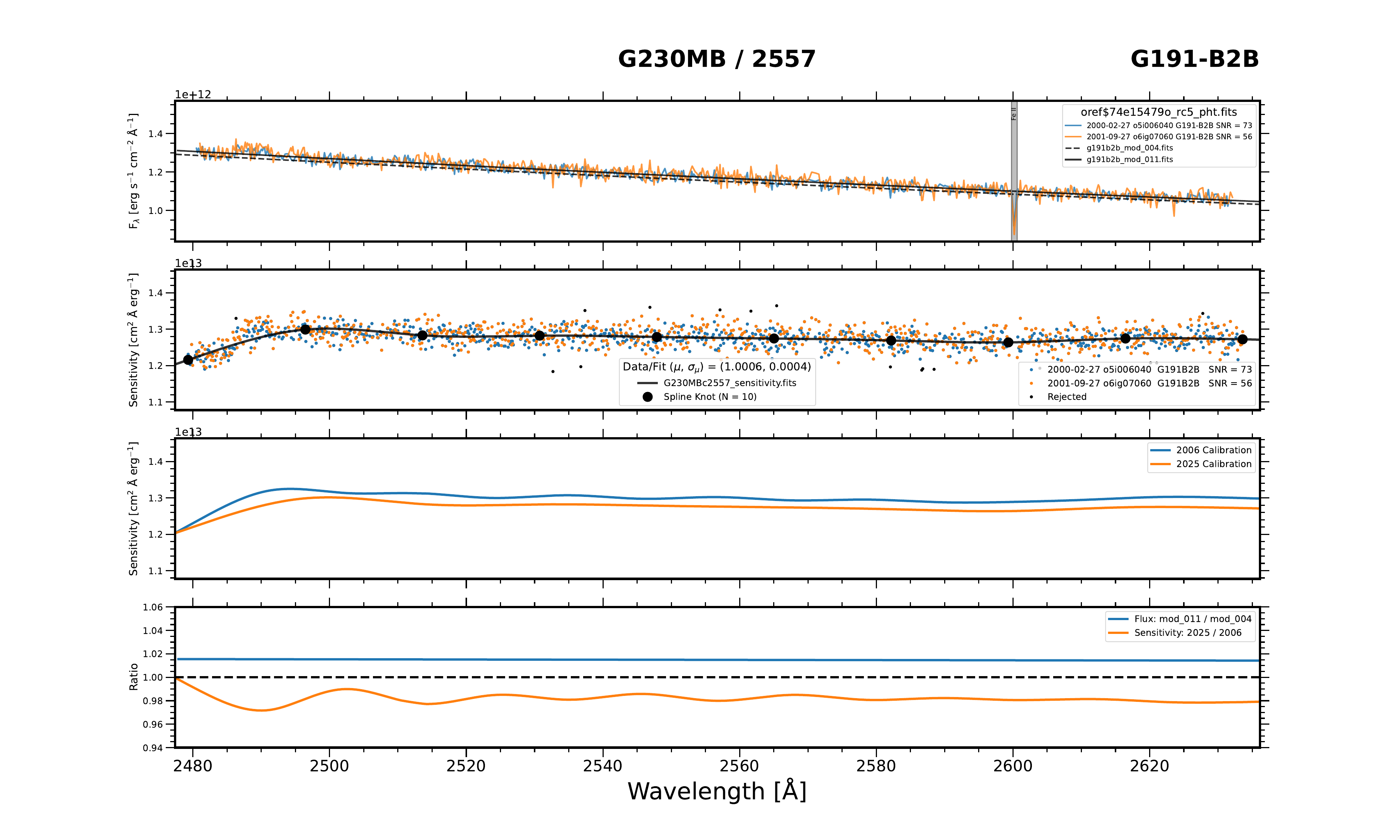}
  \footnotesize
  \caption{Calibration of G191-B2B for G230MB/2557.}
  \label{fig:G230MBC2557a}
\end{figure}
 
\begin{figure}[b]
  \hspace{-0.5in}
  \includegraphics[width=1.1\textwidth]{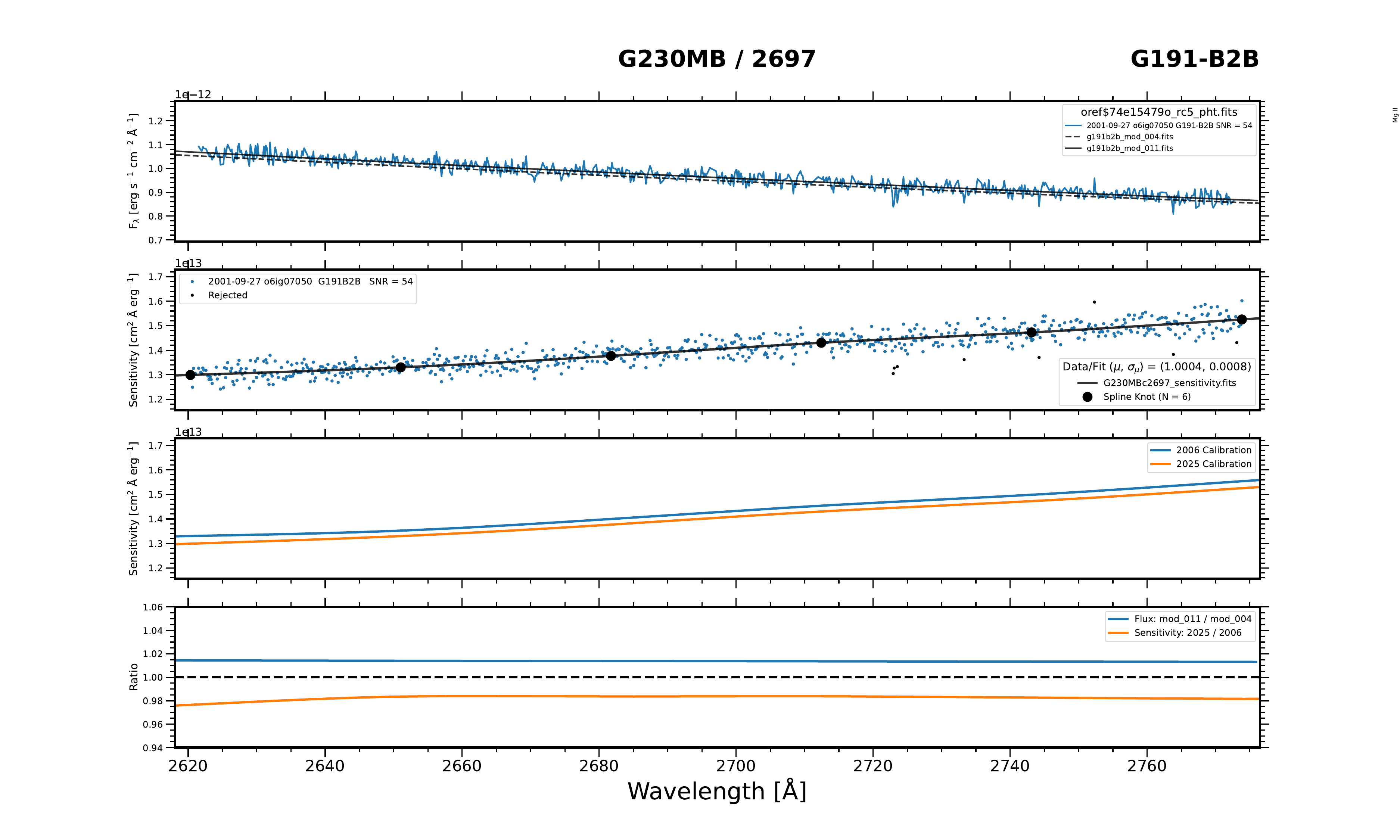}
  \footnotesize
  \caption{Calibration of G191-B2B for G230MB/2697.}
  \label{fig:G230MBC2697a}
\end{figure}
 
\clearpage
\begin{figure}[t]
  \hspace{-0.5in}
  \includegraphics[width=1.1\textwidth]{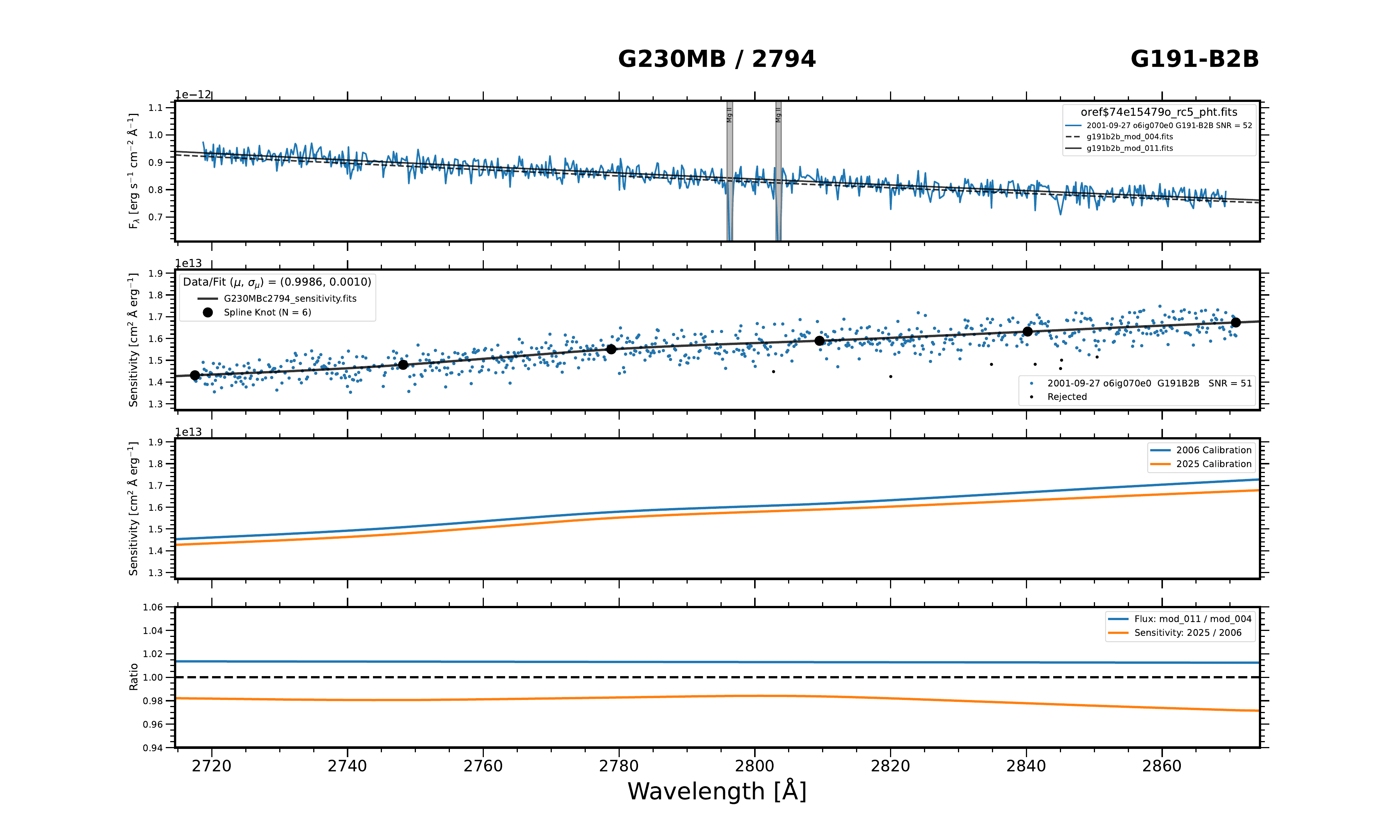}
  \footnotesize
  \caption{Calibration of G191-B2B for G230MB/2794.}
  \label{fig:G230MBC2794a}
\end{figure}
 
\begin{figure}[b]
  \hspace{-0.5in}
  \includegraphics[width=1.1\textwidth]{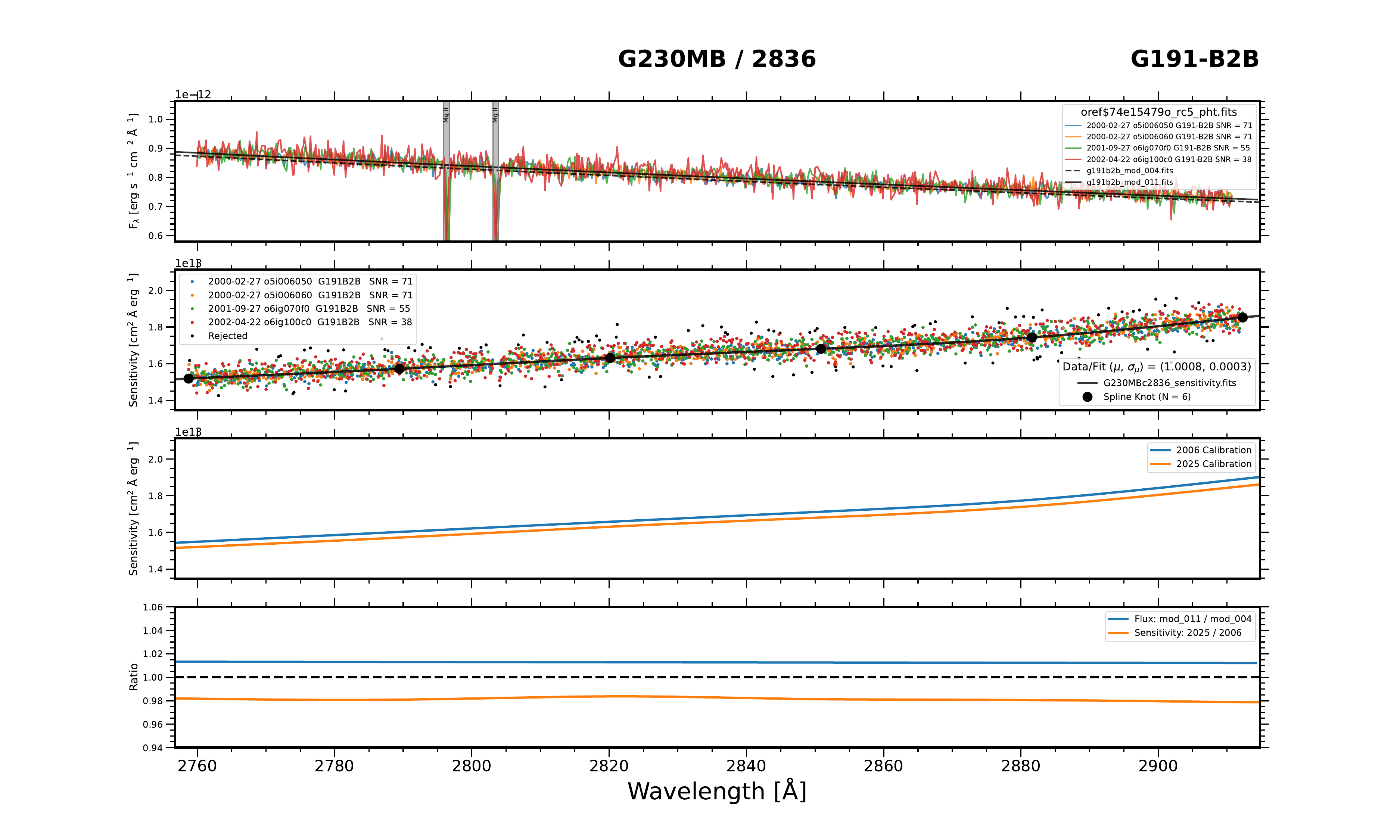}
  \footnotesize
  \caption{Calibration of G191-B2B for G230MB/2836.}
  \label{fig:G230MBC2836a}
\end{figure}
 
\clearpage
\begin{figure}[t]
  \hspace{-0.5in}
  \includegraphics[width=1.1\textwidth]{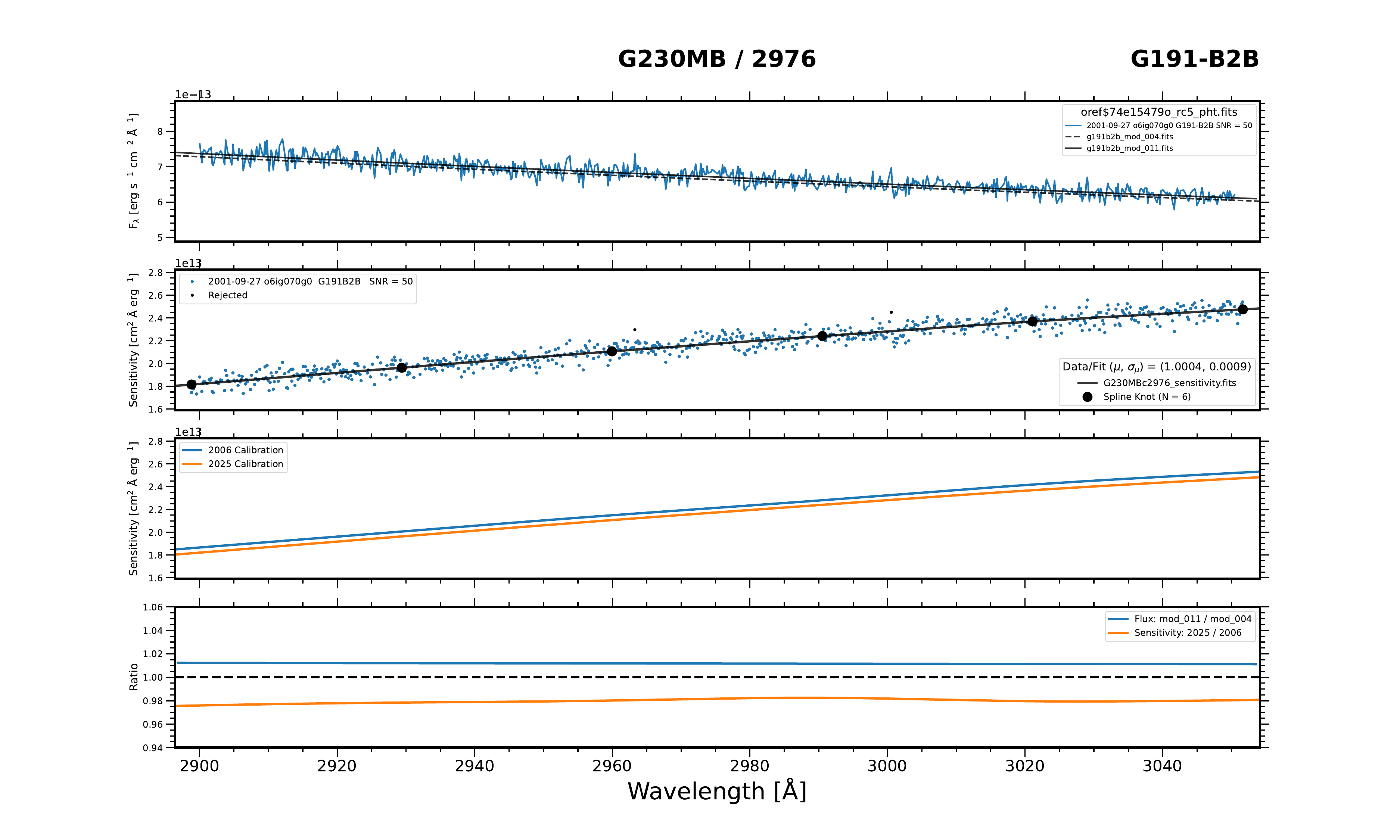}
  \footnotesize
  \caption{Calibration of G191-B2B for G230MB/2976.}
  \label{fig:G230MBC2976a}
\end{figure}
 
\begin{figure}[b]
  \hspace{-0.5in}
  \includegraphics[width=1.1\textwidth]{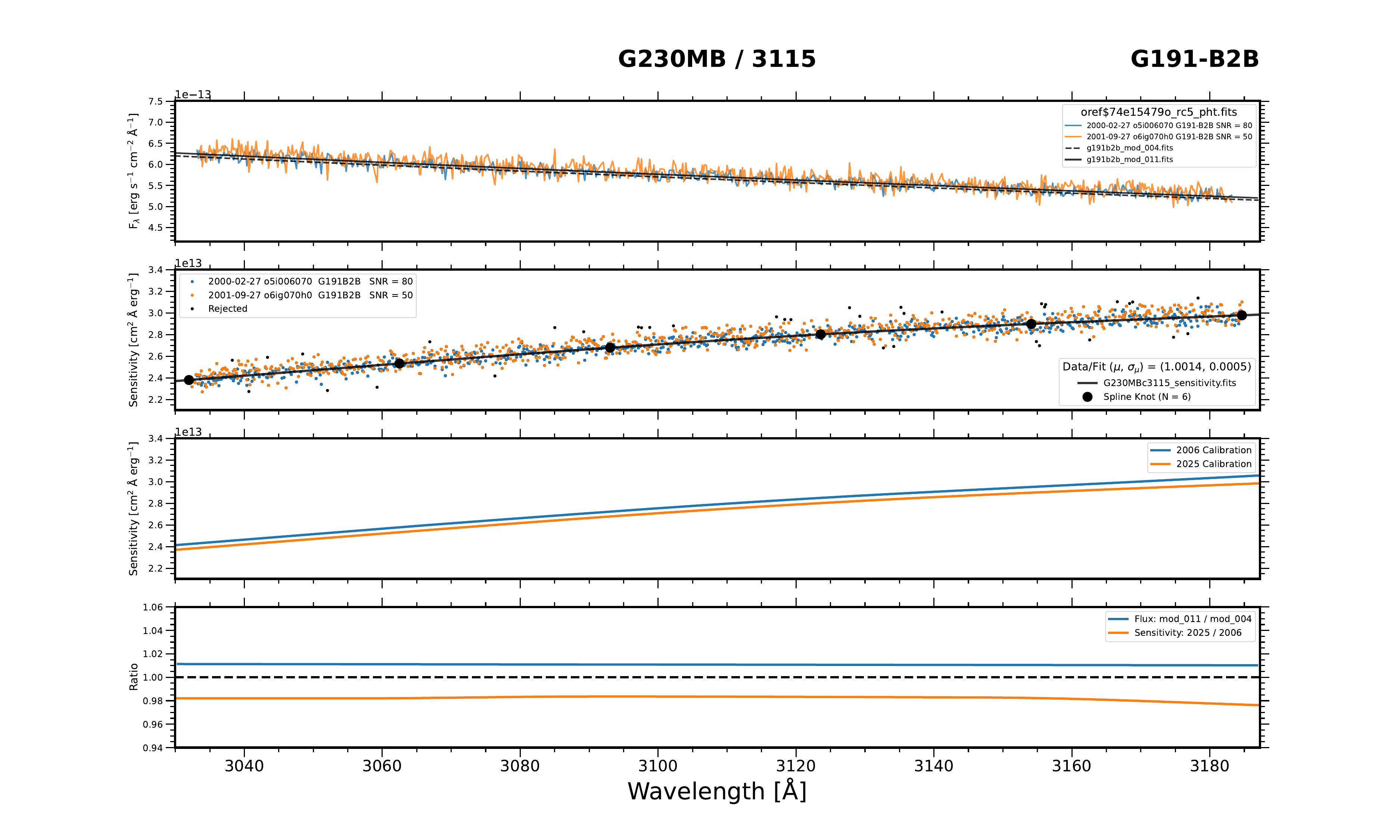}
  \footnotesize
  \caption{Calibration of G191-B2B for G230MB/3115.}
  \label{fig:G230MBC3115a}
\end{figure}
 
\clearpage
 
\clearpage
 
\clearpage
 
\clearpage
 
\clearpage
\begin{figure}[t]
  \hspace{-0.5in}
  \includegraphics[width=1.1\textwidth]{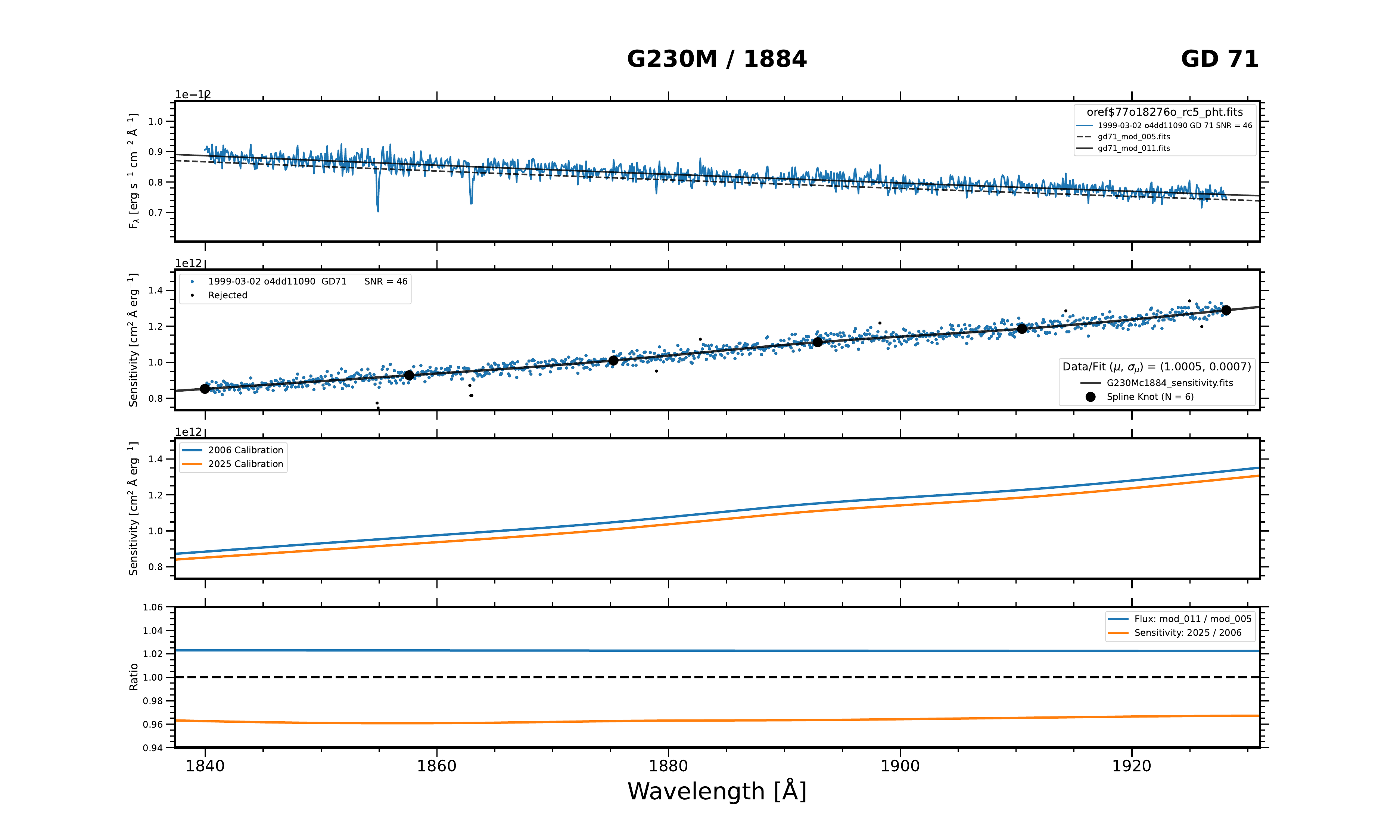}
  \footnotesize
  \caption{Calibration of GD 71 for G230M/1884.}
  \label{fig:G230MC1884b}
\end{figure}
 
\begin{figure}[b]
  \hspace{-0.5in}
  \includegraphics[width=1.1\textwidth]{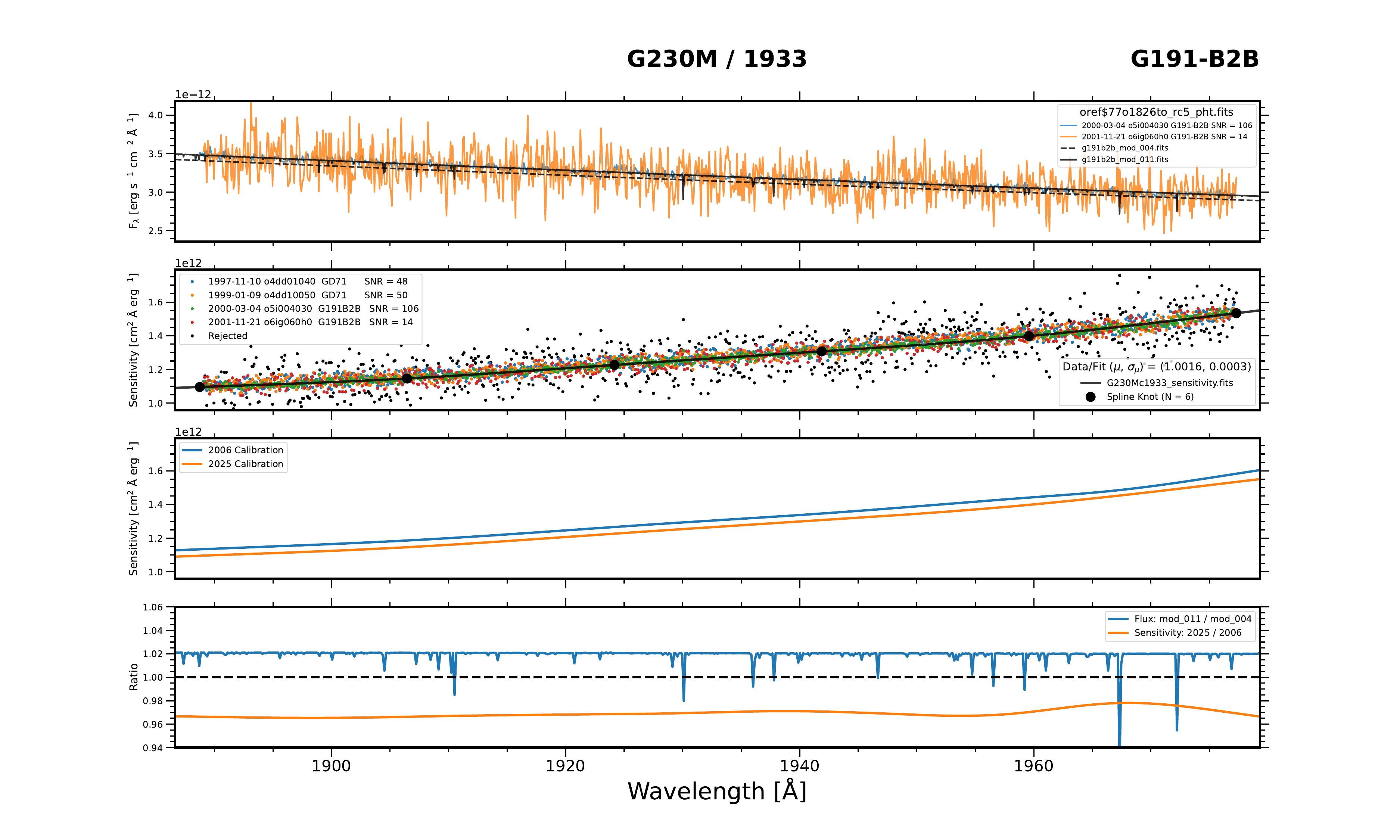}
  \footnotesize
  \caption{Calibration of G191-B2B for G230M/1933.}
  \label{fig:G230MC1933a}
\end{figure}
 
\clearpage
\begin{figure}[t]
  \hspace{-0.5in}
  \includegraphics[width=1.1\textwidth]{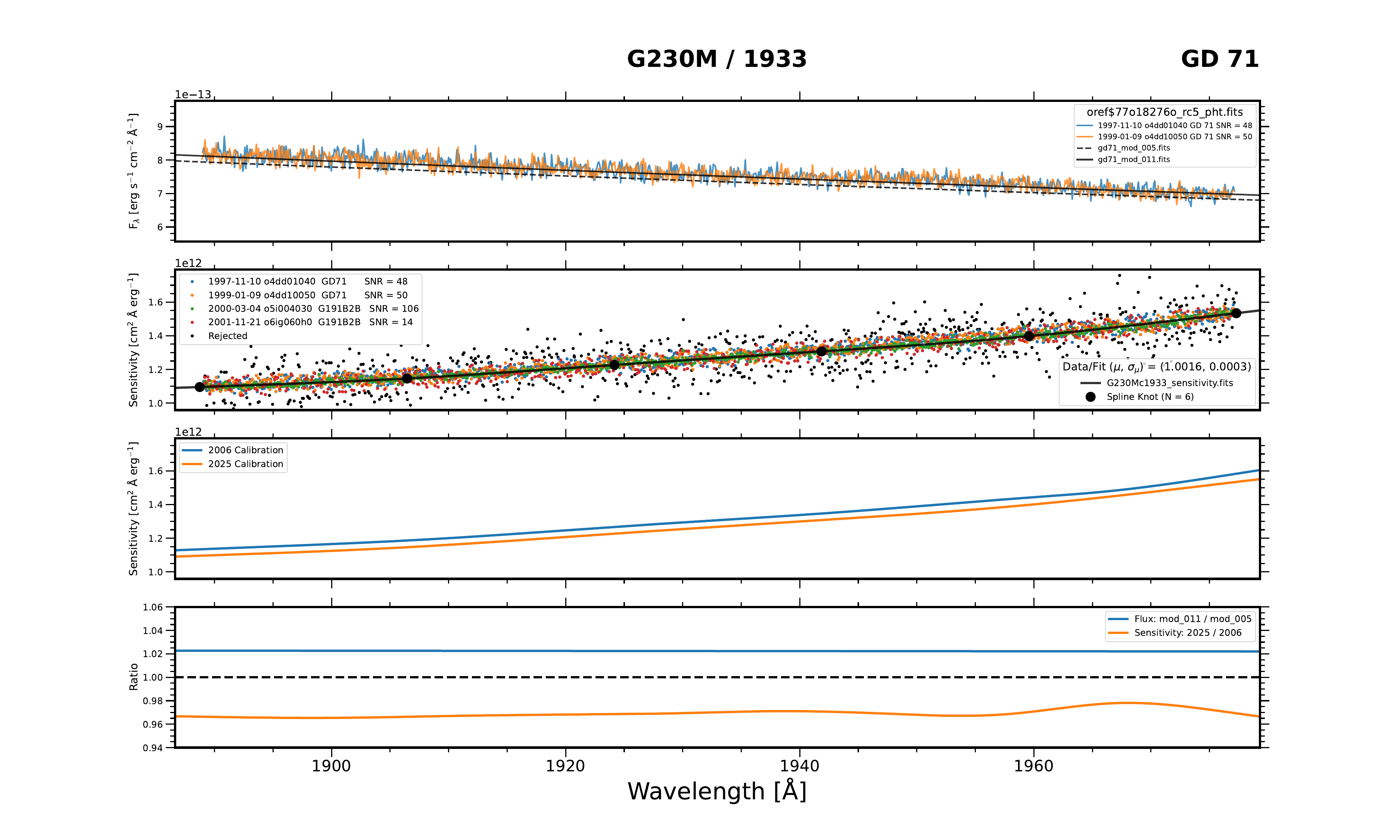}
  \footnotesize
  \caption{Calibration of GD 71 for G230M/1933.}
  \label{fig:G230MC1933b}
\end{figure}
 
\begin{figure}[b]
  \hspace{-0.5in}
  \includegraphics[width=1.1\textwidth]{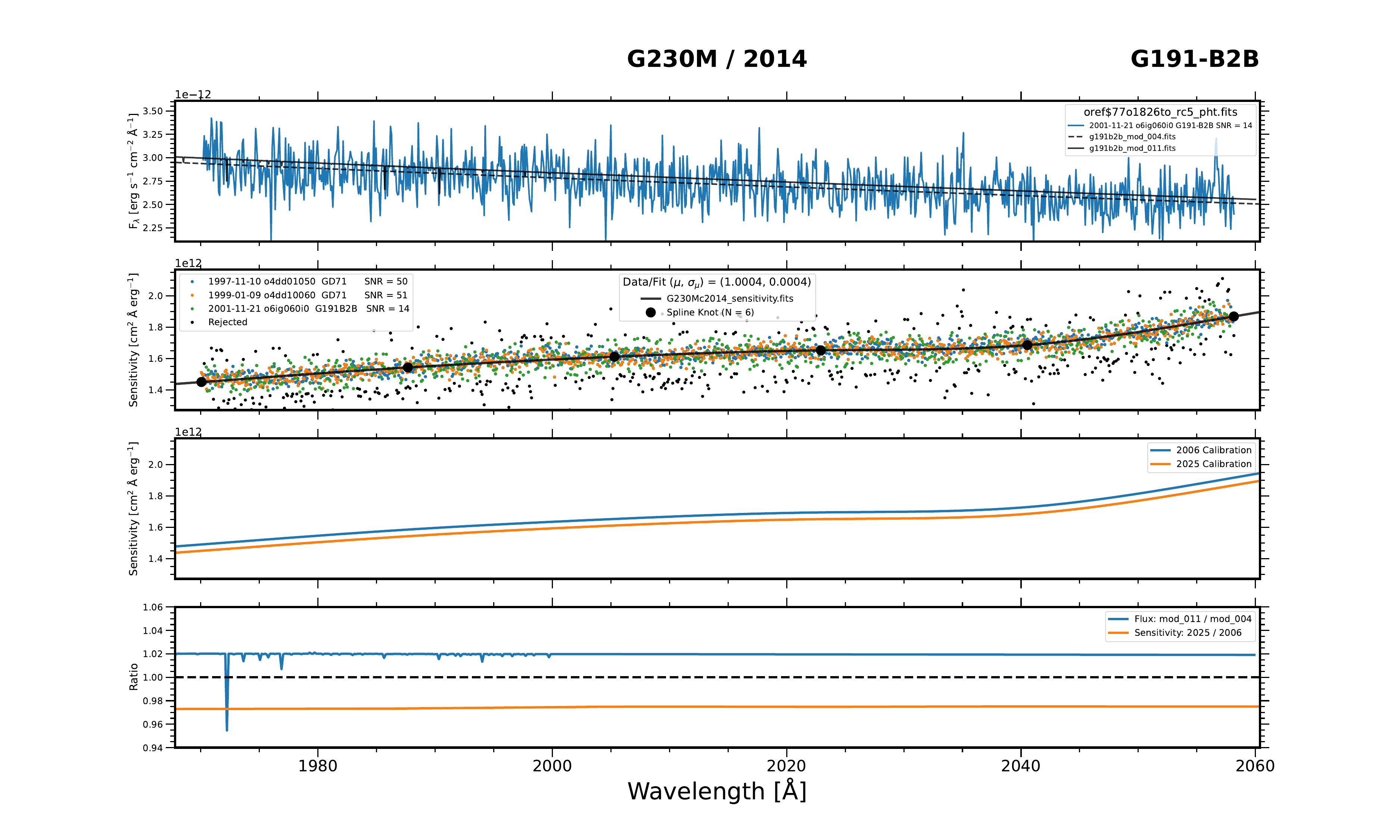}
  \footnotesize
  \caption{Calibration of G191-B2B for G230M/2014.}
  \label{fig:G230MC2014a}
\end{figure}
 
\clearpage
\begin{figure}[t]
  \hspace{-0.5in}
  \includegraphics[width=1.1\textwidth]{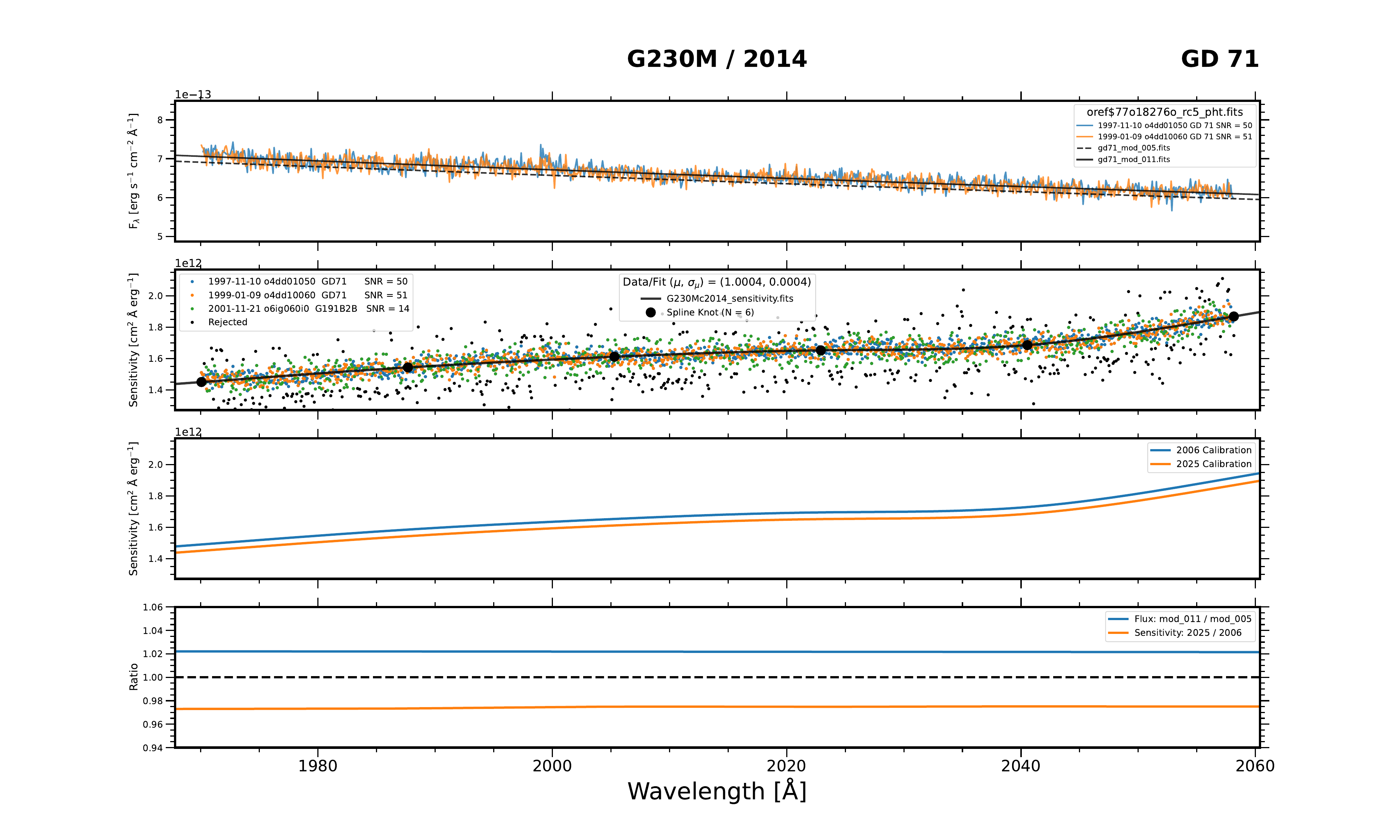}
  \footnotesize
  \caption{Calibration of GD 71 for G230M/2014.}
  \label{fig:G230MC2014b}
\end{figure}
 
\begin{figure}[b]
  \hspace{-0.5in}
  \includegraphics[width=1.1\textwidth]{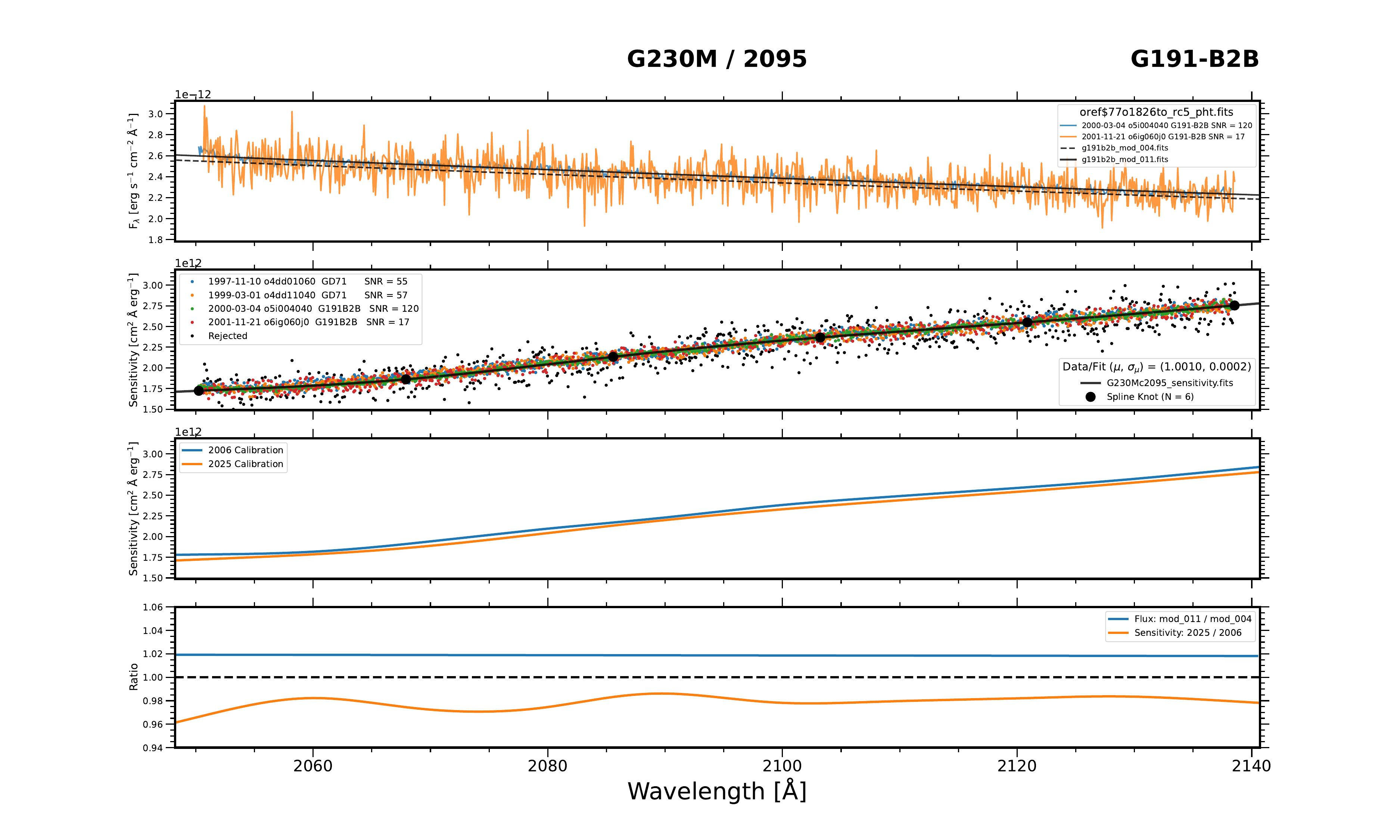}
  \footnotesize
  \caption{Calibration of G191-B2B for G230M/2095.}
  \label{fig:G230MC2095a}
\end{figure}
 
\clearpage
\begin{figure}[t]
  \hspace{-0.5in}
  \includegraphics[width=1.1\textwidth]{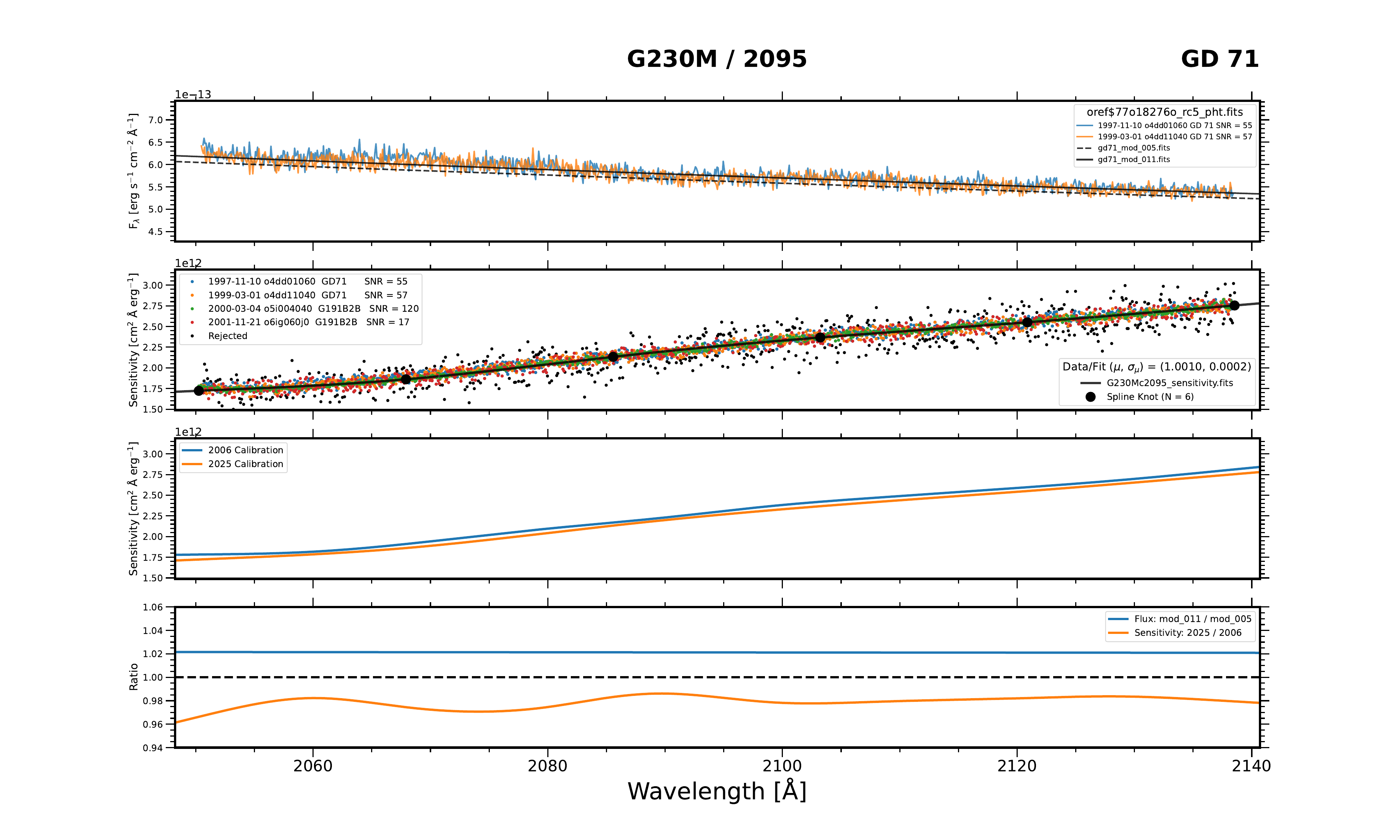}
  \footnotesize
  \caption{Calibration of GD 71 for G230M/2095.}
  \label{fig:G230MC2095b}
\end{figure}
 
\begin{figure}[b]
  \hspace{-0.5in}
  \includegraphics[width=1.1\textwidth]{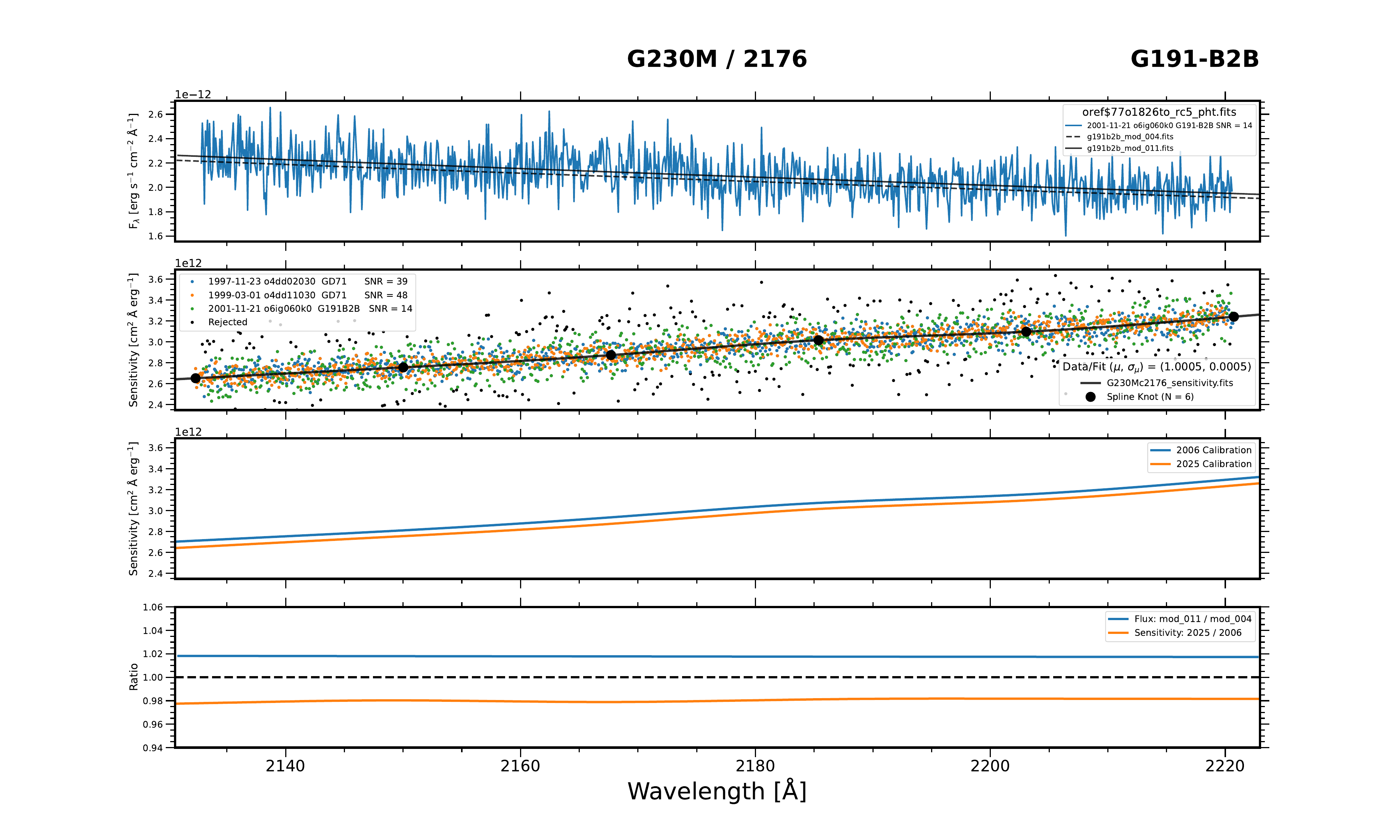}
  \footnotesize
  \caption{Calibration of G191-B2B for G230M/2176.}
  \label{fig:G230MC2176a}
\end{figure}
 
\clearpage
\begin{figure}[t]
  \hspace{-0.5in}
  \includegraphics[width=1.1\textwidth]{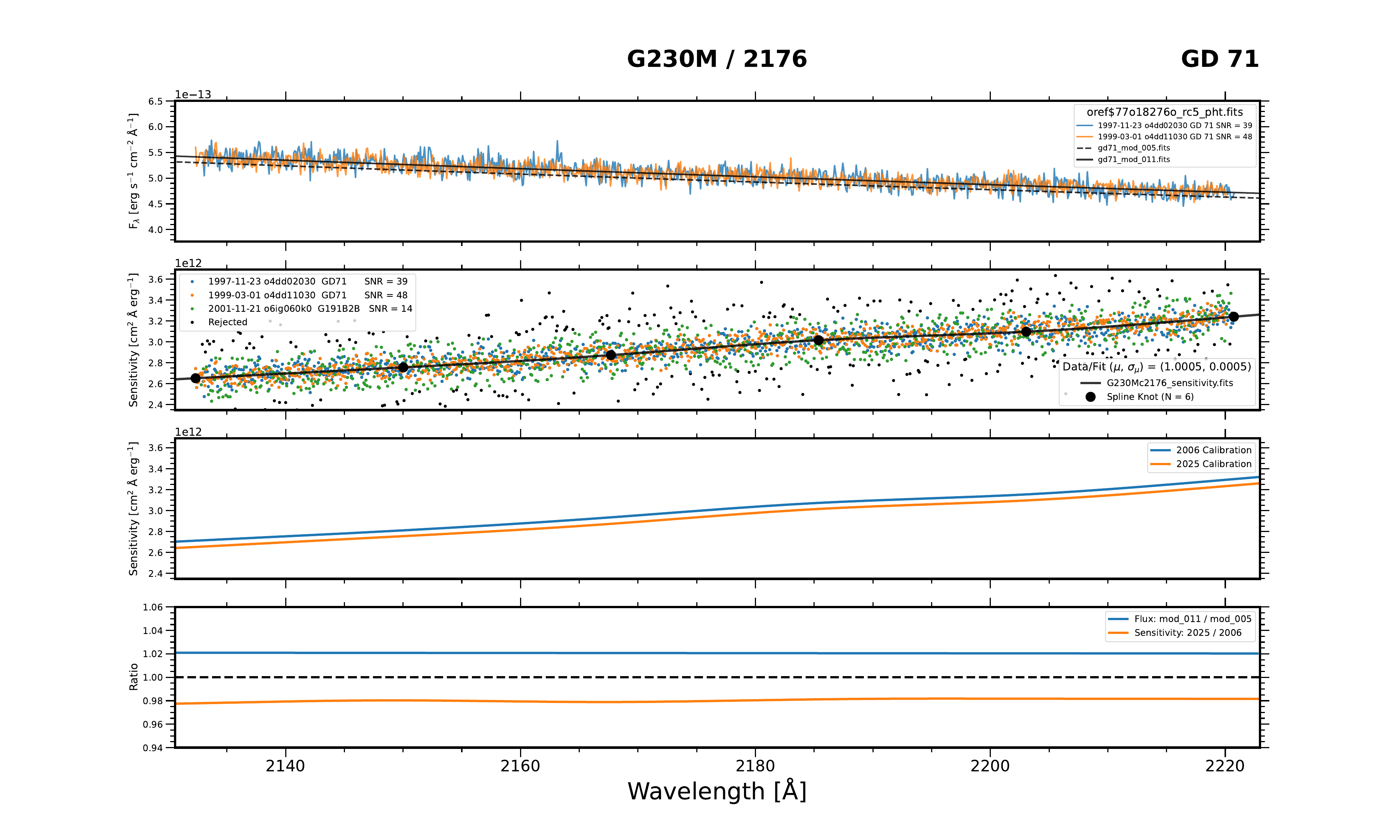}
  \footnotesize
  \caption{Calibration of GD 71 for G230M/2176.}
  \label{fig:G230MC2176b}
\end{figure}
 
\begin{figure}[b]
  \hspace{-0.5in}
  \includegraphics[width=1.1\textwidth]{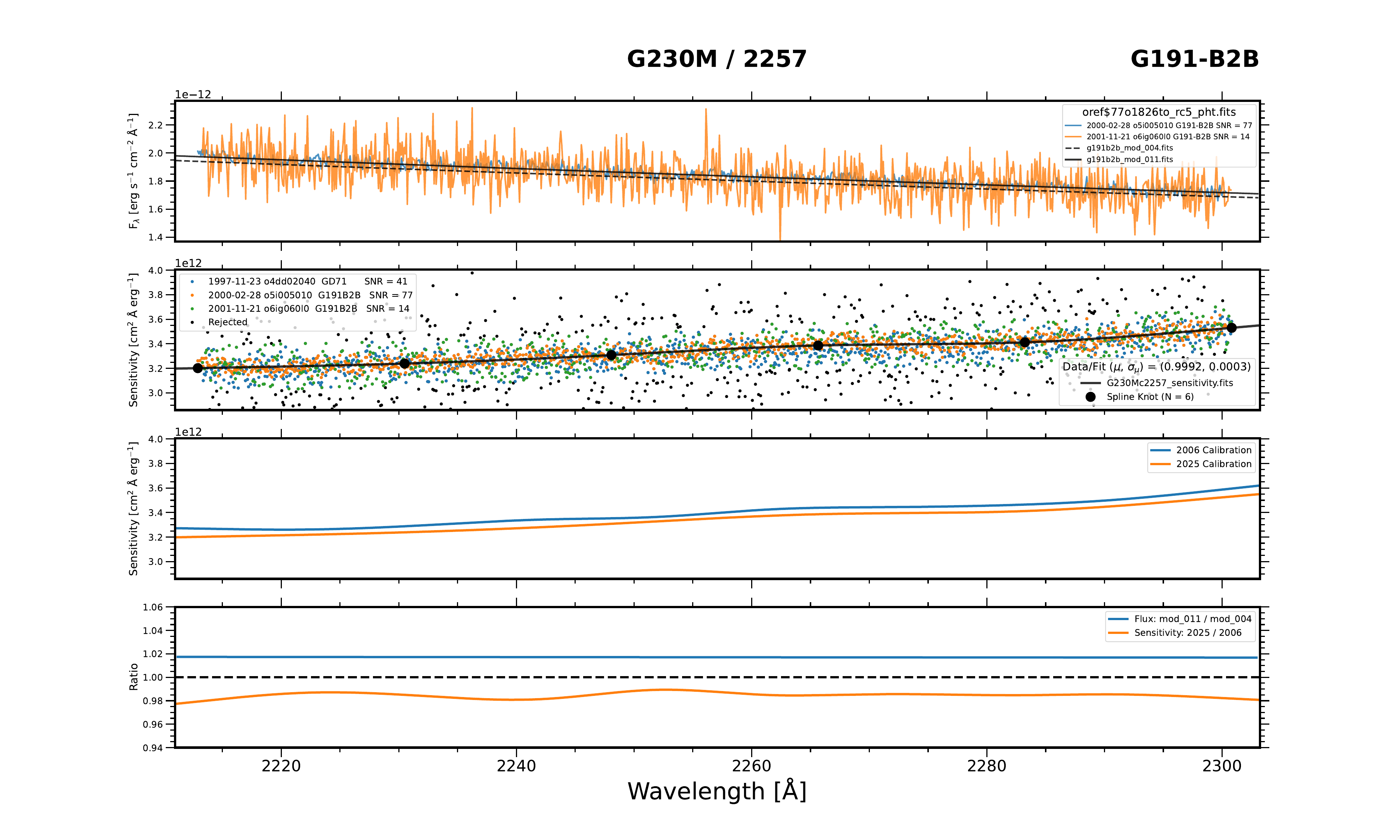}
  \footnotesize
  \caption{Calibration of G191-B2B for G230M/2257.}
  \label{fig:G230MC2257a}
\end{figure}
 
\clearpage
\begin{figure}[t]
  \hspace{-0.5in}
  \includegraphics[width=1.1\textwidth]{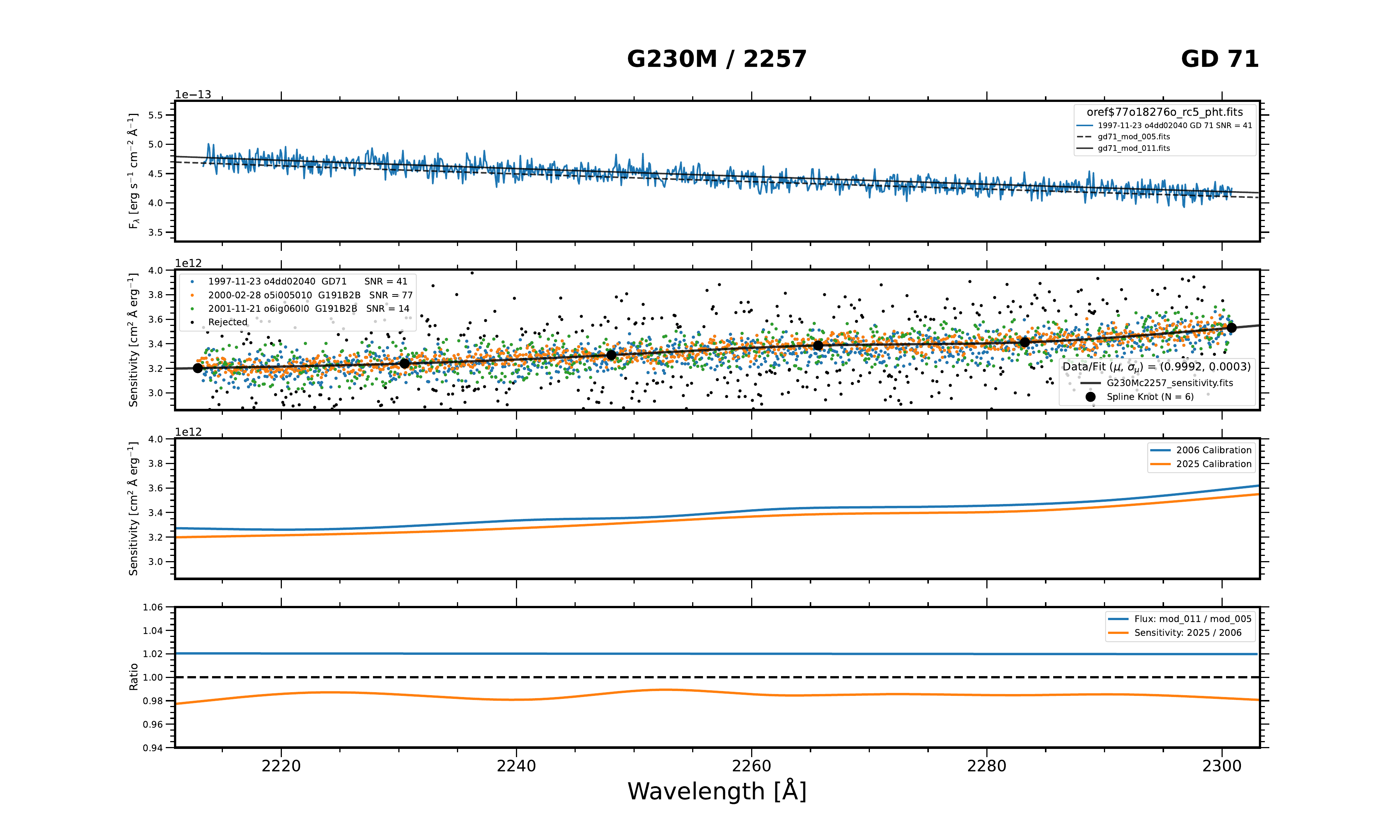}
  \footnotesize
  \caption{Calibration of GD 71 for G230M/2257.}
  \label{fig:G230MC2257b}
\end{figure}
 
\begin{figure}[b]
  \hspace{-0.5in}
  \includegraphics[width=1.1\textwidth]{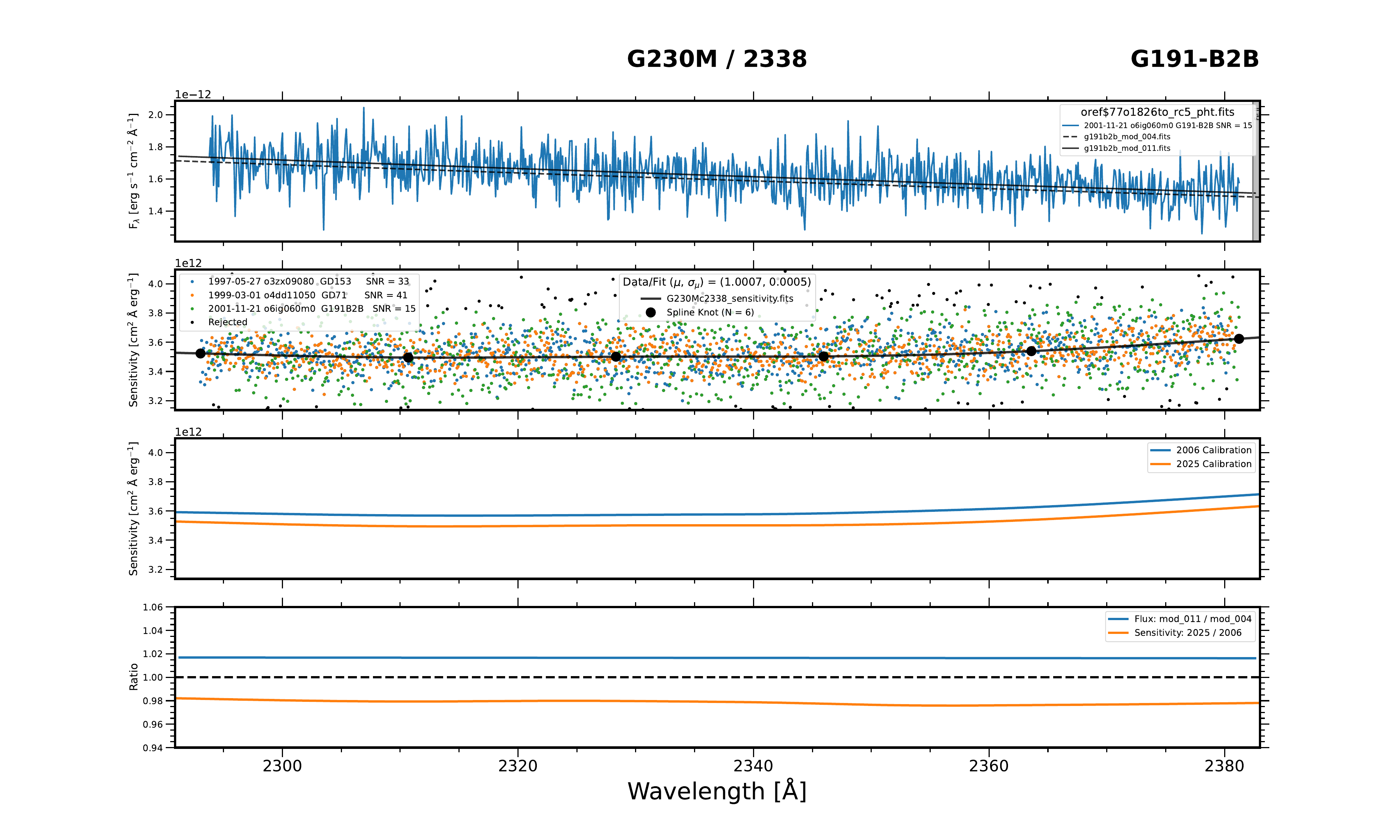}
  \footnotesize
  \caption{Calibration of G191-B2B for G230M/2338.}
  \label{fig:G230MC2338a}
\end{figure}
 
\clearpage
\begin{figure}[t]
  \hspace{-0.5in}
  \includegraphics[width=1.1\textwidth]{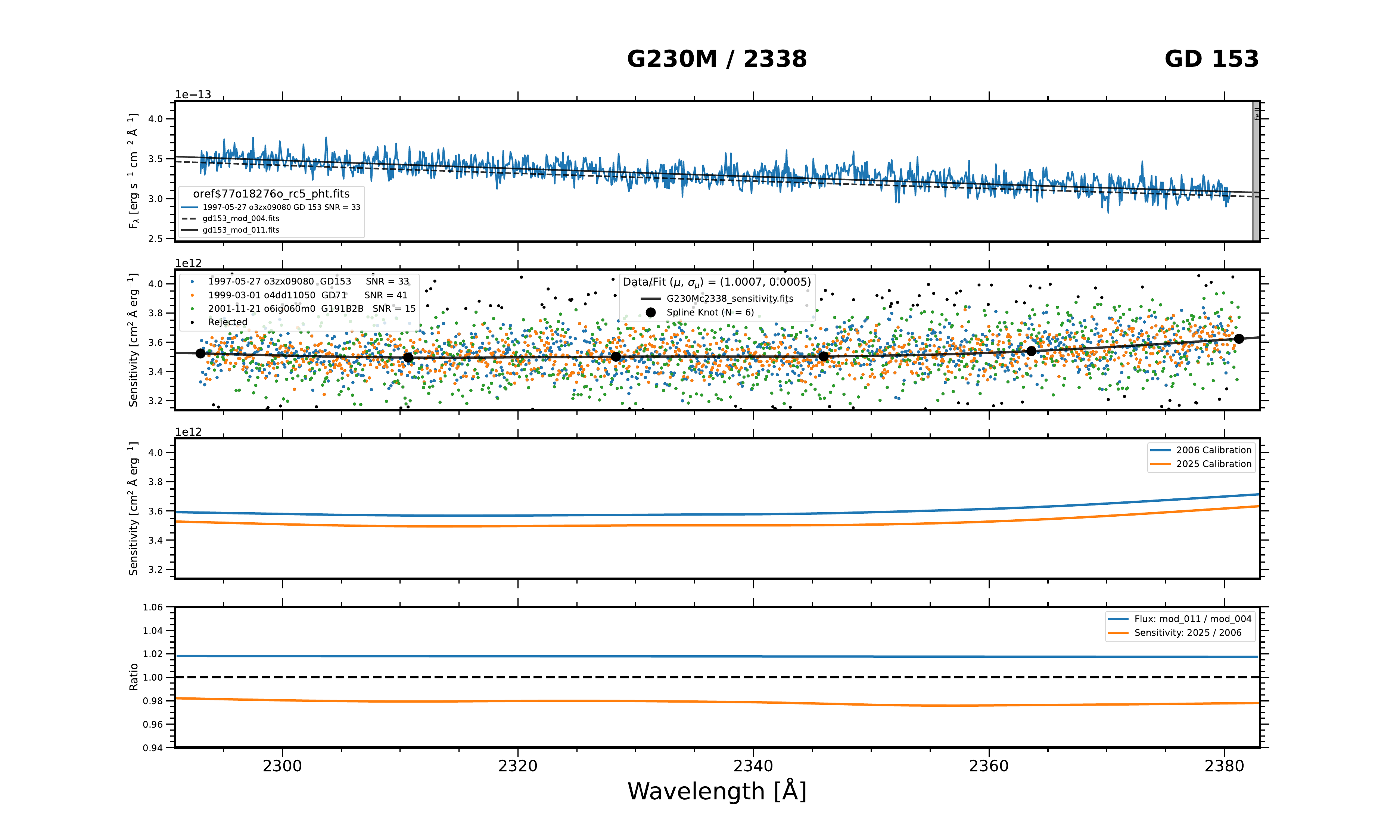}
  \footnotesize
  \caption{Calibration of GD 153 for G230M/2338.}
  \label{fig:G230MC2338c}
\end{figure}
 
\begin{figure}[b]
  \hspace{-0.5in}
  \includegraphics[width=1.1\textwidth]{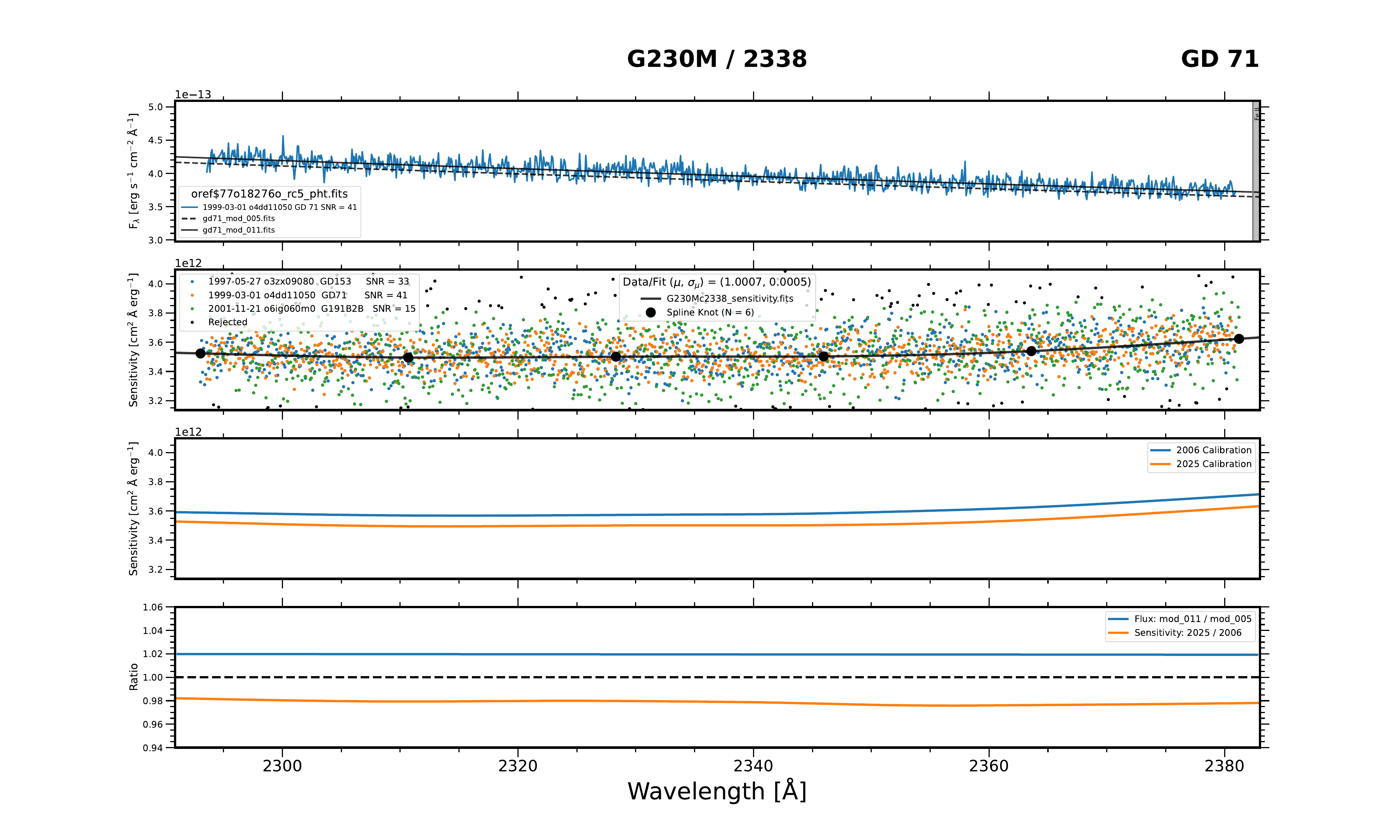}
  \footnotesize
  \caption{Calibration of GD 71 for G230M/2338.}
  \label{fig:G230MC2338b}
\end{figure}
 
\clearpage
\begin{figure}[t]
  \hspace{-0.5in}
  \includegraphics[width=1.1\textwidth]{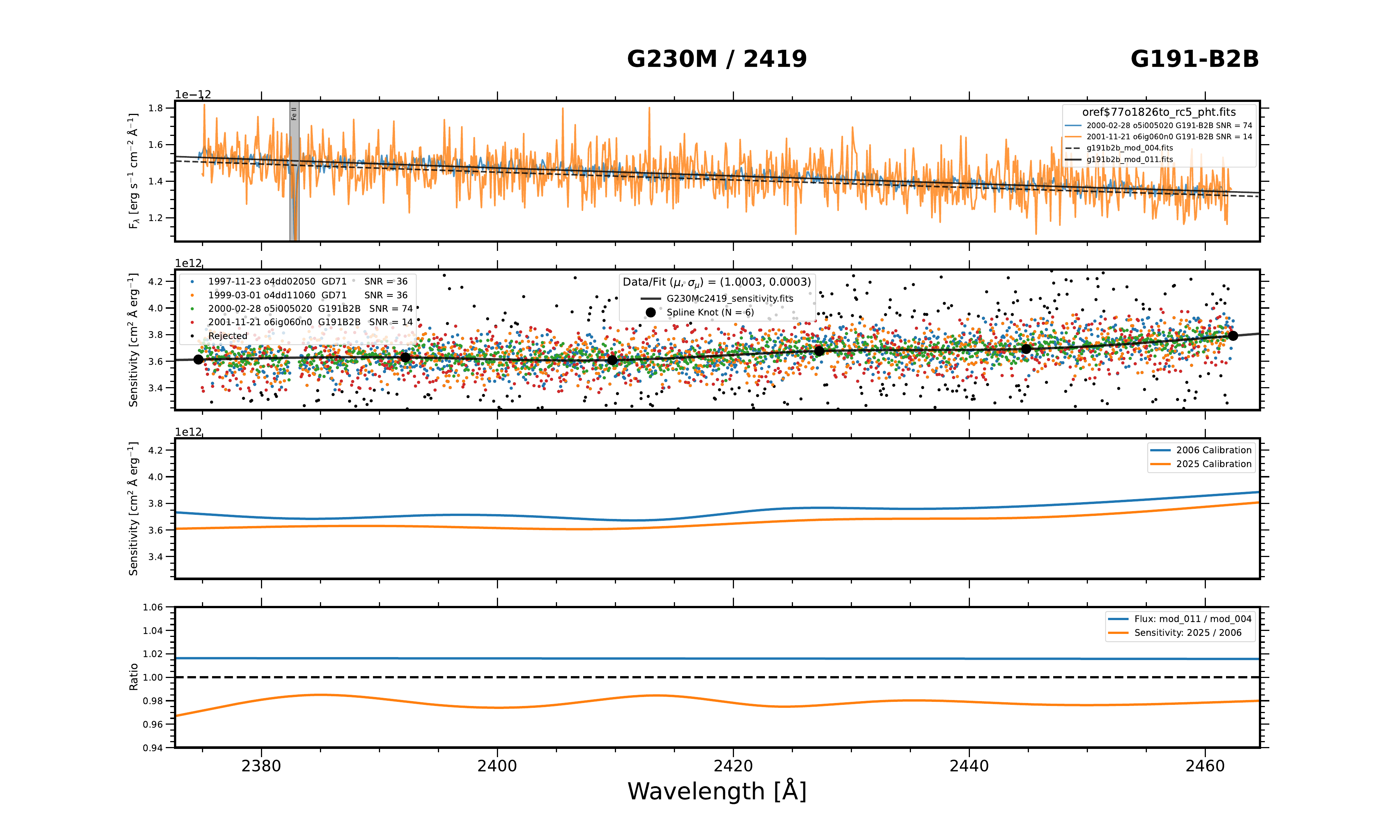}
  \footnotesize
  \caption{Calibration of G191-B2B for G230M/2419.}
  \label{fig:G230MC2419a}
\end{figure}
 
\begin{figure}[b]
  \hspace{-0.5in}
  \includegraphics[width=1.1\textwidth]{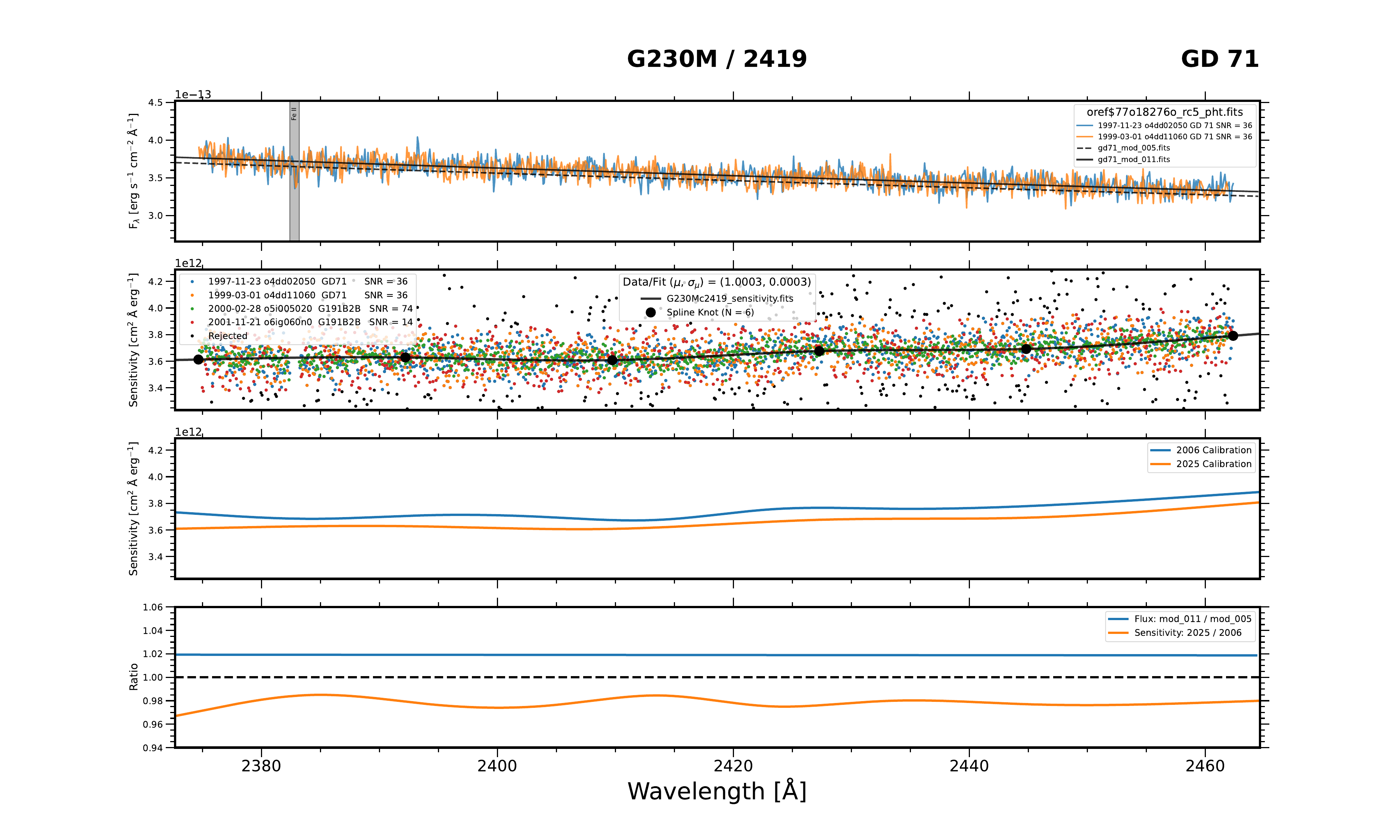}
  \footnotesize
  \caption{Calibration of GD 71 for G230M/2419.}
  \label{fig:G230MC2419b}
\end{figure}
 
\clearpage
\begin{figure}[t]
  \hspace{-0.5in}
  \includegraphics[width=1.1\textwidth]{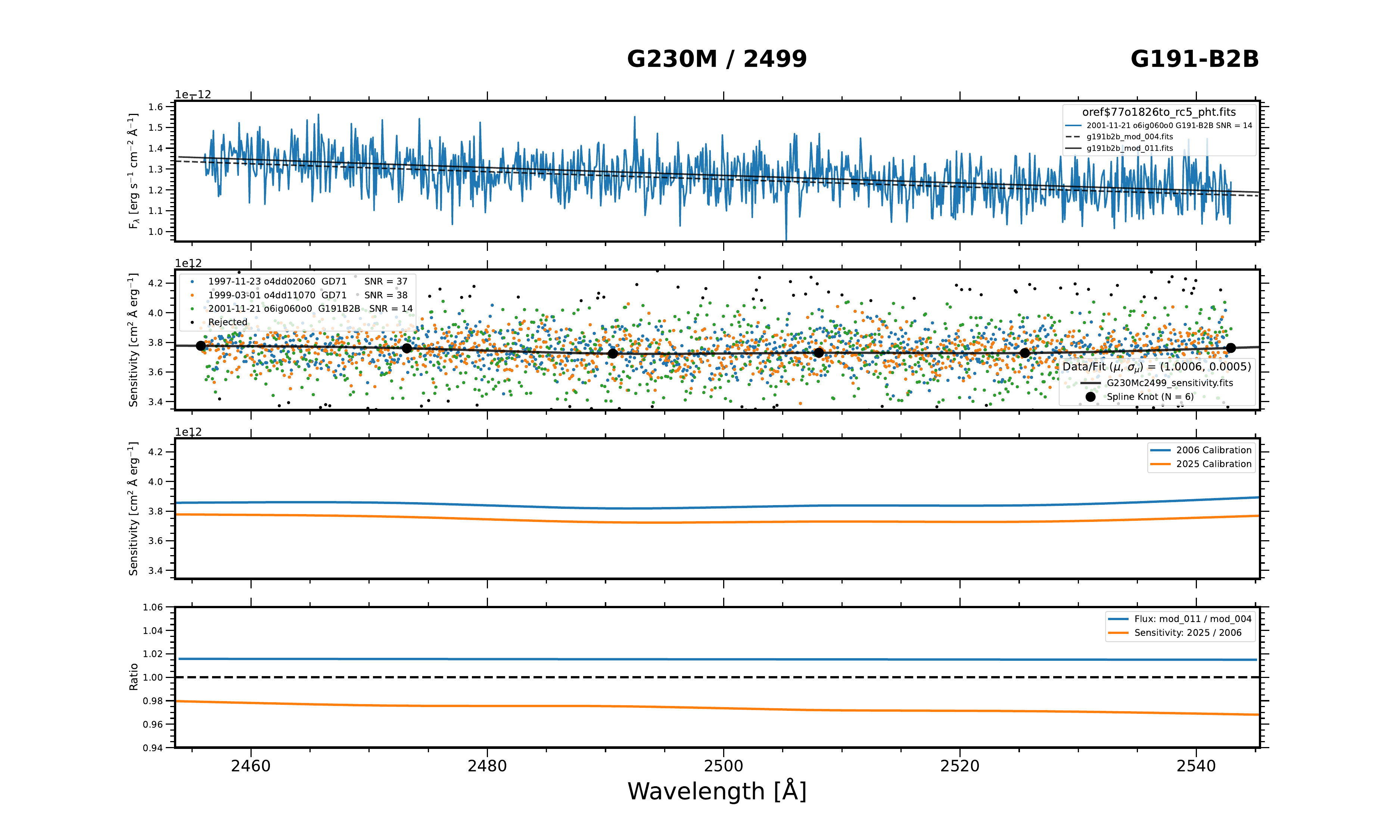}
  \footnotesize
  \caption{Calibration of G191-B2B for G230M/2499.}
  \label{fig:G230MC2499a}
\end{figure}
 
\begin{figure}[b]
  \hspace{-0.5in}
  \includegraphics[width=1.1\textwidth]{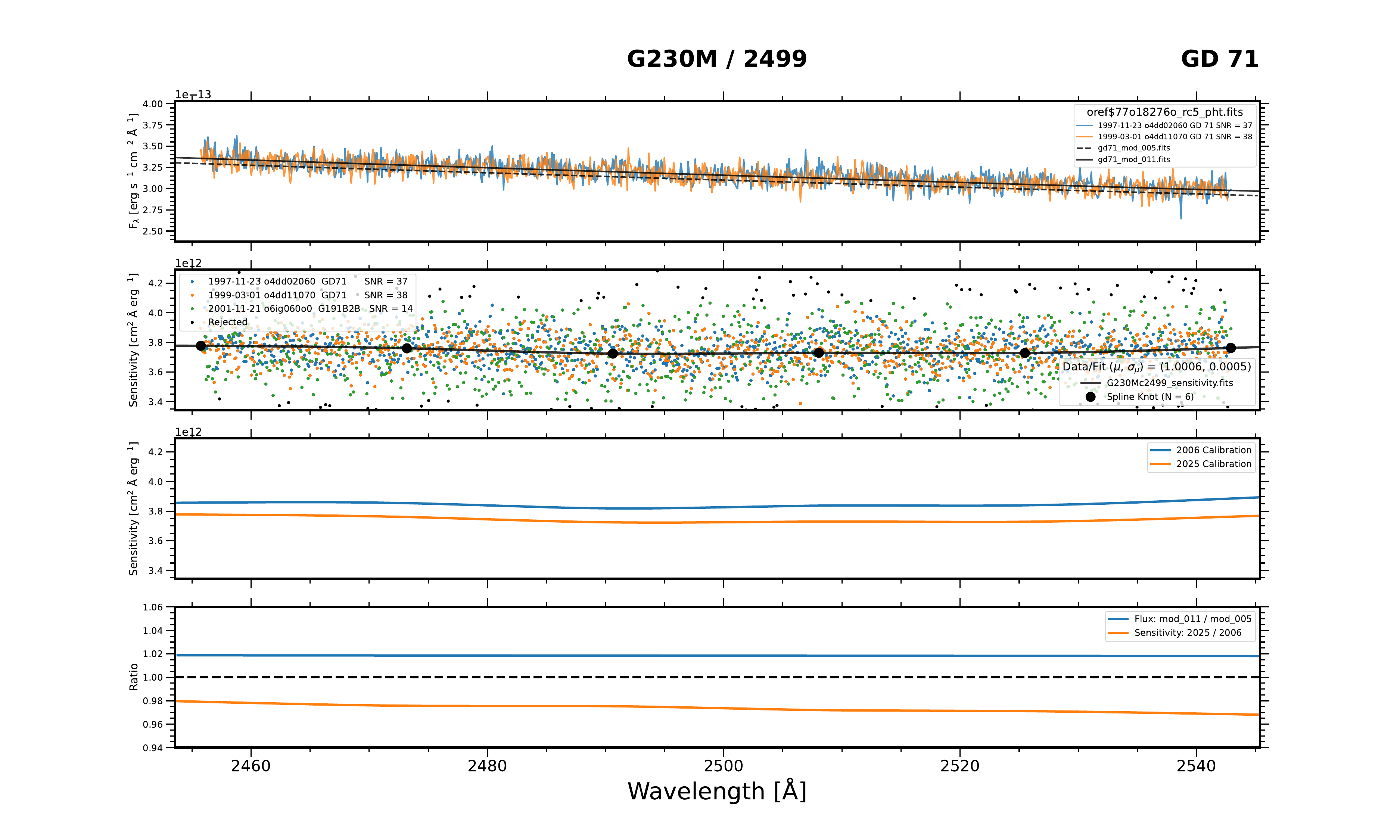}
  \footnotesize
  \caption{Calibration of GD 71 for G230M/2499.}
  \label{fig:G230MC2499b}
\end{figure}
 
\clearpage
\begin{figure}[t]
  \hspace{-0.5in}
  \includegraphics[width=1.1\textwidth]{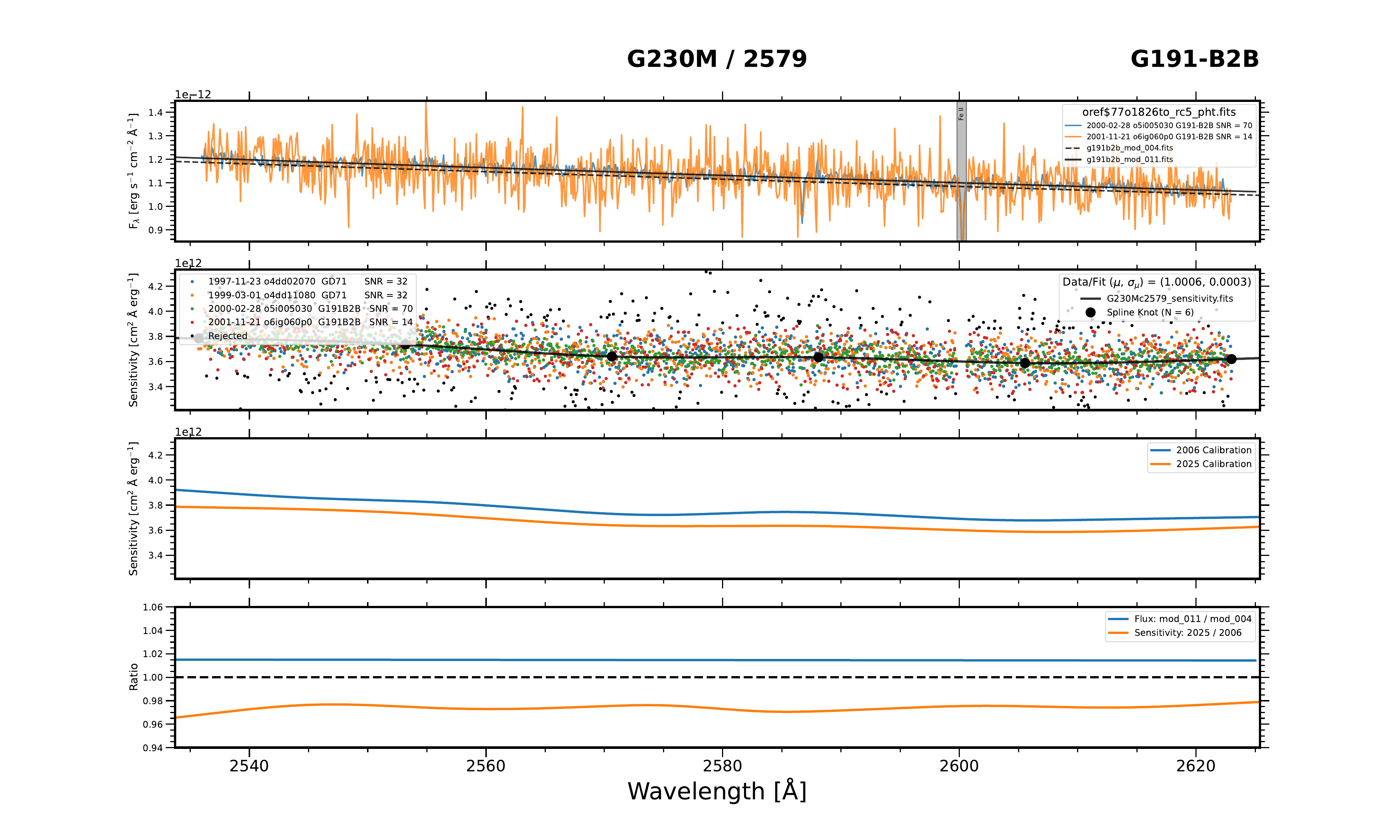}
  \footnotesize
  \caption{Calibration of G191-B2B for G230M/2579.}
  \label{fig:G230MC2579a}
\end{figure}
 
\begin{figure}[b]
  \hspace{-0.5in}
  \includegraphics[width=1.1\textwidth]{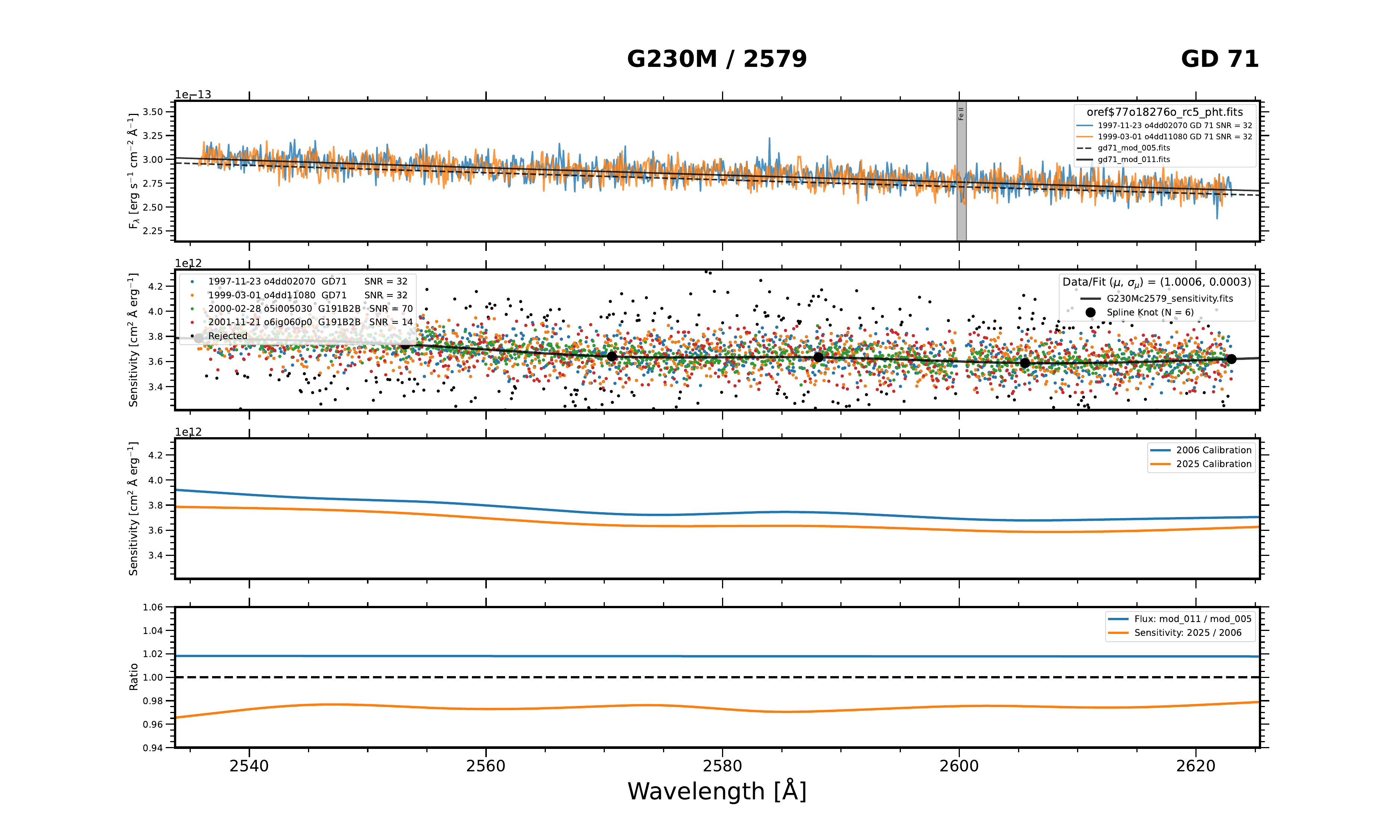}
  \footnotesize
  \caption{Calibration of GD 71 for G230M/2579.}
  \label{fig:G230MC2579b}
\end{figure}
 
\clearpage
\begin{figure}[t]
  \hspace{-0.5in}
  \includegraphics[width=1.1\textwidth]{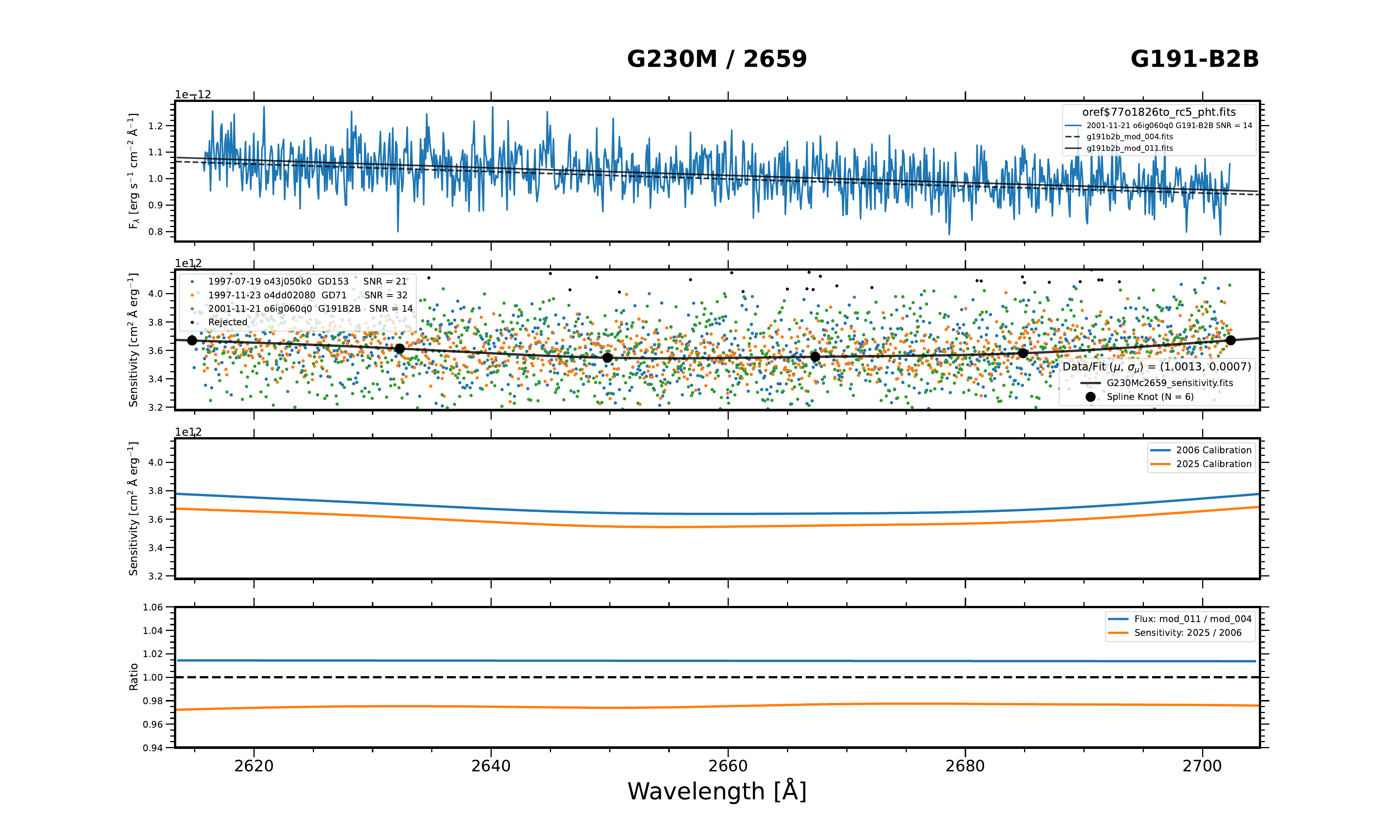}
  \footnotesize
  \caption{Calibration of G191-B2B for G230M/2659.}
  \label{fig:G230MC2659a}
\end{figure}
 
\begin{figure}[b]
  \hspace{-0.5in}
  \includegraphics[width=1.1\textwidth]{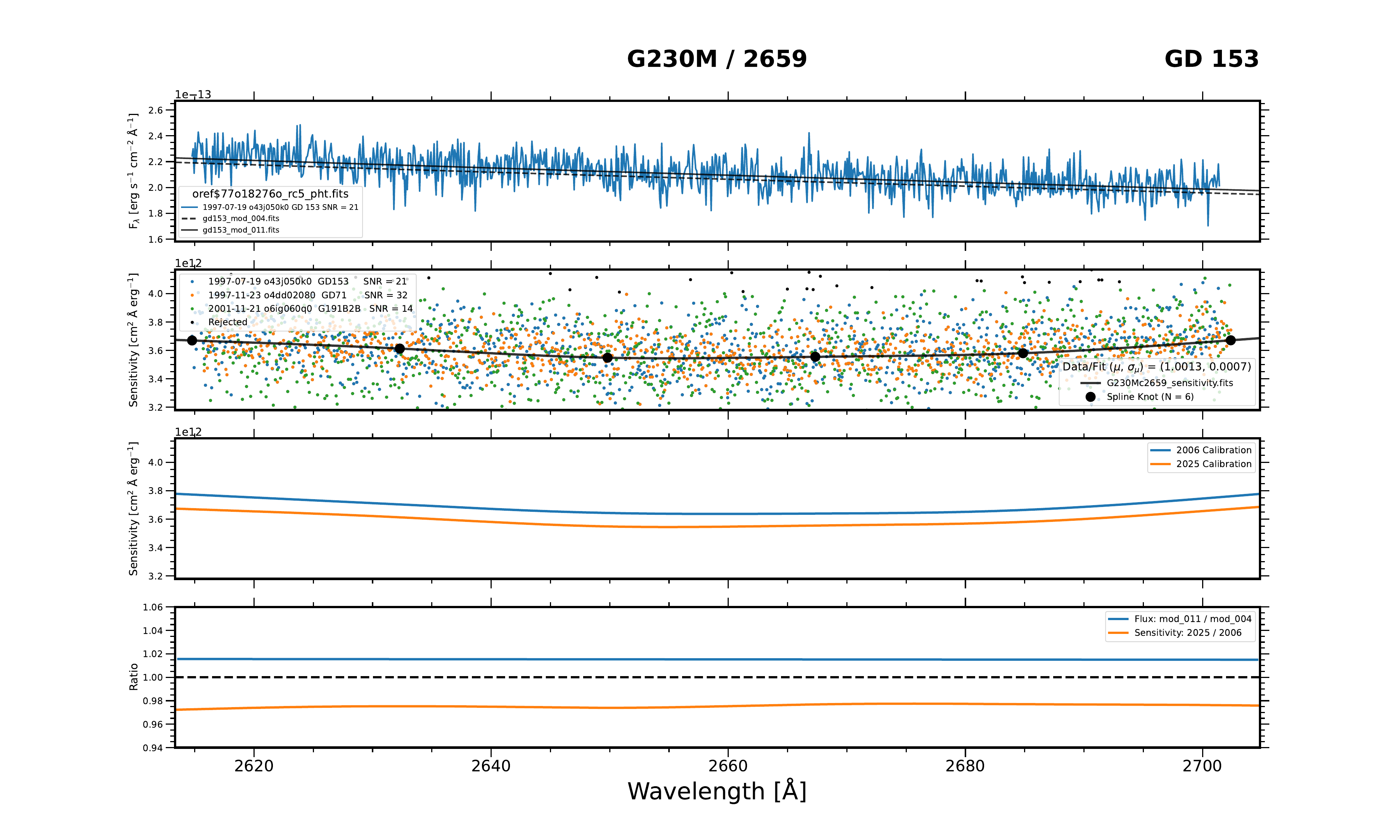}
  \footnotesize
  \caption{Calibration of GD 153 for G230M/2659.}
  \label{fig:G230MC2659c}
\end{figure}
 
\clearpage
\begin{figure}[t]
  \hspace{-0.5in}
  \includegraphics[width=1.1\textwidth]{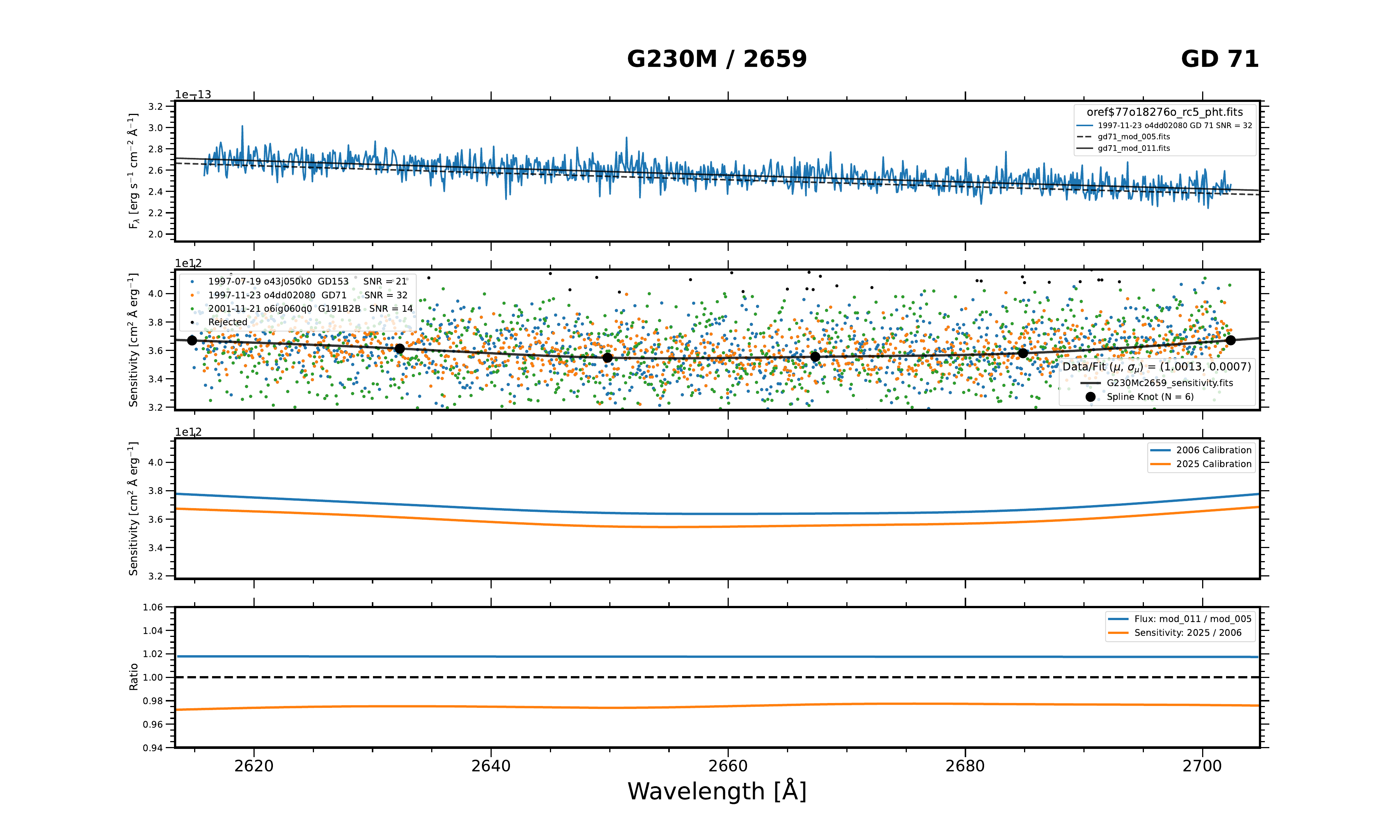}
  \footnotesize
  \caption{Calibration of GD 71 for G230M/2659.}
  \label{fig:G230MC2659b}
\end{figure}
 
\begin{figure}[b]
  \hspace{-0.5in}
  \includegraphics[width=1.1\textwidth]{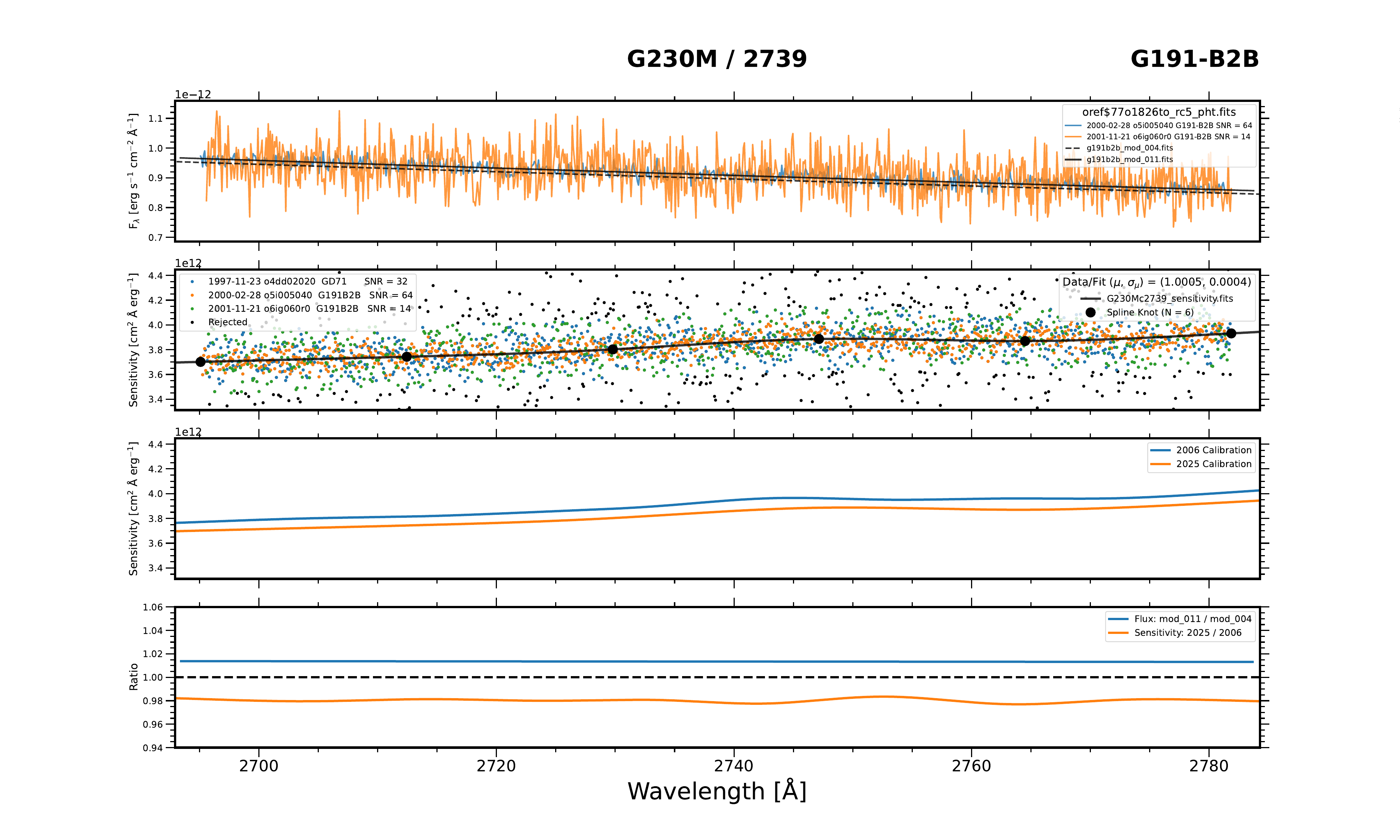}
  \footnotesize
  \caption{Calibration of G191-B2B for G230M/2739.}
  \label{fig:G230MC2739a}
\end{figure}
 
\clearpage
\begin{figure}[t]
  \hspace{-0.5in}
  \includegraphics[width=1.1\textwidth]{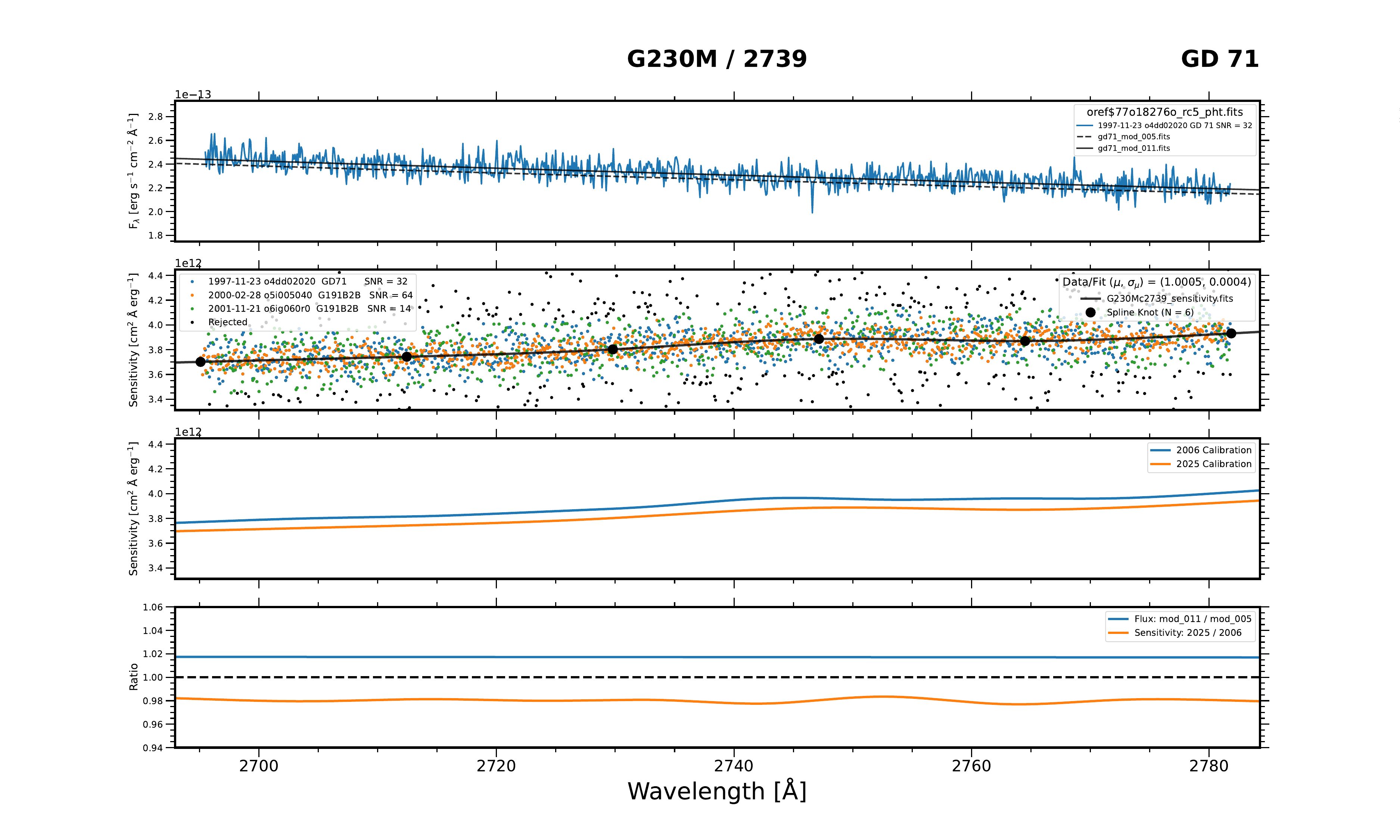}
  \footnotesize
  \caption{Calibration of GD 71 for G230M/2739.}
  \label{fig:G230MC2739b}
\end{figure}
 
\begin{figure}[b]
  \hspace{-0.5in}
  \includegraphics[width=1.1\textwidth]{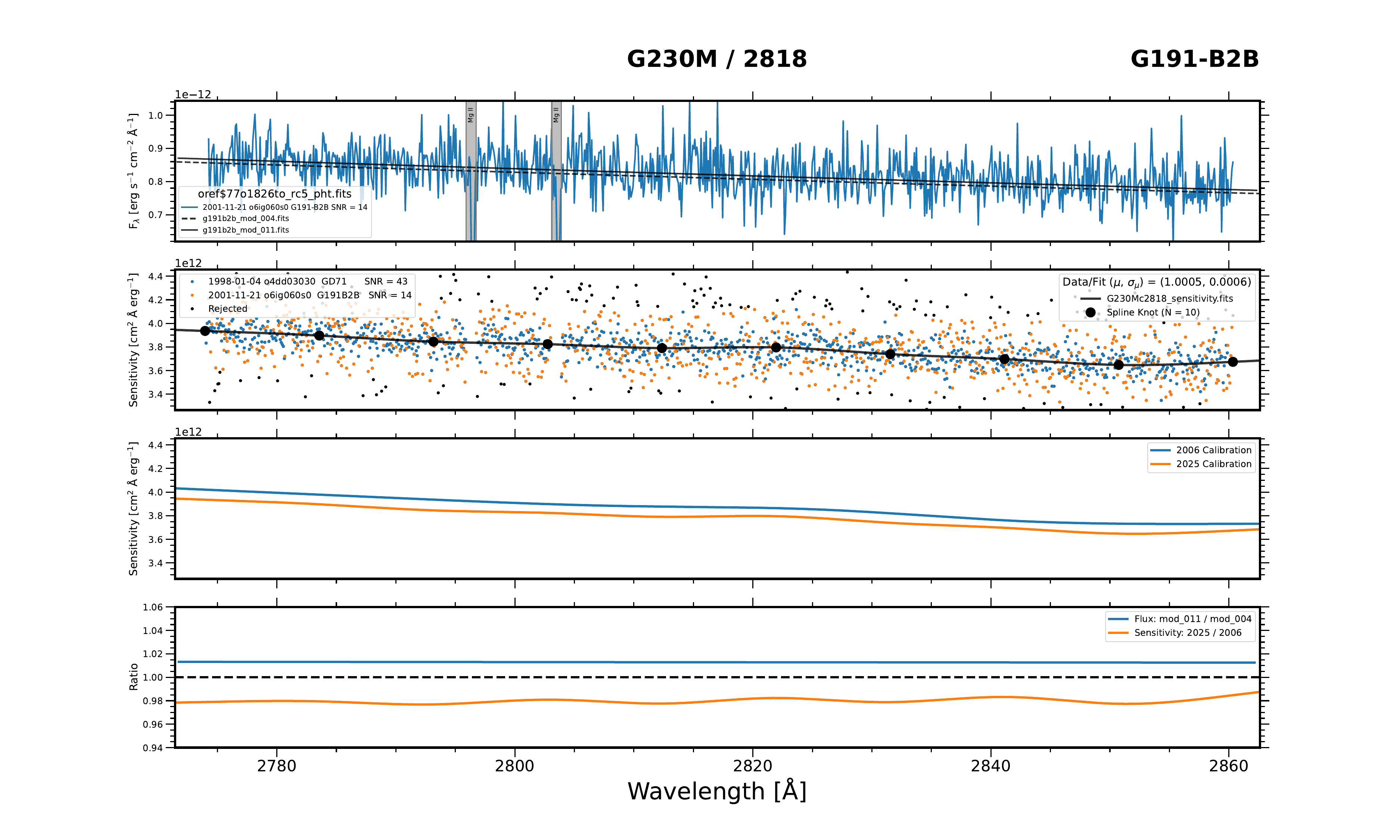}
  \footnotesize
  \caption{Calibration of G191-B2B for G230M/2818.}
  \label{fig:G230MC2818a}
\end{figure}
 
\clearpage
\begin{figure}[t]
  \hspace{-0.5in}
  \includegraphics[width=1.1\textwidth]{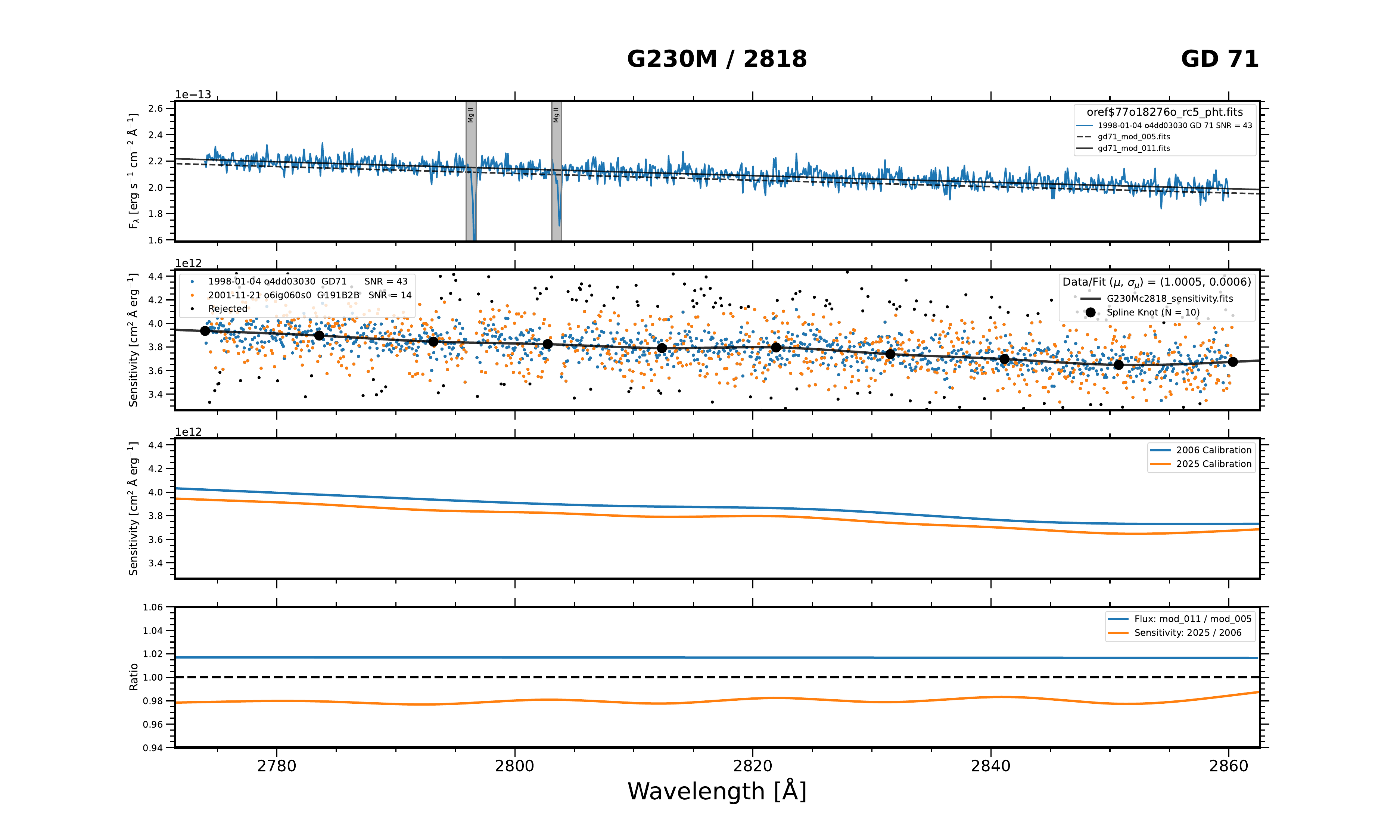}
  \footnotesize
  \caption{Calibration of GD 71 for G230M/2818.}
  \label{fig:G230MC2818b}
\end{figure}
 
\begin{figure}[b]
  \hspace{-0.5in}
  \includegraphics[width=1.1\textwidth]{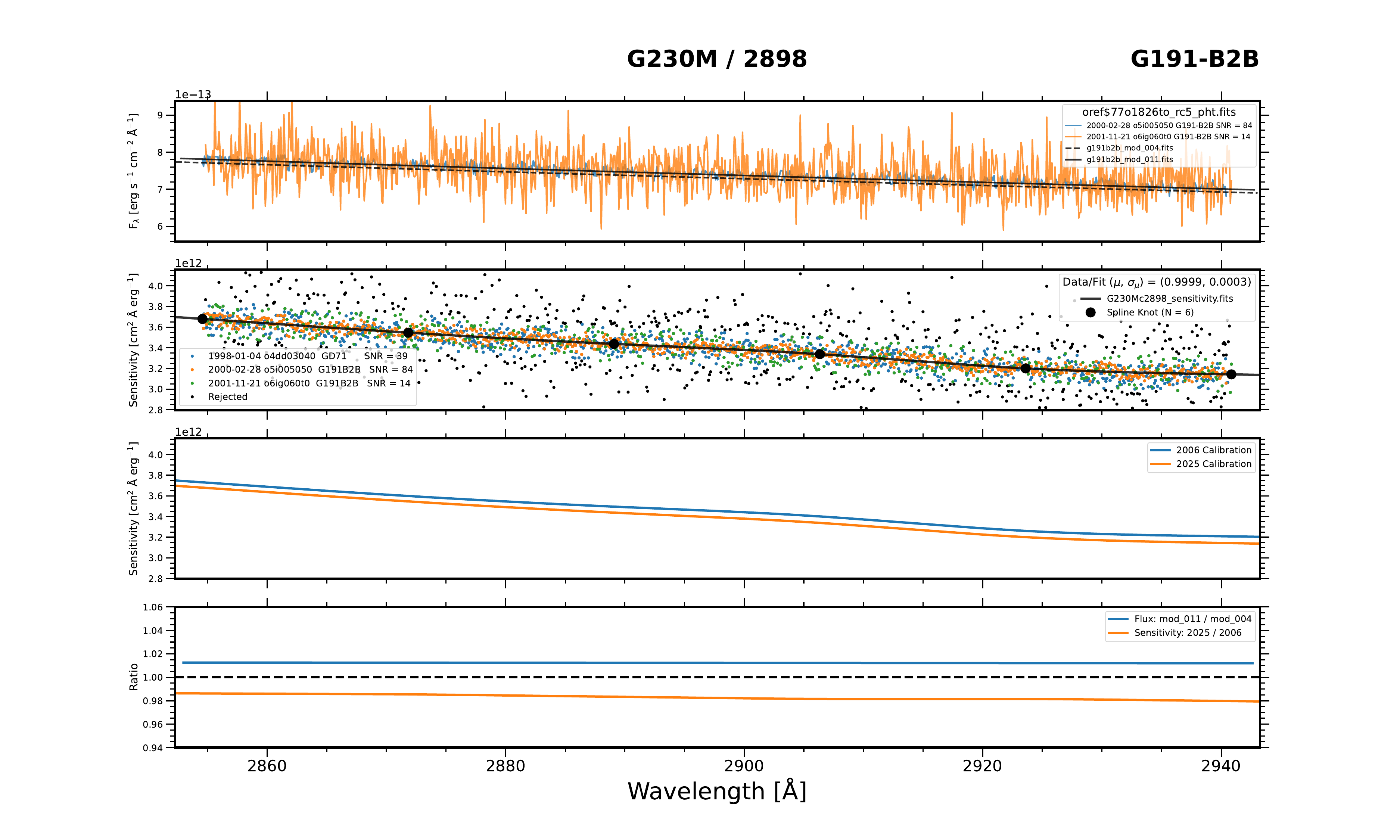}
  \footnotesize
  \caption{Calibration of G191-B2B for G230M/2898.}
  \label{fig:G230MC2898a}
\end{figure}
 
\clearpage
\begin{figure}[t]
  \hspace{-0.5in}
  \includegraphics[width=1.1\textwidth]{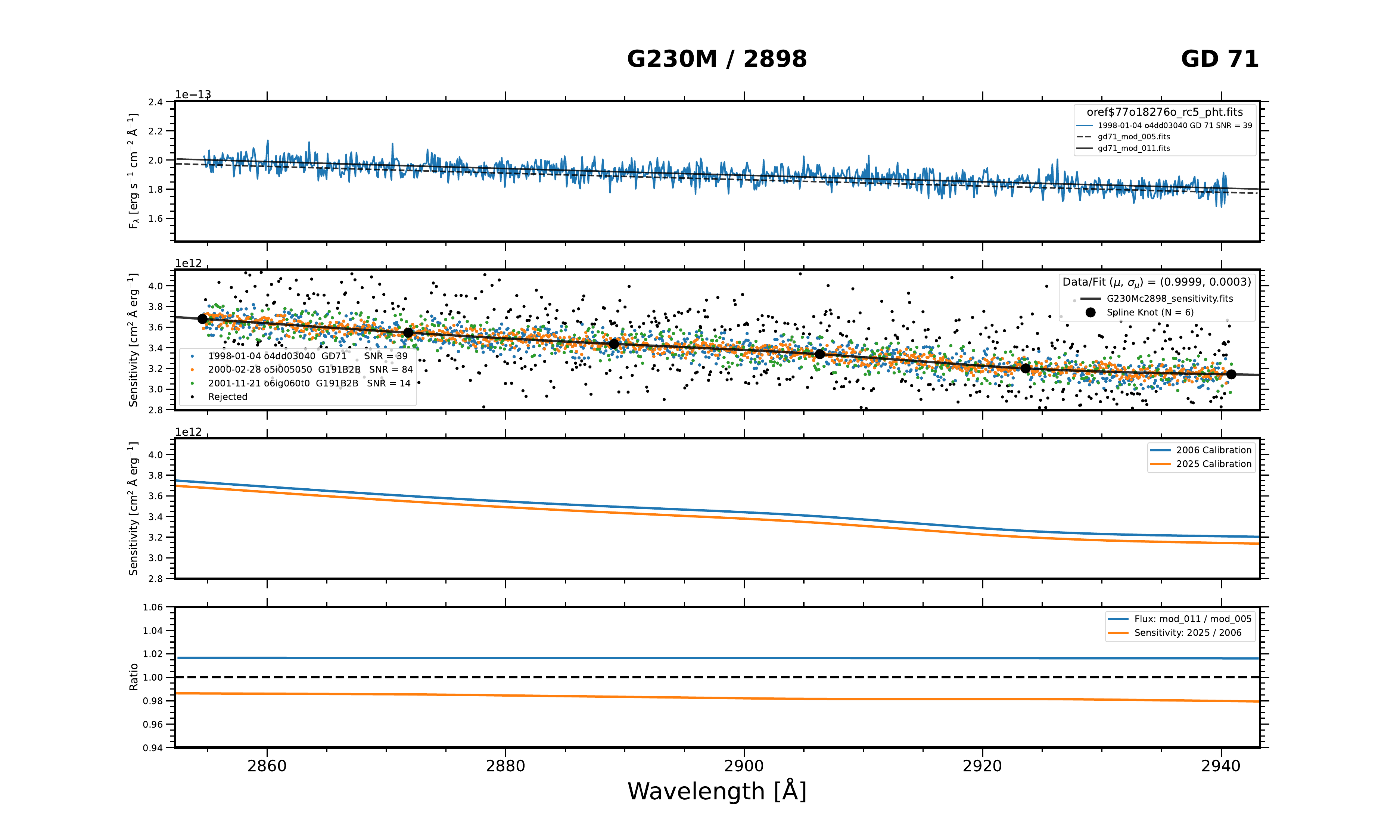}
  \footnotesize
  \caption{Calibration of GD 71 for G230M/2898.}
  \label{fig:G230MC2898b}
\end{figure}
 
\begin{figure}[b]
  \hspace{-0.5in}
  \includegraphics[width=1.1\textwidth]{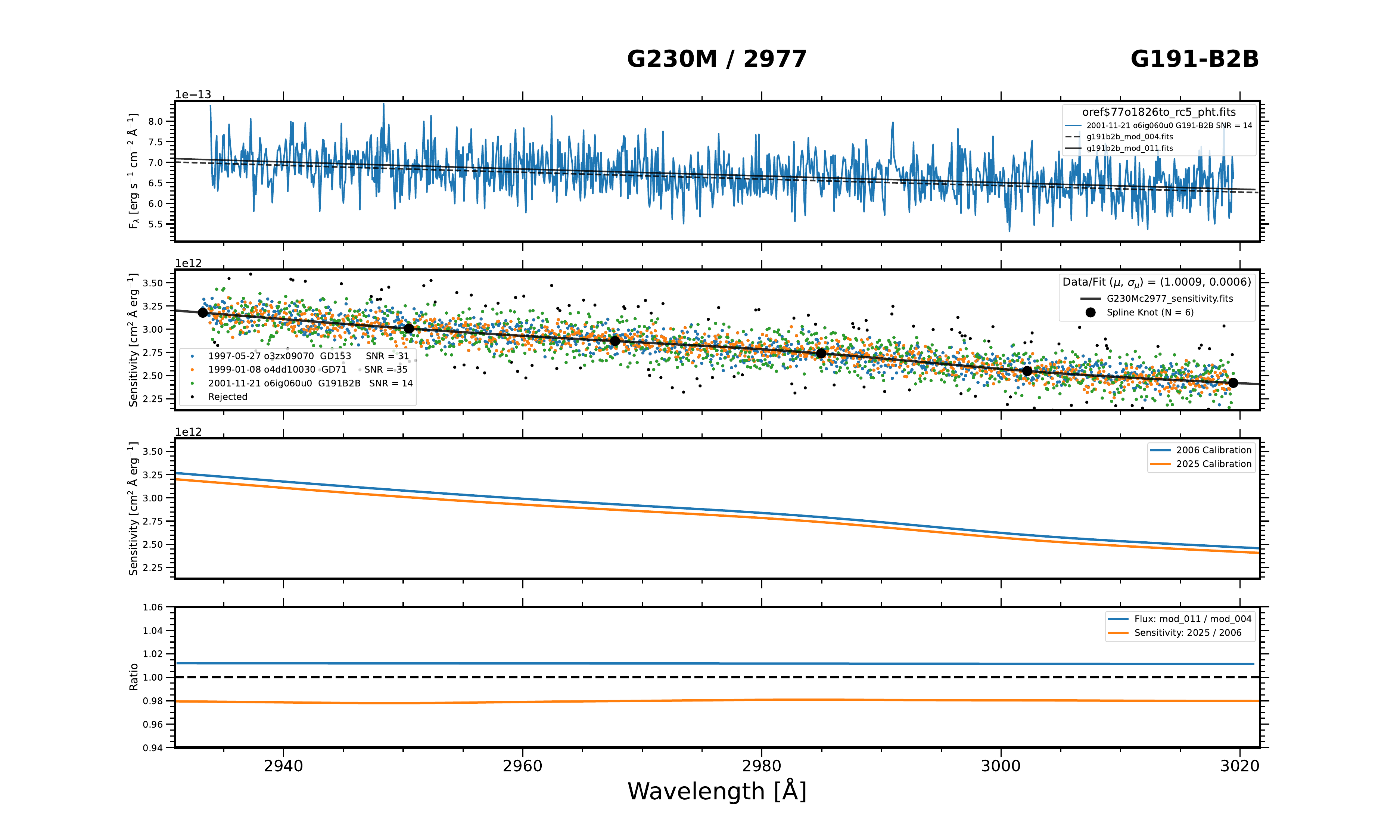}
  \footnotesize
  \caption{Calibration of G191-B2B for G230M/2977.}
  \label{fig:G230MC2977a}
\end{figure}
 
\clearpage
\begin{figure}[t]
  \hspace{-0.5in}
  \includegraphics[width=1.1\textwidth]{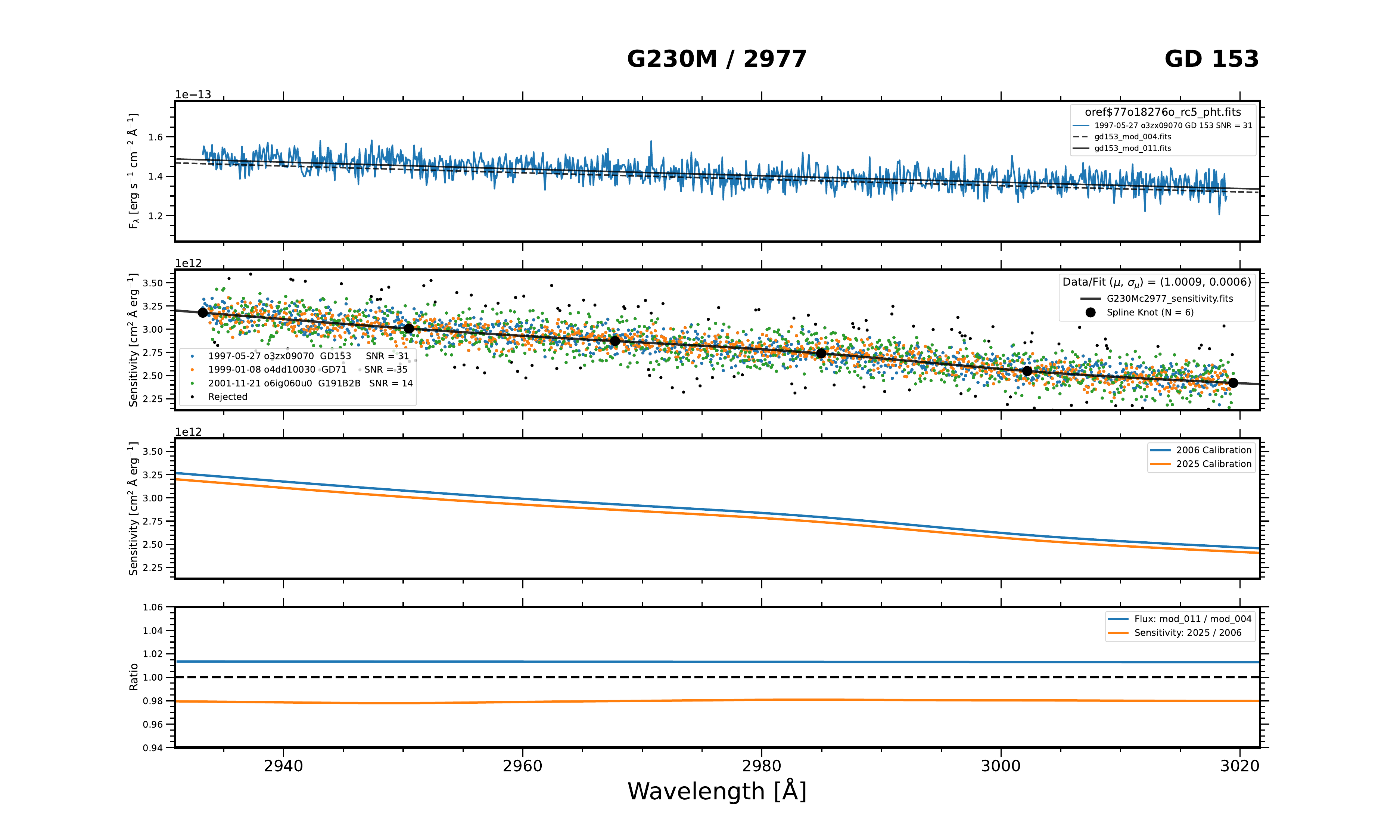}
  \footnotesize
  \caption{Calibration of GD 153 for G230M/2977.}
  \label{fig:G230MC2977c}
\end{figure}
 
\begin{figure}[b]
  \hspace{-0.5in}
  \includegraphics[width=1.1\textwidth]{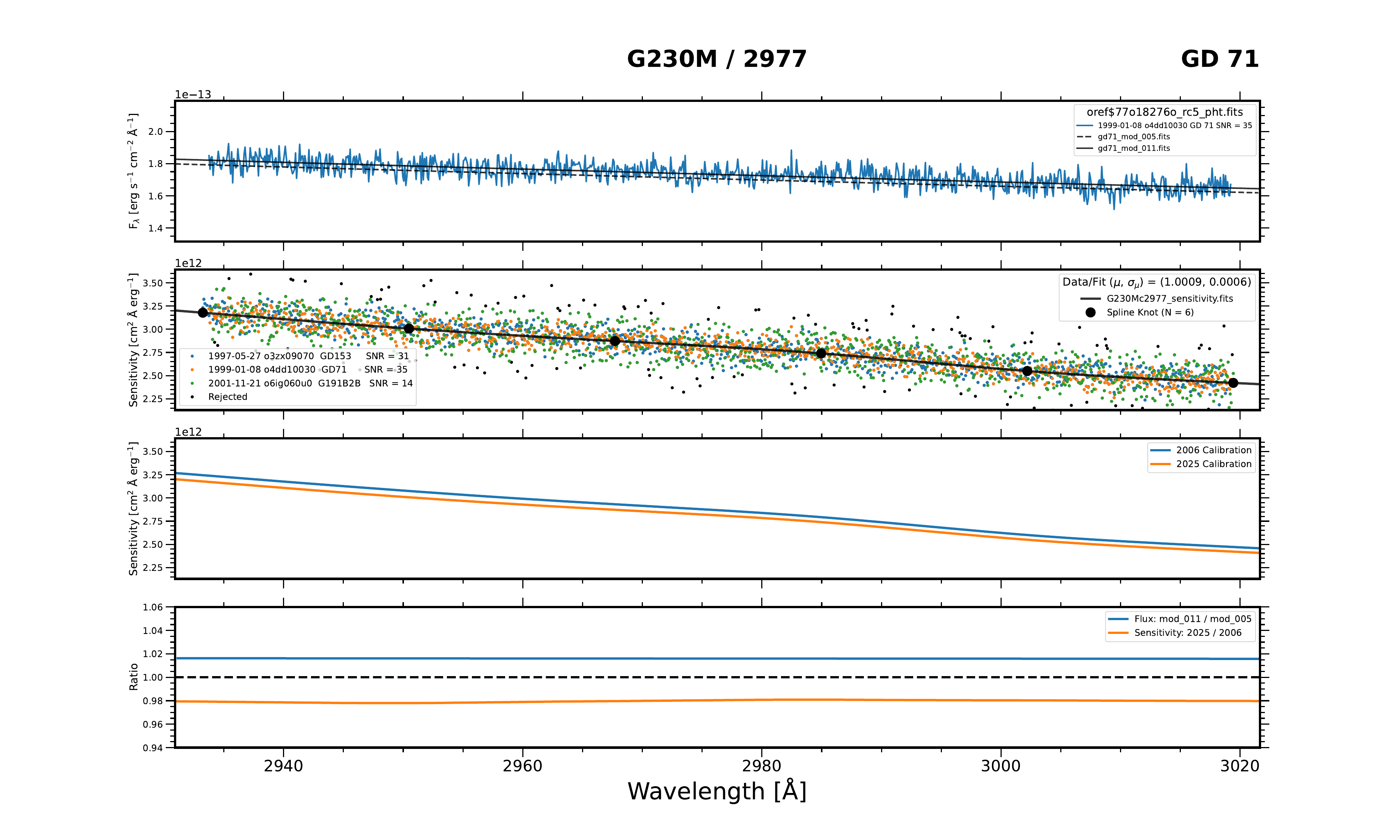}
  \footnotesize
  \caption{Calibration of GD 71 for G230M/2977.}
  \label{fig:G230MC2977b}
\end{figure}
 
\clearpage
\begin{figure}[t]
  \hspace{-0.5in}
  \includegraphics[width=1.1\textwidth]{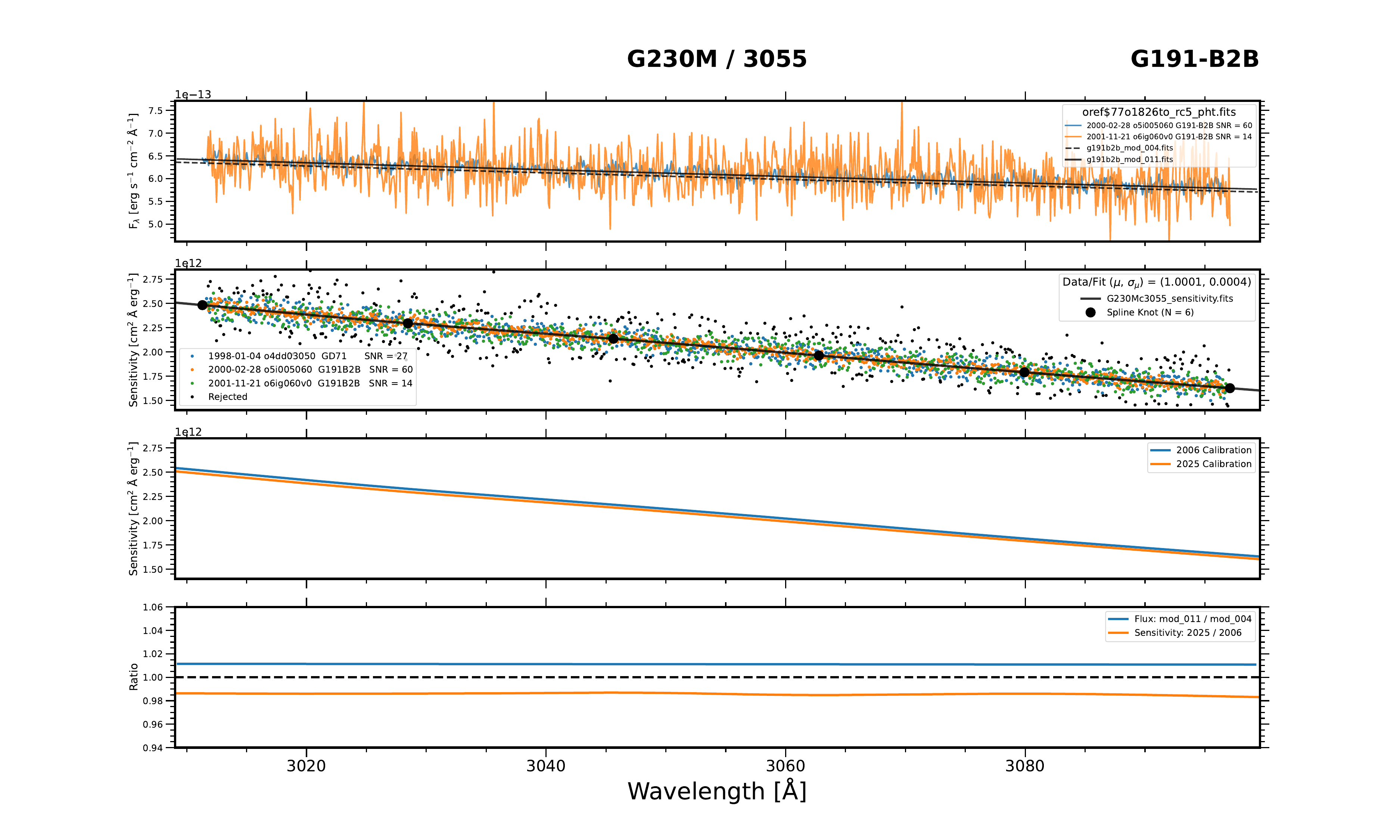}
  \footnotesize
  \caption{Calibration of G191-B2B for G230M/3055.}
  \label{fig:G230MC3055a}
\end{figure}
 
\begin{figure}[b]
  \hspace{-0.5in}
  \includegraphics[width=1.1\textwidth]{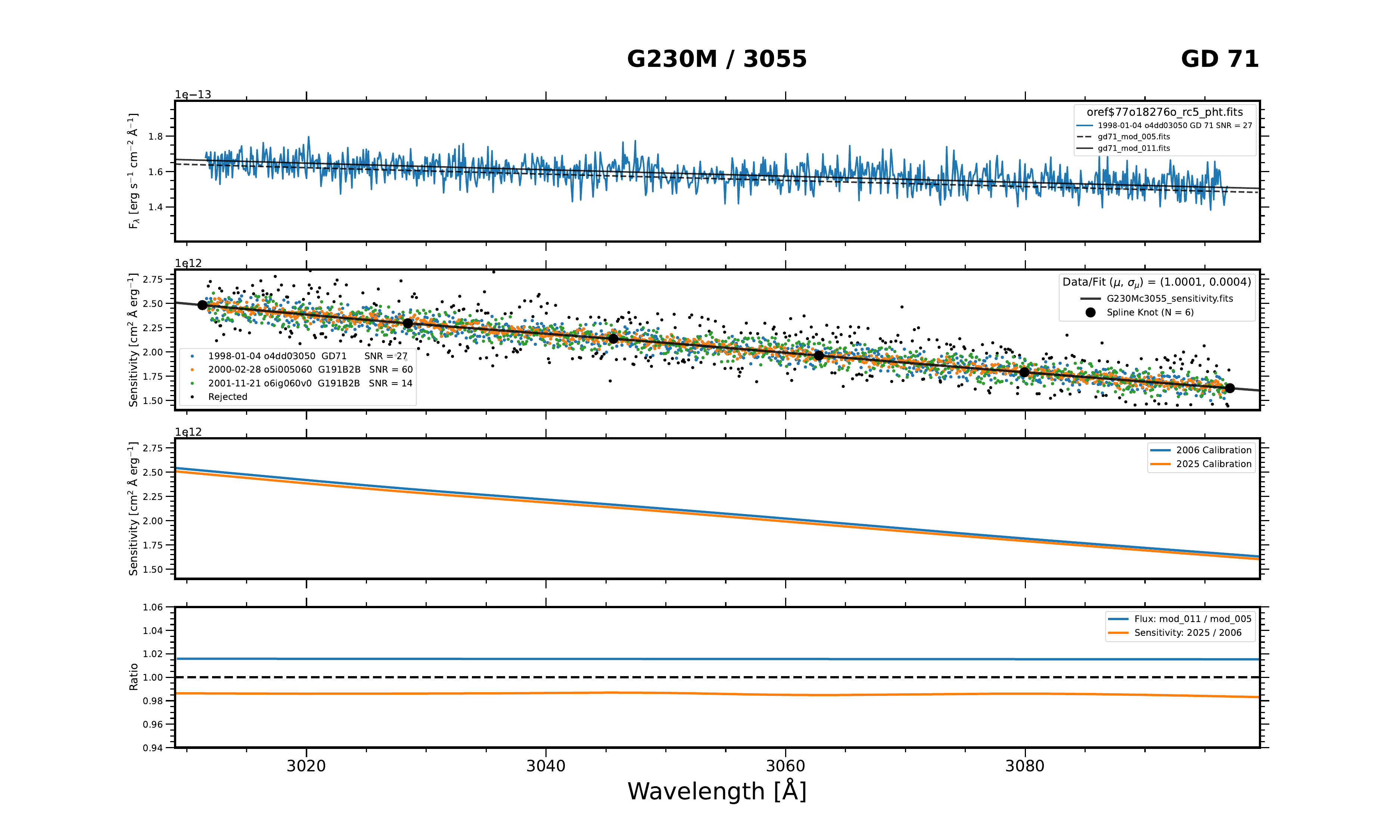}
  \footnotesize
  \caption{Calibration of GD 71 for G230M/3055.}
  \label{fig:G230MC3055b}
\end{figure}
 
\clearpage
\newpage

\begin{figure}[t]
  \hspace{-0.5in}
  \includegraphics[width=1.1\textwidth]{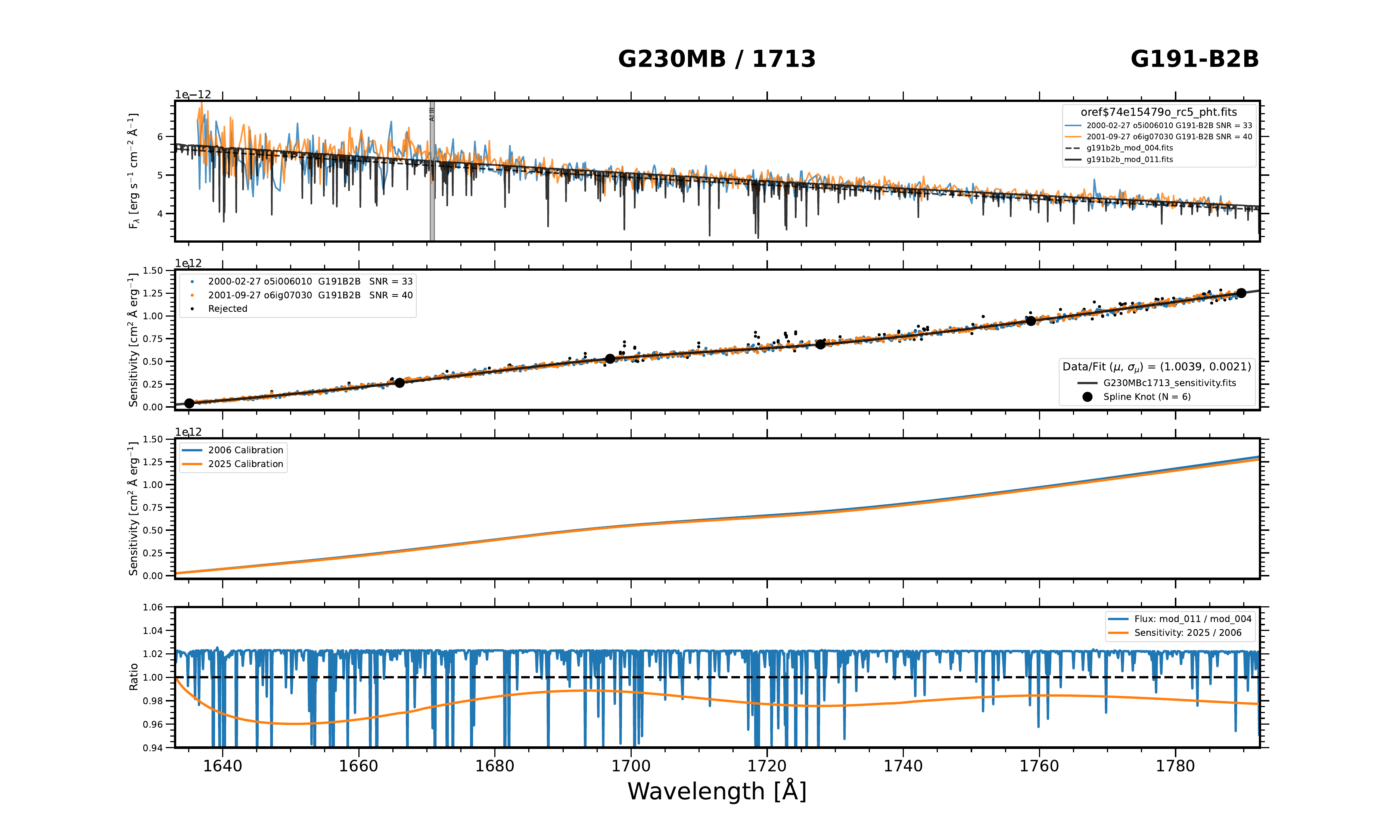}
  \footnotesize
  \caption{Calibration of G191-B2B for G230MB/1713.}
  \label{fig:G230MBC1713a}
\end{figure}
 
\begin{figure}[b]
  \hspace{-0.5in}
  \includegraphics[width=1.1\textwidth]{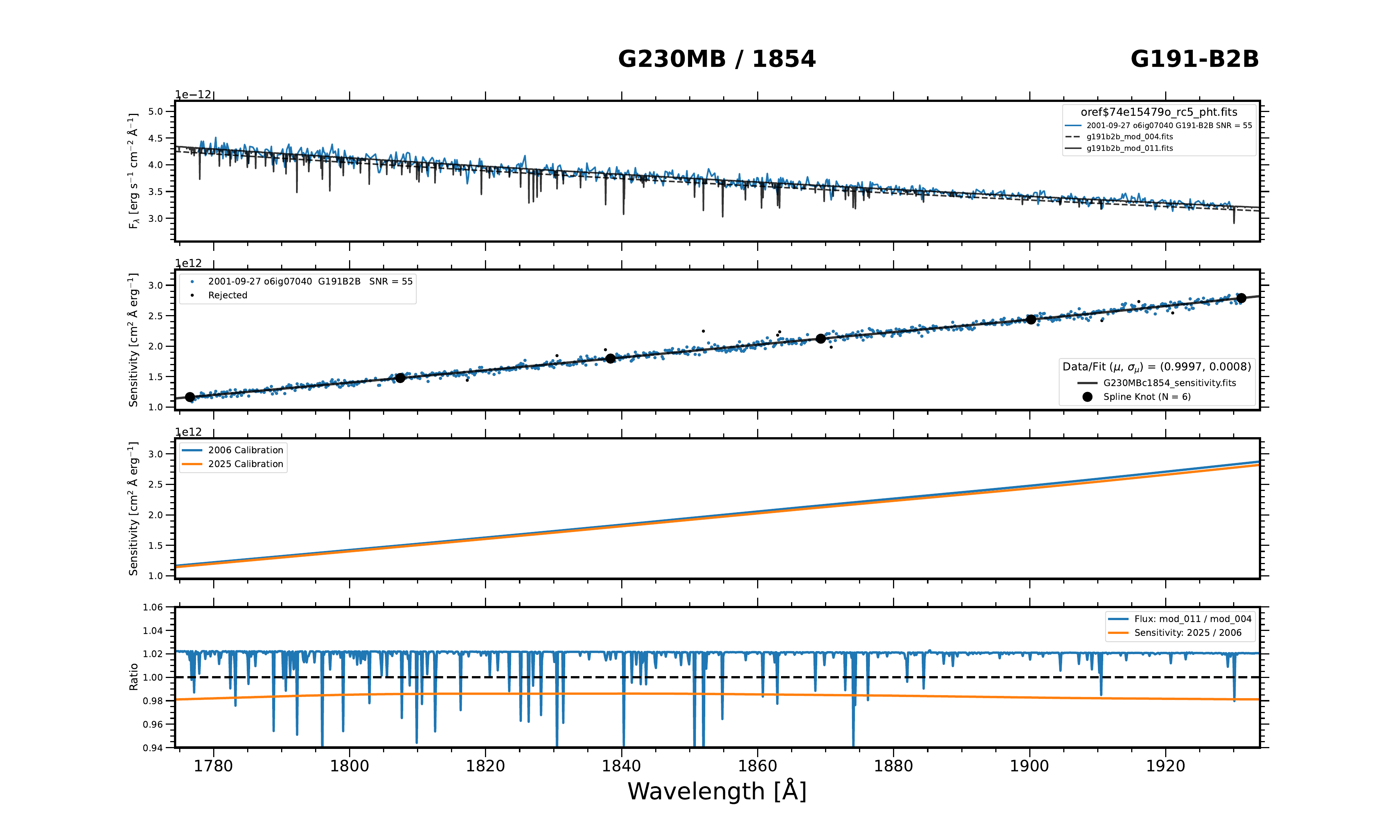}
  \footnotesize
  \caption{Calibration of G191-B2B for G230MB/1854.}
  \label{fig:G230MBC1854a}
\end{figure}
 
\clearpage
\begin{figure}[t]
  \hspace{-0.5in}
  \includegraphics[width=1.1\textwidth]{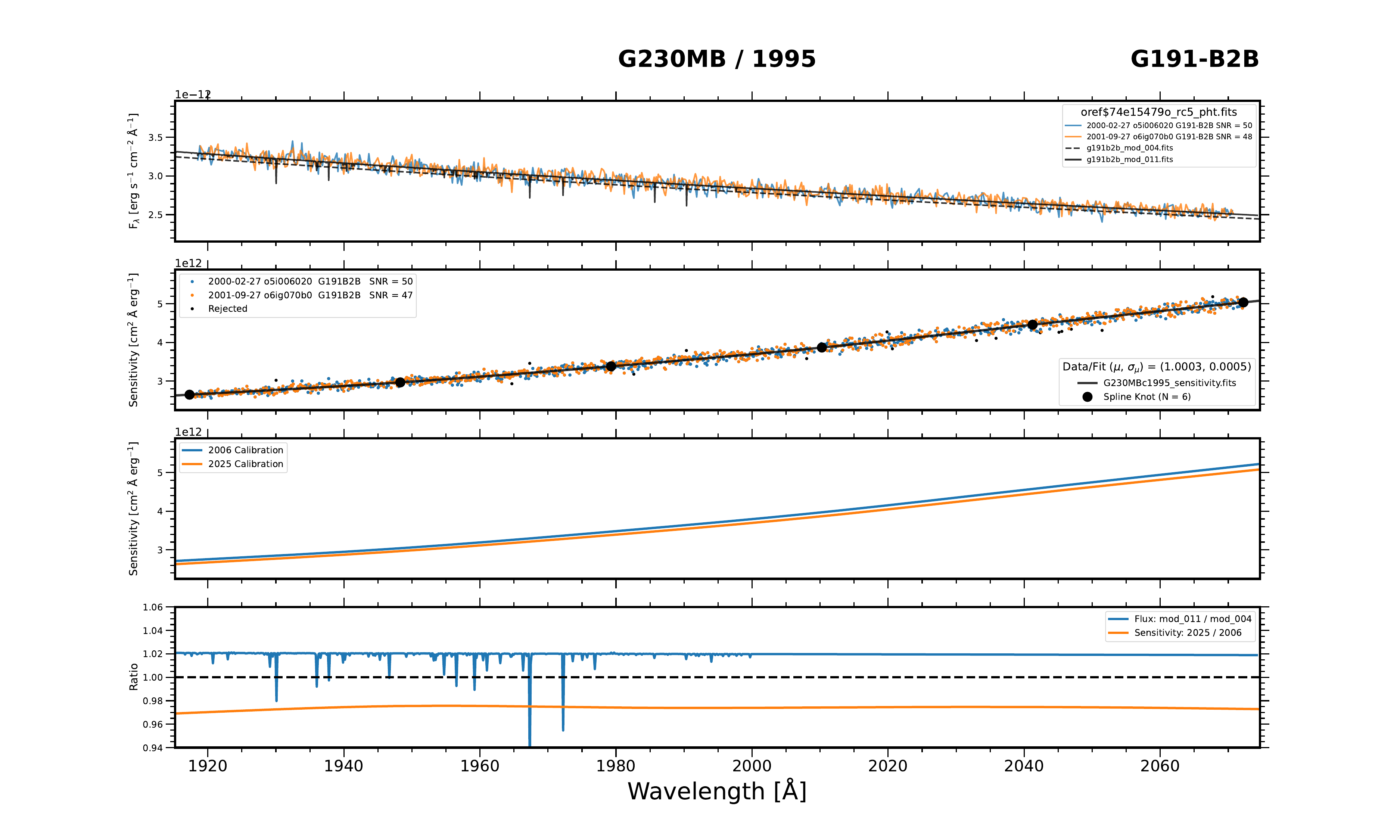}
  \footnotesize
  \caption{Calibration of G191-B2B for G230MB/1995.}
  \label{fig:G230MBC1995a}
\end{figure}
 
\begin{figure}[b]
  \hspace{-0.5in}
  \includegraphics[width=1.1\textwidth]{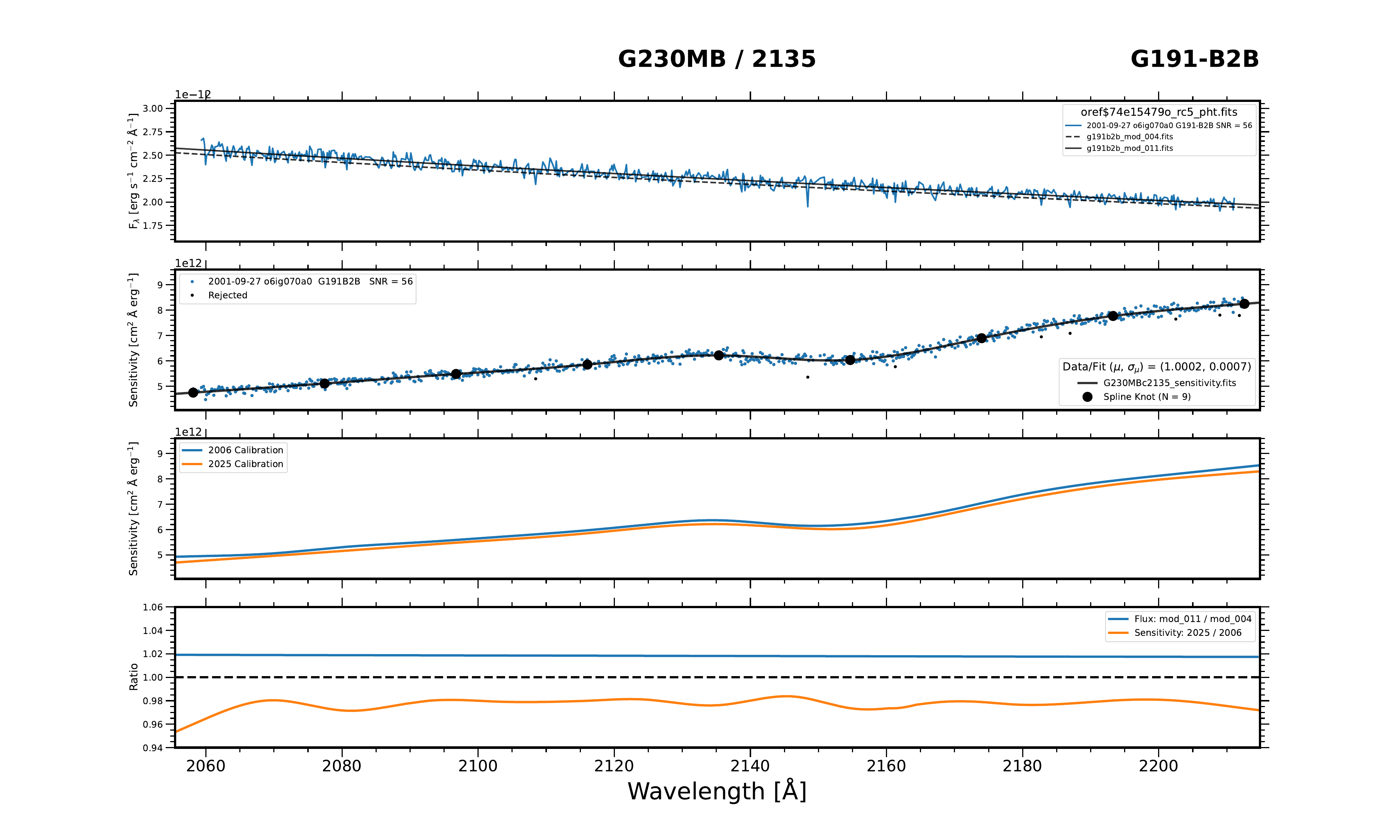}
  \footnotesize
  \caption{Calibration of G191-B2B for G230MB/2135.}
  \label{fig:G230MBC2135a}
\end{figure}
 
\clearpage
\begin{figure}[t]
  \hspace{-0.5in}
  \includegraphics[width=1.1\textwidth]{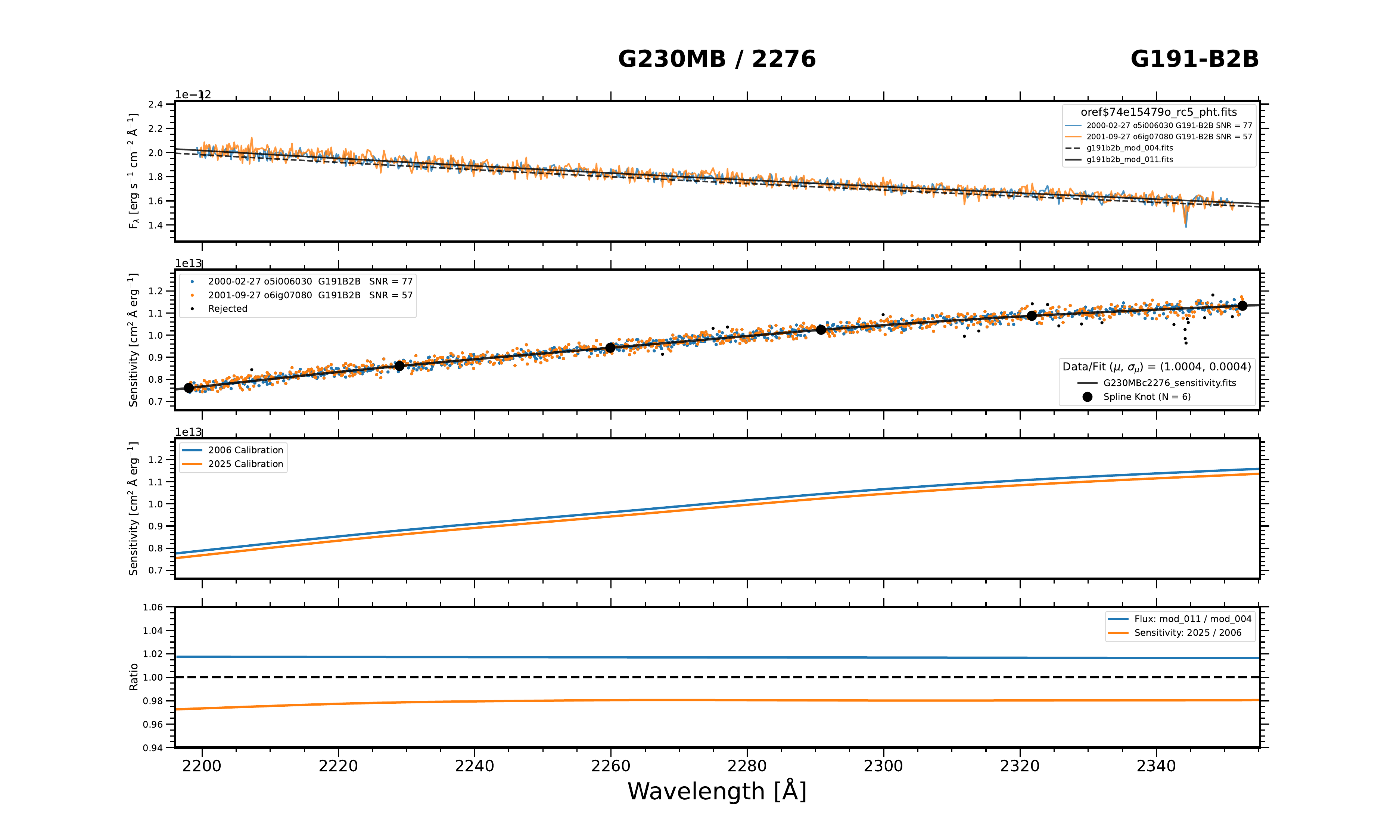}
  \footnotesize
  \caption{Calibration of G191-B2B for G230MB/2276.}
  \label{fig:G230MBC2276a}
\end{figure}
 
\begin{figure}[b]
  \hspace{-0.5in}
  \includegraphics[width=1.1\textwidth]{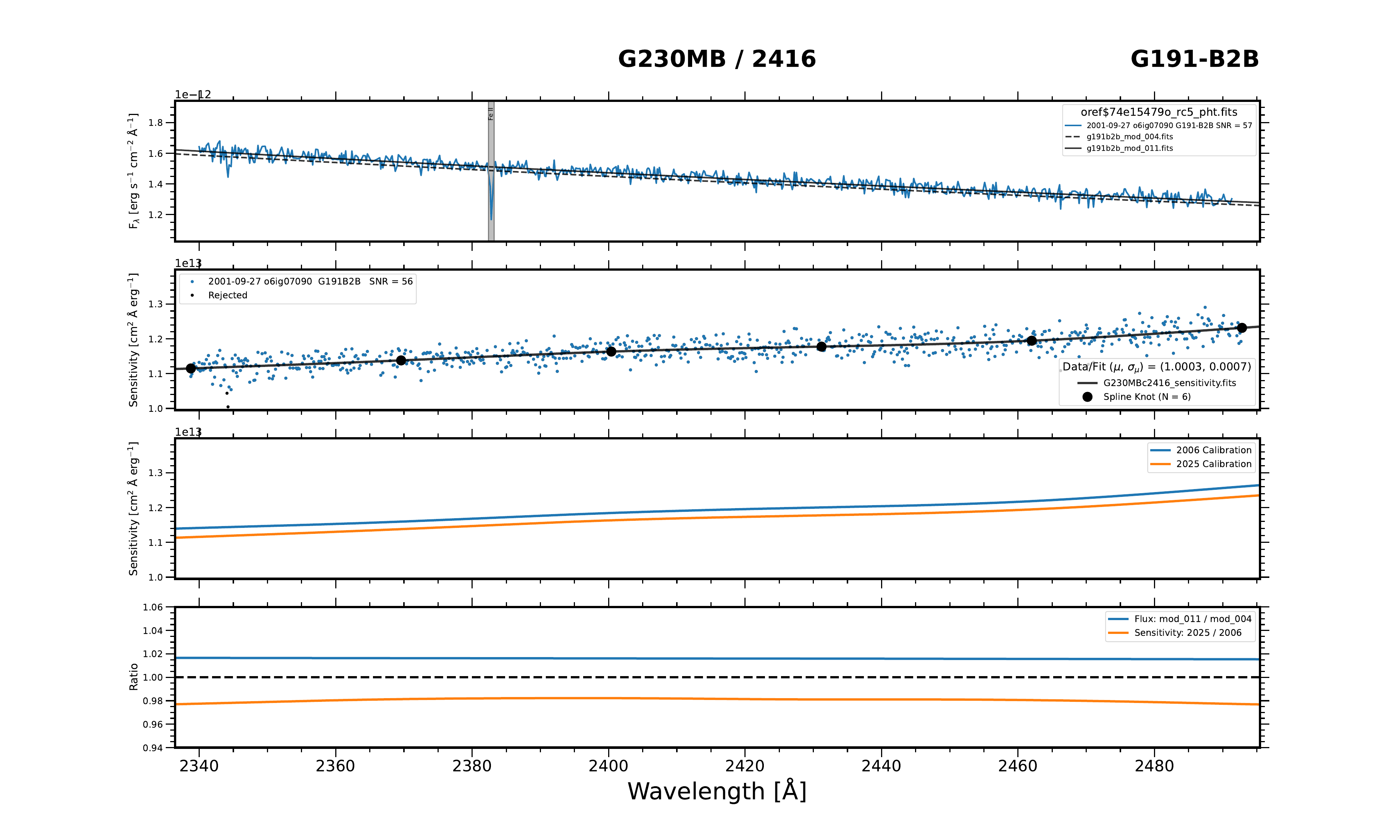}
  \footnotesize
  \caption{Calibration of G191-B2B for G230MB/2416.}
  \label{fig:G230MBC2416a}
\end{figure}
 
\clearpage
\begin{figure}[t]
  \hspace{-0.5in}
  \includegraphics[width=1.1\textwidth]{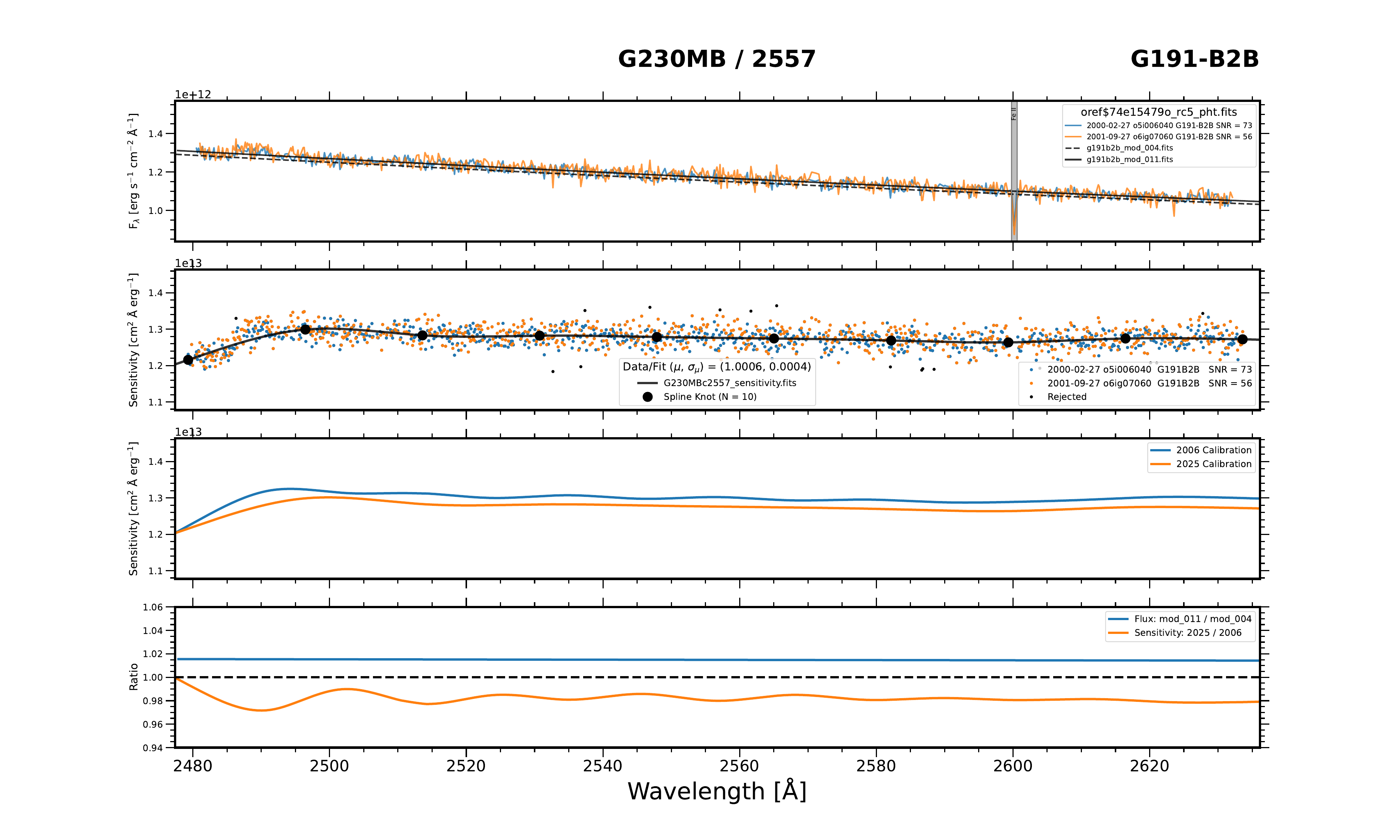}
  \footnotesize
  \caption{Calibration of G191-B2B for G230MB/2557.}
  \label{fig:G230MBC2557a}
\end{figure}
 
\begin{figure}[b]
  \hspace{-0.5in}
  \includegraphics[width=1.1\textwidth]{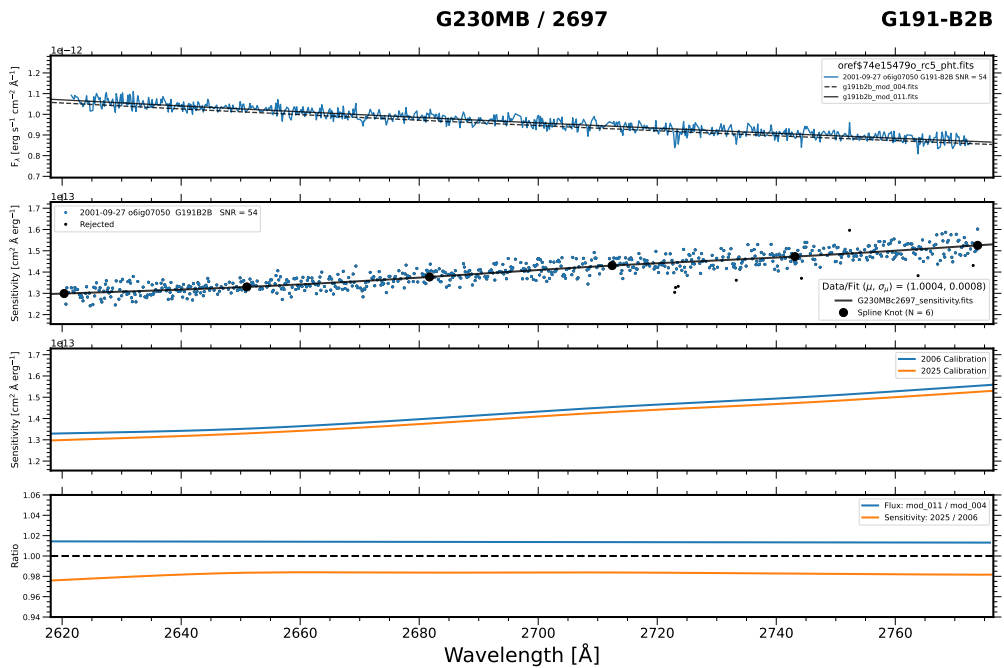}
  \footnotesize
  \caption{Calibration of G191-B2B for G230MB/2697.}
  \label{fig:G230MBC2697a}
\end{figure}
 
\clearpage
\begin{figure}[t]
  \hspace{-0.5in}
  \includegraphics[width=1.1\textwidth]{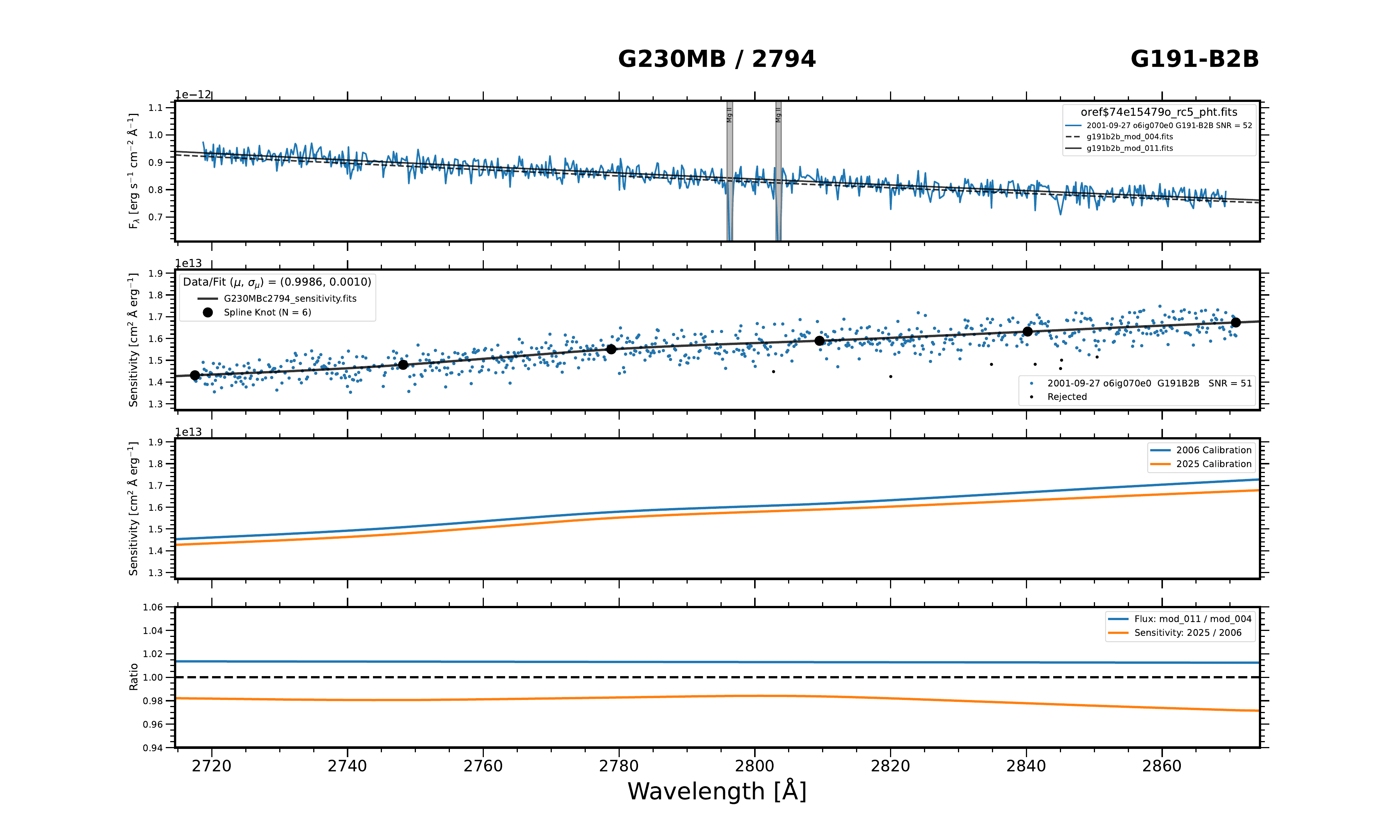}
  \footnotesize
  \caption{Calibration of G191-B2B for G230MB/2794.}
  \label{fig:G230MBC2794a}
\end{figure}
 
\begin{figure}[b]
  \hspace{-0.5in}
  \includegraphics[width=1.1\textwidth]{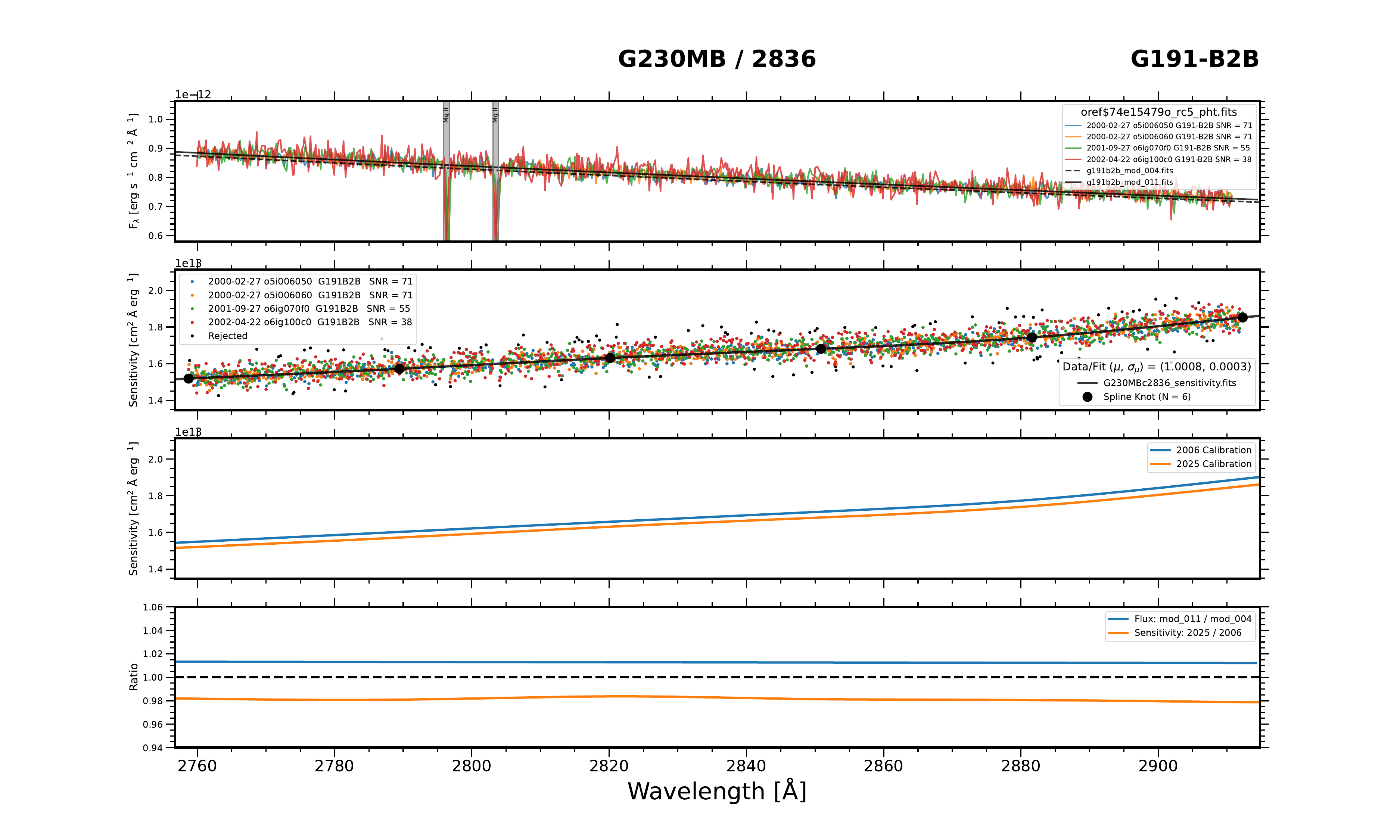}
  \footnotesize
  \caption{Calibration of G191-B2B for G230MB/2836.}
  \label{fig:G230MBC2836a}
\end{figure}
 
\clearpage
\begin{figure}[t]
  \hspace{-0.5in}
  \includegraphics[width=1.1\textwidth]{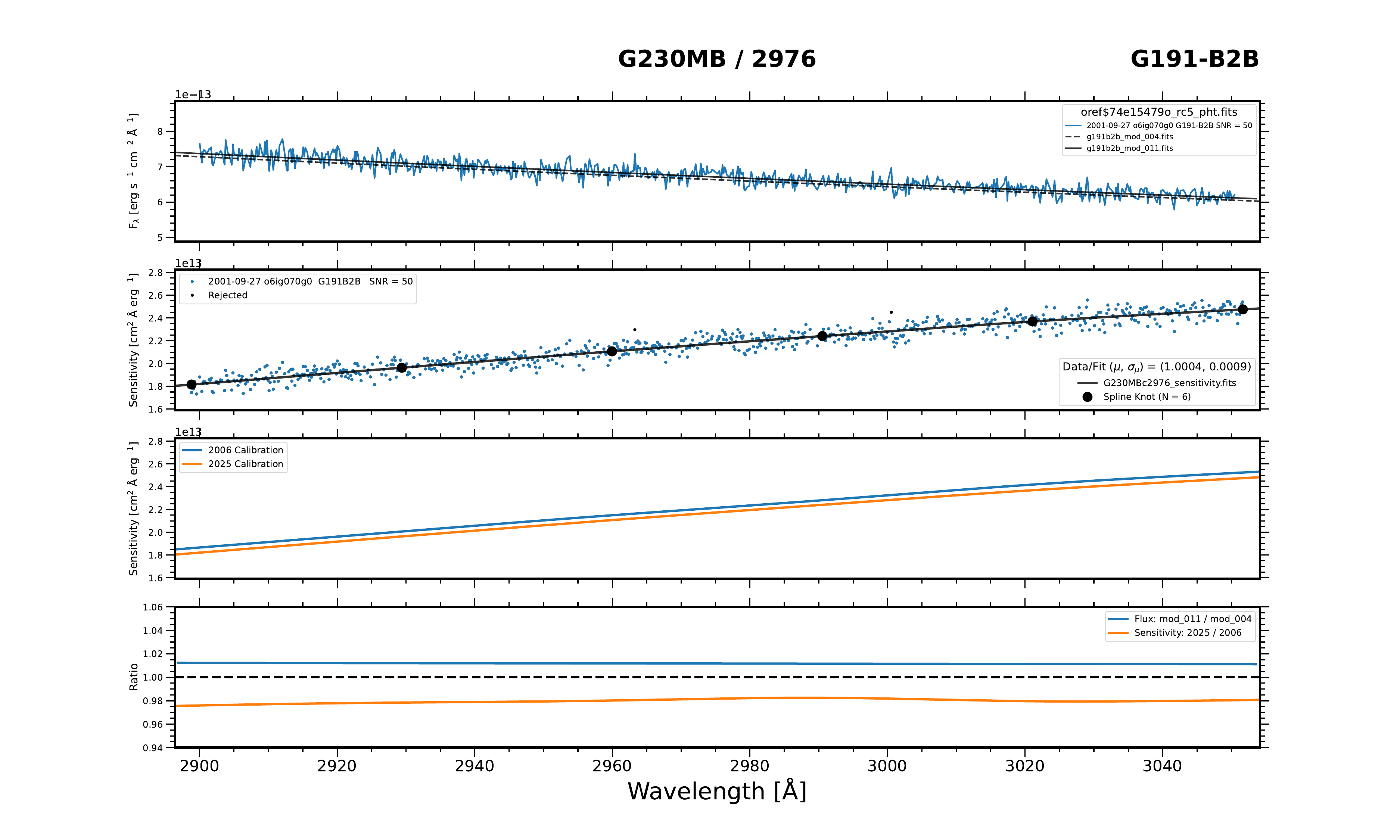}
  \footnotesize
  \caption{Calibration of G191-B2B for G230MB/2976.}
  \label{fig:G230MBC2976a}
\end{figure}
 
\begin{figure}[b]
  \hspace{-0.5in}
  \includegraphics[width=1.1\textwidth]{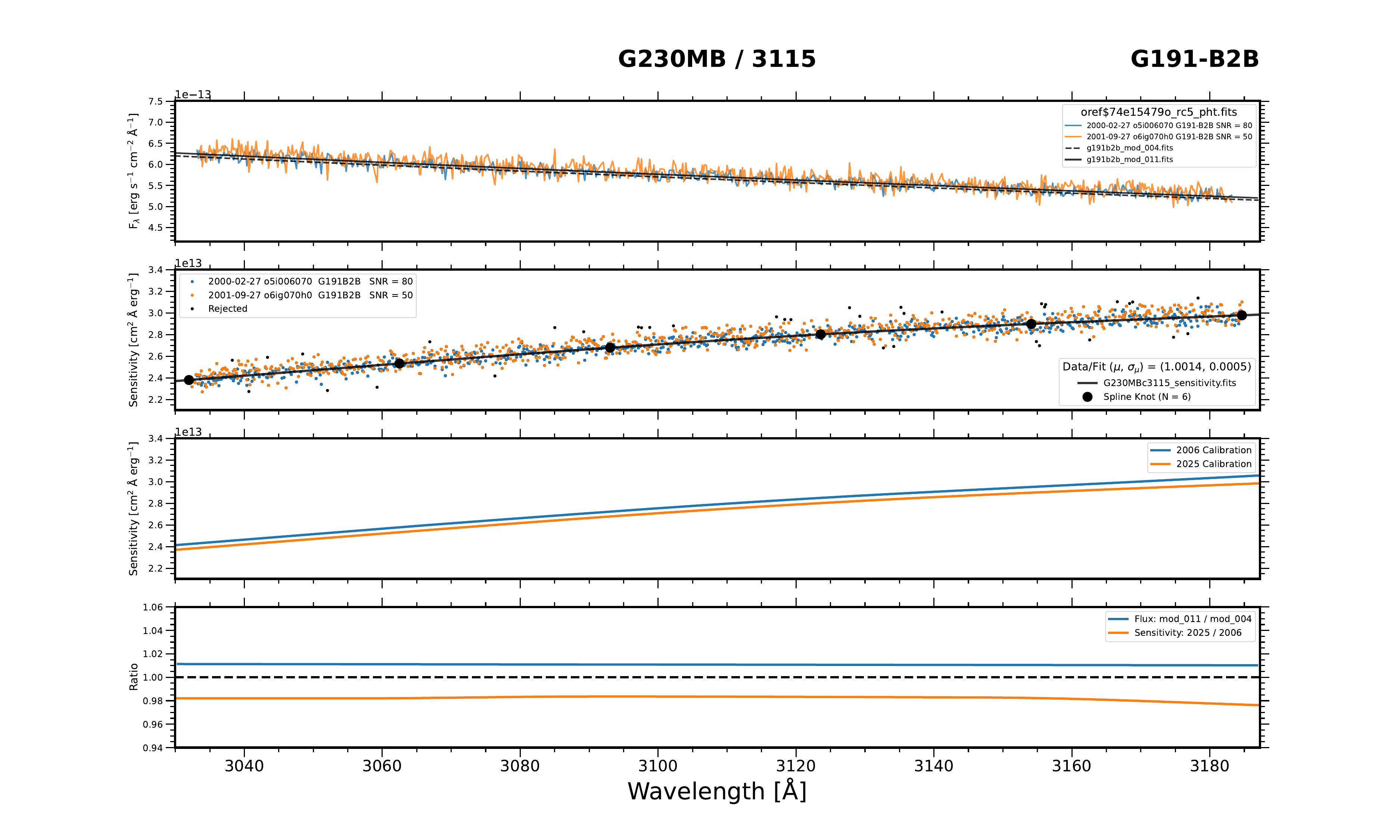}
  \footnotesize
  \caption{Calibration of G191-B2B for G230MB/3115.}
  \label{fig:G230MBC3115a}
\end{figure}
 
\clearpage
\newpage

\begin{figure}[t]
  \hspace{-0.5in}
  \includegraphics[width=1.1\textwidth]{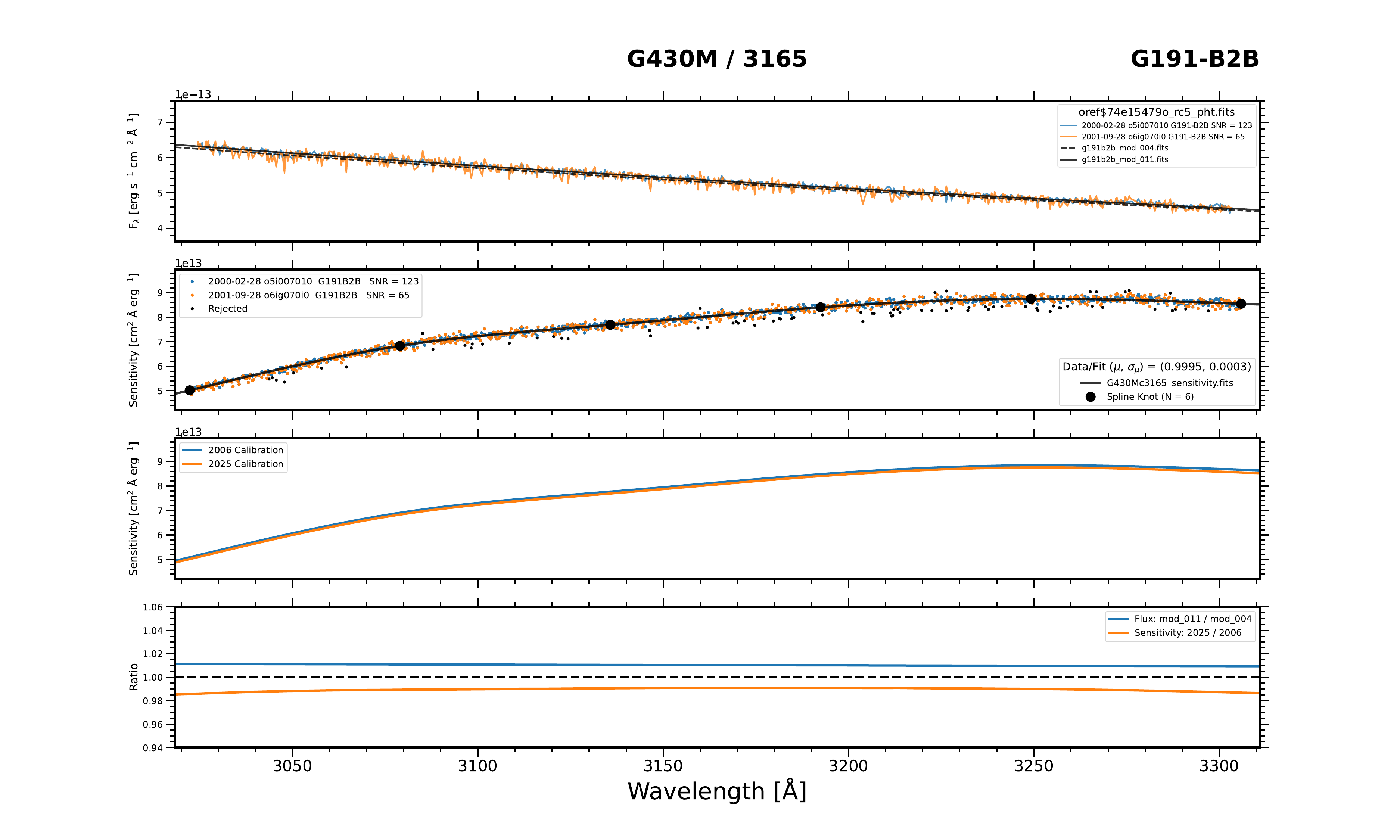}
  \footnotesize
  \caption{Calibration of G191-B2B for G430M/3165.}
  \label{fig:G430MC3165a}
\end{figure}
 
\begin{figure}[b]
  \hspace{-0.5in}
  \includegraphics[width=1.1\textwidth]{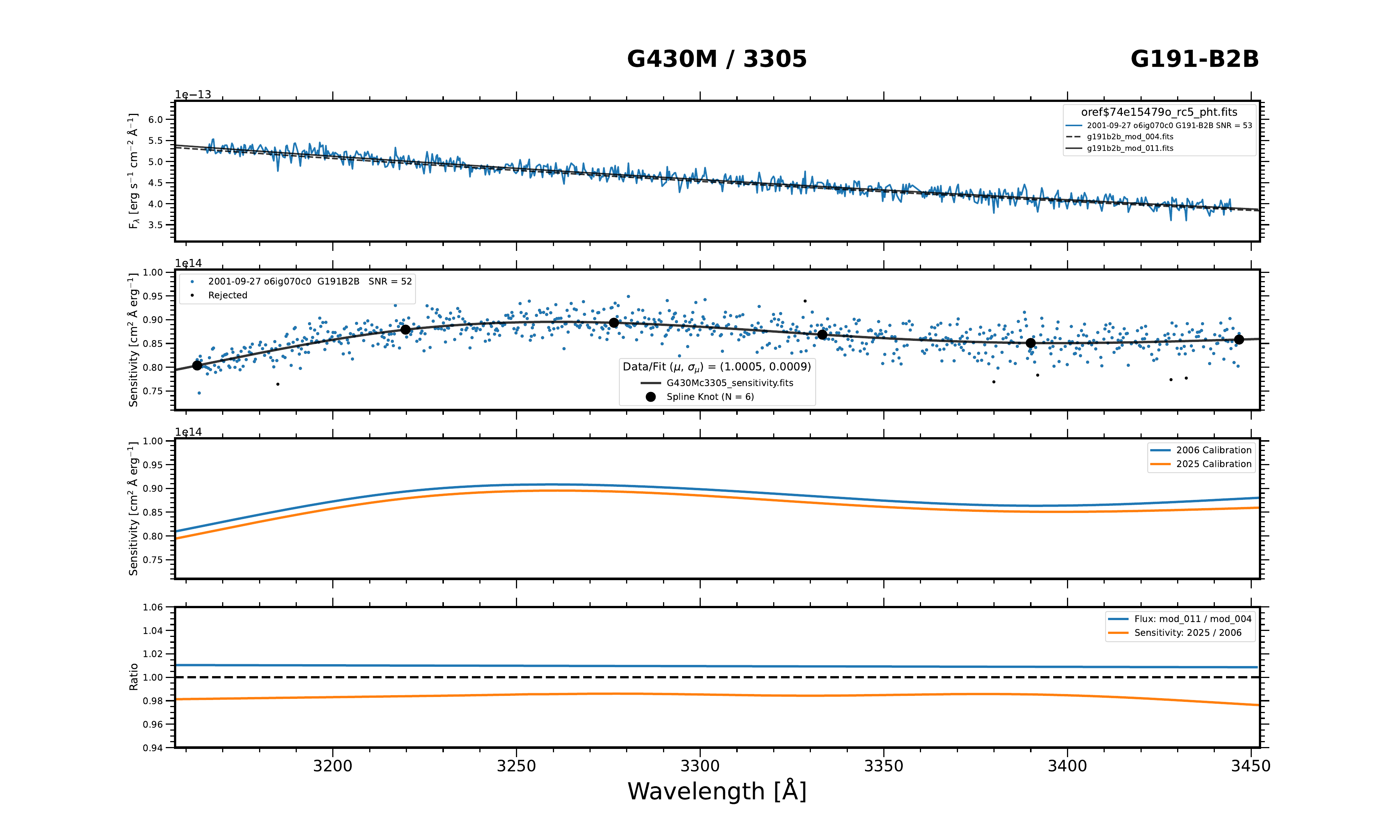}
  \footnotesize
  \caption{Calibration of G191-B2B for G430M/3305.}
  \label{fig:G430MC3305a}
\end{figure}
 
\clearpage
\begin{figure}[t]
  \hspace{-0.5in}
  \includegraphics[width=1.1\textwidth]{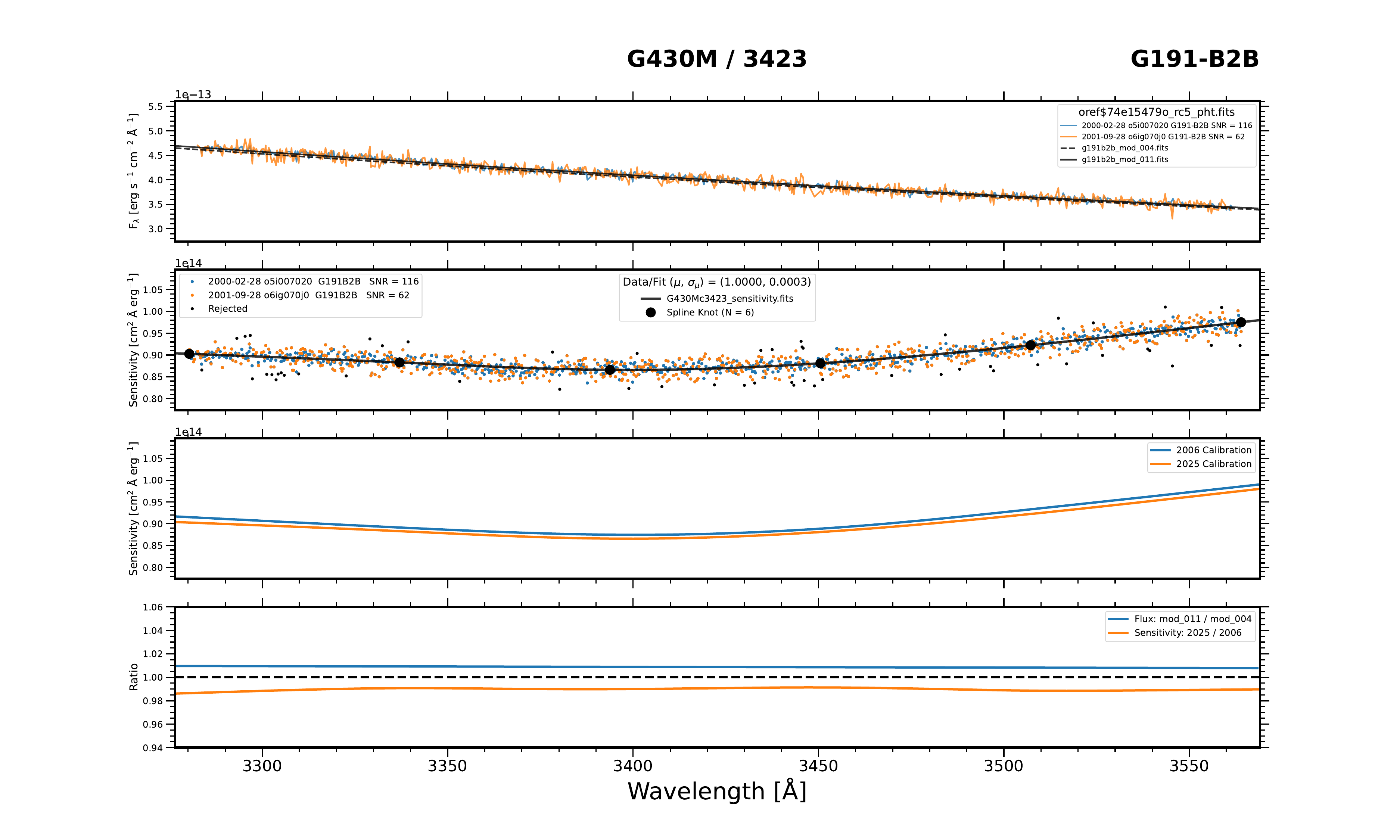}
  \footnotesize
  \caption{Calibration of G191-B2B for G430M/3423.}
  \label{fig:G430MC3423a}
\end{figure}
 
\begin{figure}[b]
  \hspace{-0.5in}
  \includegraphics[width=1.1\textwidth]{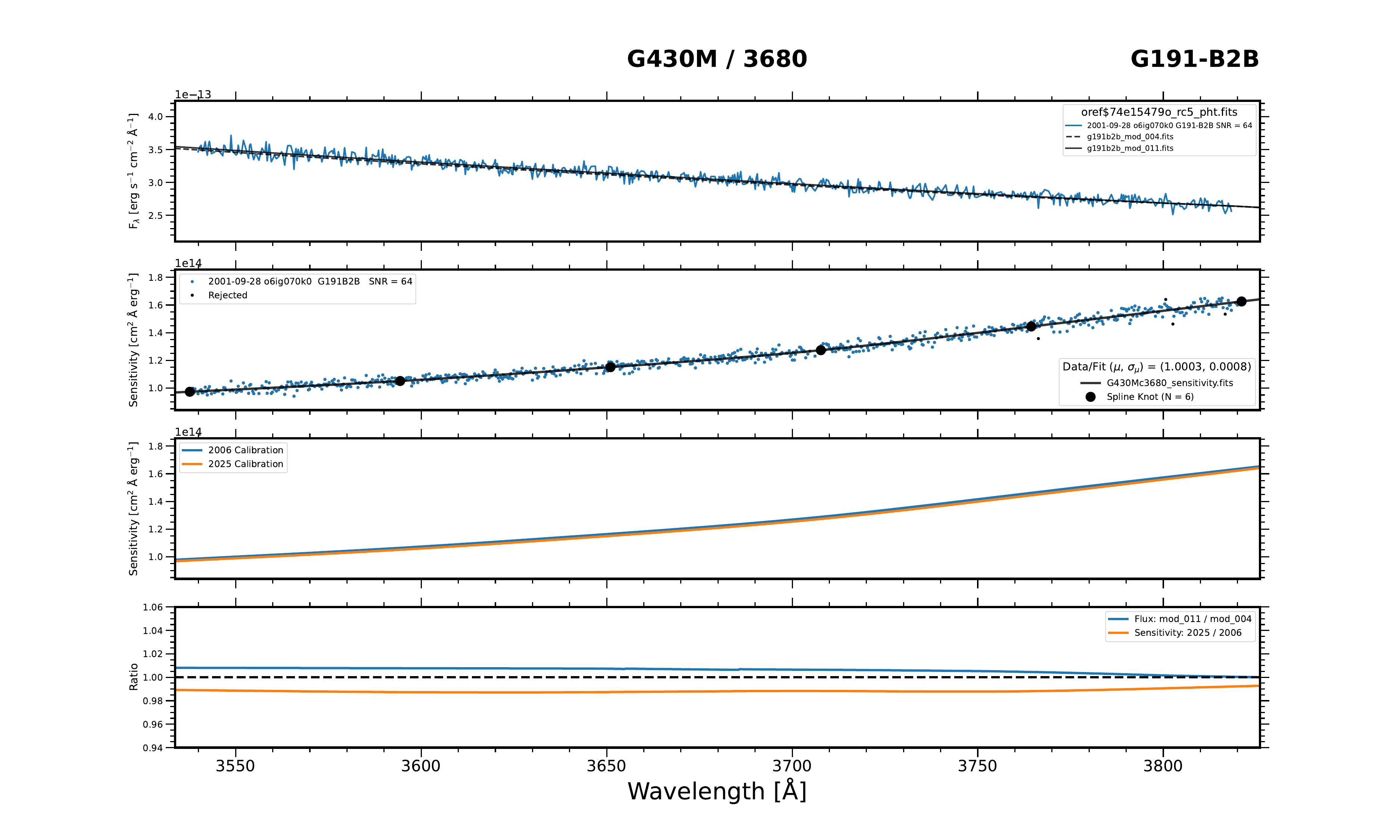}
  \footnotesize
  \caption{Calibration of G191-B2B for G430M/3680.}
  \label{fig:G430MC3680a}
\end{figure}
 
\clearpage
\begin{figure}[t]
  \hspace{-0.5in}
  \includegraphics[width=1.1\textwidth]{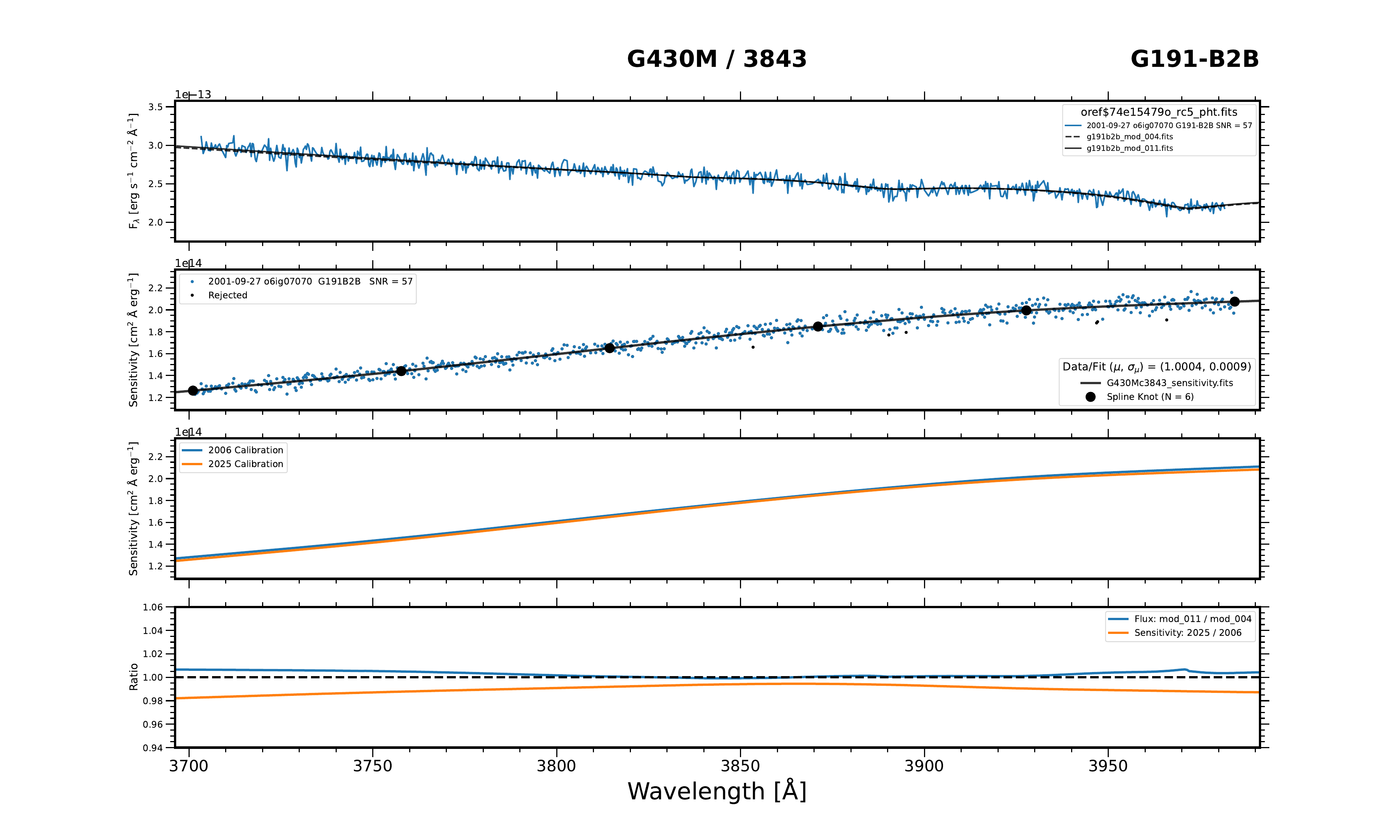}
  \footnotesize
  \caption{Calibration of G191-B2B for G430M/3843.}
  \label{fig:G430MC3843a}
\end{figure}
 
\begin{figure}[b]
  \hspace{-0.5in}
  \includegraphics[width=1.1\textwidth]{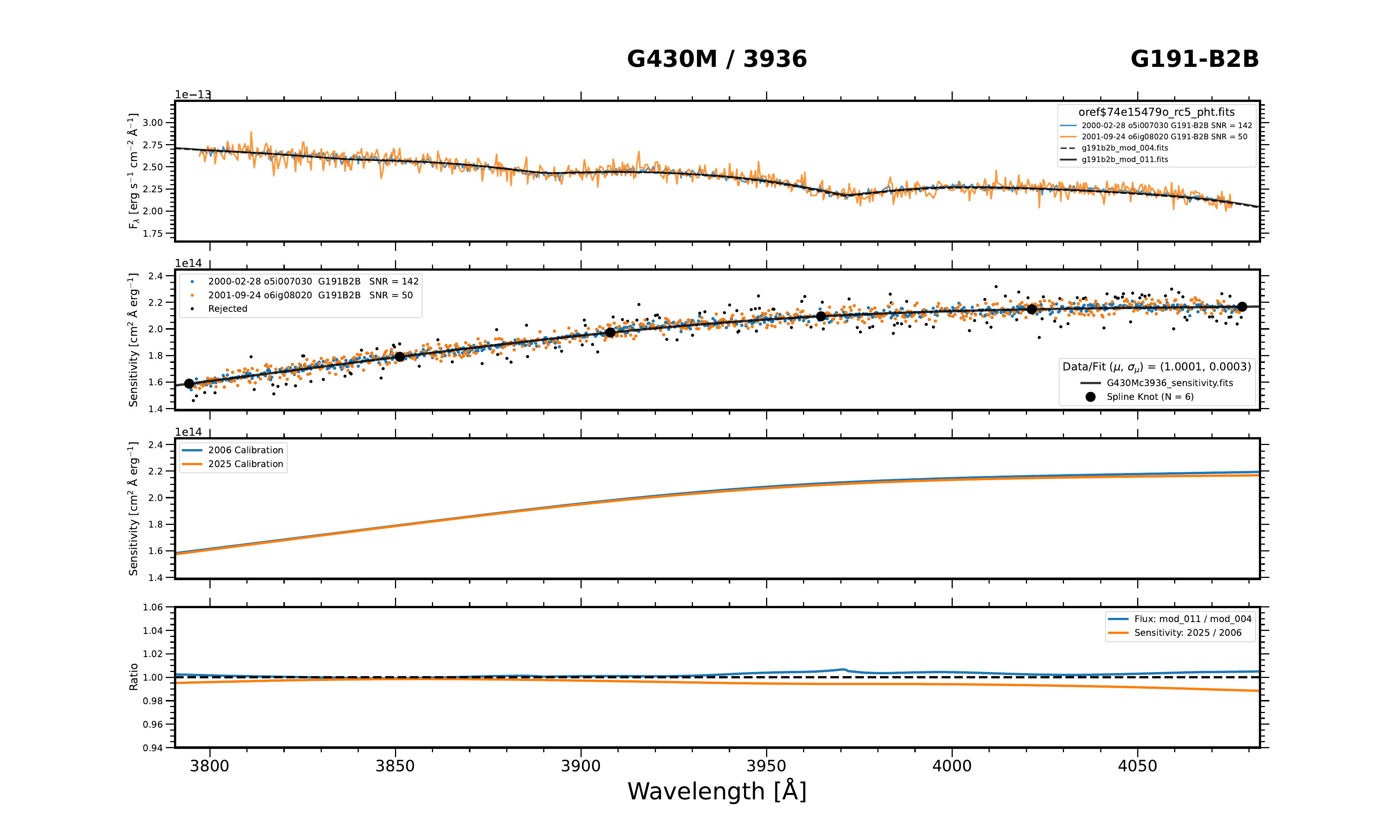}
  \footnotesize
  \caption{Calibration of G191-B2B for G430M/3936.}
  \label{fig:G430MC3936a}
\end{figure}
 
\clearpage
\begin{figure}[t]
  \hspace{-0.5in}
  \includegraphics[width=1.1\textwidth]{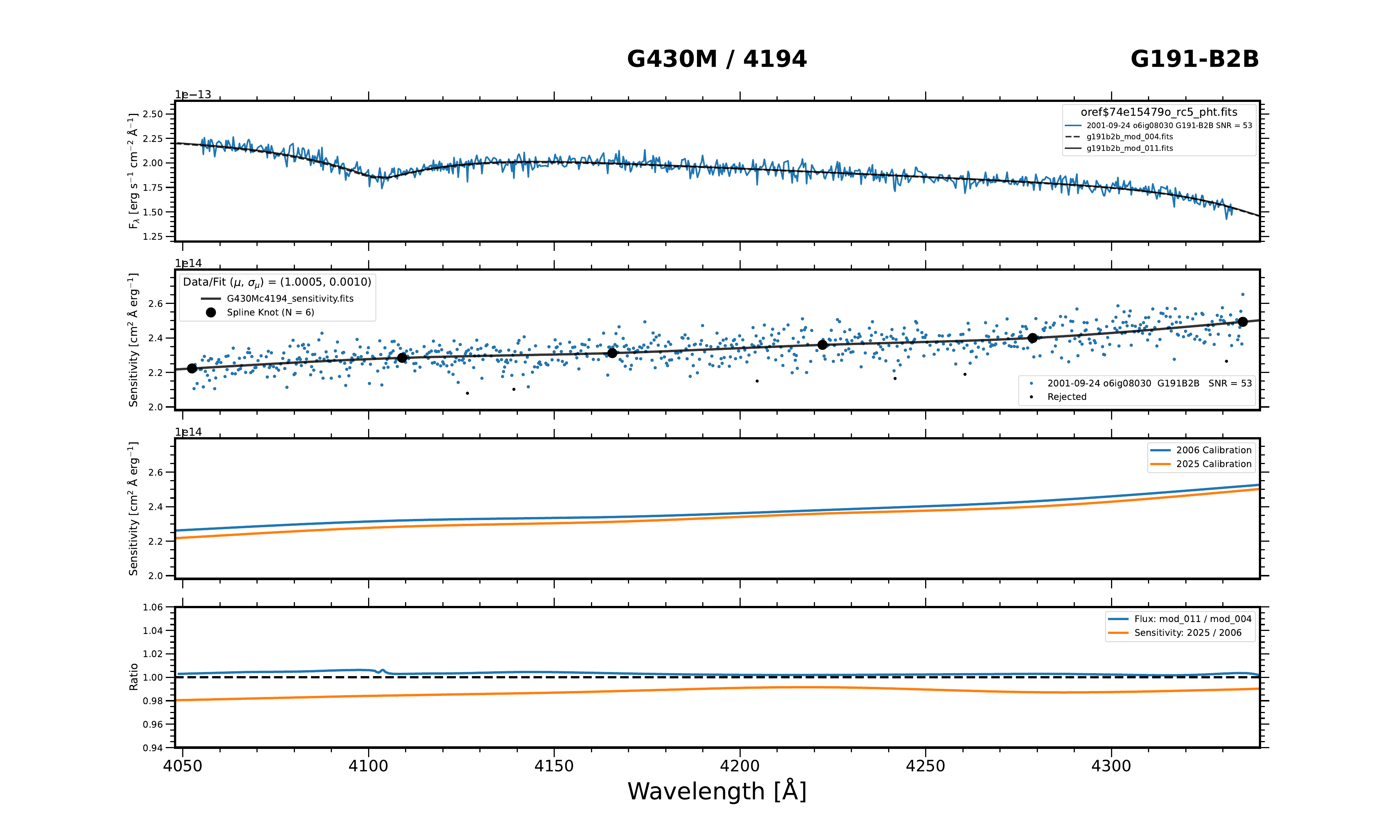}
  \footnotesize
  \caption{Calibration of G191-B2B for G430M/4194.}
  \label{fig:G430MC4194a}
\end{figure}
 
\begin{figure}[b]
  \hspace{-0.5in}
  \includegraphics[width=1.1\textwidth]{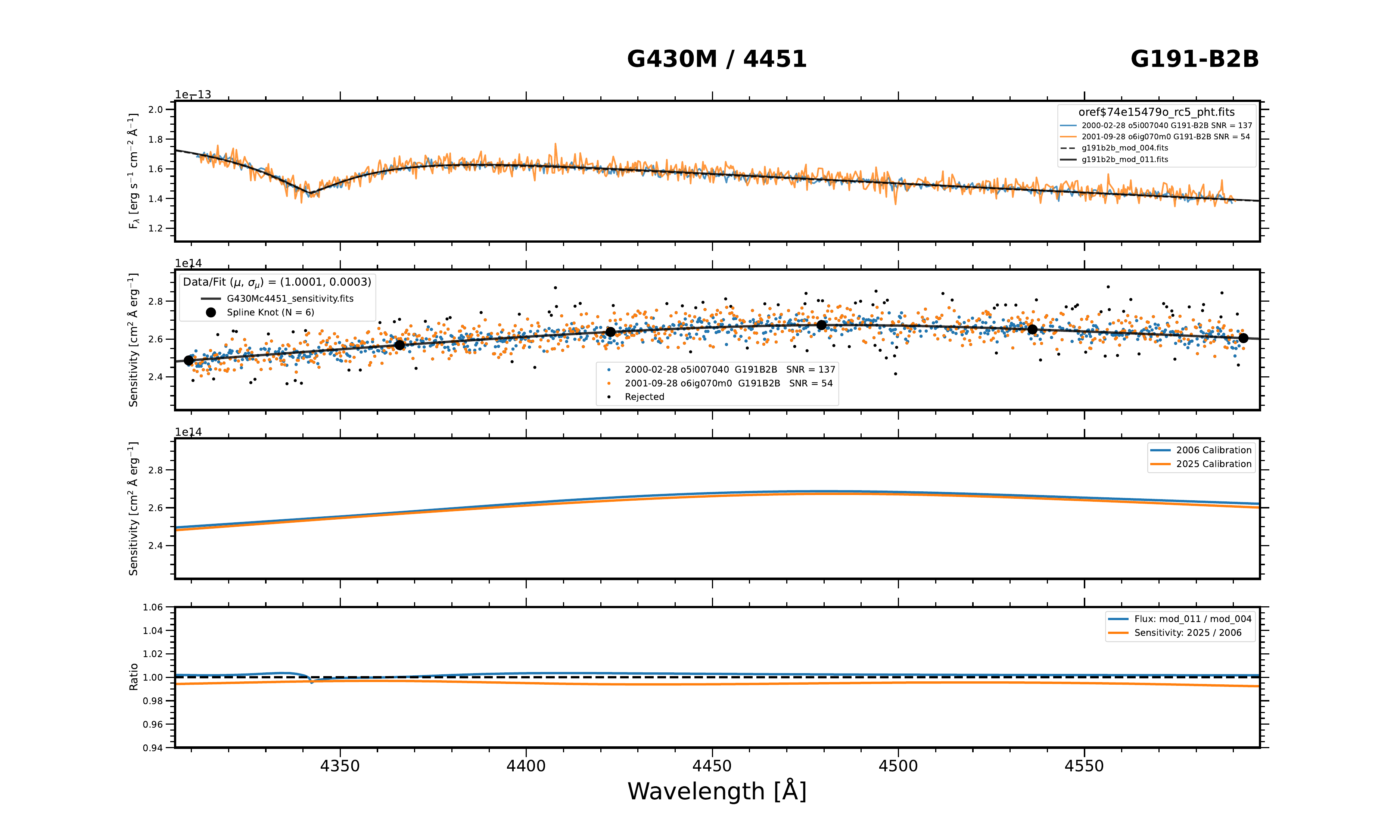}
  \footnotesize
  \caption{Calibration of G191-B2B for G430M/4451.}
  \label{fig:G430MC4451a}
\end{figure}
 
\clearpage
\begin{figure}[t]
  \hspace{-0.5in}
  \includegraphics[width=1.1\textwidth]{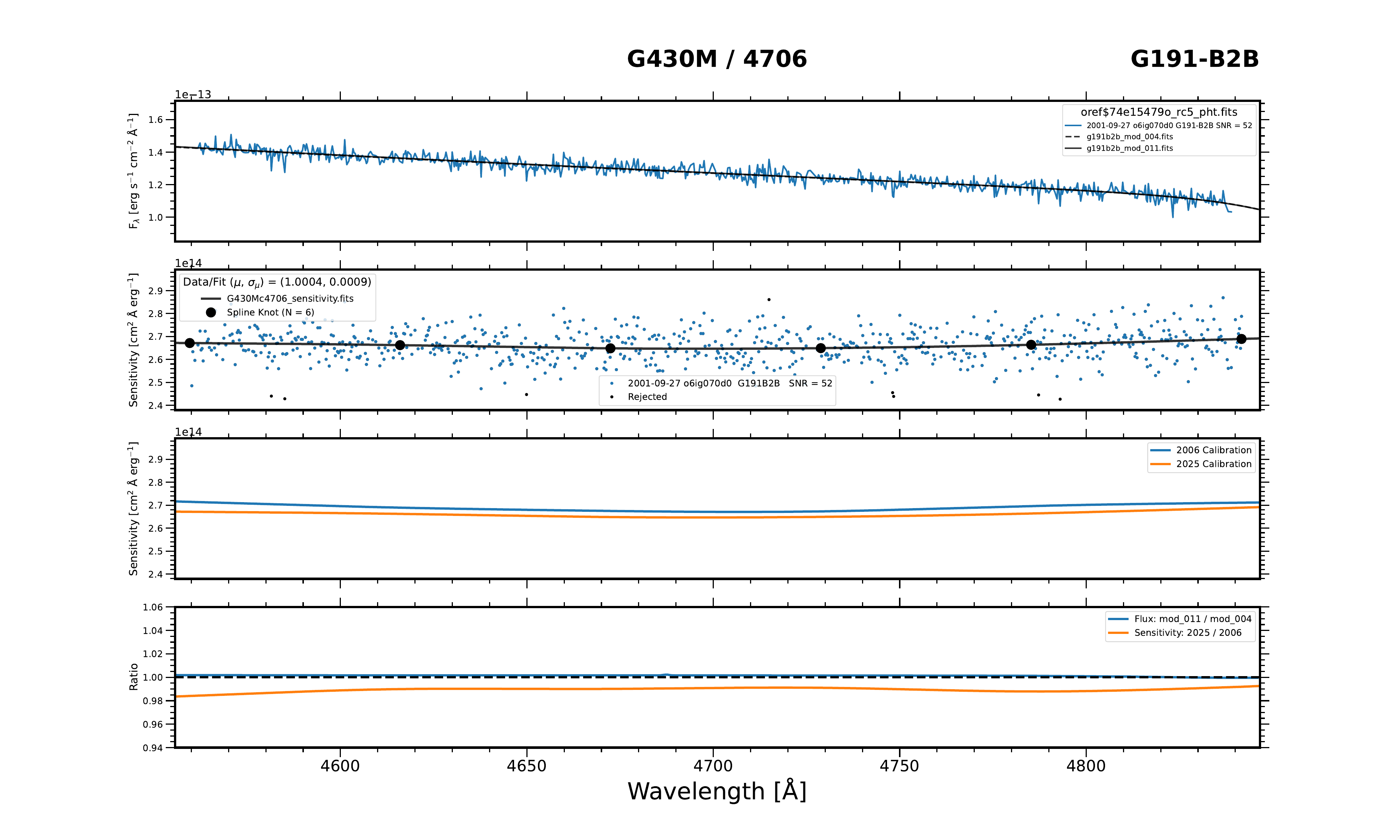}
  \footnotesize
  \caption{Calibration of G191-B2B for G430M/4706.}
  \label{fig:G430MC4706a}
\end{figure}
 
\begin{figure}[b]
  \hspace{-0.5in}
  \includegraphics[width=1.1\textwidth]{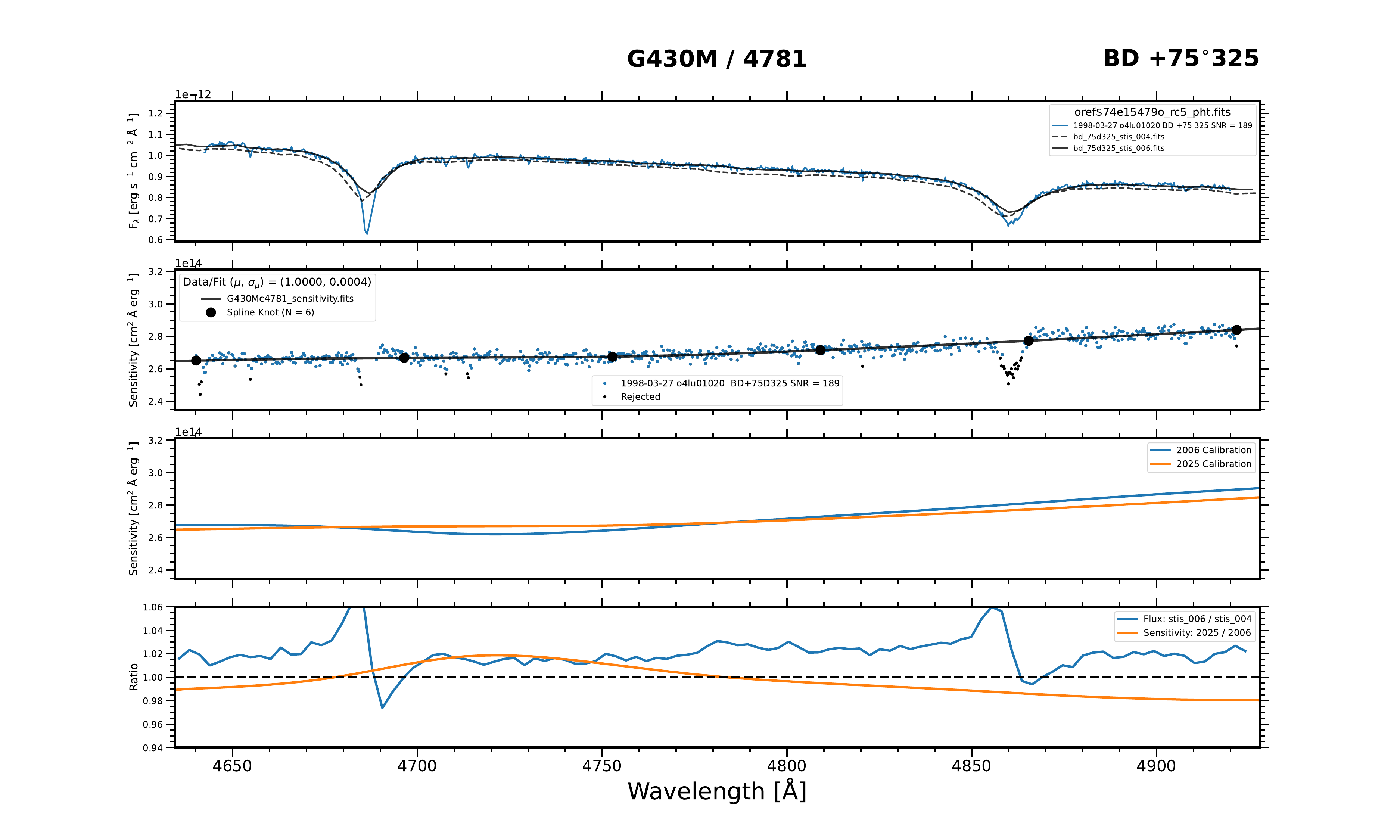}
  \footnotesize
  \caption{Calibration of BD $+$75$^{\circ}$325 for G430M/4781.}
  \label{fig:G430MC4781}
\end{figure}
 
\clearpage
\begin{figure}[t]
  \hspace{-0.5in}
  \includegraphics[width=1.1\textwidth]{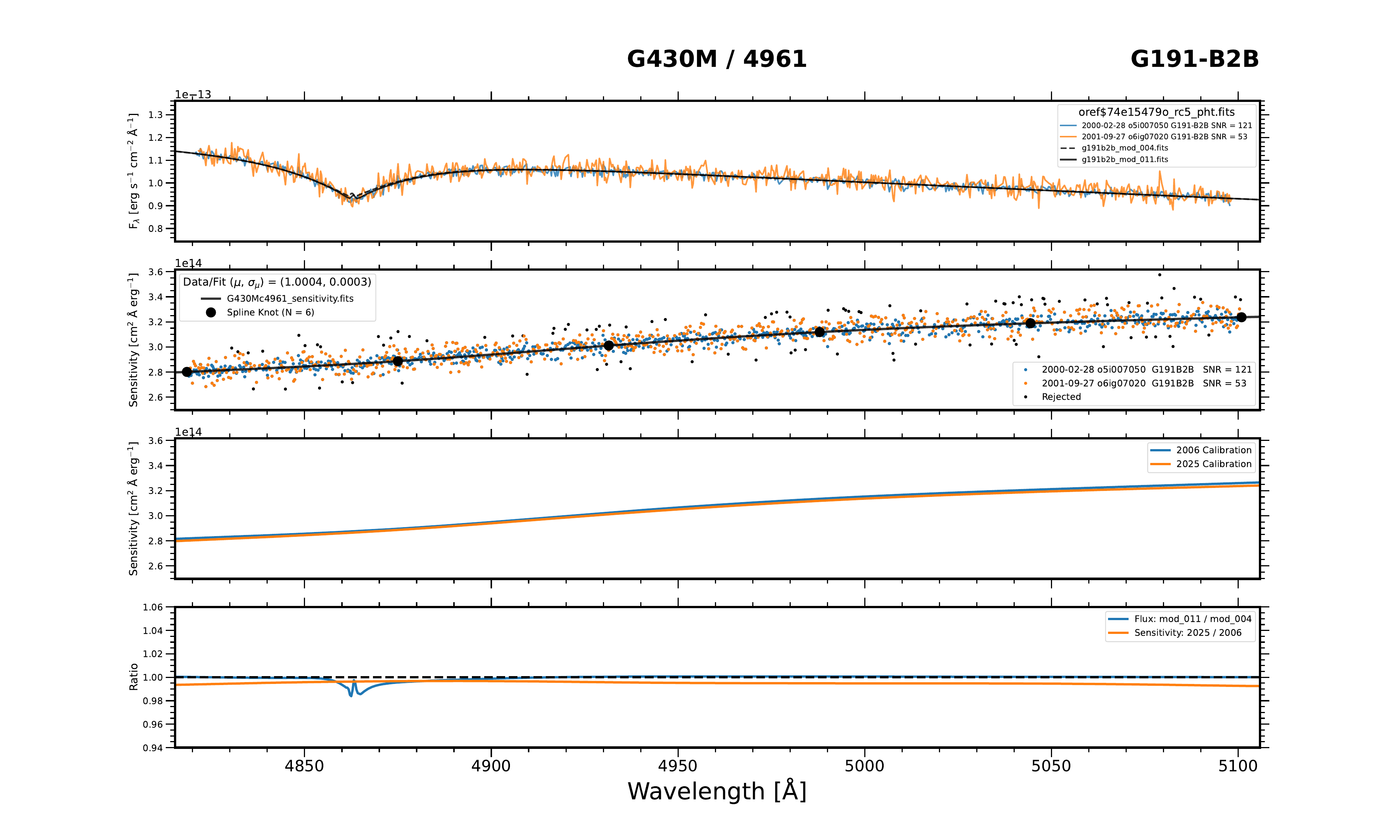}
  \footnotesize
  \caption{Calibration of G191-B2B for G430M/4961.}
  \label{fig:G430MC4961a}
\end{figure}
 
\begin{figure}[b]
  \hspace{-0.5in}
  \includegraphics[width=1.1\textwidth]{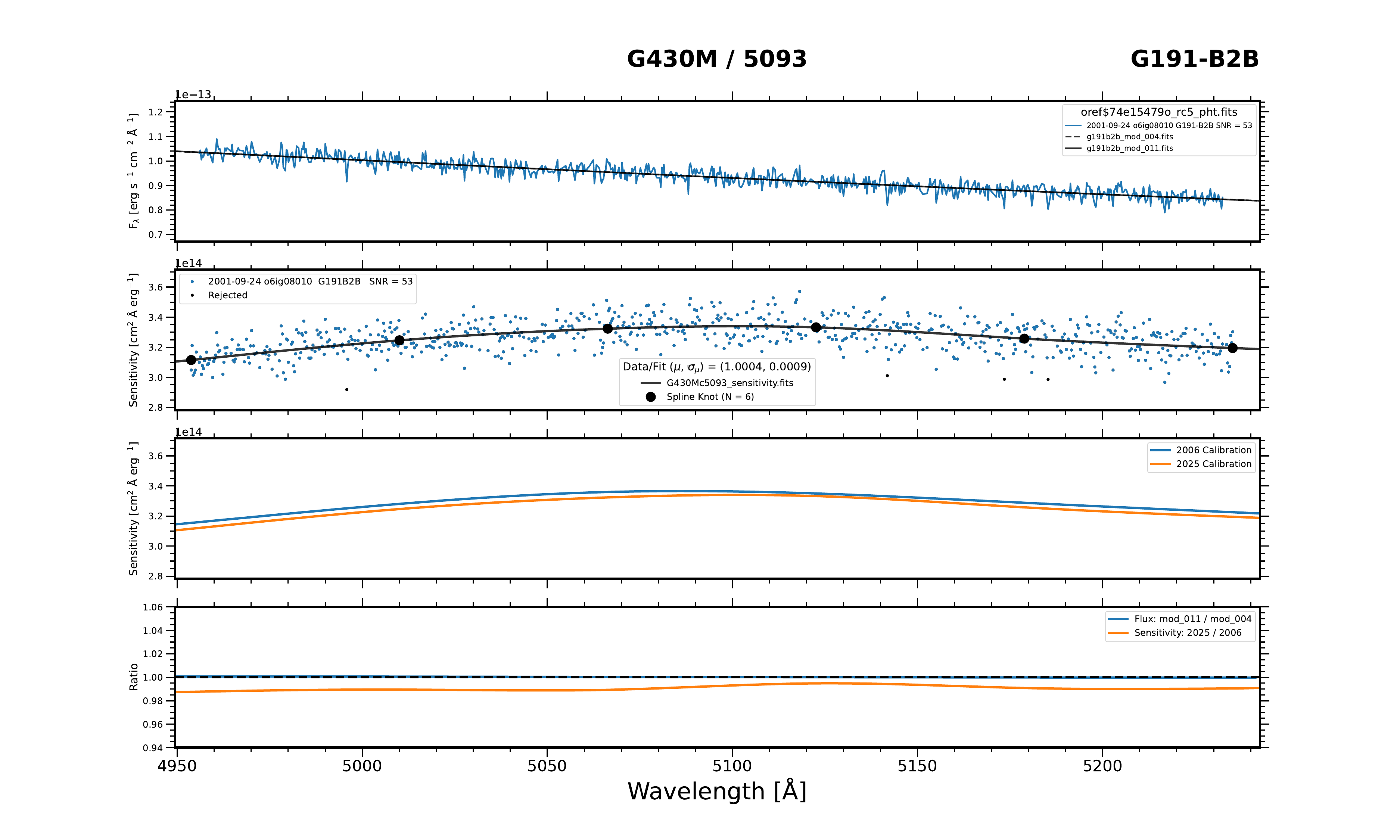}
  \footnotesize
  \caption{Calibration of G191-B2B for G430M/5093.}
  \label{fig:G430MC5093a}
\end{figure}
 
\clearpage
\begin{figure}[t]
  \hspace{-0.5in}
  \includegraphics[width=1.1\textwidth]{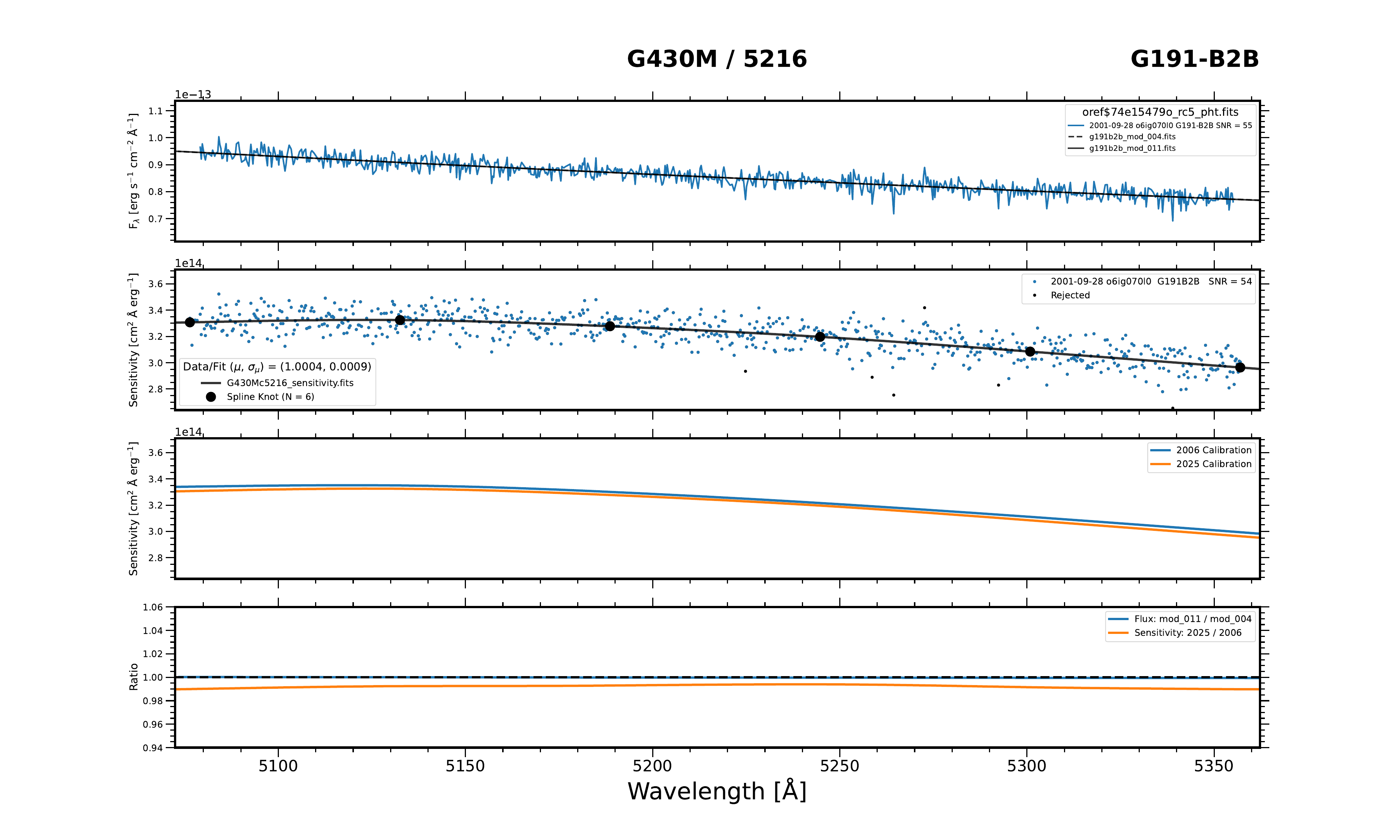}
  \footnotesize
  \caption{Calibration of G191-B2B for G430M/5216.}
  \label{fig:G430MC5216a}
\end{figure}
 
\begin{figure}[b]
  \hspace{-0.5in}
  \includegraphics[width=1.1\textwidth]{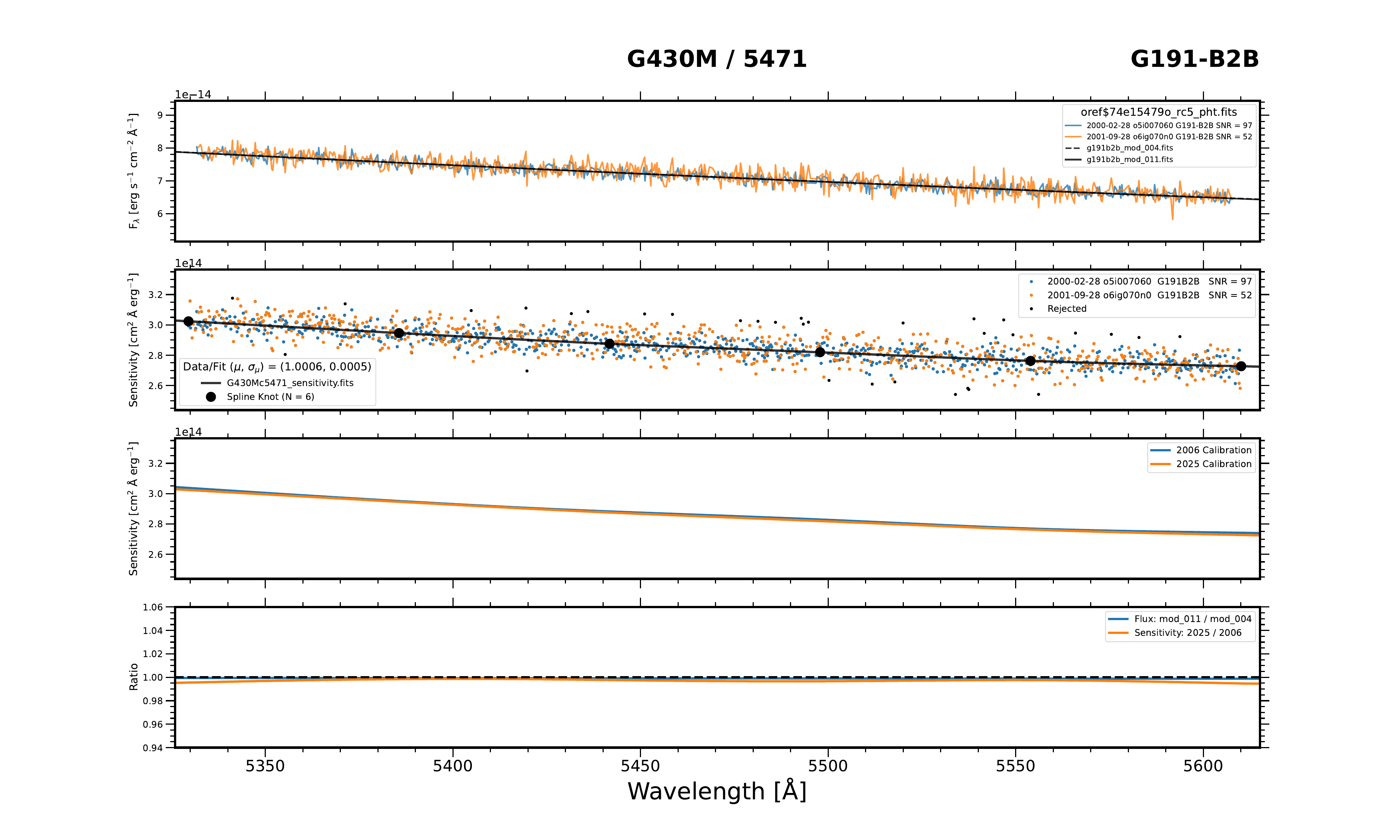}
  \footnotesize
  \caption{Calibration of G191-B2B for G430M/5471.}
  \label{fig:G430MC5471a}
\end{figure}
 
\clearpage
\newpage

\begin{figure}[t]
  \hspace{-0.5in}
  \includegraphics[width=1.1\textwidth]{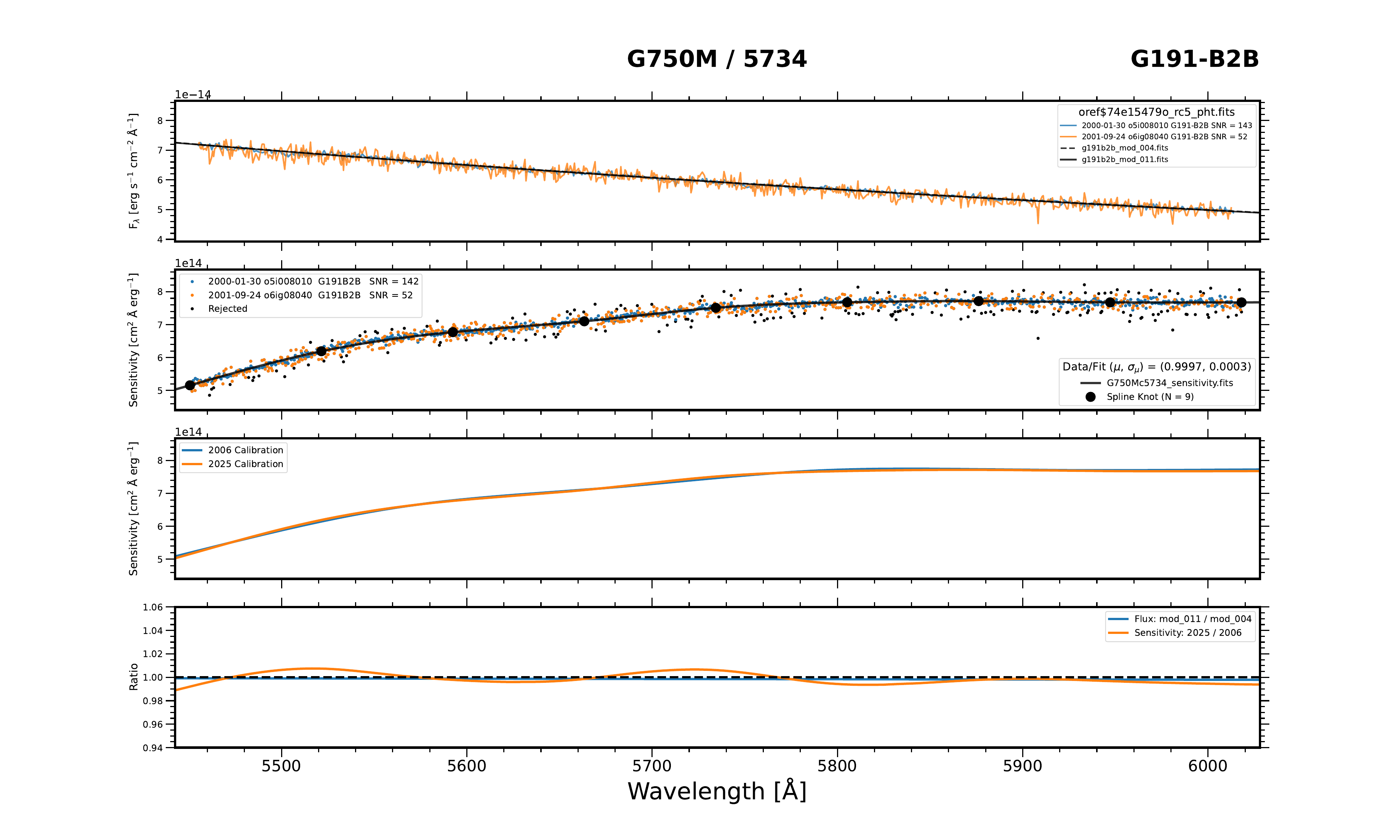}
  \footnotesize
  \caption{Calibration of G191-B2B for G750M/5734.}
  \label{fig:G750MC5734a}
\end{figure}
 
\begin{figure}[b]
  \hspace{-0.5in}
  \includegraphics[width=1.1\textwidth]{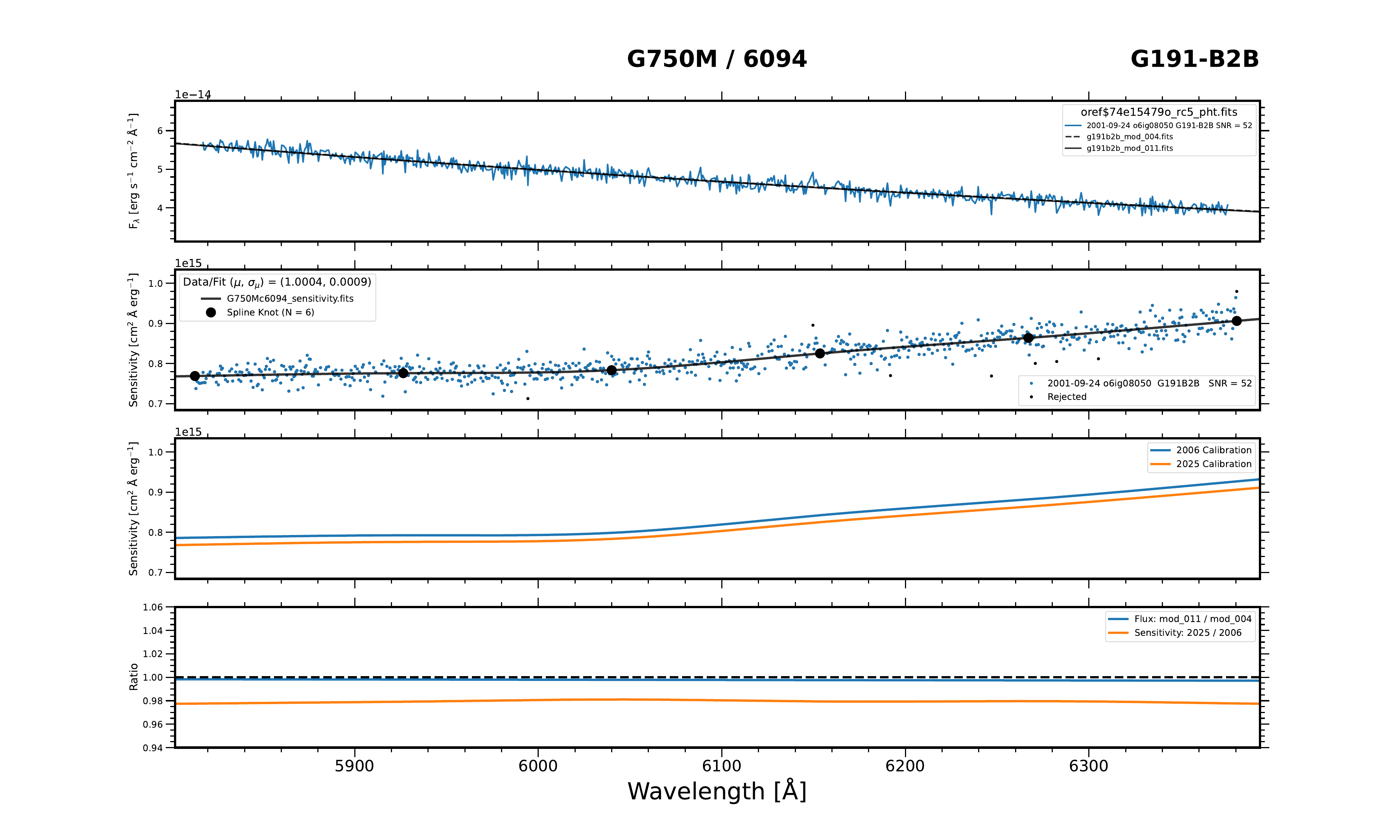}
  \footnotesize
  \caption{Calibration of G191-B2B for G750M/6094.}
  \label{fig:G750MC6094a}
\end{figure}
 
\clearpage
\begin{figure}[t]
  \hspace{-0.5in}
  \includegraphics[width=1.1\textwidth]{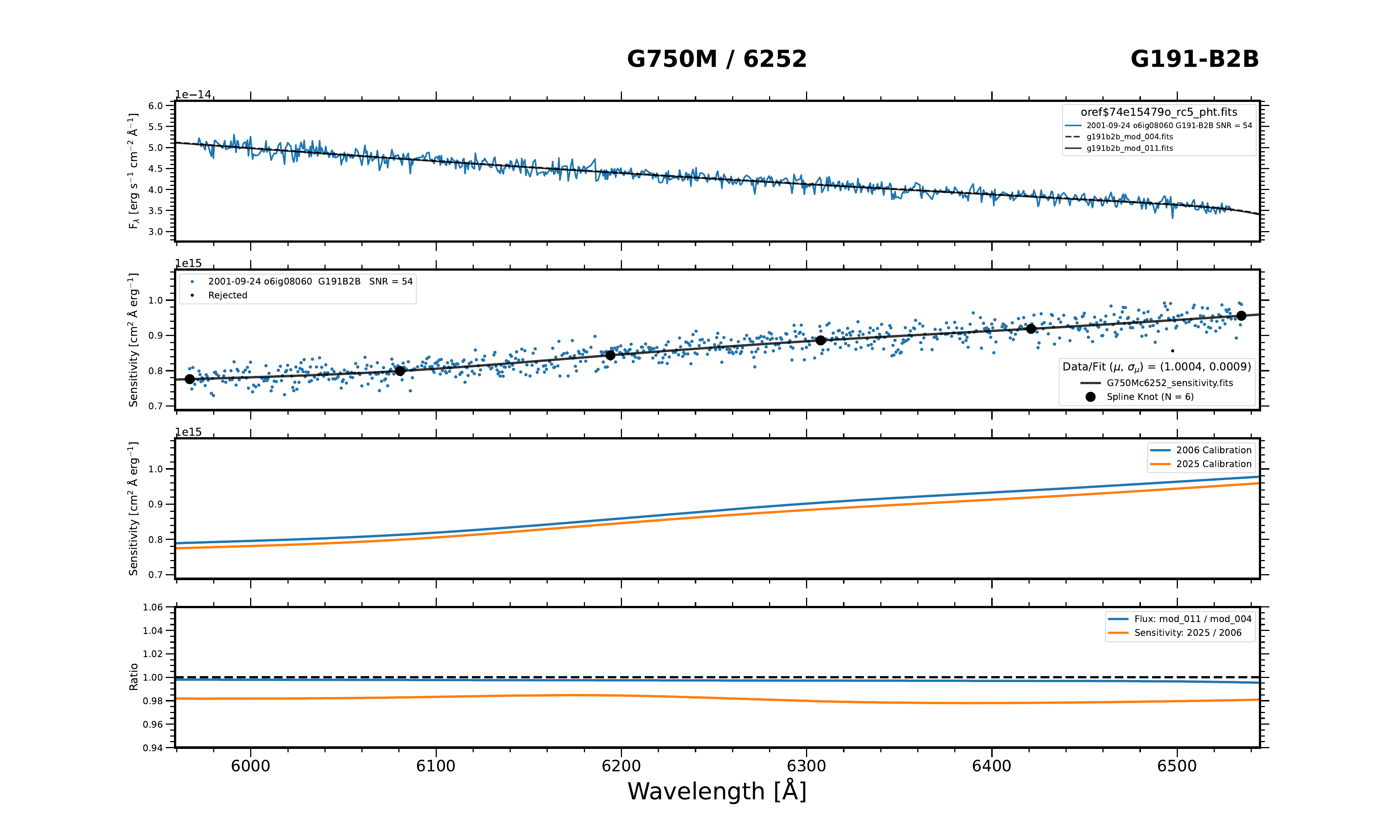}
  \footnotesize
  \caption{Calibration of G191-B2B for G750M/6252.}
  \label{fig:G750MC6252a}
\end{figure}
 
\begin{figure}[b]
  \hspace{-0.5in}
  \includegraphics[width=1.1\textwidth]{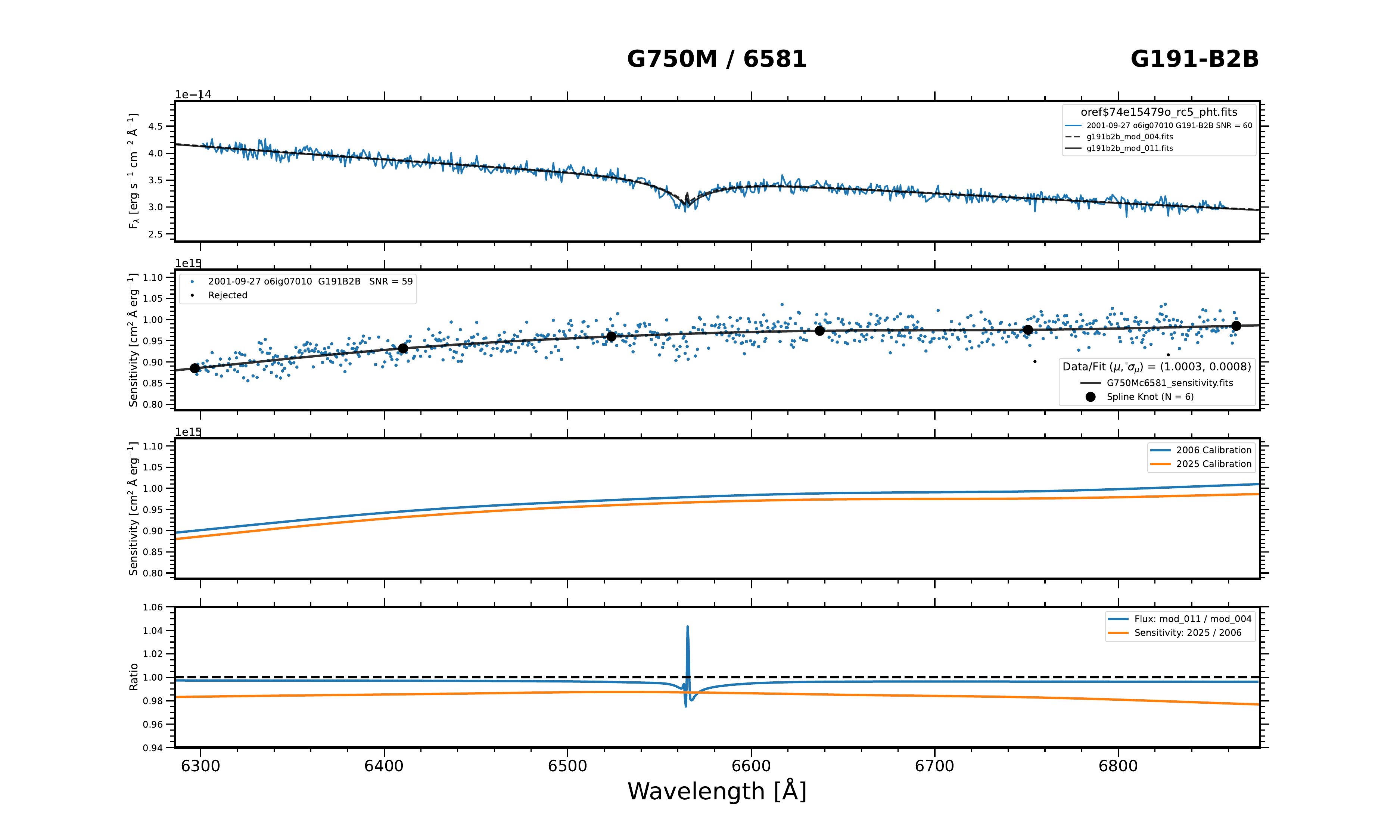}
  \footnotesize
  \caption{Calibration of G191-B2B for G750M/6581.}
  \label{fig:G750MC6581a}
\end{figure}
 
\clearpage
\begin{figure}[t]
  \hspace{-0.5in}
  \includegraphics[width=1.1\textwidth]{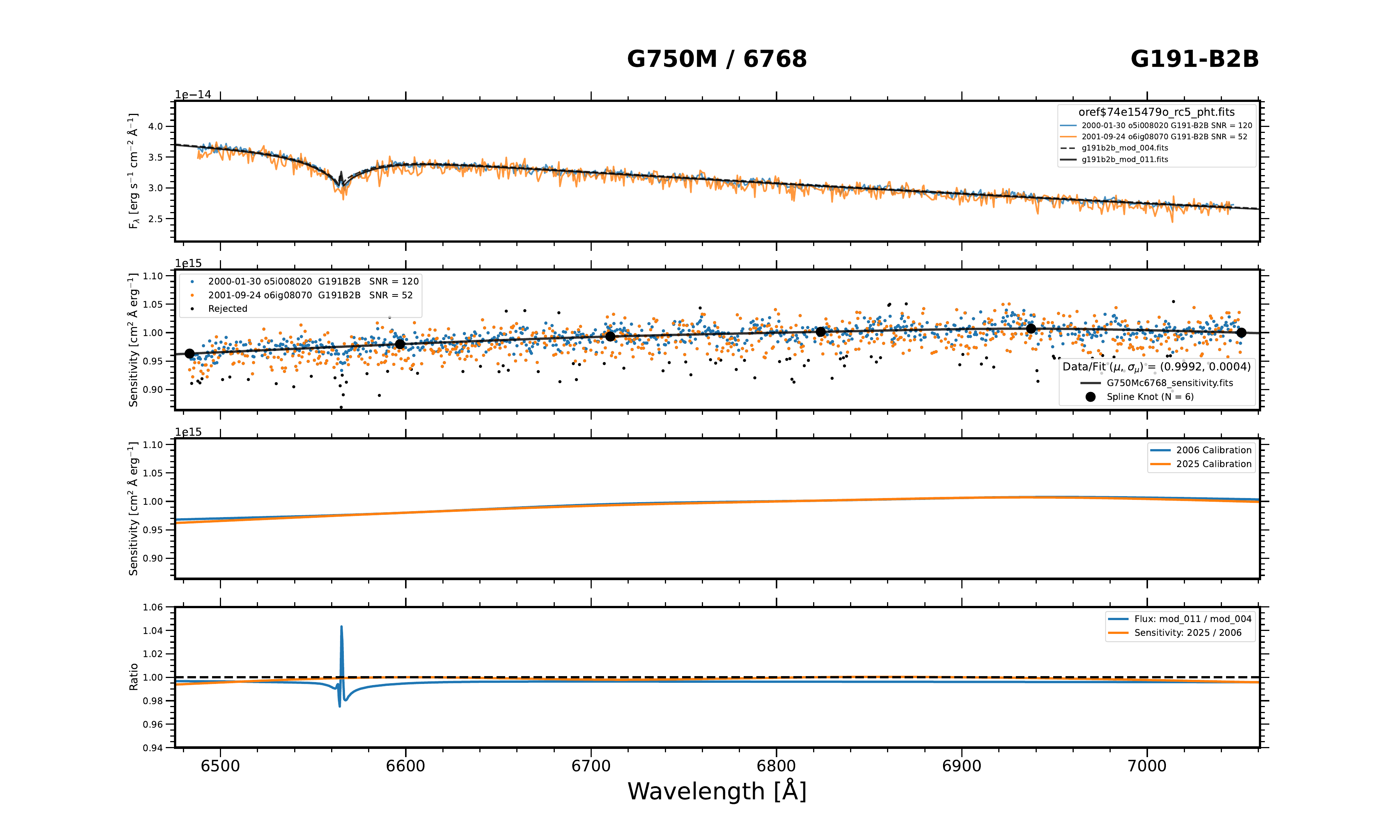}
  \footnotesize
  \caption{Calibration of G191-B2B for G750M/6768.}
  \label{fig:G750MC6768a}
\end{figure}
 
\begin{figure}[b]
  \hspace{-0.5in}
  \includegraphics[width=1.1\textwidth]{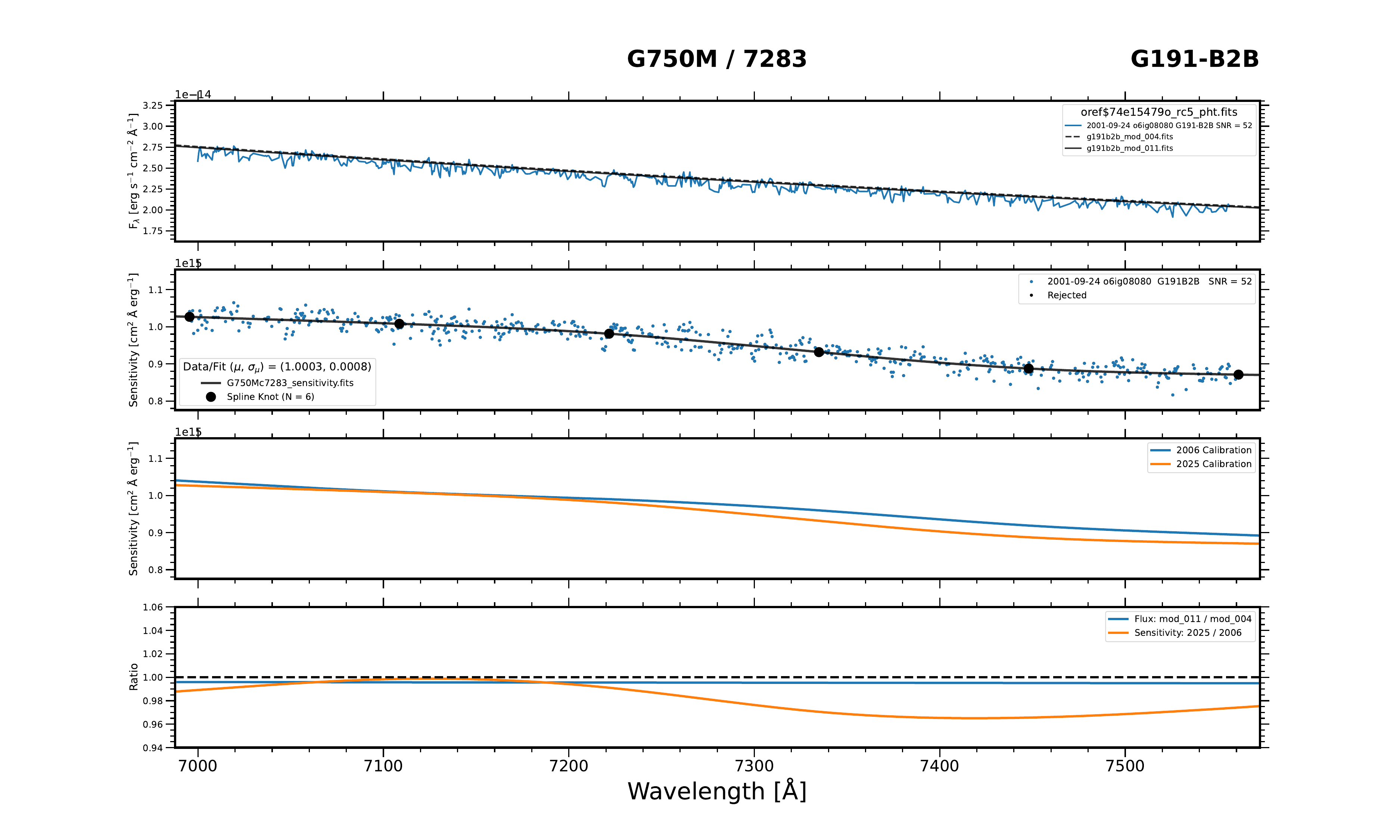}
  \footnotesize
  \caption{Calibration of G191-B2B for G750M/7283.}
  \label{fig:G750MC7283a}
\end{figure}
 
\clearpage
\begin{figure}[t]
  \hspace{-0.5in}
  \includegraphics[width=1.1\textwidth]{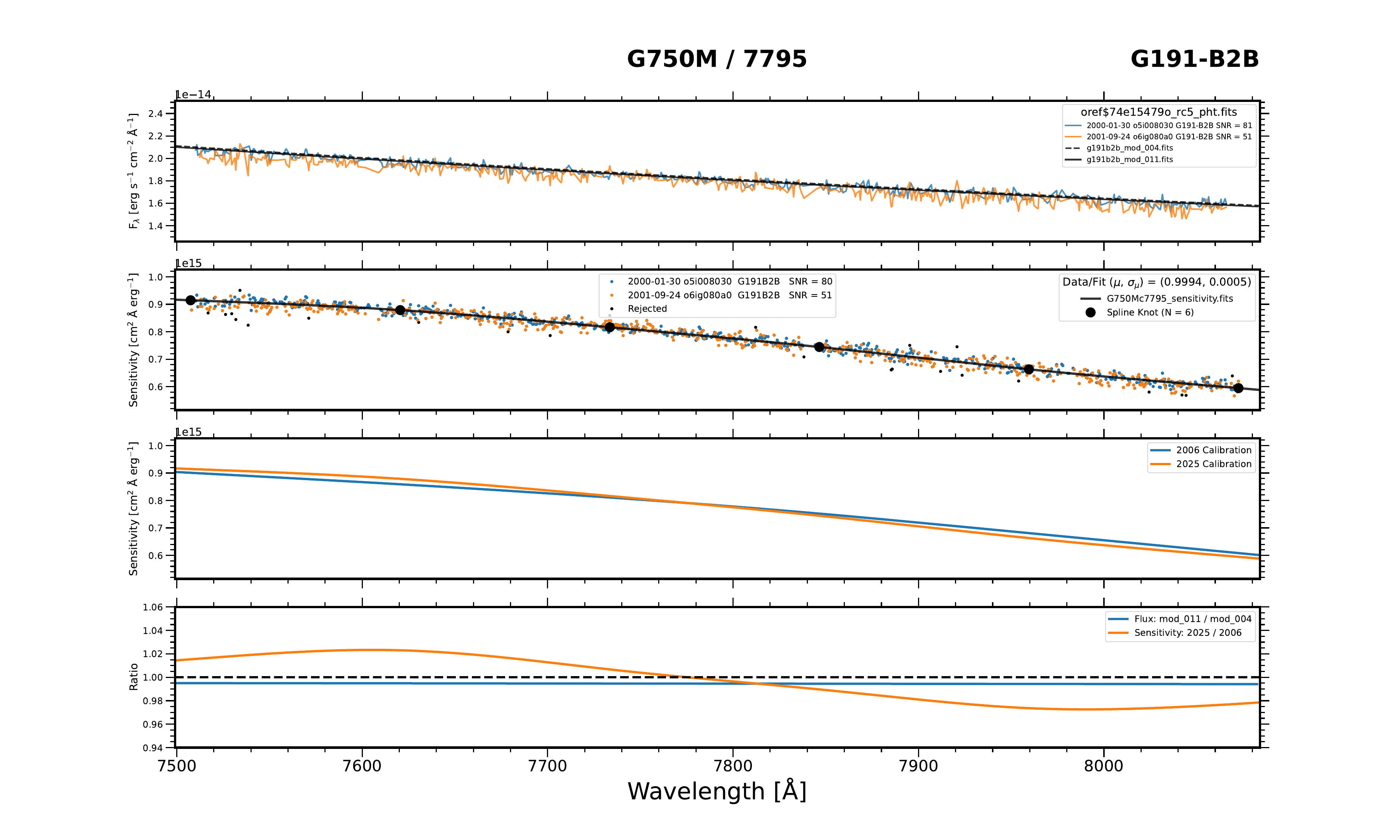}
  \footnotesize
  \caption{Calibration of G191-B2B for G750M/7795.}
  \label{fig:G750MC7795a}
\end{figure}
 
\begin{figure}[b]
  \hspace{-0.5in}
  \includegraphics[width=1.1\textwidth]{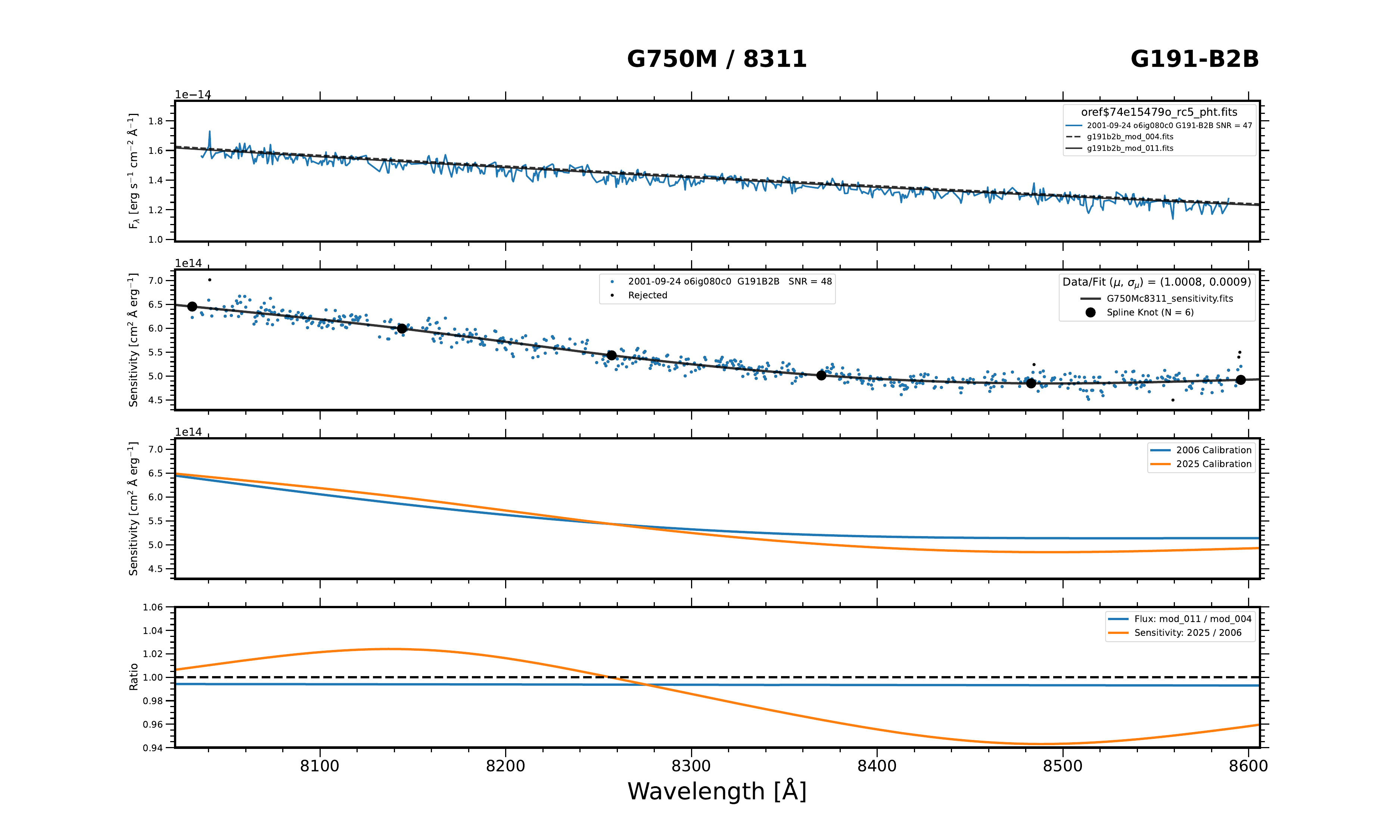}
  \footnotesize
  \caption{Calibration of G191-B2B for G750M/8311.}
  \label{fig:G750MC8311a}
\end{figure}
 
\clearpage
\begin{figure}[t]
  \hspace{-0.5in}
  \includegraphics[width=1.1\textwidth]{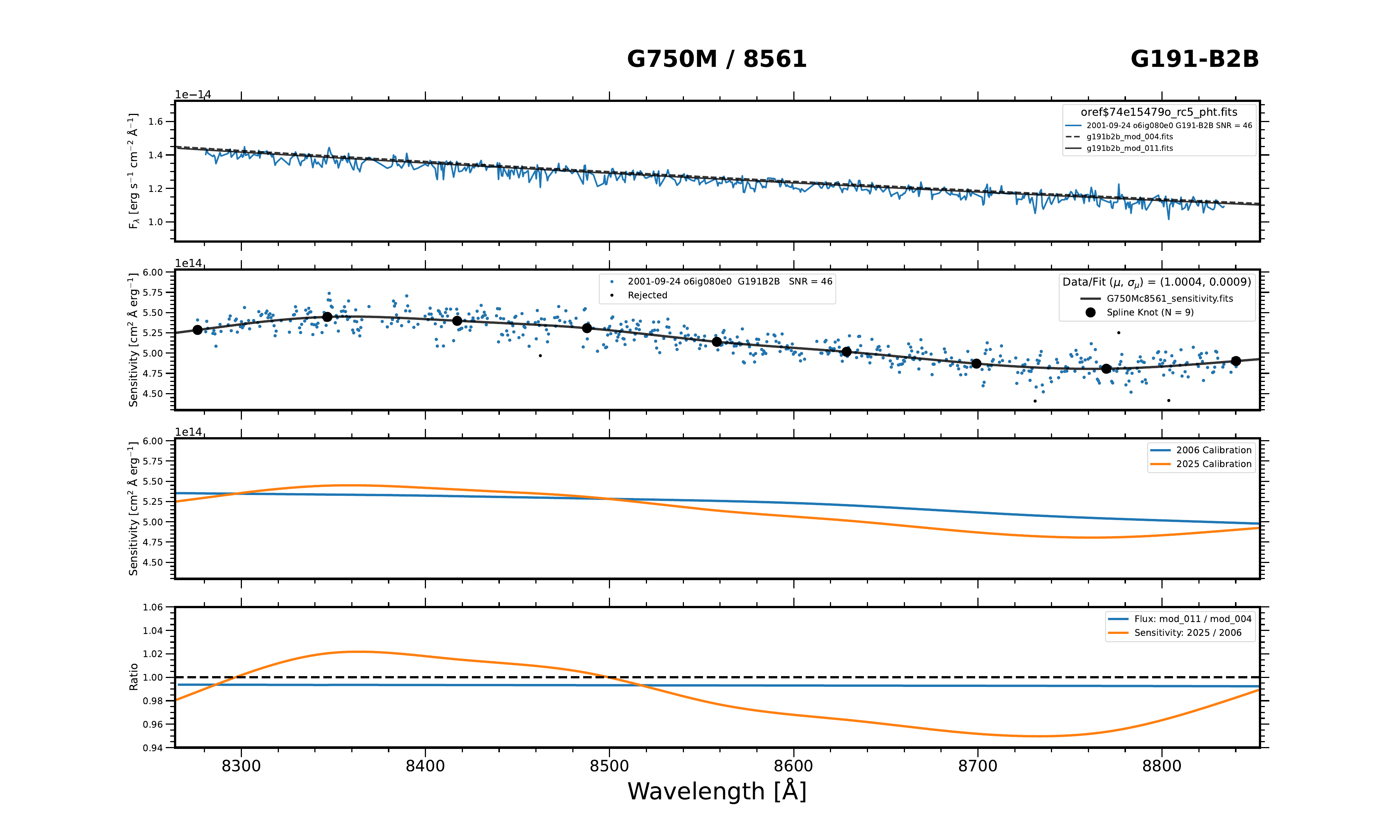}
  \footnotesize
  \caption{Calibration of G191-B2B for G750M/8561.}
  \label{fig:G750MC8561a}
\end{figure}
 
\begin{figure}[b]
  \hspace{-0.5in}
  \includegraphics[width=1.1\textwidth]{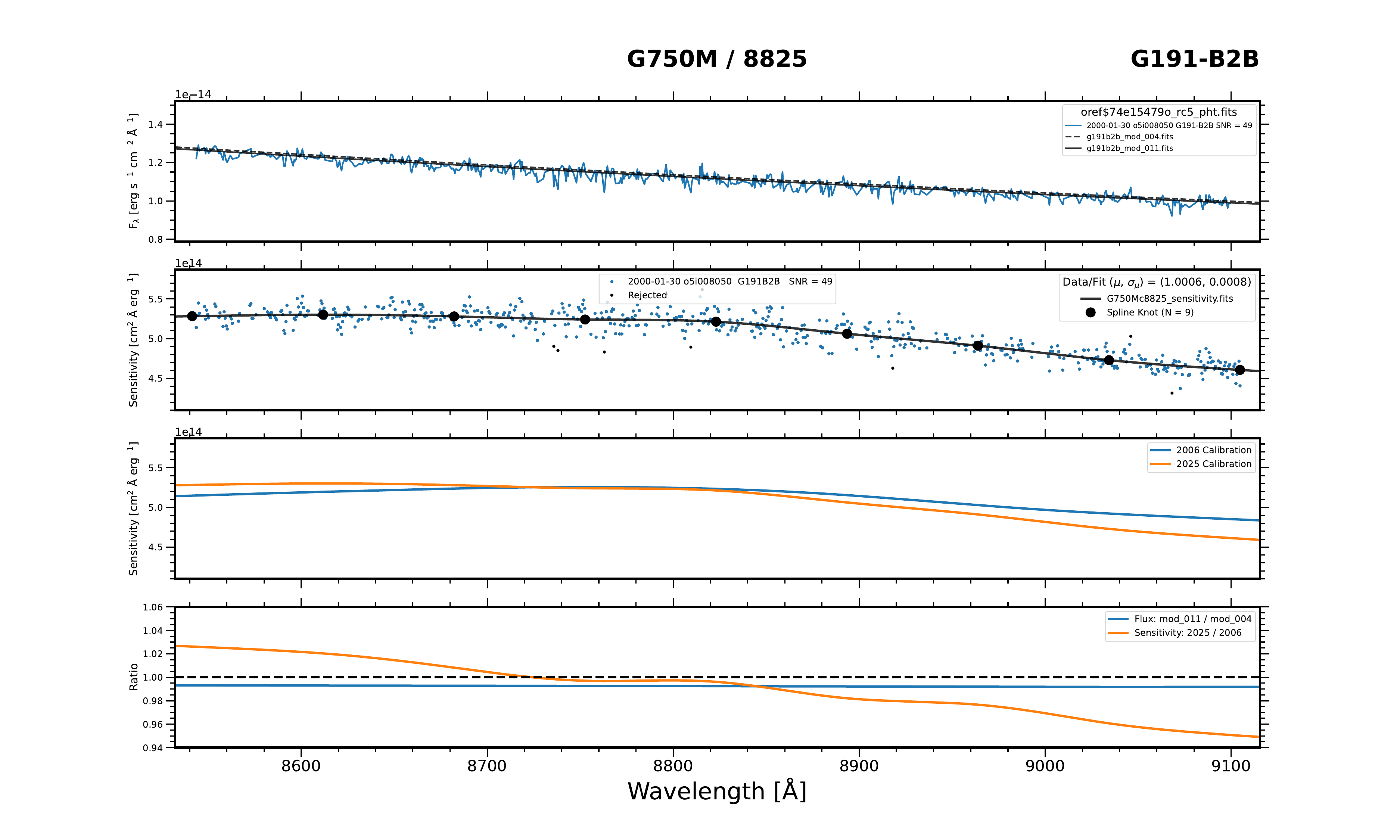}
  \footnotesize
  \caption{Calibration of G191-B2B for G750M/8825.}
  \label{fig:G750MC8825a}
\end{figure}
 
\clearpage
\begin{figure}[t]
  \hspace{-0.5in}
  \includegraphics[width=1.1\textwidth]{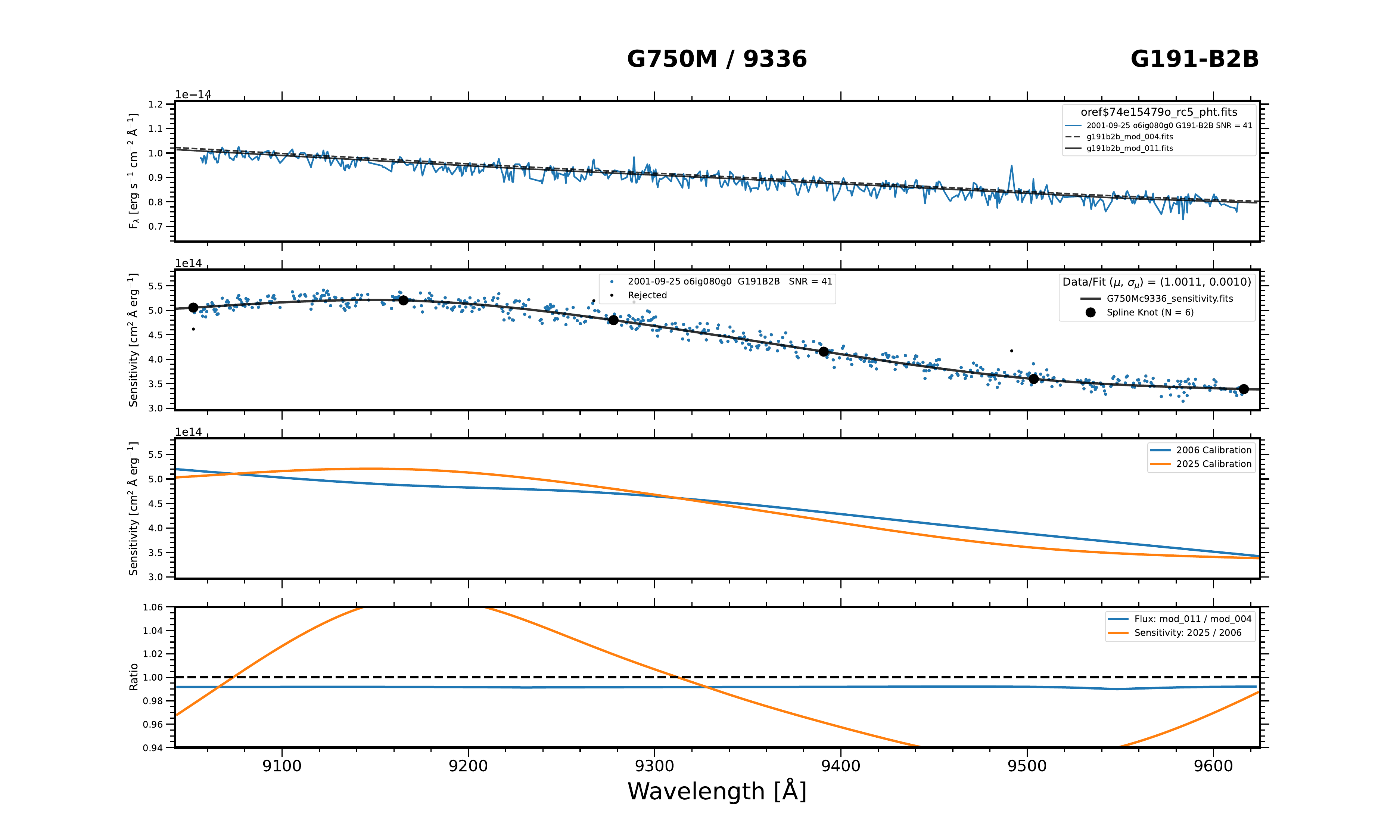}
  \footnotesize
  \caption{Calibration of G191-B2B for G750M/9336.}
  \label{fig:G750MC9336a}
\end{figure}
 
\clearpage

\end{document}